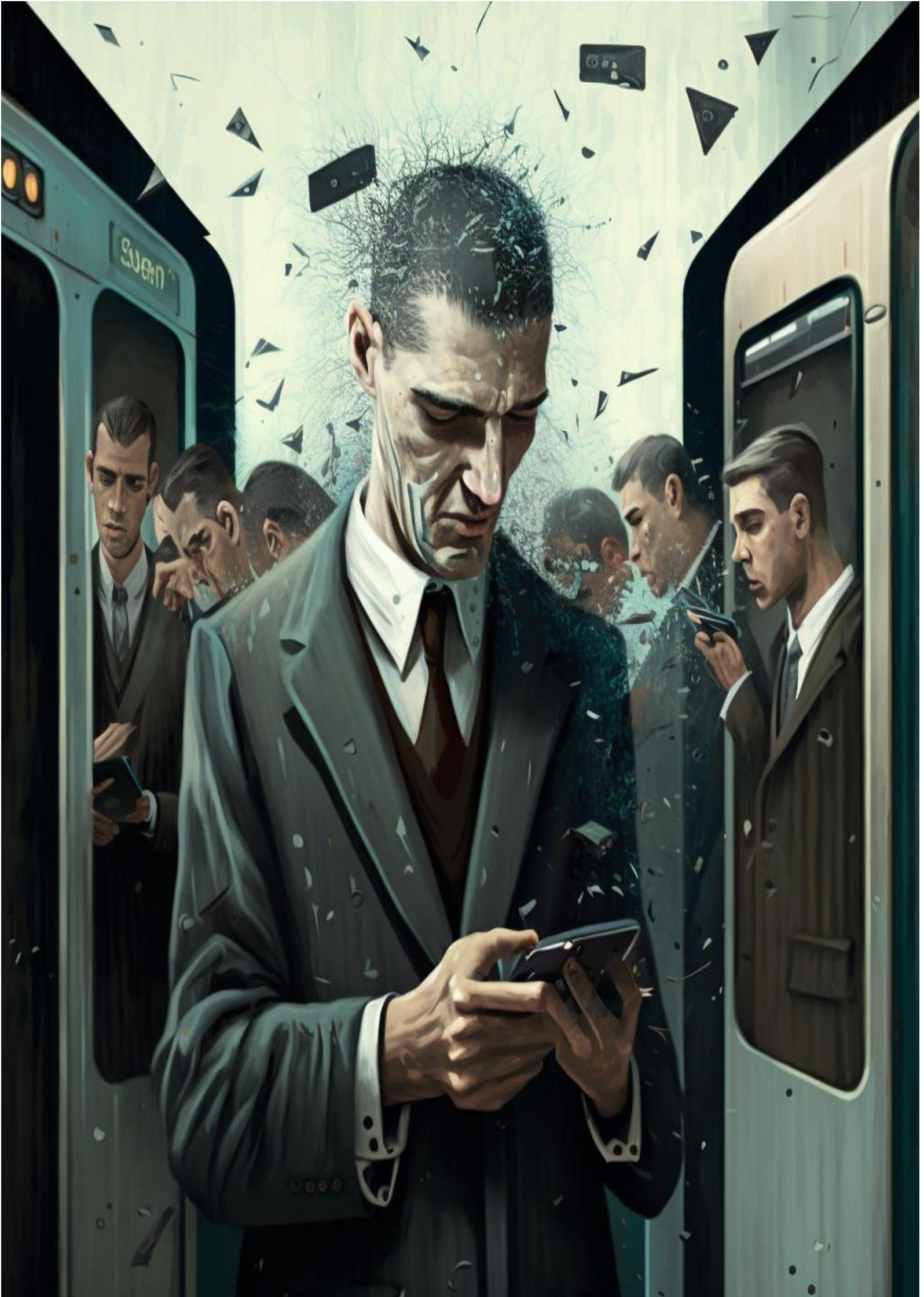

# MITIGATING THE RISK OF KNOWLEDGE LEAKAGE IN KNOWLEDGE INTENSIVE ORGANIZATIONS: A MOBILE DEVICE PERSPECTIVE

**Carlos Andres Agudelo Serna**

ORCID: 0000-0002-1627-4394

Submitted in fulfilment of the requirements of the degree of Doctor of Philosophy.

August 2023

School of Computing and Information System

The University of Melbourne

# Abstract

In the current knowledge economy, knowledge represents the most strategically significant resource of organizations. Knowledge-intensive activities advance innovation and create and sustain economic rent and competitive advantage. In order to sustain competitive advantage, organizations must protect knowledge from leakage to third parties, particularly competitors.

However, the number and scale of leakage incidents reported in news media as well as industry whitepapers suggests that modern organizations struggle with the protection of sensitive data and organizational knowledge. The increasing use of mobile devices and technologies by knowledge workers across the organizational perimeter has dramatically increased the attack surface of organizations, and the corresponding level of risk exposure.

While much of the literature has focused on technology risks that lead to information leakage, human risks that lead to knowledge leakage are relatively understudied. Further, not much is known about strategies to mitigate the risk of knowledge leakage using mobile devices – especially considering the human aspect.

Specifically, this research study identified three gaps in the current literature (1) lack of in-depth studies that provide specific strategies for knowledge-intensive organizations based on their varied risk levels. Most of the analysed studies provide high-level strategies that are presented in a generalised manner and fail to identify specific strategies for different organizations and risk levels. (2) lack of research into management of knowledge in the context of mobile devices,. And (3) lack of research into the tacit dimension of knowledge as the majority of the literature focuses on formal and informal strategies to protect explicit (codified) knowledge.

To address the aforementioned gaps, this research study adopted an exploratory and managerial practice-based perspective to investigate how knowledge intensive organizations manage their risk of knowledge leakage caused by the use of mobile devices. Hence the main research question:

- *How can knowledge intensive (KI) organizations mitigate the knowledge leakage risk (KLR) caused by the use of mobile devices?*

To answer the primary research question, the following secondary questions are also addressed:

1. *What strategies are used by knowledge-intensive organizations to mitigate the risk of knowledge leakage (KLR) caused by the use of mobile devices?*

2. *How does the perceived KLR level inform the strategies used by KI organizations?*

3. *What knowledge assets do knowledge intensive organizations protect from KL?*

The main contribution of this research study is the development of a theory-informed and empirically grounded classification framework that guides organizations in mitigating their leakage risk and improving their knowledge protection capabilities. The framework was developed through the application of a research model that was informed by a comprehensive review of the relevant literature to identify the key concepts and factors that were relevant to the research aims and questions. These concepts and factors were then organized into a conceptual research model, which served as the foundation for the classification framework. The initial development of the framework was based on theory, i.e., the knowledge-based view of the firm, and incorporated components from the mobile computing literature specifically the mobile usage contexts extending from the *social context*, *interaction framework model of context* and *the Integrative model of IT business value* *framework*. The mobile usage contexts were grouped into *human*, *enterprise*, and *technological* factors.

The research study collected qualitative data from twenty knowledge and information security professionals in managerial and executive positions from different knowledge intensive organizations within Australia which had sanctioned mobile device policies in place. The data was collected through semi-structured interviews and supplementary documentation to improve data triangulation and increase the reliability and validity of the findings. The data collection process followed the Gioia methodology that required continuous data comparison involving simultaneous data analysis and exploration (Gioia et al., 2012).

Based on the findings from the data analysis, a set of strategies were developed and organized into a hierarchical structure to form the classification framework. These constructs were arranged based on their relevance and importance to the research question, and their ability to capture the key concepts and factors identified in the conceptual research model.

After this, the collected data informed the further development and extension of the initial conceptual framework into a classification scheme of organizational strategies directed toward the protection of organizational knowledge and leakage mitigation mechanisms followed by knowledge intensive organizations based on the nature of the knowledge (tacit vs explicit) and risk level.

This study's findings also contributed to the current literature on knowledge management and knowledge protection literature:

1. By providing a synthesis of specific mitigation strategies and tactics that knowledge intensive organizations can implement categorized into enterprise, human and technological factors.

2. By proposing a classification scheme that was built on a research framework grounded on the information security, knowledge management, knowledge protection, and mobile computing literature and that can be extended to further investigate the leakage phenomenon.

3. By presenting a combination of more innovative approaches from other domains that address tacit knowledge as highlighted from the evidence.

4. By providing the adaptation of several strategies from the information security literature into the knowledge protection literature, such as zero trust, deception, active defence, active reconnaissance, and behaviour analytics.

5. By presenting protection strategies directly targeting mobility, i.e., mobile workers and mobile devices.

# DECLARATION

This is to certify that:

    I.    the thesis comprises only my original work towards the PhD except where indicated otherwise,

   II.    due acknowledgement has been made in the text to all other material used,

 III.    the thesis is less than 100,000 words in length, exclusive of tables, maps, bibliographies, and appendices.

Signed by _________________________________

Name: Carlos Andres Agudelo Serna

Date: February 2023

# PUBLICATIONS

This section includes the list of peer reviewed academic articles that I have published during the course of the PhD research. Segments of these papers are included in this thesis and have been highlighted in the relevant preface of the different chapters.

1. Agudelo-Serna, Carlos Andres; Bosua, R; Ahmad, Atif; and Maynard, Sean B. (2018) **"Towards A Knowledge Leakage Mitigation Framework For Mobile Devices In Knowledge-Intensive Organizations"** in 26th *European Conference on Information Systems (ECIS), Portsmouth, UK.*

2. Agudelo-Serna, Carlos Andres; Bosua, Rachelle; Ahmad, Atif; and Maynard, Sean B. (2017) **"Strategies to Mitigate Knowledge Leakage Risk caused by the use of mobile devices: A Preliminary Study"** in 38th *International Conference on Information Systems (ICIS), Seoul, South Korea.*

3. Serna, Carlos Andres Agudelo; Bosua, Rachelle; Maynard, Sean B; and Ahmad, Atif. (2017) **"Addressing Knowledge Leakage Risk caused by the use of mobile devices in Australian Organizations"** in 21st *Pacific Asia Conference on Information Systems (PACIS), Langkawi, Malaysia.*

4. Agudelo, Carlos A.; Bosua, Rachelle; Ahmad, Atif; and Maynard, Sean B. (2016) **"Mitigating Knowledge Leakage Risk in Organizations through Mobile Devices: A Contextual Approach"** in 27th *Australasian Conference on Information Systems (ACIS), Wollongong NSW, Australia.*

5. Agudelo, Carlos A.; Bosua, Rachelle; Ahmad, Atif; and Maynard, Sean B. (2015) **"Understanding Knowledge Leakage & BYOD (Bring Your Own Device): A Mobile Worker Perspective"** in 26th *Australasian Conference on Information Systems (ACIS), Adelaide, SA, Australia.*

# Acknowledgments

My PhD journey has been a challenging, emotional yet rewarding experience that I was able to survive thanks to the support of the following people and organizations.

First and foremost, I want to thank my supervisors, Associate Professor Sean Maynard and Associate Professor Atif Ahmad for their invaluable advice, continuous support, and patience during my PhD study, their guidance and experience have encouraged me during my academic research. I would also like to extend my gratitude to Dr. Rachell Bosua for her commitment and counselling throughout my research.

I want to express my sincere gratitude to Professor Alistair Moffat for his patience and advice during our meetings reviewing my progress and guiding me through my PhD.

A special thanks to the participants of this study for their critical contributions and insights into my research.

In addition, I would like to express my deepest appreciation to the University of Melbourne, the Australian Government, and Google for the financial support offered through the various research grants and travel scholarships over the course of this project.

I would also like to extend my deepest gratitude to my parents, Alvaro and Doris, for their unwavering support and encouragement throughout my academic journey. Their love and guidance have been an inspiration to me and have played a crucial role in my success. Their sacrifices and understanding have been invaluable in helping me achieve my goals. This thesis would not have been possible without their love and support.

I would also like to express my sincere appreciation to my little sister, Carolina, for always being there with a listening ear, a helping hand, and a word of encouragement.

Finally, I would like to express my heartfelt gratitude to my mother-in-law, Wendy, and my father-in-law, Josef, for their unwavering support and encouragement throughout this journey. Their love, guidance and understanding have been a source of inspiration and comfort to me. I would like to thank them for being such an important and loving part of my life and for being such wonderful parents to my wife and me.

# DEDICATION

This thesis is dedicated to my beloved wife, Stephanie, who somehow managed to be nothing but supportive throughout this long journey. Thanks for being a constant source of inspiration, support, encouragement and my guiding light throughout this journey.

# TABLE OF CONTENTS

# LIST OF TABLES

# LIST OF FIGURES

# LIST OF ABBREVIATIONS AND ACRONYMS

| Term | Definition |
| --- | --- |
| **APT** | Advanced Persistent Threat |
| **ASD** | Australian Signals Directorate |
| **ATT&CK** | Adversarial Tactics, Techniques, and Common Knowledge |
| **BYO** | Bring Your Own |
| **BYOD** | Bring Your Own Device |
| **CIKO** | Chief Knowledge and Information Officer |
| **CIO** | Chief Information Officer |
| **CISO** | Chief Information Security Officer |
| **CKO** | Chief Knowledge Officer |
| **COBO** | Company Owned Business Only |
| **CoP** | Communities of Practice |
| **COPE** | Company Owned / Personally Enabled |
| **CSM** | Cyber Security Manager |
| **CTI** | Cyber Threat Intelligence |
| **CTO** | Chief Technology Officer |
| **CYOD** | Choose Your Own Device |
| **DLP** | Data Loss / Leakage Prevention |
| **EDR** | Endpoint Detection and Response |
| **GDPR** | General Data Protection Regulation |
| **HRM** | Human Resource Management |
| **IAM** | Identity and Access Management |
| **ICT** | Information Communication and Technology |
| **IP** | Intellectual Property / Internet Protocol |
| **IPP** | Intellectual Property Protection |
| **IPPM** | Intellectual Property Protection Mechanism |

| Term | Definition |
| --- | --- |
| **IPR** | Intellectual Property Rights |
| **IRM** | Information Rights Management/Insider Risk Management |
| **IS** | Information Systems/ Information Security |
| **ISM** | Information Security Manager |
| **ISM** | Information Security Management |
| **ISO** | International Organization for Standardization |
| **ITM** | Insider Threat Management |
| **KBV** | Knowledge Based View |
| **KIO** | Knowledge Intensive Organization |
| **KL** | Knowledge Leakage |
| **KLR** | Knowledge Leakage Risk |
| **KM** | Knowledge Management |
| **KM** | Knowledge Manager |
| **KMS** | Knowledge Management Systems |
| **MAC** | Media Access Control |
| **MAM** | Mobile Application Management |
| **MDM** | Mobile Device Management |
| **MEM** | Mobile Enterprise Management |
| **MFA** | Multi Factor Authentication |
| **MTD** | Mobile Threat Defence |
| **NIST** | National Institute of Standards and Technology |
| **NIST - SP** | National Institute of Standards and Technology - Special Publication |
| **OSINT** | Open Source Intelligence |
| **RBAC** | Role Based Access Control |
| **RBV** | Resource Based View |
| **SCM** | Supply Chain Management |
| **SCMR** | Supply Chain Management Risk |

| Term | Definition |
|------|------------|
| **SECI** | Socialization, Externalization, Combination and Internalization |
| **SIEM** | Security Information and Event Management |
| **SM** | Security Manager |
| **SOAR** | Security Orchestration, automation and Response |
| **SWOT** | Strength Weaknesses Opportunities and Threats |
| **TPR** | Third Party Risk |
| **TPRM** | Third Party Risk Management |
| **TTP** | Tactics, Techniques and Procedures |
| **TTTP** | Tools, Tactics, Techniques and Procedures |
| **UEBA** | User Entity Behaviour Analytics |
| **VPN** | Virtual Private Network |
| **VRIN** | Valuable, Rare, Inimitable and Non-substitutable |
| **ZT** | Zero Trust |
| **ZTA** | Zero Trust Architecture |
| **ZTNA** | Zero Trust Network Architecture |

# Chapter 1. INTRODUCTION

This thesis investigates how knowledge intensive organizations can mitigate the knowledge leakage risk caused by the use of mobile devices. By studying this phenomenon, the findings from this research study suggest a number of mitigation strategies and propose a classification schema and a framework that can be used by organizations to gain an understanding of their security posture and security risk profile that can inform their knowledge protection and leakage mitigation capabilities.

The significance of this study lies in the fact that organizations in the current knowledge economy rely on knowledge intensive activities that contribute to the advancement of innovation and to sustaining their competitive advantage. Moreover, these organizations deem their intellectual capital, represented in the form of people and organizational knowledge, as the most strategically significant resource for creating and sustaining economic rent and competitive edge.

Therefore, it remains paramount for these knowledge intensive organizations to protect their knowledge resources from competitors and their environment as losing their knowledge assets may erode their competitiveness and compromise their position. In this regard, one of the major risks knowledge intensive organizations face is the knowledge leakage risk, defined as *the accidental or deliberate loss or unauthorized transfer of organizational knowledge intended to stay within a firm's boundary* *resulting in the deterioration of competitiveness and industrial position of the organization* (Frishammar et al., 2015; Mohamed et al., 2007; Nunes et al., 2006).

Additionally, to further compound this risk, the use of mobile devices by knowledge workers when conducting their work, usually outside the organization's perimeter, poses an even greater challenge and a management problem for organizations. According to the Ponemon Institute (Ponemon Institute, 2020, 2021b) the uptake of mobile devices within the workplace has grown 95% in the last five years, with employees now using more than one mobile device for work purposes (i.e., smartphone, tablet, laptop) increasing their attack surface (i.e., the sum of different points of attack). Further, smart phones, laptops and tablets remain the most vulnerable endpoints to organizations' networks and enterprise systems, decreasing their overall security posture.

It is worth mentioning that the relationship between knowledge leakage and data breaches is quite intricate. Knowledge leakage refers to the unauthorized or unintended transfer of knowledge from one entity to another. This knowledge can be in the form of data, information, or even the expertise and skills of an individual.

In the context of a data breach, the leaked data can contribute to knowledge leakage if the data contains valuable insights or proprietary information that may lead, from a business perspective to competitive advantage to the organization (Arias-Pérez et al., 2020; Bloodgood & Chen, 2021). For instance, if a data breach results in the leakage of a company's intellectual property, the breached data can be used by unauthorized individuals or competitors to gain knowledge about the company's business processes, its preferences, and behaviors. This constitutes knowledge leakage as the knowledge that was exclusive to the company is now in the hands of unauthorized individuals or competitors.

It is important to note that while all data breaches can lead to information leakage, not all of them result in knowledge leakage. The distinction lies in the value and utility of the breached data. If the leaked data fails to contribute to valuable insights from a business perspective or fails to provide a competitive advantage, it may not necessarily result in knowledge leakage (Timiyo & Foli, 2023).

Hence, the contributions of this study help to address these risks that knowledge intensive organizations face and provide organizations with guidance on selecting and implementing mitigation strategies based on their organizational risk and enable them to manage their security posture efficiently.

The rest of the Introduction chapter[1] is structured as follows; first, it starts by explaining the motivations of this study and key insights of the study in relation to the existing information security management, knowledge management, knowledge protection and mobile literature. Second, the chapter also provides an overview of the research design and key findings. The final part of this chapter outlines the structure of the remaining sections of this thesis.

## 1.1 Motivation of the Study

In recent years, knowledge leakage has come to the forefront thanks in large part to the media attention generated by high profile leakage incidents. One such example, is the infamous mobile spyware Pegasus, developed by the Israeli technology company NSO Group. This spyware employs cutting edge technology and state of the art mechanisms to infect and take total control of mobile devices allowing cyber criminals to spy on users, read text messages, gather passwords, track locations, listen on calls, access the mobile's camera and microphone, and collect information from apps (Marczak et al., 2021). The spyware has been used by both governments and criminals alike to conduct industrial espionage  campaigns and steal knowledge from competitors, adversaries and activists directly through their mobile devices

---

[1] Sections of this chapter have been published in the following publications:

- Agudelo, C. A., Bosua, R., Ahmad, A., & Maynard, S. B. (2016). Understanding knowledge leakage & BYOD (Bring Your Own Device): A mobile worker perspective. arXiv preprint arXiv:1606.01450.
- Agudelo-Serna, C. A., Bosua, R., Ahmad, A., & Maynard, S. (2017). Strategies to Mitigate Knowledge Leakage Risk caused by the use of mobile devices: A Preliminary Study.
- Agudelo-Serna, C. A., Bosua, R., Ahmad, A., & Maynard, S. B. (2018). Towards a knowledge leakage mitigation framework for mobile devices in knowledge-intensive organizations.

(Kaster & Ensign, 2022). Since the Covid-19 pandemic, news of such leakage incidents and spyware usage has continued to surge and cause major issues for organizations across the world (Lallie et al., 2021). With many people working offsite the organizational perimeter has vanished, and workers are now conducting much of their knowledge work from home, from cafes, public spaces, even from other countries whilst on holiday (Georgiadou et al., 2022). As such, knowledge intensive organizations have come to realize that the threat landscape has changed and, with the increasing risk of knowledge leakage incidents, many are now faced with a myriad of new challenges in protecting their knowledge assets.

Therefore, under the current circumstances, knowledge leakage has become a significant security risk to organizations in this new perimeter-less world where employees work from anywhere. A recent global research study on leakage conducted by the Ponemon Institute and IBM Security found each security incident in business costs an average of US$4.35 million in 2022 and that organizations spend on average US$1.2 million each on investigating and assessing information breaches (IBM, 2022; Ponemon, 2022).

In addition to this, earlier and more recent literature show how organizations struggle with leakage of sensitive organizational information across various avenues, such as social media, cloud computing and portable data devices (Ahmad et al., 2015; Ahmad, Bosua, et al., 2014; Check Point, 2022; Jiang et al., 2013; Karabacak & Whittaker, 2022; Krishnamurthy & Wills, 2010; Mohamed et al., 2006; Wu et al., 2021). Although much of the literature has focused on technical security-related aspects of leakage (i.e., data and information), scant research has been conducted on knowledge leakage specifically through mobile devices (Bloodgood & Chen, 2021; Bouncken & Barwinski, 2021; Ghosh & Rai, 2013; Ilvonen et al., 2018; Zahadat et al., 2015).

While it is true that mobile devices fall under the umbrella of 'technical' artifacts, the distinction emphasized in this study pertains to a unique aspect, the nature of interaction and the source of potential knowledge leakage risks. Rather than focusing on the technical vulnerabilities inherent in mobile devices—an extensively researched area in information security— this research concentrates on the examination of behavioral facets and the organizational strategies used to address them. It scrutinizes how the interplay between knowledge workers and their mobile devices might elicit knowledge leakage. The emphasis, therefore, shifts towards human-induced risks that emanate from worker mobility, rather than the intrinsic technical risks associated with mobile devices. This approach facilitates the exploration of a less-studied but equally significant aspect of knowledge leakage.

In today's digitally-driven landscape, it's essential to recognize that while most organizations are vulnerable to data and information breaches, the risk of knowledge leakage is predominantly faced by organizations that heavily depend on knowledge and expertise for their service or product delivery. The distinction between knowledge-intensive and non-knowledge-intensive organizations lies in their operational focus. Knowledge-intensive organizations primarily derive their value from knowledge work, relying significantly on expertise to deliver their offerings. Conversely, non-knowledge-intensive organizations depend more on manual labor, machinery, or raw materials for their service or product delivery.

Examples of knowledge intensive vs non-knowledge intensive organizations are provided below:

**Table 1-1. Examples of Knowledge Intensive vs. Non-Knowledge Intensive Organizations**

| Type of Organization | Examples | Description |
|---|---|---|
| Knowledge-Intensive Organizations | Pharmaceutical Companies | These companies heavily rely on research, development, and innovation to create new drugs and therapies. They invest significantly in scientific knowledge, clinical trials, and intellectual property to stay competitive in the market. |
| | Technology Firms | Technology companies, such as software developers, hardware manufacturers, and IT service providers, are driven by knowledge and expertise in cutting-edge technologies. They thrive on constant innovation and require highly skilled professionals to develop and deliver their products and services. |
| | Research and Development (R&D) Institutions | Organizations that focus on research and development across various fields, including biotechnology, engineering, and aerospace, are inherently knowledge-intensive. Their core activities involve creating new knowledge and applying it to address complex challenges. |
| | Consulting Firms | Management consulting firms, financial advisory firms, and other professional service providers heavily rely on the knowledge and expertise of their employees to deliver high-quality advice and solutions to clients. |
| Non-Knowledge-Intensive Organizations | Agriculture and Farming | While the agricultural sector requires some level of knowledge and expertise, it is less knowledge-intensive compared to industries heavily reliant on research and development. Traditional farming practices and some aspects of agricultural production may involve less specialized knowledge. |
| | Construction Industry | Although the construction industry requires skilled workers, it is generally considered less knowledge-intensive compared to sectors driven by technological advancements and scientific breakthroughs. |
| | Hospitality and Tourism | The hospitality and tourism sector relies more on customer service, operational efficiency, and practical skills rather than advanced knowledge-based processes. |
| | Manufacturing (Basic) | Some manufacturing industries, particularly those involved in producing basic goods using traditional processes, may be less knowledge-intensive when compared to technology-driven or research-oriented industries. |

Although the use of mobile devices (whether employee or organization owned) has shown to be convenient in the context of higher mobility, this convenience comes at a high security cost. The use of mobile devices and technologies by knowledge workers outside the organization's perimeter poses an even greater challenge and a management risk for organizations(Check Point, 2022; Janssen & Spruit, 2019; Ponemon Institute, 2020). Despite the criticality and relevance of this risk, to date, there is a paucity of literature on the knowledge leakage caused by mobile devices.

Additionally, while most of the current literature on leakage has focused primarily on explicit knowledge, that is codified knowledge into data and information (Abdul Molok et al., 2010b; T. Chen et al., 2014; D'Arcy et al., 2009; Gordon, 2007; Ilvonen et al., 2018; Krishnamurthy & Wills, 2010; Manhart & Thalmann, 2015; Morrow, 2012; Shabtai et al., 2012a; Wong et al., 2021; Yahav et al., 2014), research on the leakage of tacit knowledge, comparatively, continues to be underrepresented in the current knowledge management and protection literature (Bloodgood & Chen, 2021; Manhart & Thalmann, 2015; Mupepi, 2017; Zeiringer & Thalmann, 2022).

Furthermore, several authors have called for in-depth research on knowledge leakage mitigation strategies particularly in the context of knowledge intensive organizations as most of the existing research remains scant on individual strategies (Ahmad, Bosua, et al., 2014; Amara et al., 2008; Ilvonen et al., 2018; Kaiser et al., 2020; Manhart & Thalmann, 2015). Similarly, other authors have also highlighted that the current literature fails to address specific conditions and attributes of organizations such as type of industries and risk levels as most studies focus on rather general high-level strategies to prevent leakage (Amara et al., 2008; Bouncken & Barwinski, 2021; Foli & Durst, 2022; M. Lee et al., 2018; Päällysaho & Kuusisto, 2011).

## 1.2 Scope of the Study

Based on the motivation, the researcher identified three gaps in the current literature that informs the scope of this study:

(1) lack of in-depth studies that provide specific strategies for knowledge intensive organizations based on different risk levels. As the majority of the literature provides high-level strategies that are presented in a generalised manner and fail to identify specific strategies for different organizations and risk levels.

(2) Mobile devices and mobile knowledge management are widely neglected, particularly for the protection of tacit knowledge.

(3) the tacit dimension of knowledge remains largely understudied as the majority of the literature focuses on formal and informal strategies to protect explicit (codified) knowledge.

In response to the identified need for more specific strategies for knowledge-intensive organizations, this research delves deeper into the development of such strategies, taking into account the unique risk levels of different organizations. This involves a detailed analysis of various types of organizations, their specific knowledge management needs, and the varying levels of risk they face. The aim is to move beyond generalized strategies and provide tailored recommendations that can effectively address the unique challenges faced by different organizations.

Building on this foundation, and recognizing the increasing ubiquity of mobile devices in the workplace, the study then shifts its focus to mobile knowledge management. It explores how mobile devices contribute to knowledge leakage, particularly of tacit knowledge, and seeks to develop strategies for mitigating these risks. This involves a comprehensive review of current mobile security measures and their effectiveness in protecting tacit knowledge, thereby linking the organizational strategies with the practical implications of mobile device usage.

The research then takes a further step by delving into the largely understudied area of tacit knowledge protection. While much of the existing literature focuses on protecting explicit knowledge, this study explores the unique challenges associated with protecting tacit knowledge, given its intangible and deeply personal nature. It also investigates how tacit knowledge can be inadvertently converted into explicit knowledge through the use of mobile devices, and how this process can be managed to minimize knowledge leakage, thus bridging the gap between mobile knowledge management and tacit knowledge protection.

In the final stage, the study aims to integrate these elements into a comprehensive framework for knowledge protection in knowledge-intensive organizations. This framework considers the specific risk levels of different organizations, the unique challenges posed by mobile devices, and the need for effective protection of both explicit and tacit knowledge. By addressing these gaps, the research contributes to a more nuanced and comprehensive understanding of knowledge protection in the modern, mobile-centric, and knowledge-intensive organization.

Therefore, in this thesis, the scope is centred around the intersection of knowledge, user mobility, risk, and organizations. It aims to explore the unique challenges and risks associated with managing and protecting tacit knowledge in the context of increasing user mobility within knowledge-intensive organizations. The focus is on understanding these challenges from a business perspective, with the goal of developing strategic recommendations for mitigating knowledge leakage risks.

This study, rather than focusing on the data or information itself or delving into the technical aspects of information security, emphasizes the business implications of knowledge leakage. While these are important aspects of knowledge protection, they have been extensively addressed in the existing information security literature. This study, consequently, aims to contribute to the literature by exploring the business implications of knowledge leakage and offering strategic recommendations for knowledge-intensive organizations.

## 1.3 Research Questions

Therefore, in order to address the knowledge gap described above, the main objective of this study is to address the following research question:

*How can knowledge intensive (KI) organizations mitigate the knowledge leakage risk (KLR) caused by the use of mobile devices?*

In order to answer the main research question, the researcher adopted the Knowledge based view of the firm (KBV) as the underlying theoretical lens. The (KBV) as an extension of the resource-based view theory (RBV) considers knowledge as the most strategically significant resource of a firm used for creating and sustaining economic rent and competitive advantage (Grant, 1996b).

By answering the research question, the researcher focused on knowledge intensive organizations and highlighted the role that leakage risk plays in the interaction with mobile devices. The answer to these questions provides the steps that organizations follow to combat the risk in the context of mobility. To better understand this phenomenon, the researcher divided the overarching questions into three components: 1) Strategies 2) Risk and 3) Assets.

The following secondary research questions were defined to assist in answering the main research question and relate to the three components listed above:

1. *What strategies are used by knowledge-intensive organizations to mitigate the risk of knowledge leakage (KLR) caused by the use of mobile devices?*

   The answer to this question provides the strategies used by knowledge intensive organizations to combat the leakage risk caused by mobile devices. This is represented in the form specific strategies addressing the mobility concern and categorized in to formal and informal type of strategies, targeting tacit and explicit knowledge.

2. *How does the perceived KLR level inform the strategies used by KI organizations?*

   The answer to this question elaborates on the association between risk and strategy and explains how the level of risk relates to the type of strategy that knowledge intensive organizations employ to mitigate a potential threat.

3. *What knowledge assets do knowledge intensive organizations protect from KL?*

   The answer to this question provides the list of assets that knowledge intensive organizations protect from knowledge leakage. The understanding of the type of asset, helps to determine the type of strategy applied to the asset.

# 1.4  Research Overview

The researcher developed a research model based on the literature and drawing on the *social context interaction framework* by Chen & Nath (2008) and Bradley and Dunlop's (2005) *model of context*, that leverages the use of mobile contexts to explain the relationships between constructs that assist in answering the research questions through theoretical propositions. The author, subsequently, grouped the related mobile contexts into factors employing *the Integrative model of IT business value framework* developed by Melville et al (2004) in order to integrate and create group level constructs that are composed of individual but related contexts addressed in a similar way in the literature This is illustrated in the next section (see Table 1-2). For further detail, see section **3.2** *Mobility and Mobile Contexts approach to the Research Model Development* on page 79

## 1.4.1 Research Model

The development of the model was conducted in three stages in which the researcher:

1. Identified the mobile device usage contexts and constructs from the literature and grouped them together based on their definition, *the social context interaction framework* and *the Integrative model of IT business value framework*.

2. Indicated the relationship between the mobile contexts and the *knowledge leakage risk through mobile devices* construct.

3. Highlighted the relationship between *knowledge leakage risk through mobile devices* and *organizational knowledge mitigation capabilities* construct.

The conceptual model as depicted below in *Table 1-2,* provides a framework to understand how *the mobile device usage contexts* which is composed of *human* (personal and social), *enterprise* (environmental and organizational) and *technological factors* (device and technical) impact the leakage risk and how this risk, in turn, informs the organizational performance via mitigation capabilities which contributes to improvement of organizational information and knowledge security performance in the context of mobility and mobile devices represented in the construct of *organizational KLR mitigation strategies.*

As previously indicated the mobile device usage contexts are composed of

1. **Human Factors**: In the literature on mobile contexts, human factors encompass both personal and social elements that influence an individual's behavior, attitudes, cognitive abilities, motivations, and experiences (personal context), as well as a group's culture, values, social norms, and the influence of peers and superiors (social context). These factors have a significant impact on the effectiveness of mobile device practices and efforts.

2. **Enterprise Factors:** In the mobile context literature, enterprise factors encompass both external environmental conditions (environmental context) such as competitors, regulations, and industry, as well as internal organizational resources and capabilities (organizational context) such as policies, culture, and processes. These factors can influence the risk and practices related to mobility and mobile devices within an enterprise. Key considerations include governance, risk management, organizational structure and culture.

3. **Technological Factors:** In the literature on mobile contexts, technological factors encompass device and technological contexts, as well as the internal and external infrastructure and resources that support knowledge-sharing activities. These factors refer to the technological aspects of an enterprise that can impact the security of mobile devices and systems, including devices, network architecture, security technologies, and controls such as encryption and identity and access management.

**Table 1-2. Research Model and Propositions**

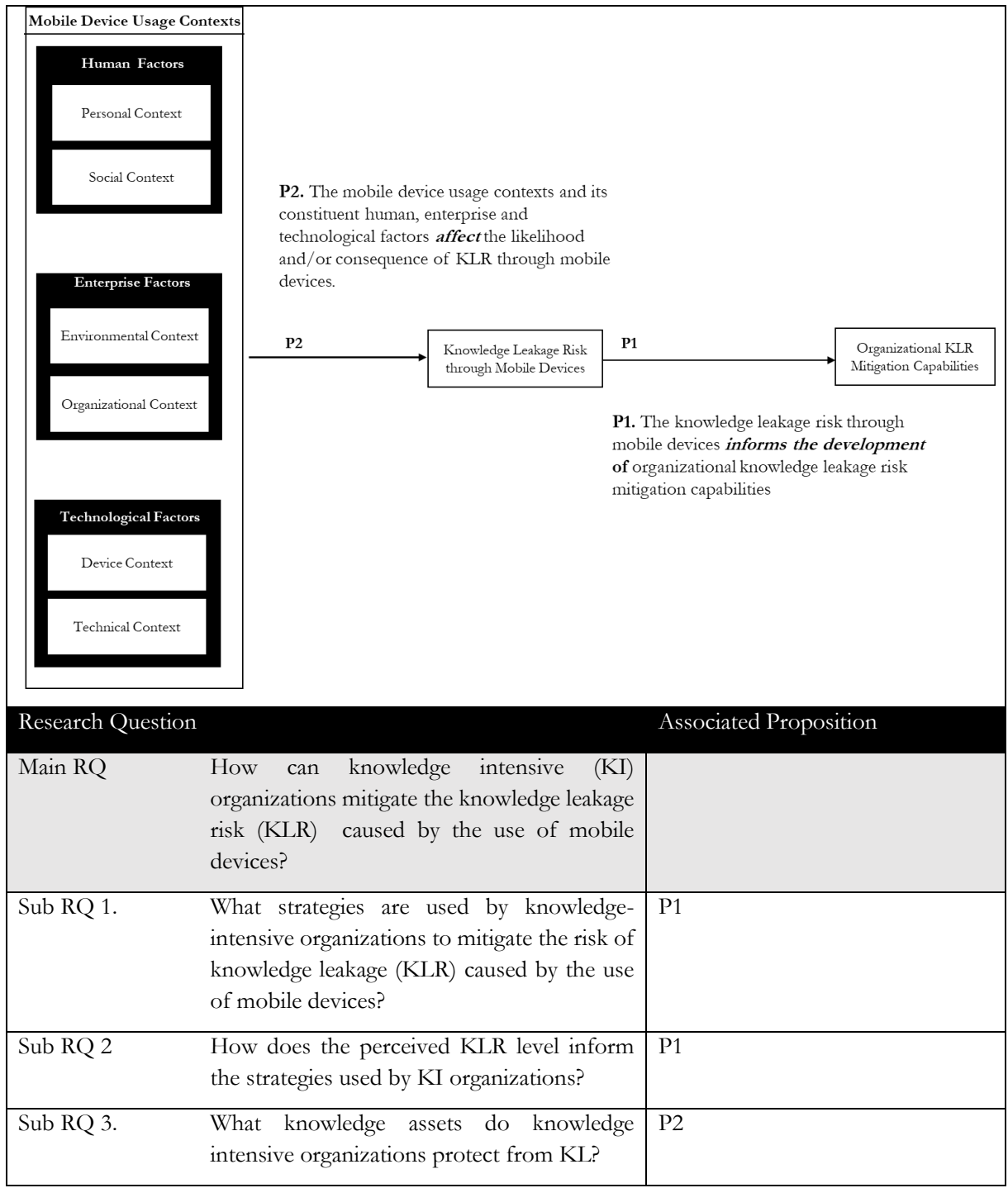

| Research Question | | Associated Proposition |
|---|---|---|
| Main RQ | How can knowledge intensive (KI) organizations mitigate the knowledge leakage risk (KLR) caused by the use of mobile devices? | |
| Sub RQ 1. | What strategies are used by knowledge-intensive organizations to mitigate the risk of knowledge leakage (KLR) caused by the use of mobile devices? | P1 |
| Sub RQ 2 | How does the perceived KLR level inform the strategies used by KI organizations? | P1 |
| Sub RQ 3. | What knowledge assets do knowledge intensive organizations protect from KL? | P2 |

Chapter 3 further describes the process of how the researcher constructed the model. See sections **3.3** *Knowledge based View Theory approach to the Research Model Development – Knowledge as an object and capability*, **3.4** *Risk Perspective on Knowledge Leakage through Mobile Devices*, and **3.5** *Propositions and constructs within the model*

## 1.4.2 Research Method

The research study collected qualitative data from twenty knowledge and information security professionals in managerial and executive positions from different knowledge intensive organizations within Australia which had sanctioned mobile device policies in place. The data was collected through semi-structured interviews and supplementary documentation to improve data triangulation and increase the reliability and validity of the findings. The data collection process followed the Gioia methodology that required continuous data comparison involving simultaneous analysis and exploration.

The collected data informed the further development and extension of the initial conceptual framework into a classification scheme of organizational strategies directed toward the protection of organizational knowledge and leakage mitigation mechanisms followed by knowledge intensive organizations based on the nature of the knowledge (tacit vs explicit) and risk level. For further detail, see section *4.3 Research Methods* on page 101.

## 1.4.3 Research Design

The research design for this study was staged in three main phases (for further detail, see section *4.4 Research Design* on page 103):

1. Contextual stage: the first stage of the research design, the contextual study, refers to the work described in chapters one and two —Introduction and Literature Review;

2. Model development stage: refers to the work contained in chapter three — Research Model Development, and;

3. Empirical study: in which the conceptual model was validated and refined (see Figure 1-1).

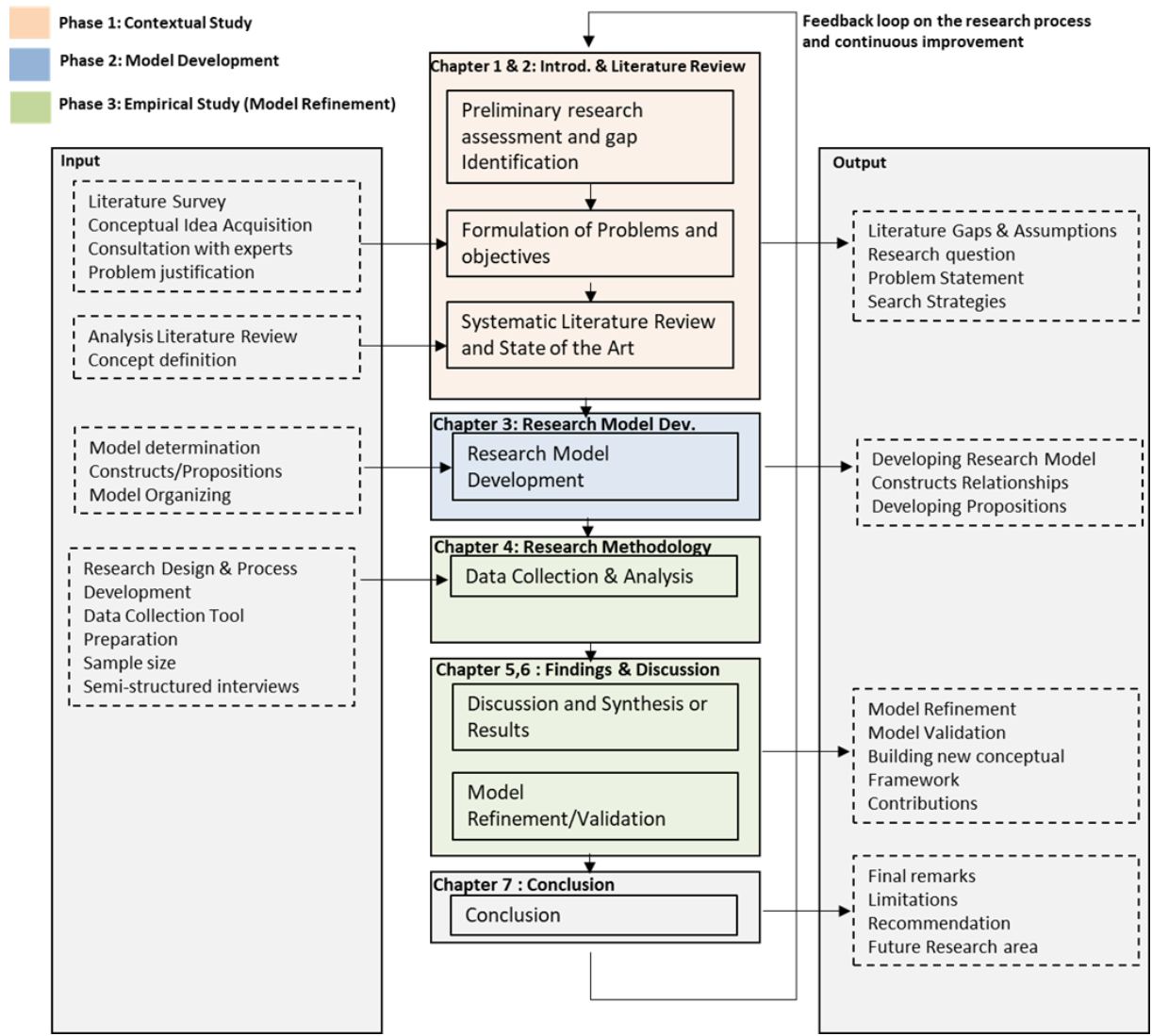

**Figure 1-1. Research Design**

The first phase, referred to as the Contextual Study, sets the foundation for the research. This phase, detailed in the first two chapters of the thesis, introduces the research topic and question. It involves a systematic review of literature spanning the fields of information security management, knowledge management, and mobile computing. The aim is to develop a nuanced understanding of the knowledge leakage phenomenon within these contexts. Throughout this phase, the researcher identified several limitations and gaps in the existing literature, particularly concerning mitigation strategies and knowledge protection in the context of mobile device usage. These identified gaps serve as the motivation for the subsequent phases of the research.

Transitioning from the Contextual Study, the second phase, Model Development, focused on constructing a research model to guide the empirical investigation. This phase, outlined in the third chapter, involved three key steps. First, the researcher identifies mobile contexts and constructs, grouping them together based on their definitions and the framework of the Integrative model of IT business value. Second, the researcher established the relationship between these mobile contexts and the construct of knowledge leakage risk through mobile devices. Finally, the researcher determined the relationship between knowledge leakage risk through mobile devices and the construct of organizational knowledge mitigation capabilities. This phase culminates in the development of propositions between the constructs, each grounded in and referenced to the relevant literature.

The third and final phase, the Empirical Study, aims to validate and refine the proposed conceptual model through empirical investigation. This phase involved conducting semi-structured interviews with industry experts in security and knowledge management. The data obtained from these interviews were then analyzed, with the researcher conducting several iterations of coding until reaching theoretical saturation. This process, constantly contrasted against the literature review and the research model, allowed the researcher to further develop a framework and classification of the leakage mitigation strategies used by organizations. This empirical study served to bridge the gaps identified in the first phase and validated the model developed in the second phase, thereby bringing the research full circle.

# 1.5 Key Findings and Contributions

The main contributions and findings of this study can be summarized as:

1.      The synthesis of specific mitigation strategies and tactics[2] that knowledge intensive organizations can implement categorized into enterprise, human and technological factors (see Table 1-3).

2.      The classification scheme that was built on a research framework grounded in the information security, knowledge management, knowledge protection, and mobile computing literature and that can be extended to further investigate the leakage phenomenon (see Figure 1-2).

3.      The combination of more innovative approaches from other domains that address tacit knowledge as highlighted from the evidence.

4.      The adaptation of several strategies from the information security literature into the knowledge protection literature, such as zero trust, deception, active defence, active reconnaissance, and behaviour analytics.

5.      The Protection strategies directly targeting mobility, i.e., mobile workers and mobile devices.

The classification scheme developed in this study is based on strategies discerned from interviews, organizational documents, and field notes. Grounded in research literature, this scheme categorizes strategies employed by knowledge-intensive organizations for knowledge management and protection. The scheme contributes to a deeper understanding of the measures used by organizations to protect their knowledge. It takes into account factors such as perceived risk profile, the formality level of the strategy (formal vs. informal), and the type of knowledge safeguarded (tacit vs. explicit). The scheme differentiates between formal strategies, such as specific policies and procedures exemplified by a Knowledge Management

---

[2] This study adopted the definition of strategy as defined by Mintzberg (1985). Please refer to section *2.7.1 Definition of Strategy*

System (KMS), and informal strategies, which manifest as ad hoc actions not formalized in policies or procedures by employees to safeguard knowledge assets, like fostering a culture of knowledge-sharing.

The classification scheme differentiates between explicit and tacit knowledge. Explicit knowledge, such as a tech firm's patented algorithm for search optimization or a software company's copyrighted source code, can be readily communicated and shared. Conversely, tacit knowledge is deeply personal and more challenging to formalize. For instance, an engineer may have a unique, intuitive approach to optimizing engine performance, developed through extensive experience with various experiments. Similarly, a seasoned aerospace engineer might possess an innate understanding of how design choices affect an aircraft's flight characteristics, a knowledge honed over years of experience and numerous design iterations. In both cases, this tacit knowledge, built on years of field experience, enables the holder to diagnose and resolve issues more efficiently than less experienced colleagues, in ways that are difficult to encapsulate in manuals or software. Refer to  2.5.1.1 and 2.5.1.2 sections in the Literature Review chapter for more detailed information.

**Table 1-3. Knowledge Leakage Mitigation Strategies**

| Factor | Context | Strategy | Type of Strategy Formal (29) | Informal (9) | Type of Knowledge Tacit (17) | Explicit (21) |
|---|---|---|---|---|---|---|
| Human (13) | Personal (8) | 1. Awareness Training and Education | ✓ | | ✓ | |
| | | 2. Mentoring | ✓ | | ✓ | |
| | | 3. Deterrence | ✓ | | | ✓ |
| | | 4. Indoctrination | ✓ | | ✓ | |
| | | 5. Individual Risk Assessment (Insider Risk) | ✓ | | ✓ | |
| | | 6. Peer Mentoring Policy | ✓ | | ✓ | |
| | | 7. Trust Development | | ✓ | ✓ | |
| | | 8. Zero Trust | | ✓ | ✓ | |
| | Social (5) | 9. Knowledge Security Culture | ✓ | | ✓ | |
| | | 10. Community of Practice | ✓ | | ✓ | |
| | | 11. Gamification | | ✓ | ✓ | |
| | | 12. Knowledge Systems (Groupware, KMS) | ✓ | | ✓ | |
| | | 13. Informal Networks | | ✓ | ✓ | |
| Enterprise (14) | Organizational (7) | 14. Risk Management (Insider Risk) | ✓ | | | ✓ |
| | | 15. Human Resource Management (HRM) | ✓ | | ✓ | |
| | | 16. Mobile Policy (BYOD, CYOD, COPE, COBO) | ✓ | | | ✓ |
| | | 17. Mobile Endpoint Management (MDM, MAM, MEM) | ✓ | | | ✓ |
| | | 18. Legal Frameworks (IPR, Copyright, Patents, Trademark, NDA) | ✓ | | | ✓ |
| | | 19. Industry Cyber Security Frameworks | ✓ | | | ✓ |
| | | 20. Ad-hoc Informal Processes (Tacitness, Secrecy, Lead time advantage, Complexity) | | ✓ | | ✓ |
| | Environmental (7) | 21. Market/ Environment Analysis | ✓ | | | ✓ |
| | | 22. Deception/ Misinformation/Disinformation | | ✓ | | ✓ |
| | | 23. Inter Organizational Liaisons | ✓ | | ✓ | |
| | | 24. Competitor Analysis | | ✓ | | ✓ |
| | | 25. Supply Chain Risk Management | ✓ | | | ✓ |
| | | 26. Active Defence | | ✓ | ✓ | |
| | | 27. Active Reconnaissance | | ✓ | ✓ | |
| Technological (11) | Device (5) | 28. User and Behaviour Analytics (UEBA) | ✓ | | ✓ | |
| | | 29. Device Profiling | ✓ | | | ✓ |
| | | 30. Whitelisting | ✓ | | | ✓ |
| | | 31. Compartmentalization | ✓ | | | ✓ |
| | | 32. Monitoring and Endpoint Detection & Response (EDR) | ✓ | | | ✓ |
| | Technological (6) | 33. Prevention | ✓ | | | ✓ |
| | | 34. Surveillance | ✓ | | | ✓ |
| | | 35. Deception (Decoy systems, honeypot, honey files) | ✓ | | | ✓ |
| | | 36. Perimeter Defence | ✓ | | | ✓ |
| | | 37. Conditional and Context based access | ✓ | | | ✓ |
| | | 38. Layering (Defence in Depth, MFA, Encryption, DLP) | ✓ | | | ✓ |

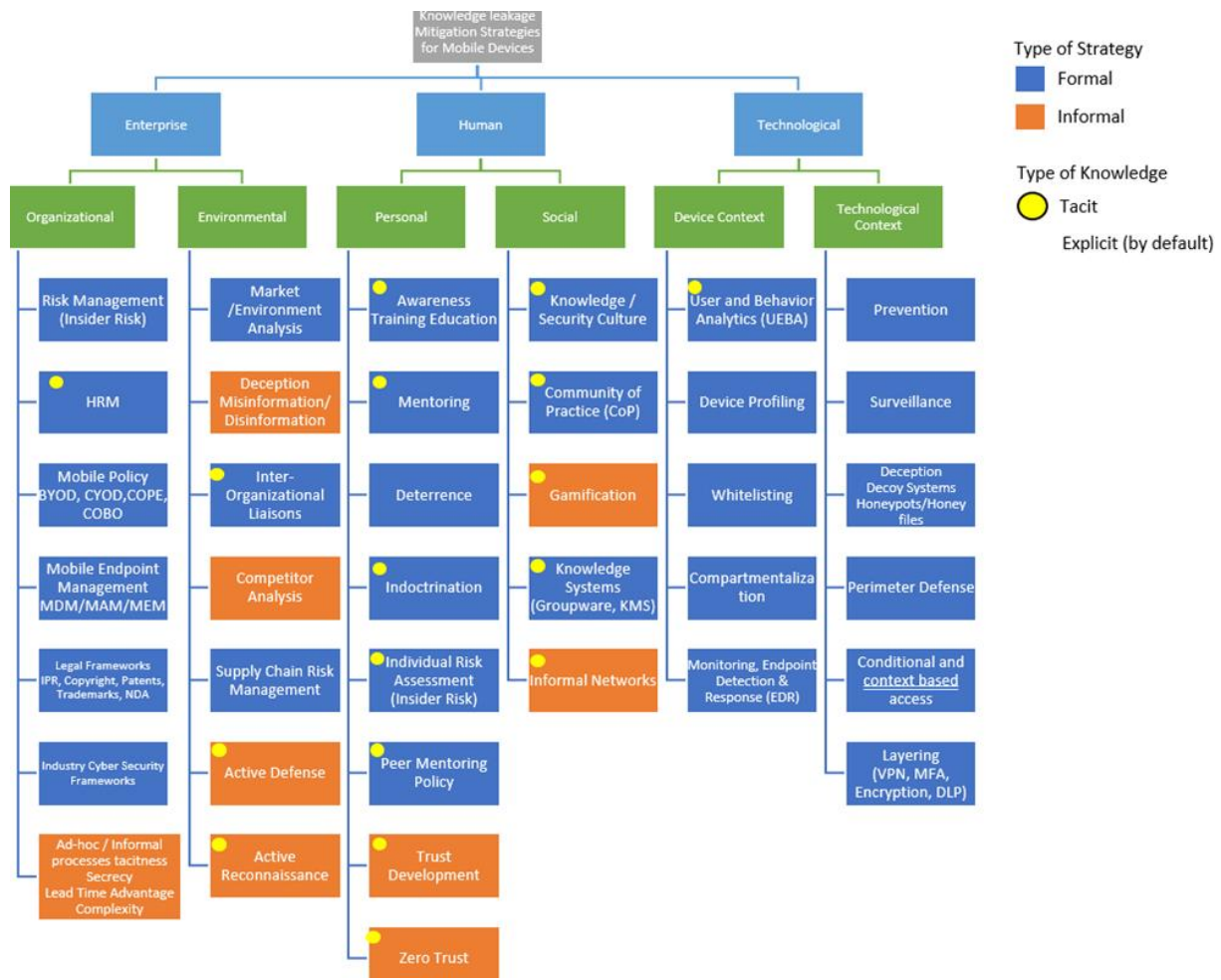

**Figure 1-2. Knowledge Leakage Mitigation Strategies Classification Scheme**

The proposed mitigation framework depicted in Table 1-3 above and Figure 1-2 above is a comprehensive structure that categorizes strategies to mitigate knowledge leakage into three main factors: Human Factors, Enterprise Factors, and Technological Factors. Each of these factors is further divided into two contexts, and each context encompasses a set of strategies. The strategies are characterized by two dimensions formality (either Formal or Informal) and the type of knowledge they protect (either Tacit or Explicit).

Human Factors are divided into Personal and Social Contexts. Personal Context includes strategies such as Awareness Training and Education, Mentoring, Deterrence, Indoctrination, Individual Risk Assessment, Peer Mentoring Policy, Trust Development, and Zero Trust. Social Context, on the other hand, includes strategies like Knowledge Security Culture, Community of Practice, Gamification, Knowledge Systems, and Informal Networks.

Enterprise Factors are divided into Organizational and Environmental Contexts. Organizational Context includes strategies such as Risk Management, Human Resource Management, Mobile Policy, Mobile Endpoint Management, Legal Frameworks, Industry Cyber Security Frameworks, and Ad-hoc Informal Processes. Environmental Context includes strategies like Market/Environment Analysis, Deception/Misinformation/Disinformation, Inter-Organizational Liaisons, Competitor Analysis, Supply Chain Risk Management, Active Defence, and Active Reconnaissance.

Technological Factors are divided into Device and Technological Contexts. Device Context includes strategies such as User and Behaviour Analytics, Device Profiling, Whitelisting, Compartmentalization, and Monitoring and Endpoint Detection & Response. Technological Context includes strategies like Prevention, Surveillance, Deception, Perimeter Defence, Conditional and Context-based access, and Layering.

# 1.6 Structure of this Thesis

The remaining of this thesis, summarized in *Figure 1-3. Thesis Structure* is structured as follows:

**Chapter 2**: **Literature Review** describes the methodology and steps taken to conduct the literature review, as well as the literature on relevant and key topics for this research study such as knowledge, knowledge-based view theory of the firm (KBV), knowledge intensive organizations, knowledge leakage, knowledge leakage risk, mobile contexts, knowledge management, knowledge security, information security management and knowledge protection mechanisms.

**Chapter 3**: **Research Model** outlines the methodology and steps taken to develop the research model based on the literature review and the synthesis of salient topics resulting from the review and analysis of different streams of literature such as information security management, knowledge management, and mobile/mobility literature.

**Chapter 4: Research Methodology** presents and justifies the research paradigm, the research design, research method, as well as the data collection techniques, and the qualitative data analysis conducted based on the data gathered. Moreover, it also summarizes the steps and measures taken to ensure research rigour as well as data validity and reliability.

**Chapter 5: Findings** presents the results and key findings based on the qualitative data analysis conducted, as well as the results of the empirical study and validates the conceptual research model.

**Chapter 6: Discussion** highlights the contributions that the findings bring to IS research and practice in general. The discussion chapter starts by explaining the findings and key insights of the study in relation to the existing knowledge leakage and mobile computing and knowledge management literature. The chapter also highlights the connection of this research to broader topics in the IS literature. The final part of the chapter outlines the implications and conclusions of this study for both IS research and practice.

**Chapter 7: Conclusion** provides the summary of the key research findings in relation to the research aims and research questions, as well as the value and contribution thereof. The conclusion chapter also reviews the limitations of the study and proposes opportunities for further research. The conclusion chapter first, starts by outlining the research questions and answering them based on the findings. Second, the chapter also highlights both the theoretical and practical contributions of the research study in relation to the research questions and research aim. Third, the next section thereafter presents the limitations of the study. The final part of the chapter describes the future avenues for research based on the work developed in the study and its limitations.

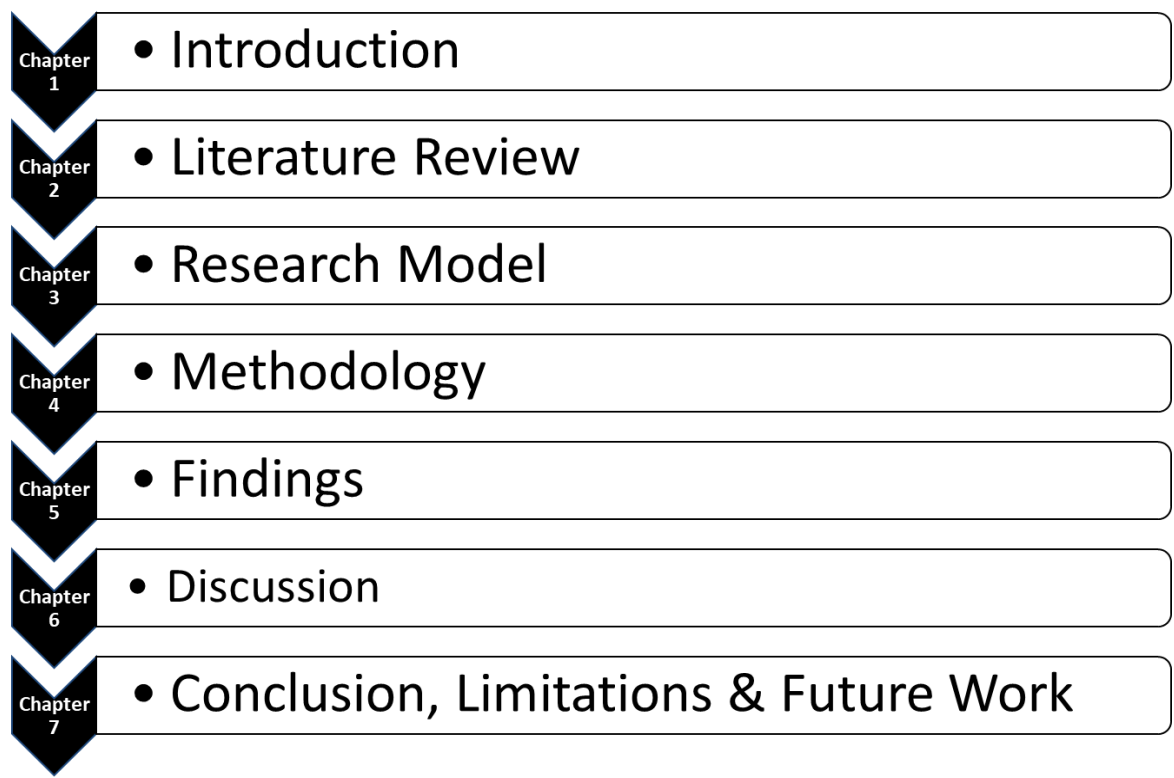

**Figure 1-3. Thesis Structure**

## 1.7 Summary

This chapter presented the introduction of this study as well as the motivation, the focus, an overview of the research design and summary of the key findings and an overall structure of the thesis. The following chapters develop further on each of these sections, The next chapter describes the methodology and steps taken to conduct the literature review.

# Chapter 2. LITERATURE REVIEW

This chapter[3] illustrates the methodology and steps taken to conduct the literature review, as well as the literature on relevant and key topics for this research study such as knowledge, knowledge-based view theory of the firm (KBV), knowledge intensive organizations, knowledge leakage, knowledge leakage risk, mobile contexts, knowledge management, knowledge security, and knowledge protection mechanisms.

## 2.1 Literature Methodology

Following a methodology extensively used in information systems research (Watson, 2015; Webster & Watson, 2002), This research study utilized a systematic approach to categorize and thoroughly examine relevant literature in order to provide a reliable and consistent assessment of the current state of the research field. The researcher conducted an *a priori* analysis of the literature, which was designed to reduce the possibility of biases and ensure the study's repeatability (Webster & Watson, 2002). The literature review required for this study was interdisciplinary in nature,

---

[3] Sections of this chapter have been published in the following publications:

- Agudelo, C. A., Bosua, R., Ahmad, A., & Maynard, S. B. (2016). Understanding knowledge leakage & BYOD (Bring Your Own Device): A mobile worker perspective. arXiv preprint arXiv:1606.01450.
- Agudelo-Serna, C. A., Bosua, R., Ahmad, A., & Maynard, S. (2017). Strategies to Mitigate Knowledge Leakage Risk caused by the use of mobile devices: A Preliminary Study.
- Agudelo-Serna, C. A., Bosua, R., Ahmad, A., & Maynard, S. B. (2018). Towards a knowledge leakage mitigation framework for mobile devices in knowledge-intensive organizations.

encompassing multiple domains within information systems, including knowledge management, mobile computing and information security management.

Webster & Watson (2002) mainly identify two types of literature reviews. The first type of literature review deals with a mature topic where an accumulated body of research exists that needs analysis and synthesis and in which authors should conduct a thorough literature review and then propose a conceptual model that synthesises and extends existing research.

The second type of literature review focuses on emerging topics where authors can benefit from identifying potential theoretical foundations. As a result, the review of current literature on the emerging issue is often more concise. The primary contribution to the field in this type of literature review comes from the development of a new conceptual model based on fresh theoretical foundations. Such a model can provide a framework for future research and ultimately advance our understanding of the emerging issue.

Based on this distinction, this research study falls under the second category of literature review, and as such, studies the current literature on the emerging topic of knowledge leakage caused by the use of mobile devices in knowledge intensive organisations and seeks to develop a conceptual model that explains the leakage phenomenon in the context of mobile devices.

## 2.1.1 Research Keywords and Terms

In order to examine the current literature including peer-reviewed conceptual, empirical and reviewed articles related to the study domains of this research from key information systems journals, conferences, white papers and books from different outlets and popular literature databases such as Science Direct, ABI/INFORM, ProQuest, JSTOR, Google Scholar, and AIS Electronic library, the researcher defined a series of keywords and Boolean operators to conduct a systematic literature review. As shown in Table 2-1, the table displays the list of research terms and operators which follow a structural approach in which each column represents a different topic or field, analysed during the literature survey, and the use of Boolean operators that show the relationship amongst topics or fields. **AND** operators are used to indicate the intersection of terms; **OR** operators are used to include similar terms; **NOT** operators are used to exclude terms not relevant to the search. The researcher followed a rigorous scientific approach to conducting the literature review and ensuring to a greater extent that the search process is reproducible based on guides and frameworks in Information Systems Research (Okoli & Schabram, 2010; Watson, 2015; Webster & Watson, 2002).

## 2.1.1.1 Exclusion of the research term *Data* in the literature search

The inclusion of the Boolean operator NOT with the term DATA in the search strategy serves a critical purpose in aligning the literature review with the research objectives and the theoretical lens of this study. The primary focus of this research is to explore the phenomenon of knowledge leakage from a business perspective, specifically through the lens of knowledge theory of the firm. This perspective emphasizes the human elements of knowledge creation, sharing, and protection, and is distinct from the technical or engineering perspective that often dominates discussions of data breaches or data leakage.

By excluding the term DATA, the search strategy effectively filters out literature that primarily addresses technical aspects of information security, such as data encryption, firewall configurations, or other engineering solutions to data breaches. While these topics are undoubtedly important in the broader context of information security, they do not align with the research objectives of this study, which are centered on the human, organizational, and strategic aspects of knowledge leakage.

Moreover, the exclusion of DATA helps to maintain the focus on knowledge as a distinct concept. In the context of this study, knowledge refers to the insights, expertise, and know-how possessed by individuals and organizations, which is often tacit and deeply personal in nature. This is distinct from data, which can be easily codified, stored, and transmitted. By excluding DATA, the search strategy ensures that the literature review remains focused on this more nuanced and complex concept of knowledge, rather than the more straightforward and technical concept of data.

Finally, the use of **NOT DATA** in the search strategy also helps to ensure the reproducibility and rigor of the literature review. By clearly defining the scope of the literature review and the criteria for inclusion and exclusion, the search strategy allows other researchers to replicate the literature review and verify the findings. This is a key aspect of scientific rigor and contributes to the overall credibility and validity of the research (Watson, 2015).

**Table 2-1. Research Keyword and Terms**

| Table Research Keywords and Terms | | | | | | |
|---|---|---|---|---|---|---|
| Research Terms – Boolean AND applied across each row to include multi-disciplinary fields and OR operator down each column to cover synonyms and similar terms | | | | | | |
| Knowledge (AND) | Leakage (AND) | Risk (AND) | Mobile (AND) | Devices (AND) | Knowledge Protection (AND) | Strategy |
| (OR) Know-how | (OR) Spill over | OR (Danger) | (OR) Mobility | (OR) Smartphone | (OR) Knowledge Security | (OR) Policy |
| (OR) Knowledge assets | (OR) Spill | OR (Exposure) | (OR) portable | (OR) Laptop | (OR) Knowledge Retention | (OR) Control |
| (OR) Intellectual Property | (OR) Loss | | (OR) BYOD | (OR) Tablet | | |
| (OR) Intellectual capital | (OR) disclosure | | (OR) Bring your own device | | | |
| (OR)Trade secrets | (OR) absorption | | (OR) CYOD | | | |
| Industrial Secrets | (OR) appropriation | | (OR) Choose your own device | | | |
| Copyright | | | (OR) Bring your own | | | |
| OR Design | | | OR BYO | | | |
| OR Patent | | | | | | |
| **OR** Information | | | | | | |
| **NOT** Data | | | | | | |
| OR Insight | | | | | | |

## 2.1.2 Literature Flow Chart

Following a reproducible method for identifying, evaluating, and synthesising the existing body of knowledge initially rooted in the health sciences and consequently adopted into the Information Systems Research (Okoli & Schabram, 2010) and based on the PRISMA (Preferred Reporting Items for Systematic Reviews and Meta-Analyses) flow diagram to depict the flow of information through the different phases of the systematic literature review (Moher et al., 2010; Watson, 2015), the researcher mapped out the number of sources identified, included and excluded, and the reasons for exclusions (see Figure 2-1). This approach helps to illustrate the process followed and to gauge the completeness and transparency of a systematic review protocol submitted for publication in a journal or other publication medium (Moher et al., 2010; Okoli & Schabram, 2010). The complete list of included articles are located in *Appendix A – Papers and reports analyzed for Literature Review.*

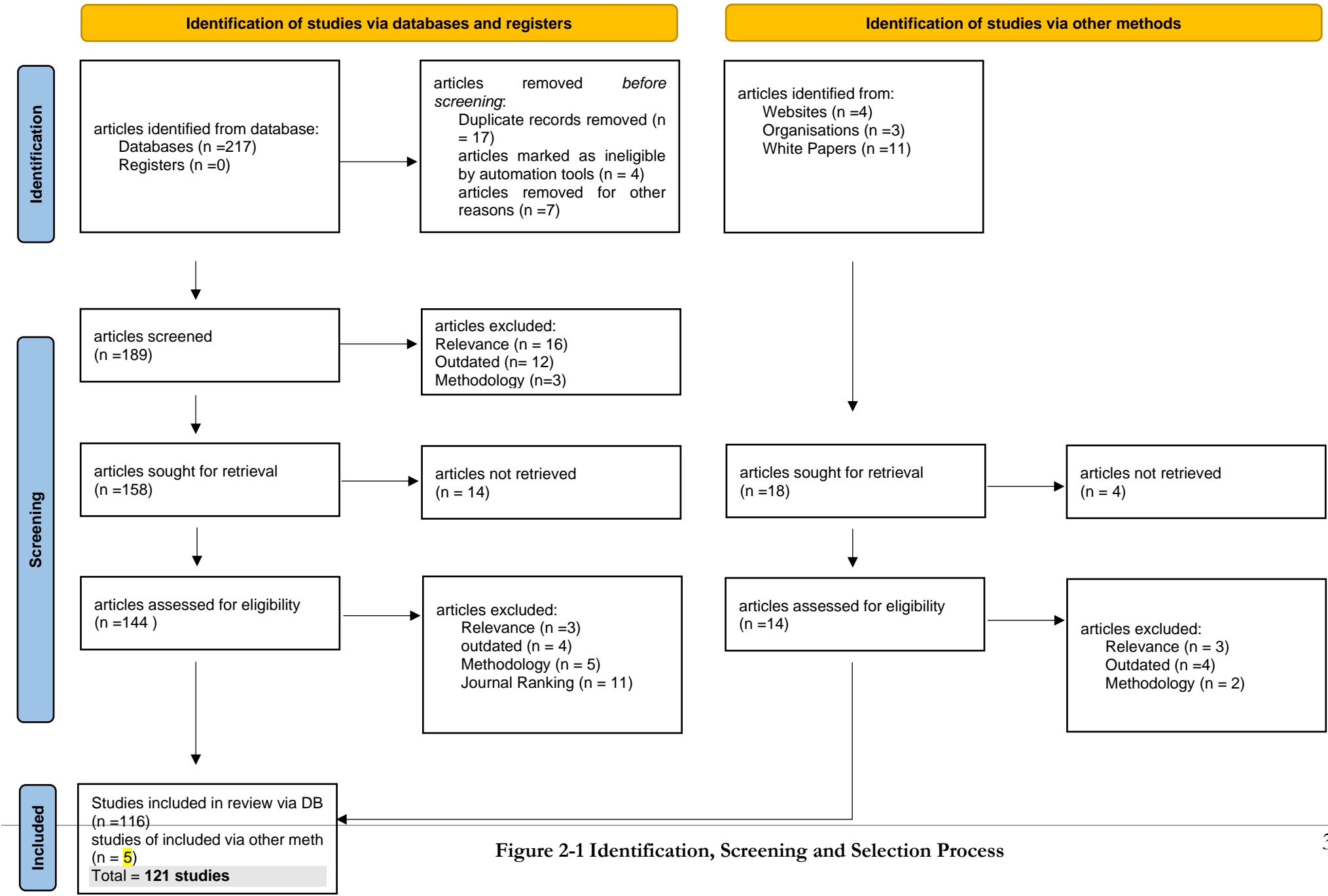

Figure 2-1 Identification, Screening and Selection Process

## 2.2 Research themes

In accordance with Webster & Watson, (2002) and Watson (2015), the researcher categorised the body of literature based on concepts or themes and organised such concepts using the conceptual framework presented in Figure 2-2. As depicted in the figure below, this study used the Knowledge Based view Theory of the firm to underpin the different salient themes of the literature and to integrate and analyse them conceptually from the knowledge perspective.

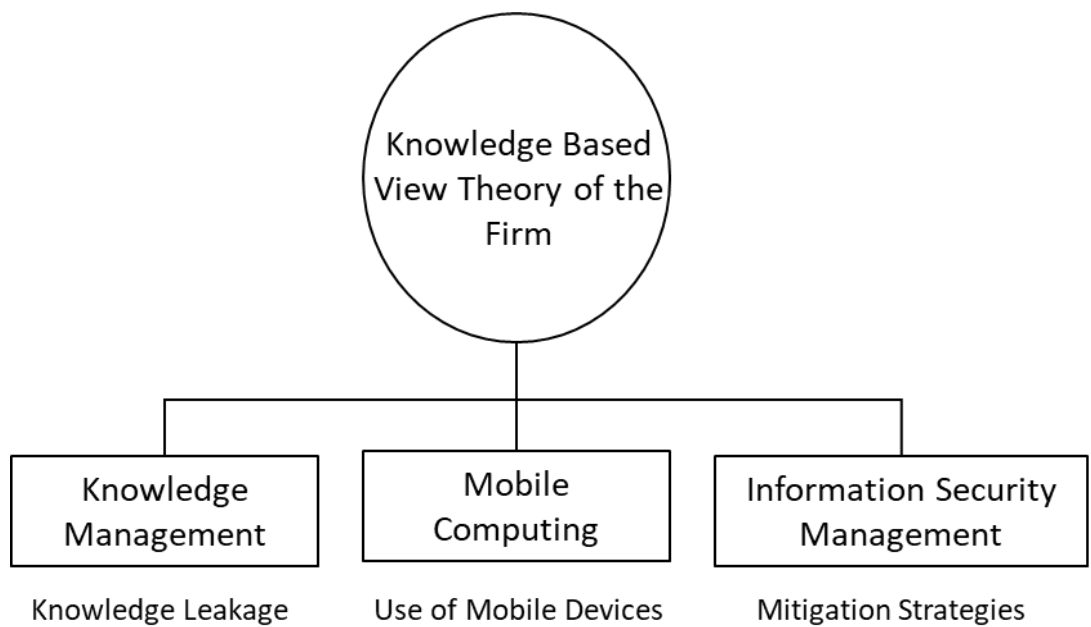

**Figure 2-2. Conceptual Framework for Literature Integration, Synthesis and Analysis**

In the subsequent sections the different streams of the literature are further analysed and detailed. By starting with the knowledge based view theory of the firm as a first topic, the researcher helps the reader to understand the perspective and the lens from which the remaining literature concepts are reviewed. Following this, the author continues to analyse the literature streams moving from a broader to a narrower view and critically analysing the research pertinent to the research topic, and subsequently indicating the insights and gaps found in the current body of literature.  As shown in Figure 2-3, the literature review funnel illustrates how the researcher conducted the reviewing process and remains a common approach in analysing heterogenous literature (Berthon et al., 2003; De Moya & Pallud, 2017).

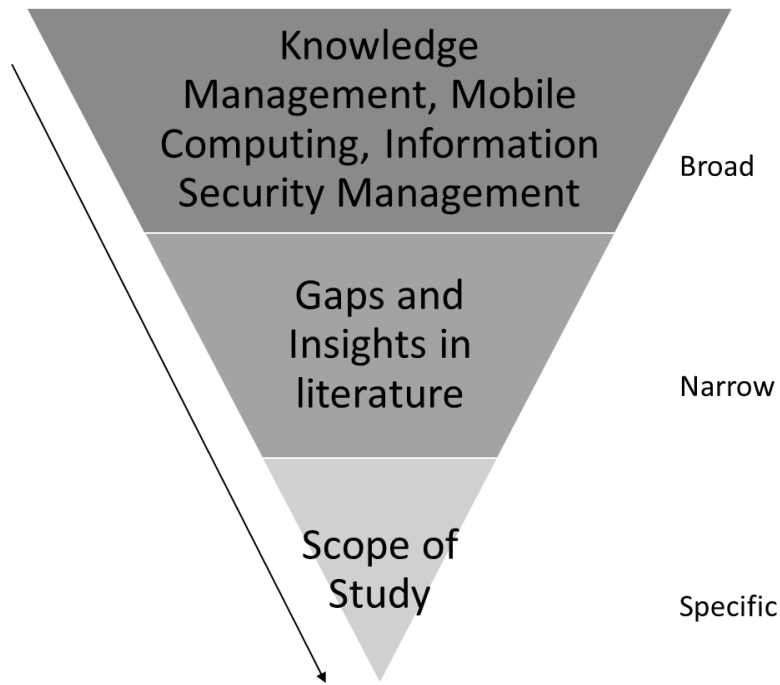

**Figure 2-3. Literature Review Funnel**

## 2.3 Knowledge Based View Theory of the Firm

The knowledge-based view of the firm, which originated from the strategic management literature, offers a perspective that builds upon and extends the resource-based view of the firm (RBV). The RBV was first introduced by Penrose (1960) and subsequently expanded by other scholars (J. Barney, 1991; Conner, 1991; Wernerfelt, 1984). The knowledge-based view theory of the firm (KBV) as an extension of the resource-based view theory (RBV) considers knowledge as the most strategically significant resource of a firm used for creating and sustaining economic rent and competitive advantage (Grant, 1996b).

While RBV argues that successful firms will achieve their competitiveness through the development of *distinctive* and unique *capabilities* often in the form of implicit or intangible assets (D J Teece et al., 1991), it fails to properly incorporate and acknowledge the relevance and significance of human, social, and organizational resources, such as stock of organizational knowledge that could be described as rare and an important source capable of generating a sustaining competitive advantage and a high economic rent (Ranft & Lord, 2002). Instead, RBV mostly focuses on economic and technical resources (Theriou et al., 2009). While the resource-based view of the firm acknowledges the significance of knowledge in enabling firms to gain a competitive advantage, proponents of the KBV contend that the resource-based perspective fails to address specific knowledge features, particularly, the way RBV characterizes knowledge as a generic resource, neglecting to differentiate between various types of knowledge-based capabilities and failing to recognize their unique characteristics (Nickerson & Zenger, 2004). In addition to this, information technologies may play a further critical role in the KBV of the firm in the sense that information systems have the potential to combine, improve, and expedite the management of vast quantities of knowledge within and between firms (Alavi & Leidner, 2001).

In this regard, previous studies (Alchian & Demsetz, 1972; Amit & Schoemaker, 1993; J. B. Barney, 1986) reported that economical and effective  production with diverse resources were not always the result of technical and economic resources, as RBV argued,  but rather the result of  *knowing* more precisely the efficient performance of such resources. This new approach resulted in the development of the KBV.

Importantly, the KBV theory primarily contends that firms represent entities that create, integrate and share knowledge and their ability to generate value relies on intangible knowledge based capabilities. Further, firm's competitive success depends on knowledge based assets that create core competencies. (Pemberton & Stonehouse, 2000). Therefore, in KBV, the most important premise is that the crucial input in production and principal source of value is knowledge (Grant, 1996a; J.-C. Spender & Grant, 1996a).

Furthermore, KBV also argues that knowledge-based resources that are challenging to replicate and involve social complexity have the potential to generate a sustainable competitive advantage over the long term (Alavi & Leidner, 2001). In fact, as other authors have mentioned, knowledge based resources should also meet the **VRIN** criteria, that is, *valuable*, *rare*, *inimitable* (imperfectly imitable) and *non-substitutable* to provide sustainable competitive advantage. Moreover knowledge and resources can be viewed as distinct from one another and both can exist inside as well as outside of the firm, while capabilities remain inside the firm and connect knowledge with organizational performance (Kaplan et al., 2001). In this integrated view by Kaplan et al.(2001), the firm is surrounded by two boundaries (i.e., absorption boundary and integration boundary) that illustrates the idea that the firm can possess a porous boundary where knowledge can be integrated without being incorporated (absorbed); and reflects the view that knowledge can be dynamic and therefore can be misappropriated and replicated by competitors as well. This notion is important because it represents the basis of knowledge leakage and will be elaborated further in the chapter. This integrated view is represented in Figure 2-4.

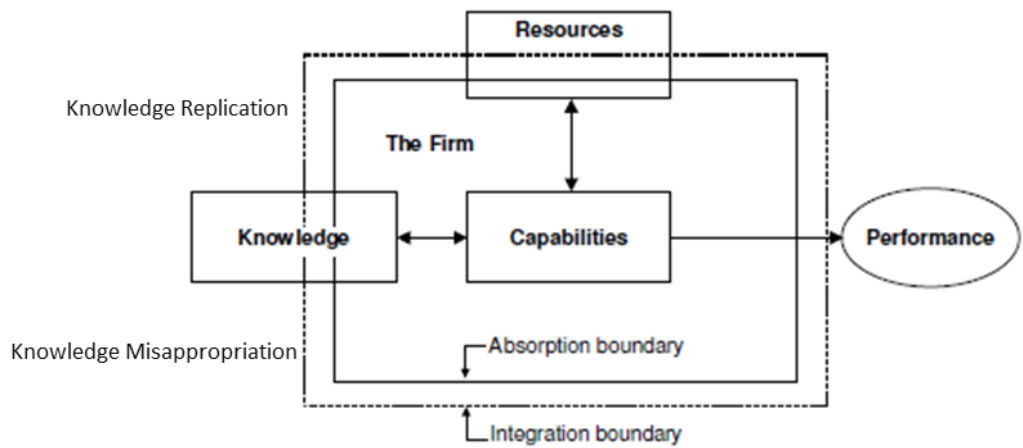

**Figure 2-4. Integrated Knowledge-based view of the firm. Adapted from Kaplan et al. (2001)**

## 2.3.1 Knowledge Intensive Organizations

As mentioned previously, the knowledge-based view of the firm considers a firm as a knowledge-creating entity, and posits that knowledge and the capability to generate and use it, remains the key source of a firm's sustainable competitive advantage which is extensively supported by researchers in the knowledge management and strategic management literature. (Melville et al., 2004; Ikujirō Nonaka et al., 1995; Ikujiro Nonaka & Toyama, 2003; Parker, 2012; Penrose, 1960; J. Spender, 1996).

This knowledge based perspective led to the emergence of the concept of **knowledge intensive organizations**, that is, firms whose main activity relies on the employment and leverage of knowledge based resources (Boisot & Canals, 2004; Ikujirō Nonaka et al., 1995). In organizations, these knowledge based resources are represented as both **individual knowledge** (i.e., human capital) in the mind of employees, as well as **organizational knowledge** (i.e., relational and structural capital) such as organisational routines, strategies, processes, databases, hierarchies, knowledge repositories and knowledge management systems. Collectively, all of these knowledge based resources (i.e., human, relational and structural) represent the intellectual capital of the firm and grant organizations their competitive advantage (Bolisani et al., 2013; MacDougall & Hurst, 2005; Ritala et al., 2015)

Further, these knowledge based resources constitute unique assets, also known as **knowledge assets**, difficult to imitate due to their socially complex nature. Such heterogeneous organizational knowledge bases and capabilities among firms represent the major determinants of sustained competitive advantage and improved organisational performance.

This organizational knowledge base originates and is ingrained in various components of the firm, including its organizational culture and identity, policies, routines, documents, systems, and personnel. This knowledge is distributed throughout the organization and can be difficult to codify and transfer, making it a valuable but complex resource (J.-C. Spender & Grant, 1996b).

## 2.3.1.1 The Importance of Knowledge-Intensive Organizations

Knowledge-intensive organizations play a crucial role in the modern economy and are often at the forefront of innovation. These organizations, which include sectors such as technology, finance, healthcare, and professional services, rely heavily on the knowledge and expertise of their employees to create value. They are characterized by a high degree of expertise, a strong focus on research and development, and a reliance on intellectual capital (Vafaei-Zadeh et al., 2019).

In the knowledge-based economy, these organizations are key drivers of economic growth and competitiveness. They contribute to the economy not only through their direct economic output but also by driving innovation, creating high-quality jobs, and contributing to productivity growth. Their activities often lead to the development of new products, services, and technologies that can have far-reaching impacts across various sectors of the economy (Figueiredo & de Matos Ferreira, 2020).

Due to their unique characteristics and the inherent challenges they face in managing and protecting their knowledge assets, knowledge-intensive organizations provides insights into the management and protection of critical knowledge assets, which are key drivers of innovation, economic growth, and societal advancement (Ritala et al., 2015). These organizations, by their very nature, rely heavily on the knowledge and expertise of their employees to drive innovation, maintain competitive advantage, and achieve their strategic objectives. This makes the management and protection of their knowledge assets a critical concern (Durst et al., 2015).

In knowledge-intensive organizations, knowledge is not just an ancillary resource; it is the core asset that fuels their operations and underpins their value proposition. The knowledge held within these organizations is often complex, specialized, and deeply embedded in their processes and people. This makes it both incredibly valuable and particularly vulnerable to leakage (Ritala et al., 2018).

Moreover, knowledge-intensive organizations often operate in highly competitive and rapidly evolving industries where the loss or leakage of knowledge can have significant strategic implications. In such contexts, the protection of knowledge is not just about preventing loss; it is about safeguarding the organization's competitive position and future viability (Wu et al., 2021).

## 2.3.2 Transferability

According to the resource-based view of the firm, a firm's ability to achieve sustainable competitive advantage is largely determined by the *transferability* of its resources and capabilities (Barney, 1986). In the context of knowledge, transferability is critical not only between firms, but also within the firm. The management literature has recognized a fundamental epistemological distinction between "knowing how" and "knowing about", as captured by the contrasting concepts of subjective vs. objective knowledge, implicit or tacit vs. explicit knowledge, personal vs. propositional knowledge, and procedural vs. declarative knowledge. The crucial difference between these two types of knowledge lies in their transferability and the mechanisms by which they can be transferred across individuals, space, and time. Explicit knowledge can be communicated and shared relatively easily, since it can be codified and transferred through language or other means. In contrast, tacit knowledge is revealed through its application, and its transfer between individuals is often slow, costly, and uncertain, since it cannot be easily codified and must be acquired through practice and observation (Kogut and Zander, 1992).

## 2.3.3 Capacity for Aggregation

The effectiveness of knowledge transfer is partly dependent on the potential for knowledge *aggregation*. Knowledge transfer involves both transmission and reception, with the recipient's absorptive capacity analyzed in terms of their ability to incorporate new knowledge into existing knowledge, at both individual and organizational levels. This requires *additivity* across different knowledge elements.

The efficiency of knowledge aggregation and transferability is enhanced when knowledge can be expressed in a common language. The use of statistics to summarize and aggregate explicit knowledge, such as figures and indicators, is an example of this. Information systems play an important role in facilitating the transfer of such knowledge across different parts of the organization. Conversely, *idiosyncratic* knowledge, such as individual skill sets and quirks of specific tools or technology, cannot be aggregated at a single location. This is referred to as "knowledge of the *particular* circumstances of *time and place*" by Hayek (1945: 521) and "*specific* knowledge" by Jensen and Meckling (1992). The ability to transfer and aggregate knowledge is a critical factor in determining the optimal location of decision-making within the firm.

## 2.3.4 Appropriability

Appropriability refers to the owner's ability to receive a return on a resource equal to the value it generates (Teece, 1987; Levin et al., 1987). Knowledge is a resource that faces unique challenges in terms of *appropriability*. Tacit knowledge cannot be directly *transferred* and can only be appropriated through productive application.

Explicit knowledge presents two key appropriability issues: firstly, as a public or non-rivalrous good, once acquired it can be resold without loss (Arrow, 1984); secondly, the act of marketing knowledge makes it accessible to potential buyers (Arrow, 1971: 152). Therefore, except for situations involving legally established property rights such as patents and copyrights, knowledge is often in-appropriable through market transactions. Ambiguity over knowledge ownership can arise from a lack of clear property rights. While most explicit knowledge and all tacit knowledge is stored within individuals, much of this knowledge is created within the firm and is firm-specific, resulting in challenges over the allocation of returns and achieving optimal investment in new knowledge (Rosen, 1991).

### 2.3.4.1 Specialization in Knowledge Acquisition

The human brain has limited capacity to acquire, store and process knowledge. The result is that efficiency in knowledge production (the creation of new knowledge, the *acquisition* of existing knowledge, and storage of knowledge) requires that individuals specialize in particular areas of knowledge. This implies that experts are (almost) invariably specialists.

The human brain has a finite capacity for acquiring, storing, and processing knowledge, which places limitations on the efficiency of knowledge production. To achieve optimal results in knowledge production, individuals are required to specialize in particular areas of knowledge, and this often leads to the emergence of experts who possess deep expertise in specific domains (J.-C. Spender & Grant, 1996b).

Expertise is characterized by a high level of knowledge and skill in a narrow area, and this specialization allows individuals to efficiently process and utilize knowledge in their respective fields. However, as the scope of knowledge continues to expand, specialization can be challenging, and experts may need to broaden their knowledge base to remain relevant in their fields. To address this challenge, experts may engage in interdisciplinary learning and collaboration to expand their knowledge and skills beyond their primary area of expertise. This approach can facilitate knowledge integration and innovation, as experts bring different perspectives and ideas from their respective domains to tackle complex problems. Furthermore, interdisciplinary collaboration can help experts to recognize and appreciate the value of different types of knowledge, including tacit and explicit knowledge, which can ultimately lead to more effective knowledge production and acquisition. Therefore, while expertise in a particular area is essential for efficient knowledge production, collaboration and learning across domains can help to foster innovation and ensure the continued relevance of experts in an ever-changing knowledge landscape (Levin et al, 1987).

## 2.3.4.2 The Knowledge Requirement of Production

In the context of a knowledge-based theory of the firm, *production* is viewed as the process of transforming inputs into outputs. Central to this theory is the premise that knowledge is the critical input and the primary source of value in this process.

All human *productivity*, regardless of the industry or sector, is fundamentally dependent on knowledge. Even machines, which are often considered the epitome of technological advancement, are simply embodiments of knowledge, designed and built by individuals who possess specialized knowledge and expertise. Therefore, a firm's ability to leverage and manage its knowledge resources effectively is essential to achieving sustainable competitive advantage in today's knowledge-driven economy (Teece, 1987).

## 2.4 Knowledge

There is abundant literature addressing the difference between data, information and knowledge in sources such as Boisot and Canals (2004) and Dahlbom and Mathiassen.(1993). These authors state that raw data becomes information through processes that add meaning from such data, and, adding the contextual understanding in conjunction with the background of such data allows knowledge to be inferred. Therefore, knowledge is intertwined with data and information. Consequently, knowledge leakage is also related to data and information leakage. This distinction is important for our study because the leakage of data/information may also result in knowledge leakage just by drawing on inferences, that is, humans gain knowledge by inferences – the process of inferring things based on what is already known (Dahlbom and Mathiassen 1993).

According to Schwartz (2006) as a way of circumventing to a certain extent this debate about knowledge, information and data (KID) regarding the granularity of knowledge, we should take the perspective of knowledge management (KM) process and focus on praxis, "taking as a starting point the question, What do we need to do with knowledge in order to make it viable for an organization to use, reuse, and manage it as a tangible resource, and apply it toward specific actions?" (p. 11). Schwartz (2006) also argues that such KID distinction although important can be conveniently ignored as it is not essential to the fundamental mission of KM. In taking this approach, this study acknowledges that knowledge leakage remains a KM issue and therefore posits that knowledge should be analysed from an applied pragmatic and holistic view (David G. Schwartz & Te'eni, 2011).

Similarly, Davenport and Prusak (1998) defined knowledge as the following:

"Knowledge is a fluid mix of framed experience, values, contextual information, and expert insight that provides a framework for evaluating and incorporating new experiences and information. It originates and is applied in the minds of knowers. In organizations, it often becomes embedded not only in documents or repositories but also in organizational routines, processes, practices, and norms."

According to this definition, knowledge is complex, a mixture of various elements, intuitive and therefore, hard to capture. Moreover, knowledge is embedded in people, and as such, may be unpredictable and intangible. Knowledge derives from information and to turn information into knowledge, human mediation is required (Davenport and Prusak 1998). Although knowledge is further divided into tacit (present in employee's minds) and explicit (knowledge that has been codified into artefacts) (Nonaka 1991), Nonaka and Takeuchi (1995) also suggested that knowledge can be created as a result of a dynamic process and constitutes a epistemological cycle whereby knowledge is continuously converted from tacit form to explicit form and vice-versa. This continuous and dynamic process complements the interaction at different ontological levels, that is to say, the knowledge dynamic transcends from interaction between individuals, to groups, to the organization as one entity. In his seminal work, where Nonaka (1994) proposed the SECI model, as shown in Figure 2-5, he illustrated this process and presented the four conversion modes that result from these interactions, and the spiral of knowledge created by the continuous switch from one knowledge type to the other. These four modes comprised the following phases: Socialization, Externalization, Combination, and Internalization.

The spiral of knowledge starts with the *socialization* phase in which the tacit knowledge is exchanged amongst individuals through interactions and experiences, then the tacit knowledge is converted into new explicit knowledge via the *externalization* phase in the form of documents, concepts, images and models. After this, the new explicit knowledge is mixed with existing intra- or inter- organizational explicit knowledge during the *combination* phase to create more complex and systematic explicit knowledge which may result in further technological artefacts such as communication and electronic systems, databases, information systems and intranets. Finally, in the *internalization* phase, individuals absorb the explicit knowledge complementing their existing tacit knowledge base, and as a result, the new formal knowledge is linked to their personal experiences which then can be and used in other circumstances and transferred and replicated to other individuals, groups, and the organisation or other organisations, as required.

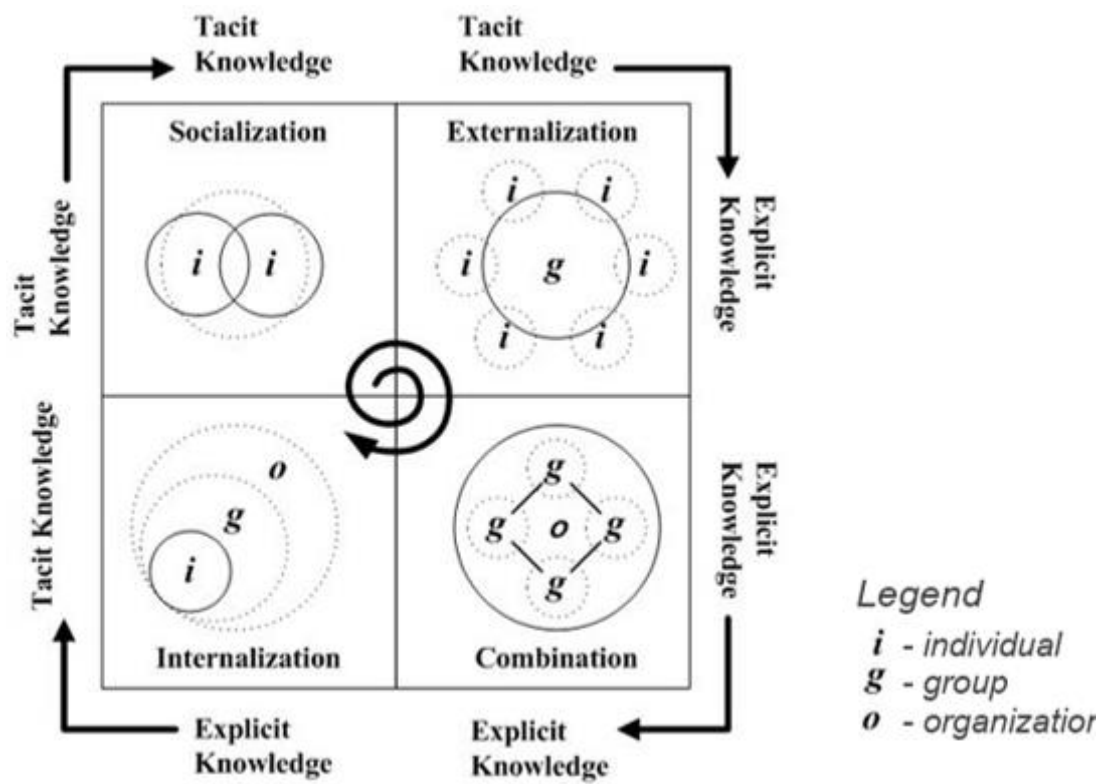

**Figure 2-5. SECI Model adopted from Nonaka and Takeuchi (1995)**

According to a number of researchers (T. Chen et al., 2014; Choi et al., 2015; Crager et al., 2007) from the perspective of mobile devices, explicit knowledge leakage constitutes the main concern, as knowledge which is represented as storage of data on the device itself is exposed to confidentiality issues and represent a more serious concern since its disclosure is more likely to occur in mobile device (and mobile technology) settings than tacit knowledge leakage, such as when key personnel leave the organization to a competitor (Frishammar et al. 2015). In contrast, Yu et al.

(2018) found that leakage of tacit knowledge through mobile devices remains a possibility when voice communications are intercepted and analysed.

Information and knowledge have become key strategic assets (Bollinger and Smith 2001; Grant 1996; Spender and Grant 1996) for knowledge-intensive organizations to achieve sustained competitive advantage, innovation and value creation (Nonaka and Toyama 2003; Sveiby 1997). Similarly, MacDougall and Hurst (2005) contend that the adoption of knowledge workers, employees who produce value by utilizing their knowledge rather than physical labor, allows organizations to develop their knowledge assets. These individuals perform work based on their information assets for the coordination and management of organizational activities (Sorensen et al. 2008). Ristovska et al. (2012) also focus on the importance of knowledge embedded in knowledge workers as it is an organizational asset for achieving sustainable competitive advantage which can be materialized into documentation and organizational processes. The importance of expertise in organizations relies heavily on exercising specialist knowledge and competencies, or alternatively, the management of organizational competencies and capabilities which belong to employees or knowledge workers (Blackler 1995; Thompson and Walsham 2004).

Accordingly, knowledge, in this sense represents the mix of contextual information residing in the mind of the knowledge worker, personalized by the individual, based on facts, procedures, concepts, interpretations, ideas, observations and judgments which is codified into artefacts such as documentation, processes and guidelines in organizations (Alavi & Leidner, 1999; Ikujiro Nonaka & Toyama, 2003).

## 2.5 Knowledge Leakage – Definition and gaps

Based on the previous sections, knowledge is a complex concept that involves multiple facets and as such can represent different entities in different contexts. Utilising the knowledge based view theory of the firm as a lens to qualify knowledge as an organisational asset, the definition of knowledge leakage consequently involves the point of view of the organisation. As reported by a number of researchers knowledge leakage (KL) can be defined as the **accidental** or **deliberate** loss or unauthorized transfer of organizational knowledge intended to stay within a firm's boundary resulting in the deterioration of competitiveness and industrial position of the organization (Frishammar et al., 2015; Mohamed et al., 2007; Nunes et al., 2006).

According to this knowledge leakage definition, KL can occur from the disclosure of sensitive details, information or data as meaning can be inferred by a competitor based on understanding of context and leveraged even further to generate insights and advance their own competitiveness to the detriment of the organization's competitive advantage (Abdul Molok et al., 2010a; Ahmad et al., 2015; Annansingh, 2012; Bouncken & Barwinski, 2021; Davenport & Prusak, 1998; Wu et al., 2021).

As mentioned above, knowledge leakage, in the meaning of knowledge leaking away from its origin, can occur in different situations, however, recent research has also shown the dichotomy of leakage as it can be considered to be positive, when the organization benefits from it, or negative, when it is detrimental to the organization (Cohen & Levinthal, 1990; Grillitsch & Nilsson, 2017; Jiang et al., 2016).

This clarifies that knowledge leakage can have a positive connotation in other contexts. For example, in collaboration, a positive knowledge leakage can occur in the form of knowledge spill over between cooperation partners (Arias-Pérez et al., 2020; Mupepi, Mambo Governor Modak et al., 2017). On the other hand, a negative knowledge leakage occurs when an actor, consciously or inadvertently, leaks knowledge about the focal firm to a competitor or partner to the detriment of the firm's competitive advantage (Bouncken & Barwinski, 2021; Cheng & Kam, 2008; Durst et al., 2015).

Despite the criticality that knowledge leakage can have on organizations in either direction and the fact that knowledge management practices, such as knowledge transfer or acquisition have been studied extensively, the study of knowledge leakage appears to be insufficiently researched and warrants more investigation (Bloodgood & Chen, 2021; Kang & Lee, 2017).

Therefore it remains paramount to expand and build upon the current understanding of knowledge and focus on the leakage in organisational contexts by exploring its factors and determinants and propose mitigation strategies that are necessary to achieve the protection of organisational knowledge (Mupepi, 2017; Wu et al., 2021).

Furthermore, the lack of studies reviewing the knowledge leakage phenomenon caused explicitly by the use of mobile devices highlights the current gap in the literature, especially, as stated by different knowledge management authors specialised in mobile knowledge work such as Vishwanath (2016), Nelson et al (2017), Povarnitsyna (2020) and Tavares (2020), recent research shows that the role that mobile devices play for knowledge work remains understudied and given the importance and mobility affordance that such tools offer to work '*anytime*', '*anyplace*' and '*far beyond the traditional, centralized office*' remains of utmost importance that future research addresses such limitations in the knowledge management body of literature.

## 2.5.1 How Knowledge Leakage occurs

Based on the distinctions of tacit and explicit knowledge in the previous section, each type of knowledge posits its own set of challenges and mitigation strategies. Knowledge assets normally flow from different boundaries or containers within or outside the organization, that is, changing zones (see Figure 2-6). Under such circumstances, the transition amongst zones may result in knowledge leakage incidents when controls fail to address and ensure its confidentiality. (Bouncken & Barwinski, 2021; K. C. Desouza, 2011; K. C. Desouza & Vanapalli, 2005; Tseng et al., 2011; Ahmad & Ruighaver, 2005; Ahmad, 2002).

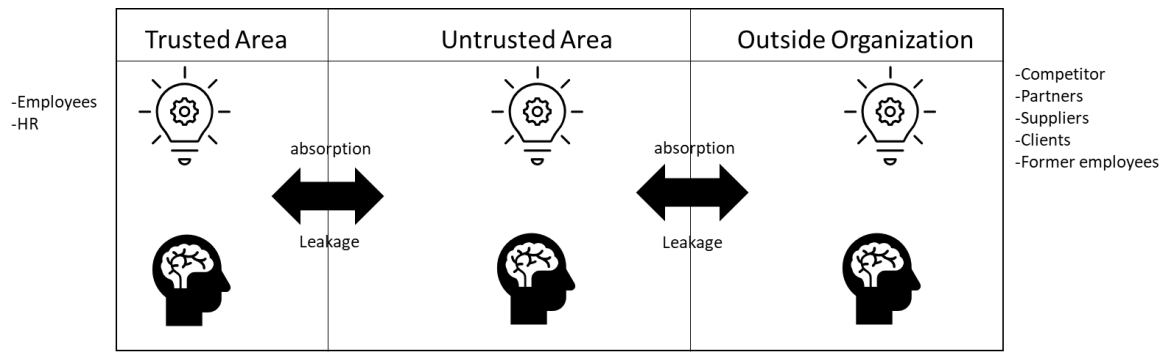

**Figure 2-6. Knowledge zones adapted from Vanapalli, 2005**

## 2.5.1.1 Explicit Knowledge

Explicit knowledge, as previously defined, refers to the type of knowledge that can be readily accessed, codified, articulated, and stored (Ikujiro Nonaka, 1991; Ikujirō Nonaka et al., 1995; Ikujiro Nonaka & Konno, 1998). This knowledge type remains the most common type reviewed and analysed in the scant leakage literature due to the fact that it is often conflated with data and information (Abdul Molok et al., 2010a; Ahmad, Bosua, et al., 2014; Shedden et al., 2011, 2012). This type of knowledge can leak as a result of different technological attacks, for example as a result of data breaches, virus, malware and common phishing attacks (Bloodgood & Chen, 2021; Ponemon Institute, 2018, 2021a).

## 2.5.1.2 Tacit Knowledge

Tacit knowledge, unlike explicit knowledge, refers to the type of knowledge that is more difficult to articulate and to codify and therefore more difficult to transfer to others by means of verbalising or writing it down. Tacit knowledge conventionally includes personal experience, heuristics, intuition, and insights (Ancori et al., 2000; Bosua & Scheepers, 2007; Brown & Duguid, 2001; K. C. Desouza, 2003b; Polanyi & Sen, 1997). However, although more difficult to leak, tacit knowledge can still be extracted and inferenced through means of social engineering (eavesdropping on conversations over the phone, recording mobile calls, pictures and photos) and espionage techniques that allow inferences to be made and pieces of seemingly disjointed knowledge to be connected together and recreated once the knowledge base shared is sufficient between the parties involved (Khatib, 2021; Thorleuchter & Van Den Poel, 2013), for example in the case of competing organisations in the same field working on developing similar products.

## 2.5.2 Knowledge Leakage Risk

Although knowledge loss due to a lack of knowledge management procedures is also defined as knowledge leakage (Bloodgood & Chen, 2021; Norman, 2001; Nunes et al., 2006), negative knowledge leakage directly or indirectly caused by knowledge workers when performing knowledge work represents the most common route of leakage and thus underlines the criticality of managing the associated risks that these workers pose (Mohamed et al., 2007, 2006; Wu et al., 2021). More recent research have emphasized the risk that mobility and **mobile devices** further represent to organizations due to the lack control, connectivity features, and increased leakage exposure (Agbo & Oyelere, 2019; Franklin et al., 2020; Janssen & Spruit, 2019; Morrell, 2020; Pratama & Scarlatos, 2020; Vishwanath, 2016; Weichbroth & Łysik, 2020). Likewise, the accidental knowledge leakage/ loss arising from misbehaviours (failing to comply with policy and procedures) constitutes yet another challenging channel of leakage for organizations to control and risk manage (Durst et al., 2015; Li et al., 2019; Mupepi, 2017; Nunes et al., 2006; Thorleuchter & Van Den Poel, 2013; Vafaei-Zadeh et al., 2019).

In this regard, the inadvertent leakage/ loss caused by insiders can be influenced by addressing human behaviour, habits through policy, culture, training and awareness as opposed to malicious insiders who deliberately seek to leak knowledge/ information (Bishop et al., 2010; Colwill, 2009; Crossler et al., 2013; D'Arcy et al., 2009; Dtex, 2022; Toelle, 2021) and purposefully disregard such controls. Therefore, different strategies should be applied to address unintentional leakage caused by non-malicious insiders and intentional leakage caused by malicious insiders, which, as mentioned before, in the context of mobility, more specifically, mobile devices exacerbates this issue further (Allam et al., 2014; Janssen & Spruit, 2019; Ortbach et al., 2015; Scarfo, 2012; Zahadat et al., 2015).

In the context of risk, drawing upon the standard definition of risk based on the ISO 31000 standard (27005:2011, 2011), risk as a function of impact versus probability, knowledge leakage risk (KLR), can be defined as the probability that KL occurs multiplied by the impact of the KL to the organization (Li et al., 2019; Tsang et al., 2016). However, previous researchers (Baskerville, 2005; Baskerville et al., 2014; Ghosh & Rai, 2013; Janssen & Spruit, 2019; Tsang et al., 2016) have highlighted the fact that these assessments of risk remain, for the most part, subjective exercises within organizations.

$$KLR = Knowledge\ Leakage\ Probability\ x\ Knowledge\ Leakage\ Impact$$

## 2.5.3 Knowledge Leakage Vectors

Based on the previous definition of risk, each assessment of risk is related to the different vectors that cause it. Knowledge leakage risk is often associated with collaborations and alliances in the supply chain such as partners, customers and competitors (Figure 2-7). Collaborative activities such as inter-firm partnerships, licensing, franchising, and outsourcing can provide firms with access to external knowledge resources and opportunities to generate new ideas. However, such activities can also increase the risk of unintended knowledge outflows or "knowledge leakage". Knowledge leakage can occur when firms inadvertently share their intellectual property, trade secrets, or other sensitive resources with their collaborators, or when their collaborators develop similar products or services using the knowledge gained from the partnership.

To address the issue of knowledge leakage, firms seek to implement effective governance mechanisms that facilitate knowledge acquisition while mitigating the risk of knowledge loss. Previous research has suggested that different governance mechanisms may be needed to address the diverse knowledge management practices of different firms. For example, Parker's (2012) study of new technology-based firms involved in inter-firm collaborative new product development projects found that relational governance mechanisms promote knowledge acquisition, while contractual governance mechanisms better address knowledge loss and leakage. Thus, a combination of governance approaches may be required to effectively manage knowledge in collaborative settings.

Another approach to reducing the risk of knowledge leakage is to limit the scope of the alliance or partnership. Oxley and Sampson (2004) study found that reducing the scope of the alliance can be an effective way to address knowledge leakage in competitive environments. However, this approach also has limitations, as it may restrict the potential benefits of collaboration and may not be feasible in all situations.

The risk of knowledge leakage is also influenced by the partners' absorptive capacity, which refers to their ability and capacity to assimilate and apply new knowledge. Partners with higher absorptive capacity are more likely to benefit from the knowledge gained from the partnership, but also more likely to engage in knowledge leakage (Durst et al., 2015). The number and type of partners involved in an alliance can also affect the risk of knowledge leakage, as more partners and partners with competing interests can make it more difficult to monitor and control the flow of knowledge. Therefore, firms need to carefully consider the potential risks and benefits of collaborative activities and implement effective governance mechanisms to manage knowledge leakage and promote successful collaborative outcomes (Durst et al., 2015; Lau et al., 2010; J. E. Oxley & Sampson, 2004; J. Oxley & Wada, 2009).

Similarly, a number of studies have examined the risk posed by knowledge workers within knowledge intensive firms which constitutes one of the most common vector of knowledge leakage (Arias-Pérez et al., 2020; Bloodgood & Chen, 2021; Bouncken & Barwinski, 2021; Galati et al., 2019; Jiang & Li, 2008; Li et al., 2019; Mohamed et al., 2006; Olander et al., 2011; Ritala et al., 2015; Vafaei-Zadeh et al., 2019). Either intentionally or accidentally employees often cause knowledge to leak when conducting their work through sharing, exchanging and capturing knowledge stemming from improper habits and behaviours (Bishop et al., 2010; Homoliak et al., 2019; Toelle, 2021). Additionally, it is often the case that case when knowledgeable employees leave their organizations, firms fail to follow a rigorous offboarding process to ensure the organizational knowledge remain protected(Durst et al., 2018; Mupepi, 2017). Whether through legal agreements or rigorous knowledge management processes such as capturing and transferring to other employees or into knowledge repositories, knowledge leakage can be contained and even prevented, particularly, in the case of inadvertent leakage through organizational cultural endeavours such as reinforcement, mentoring, communication, training, education, and awareness. Intentional knowledge leakage, however, remains a challenge and represents an on-going avenue for future research (Ahlfänger et al., 2021; Bloodgood & Chen, 2021; Thorleuchter & Van Den Poel, 2013).

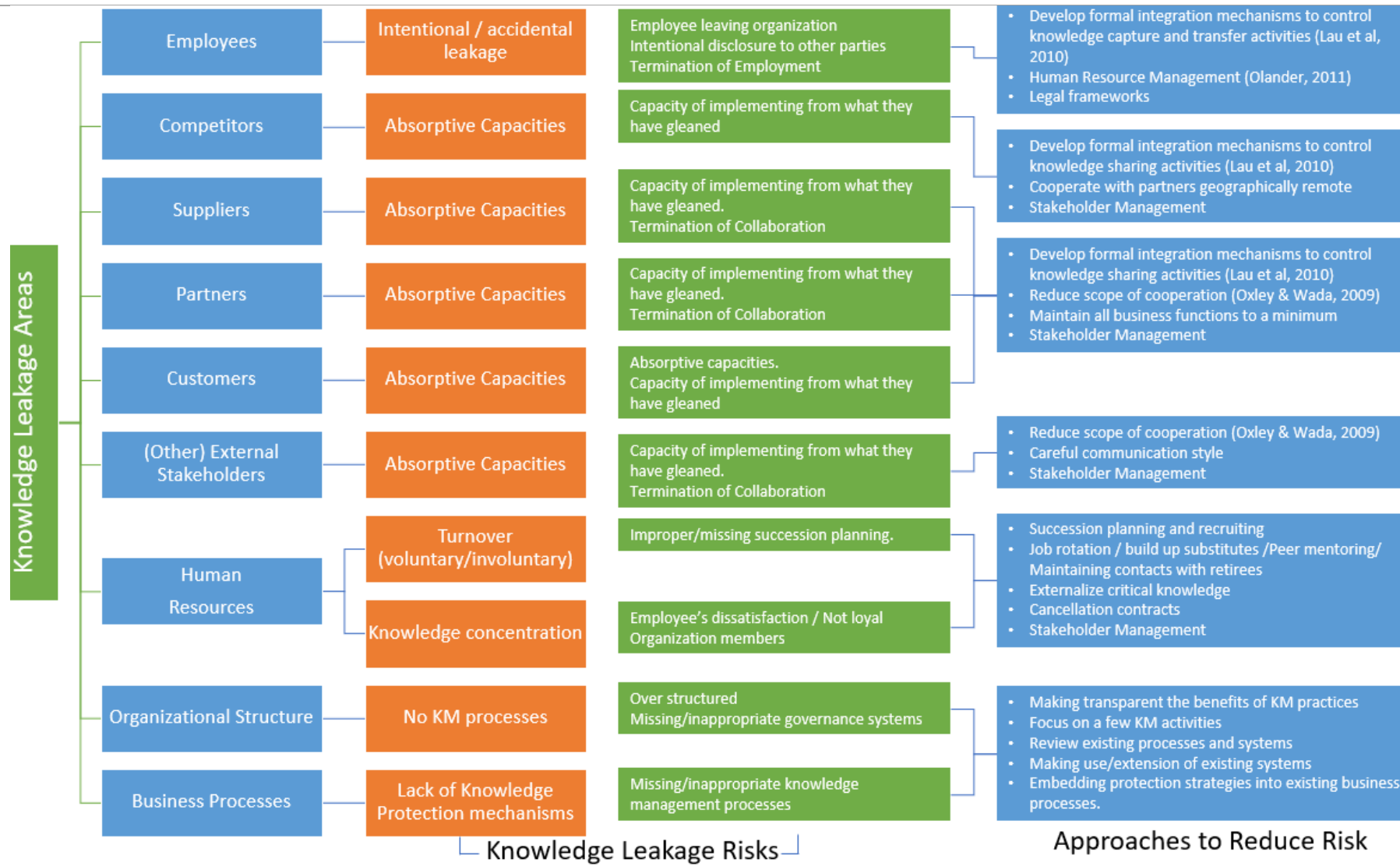

Figure 2-7. Knowledge Leakage Areas identified in the literature from different sources Durst et al (2014), Olander (2011), Oxley (2009), Lau (2010)

## 2.6 Knowledge Leakage and Mobile Contexts – Understanding Mobility and Mobile Devices

In order to address the issue of knowledge leakage risk through mobile devices and mobility in organizations, in line with the current research in mobile knowledge management, (Agbo & Oyelere, 2019; Aliannejadi et al., 2021; Benítez-Guerrero et al., 2012; Luo et al., 2020; Nelson et al., 2017; B. Schilit et al., 1994; Wakefield & Whitten, 2006) this study took a context-specific perspective to understand how risk changes according to the circumstances and factors within which leakage occurs.

Knowledge management in the context of device mobility can be viewed in terms of several qualities for the mobile devices (Rieger & Majchrzak, 2017):

- Physical dimensions and weight
- Whether the device is mobile or the host to which it is attached is mobile
- To what kind of host devices can it be bound
- How devices communicate with a host
- When the mobility occurs

Strictly speaking, most of the mobile devices are not mobile. It is the host that is mobile, i.e., a mobile human host carries a non-mobile device. Mobile knowledge workers or employees differ from conventional knowledge workers in that their work is conducted outside of traditional, centralized offices in independent, modular, and mobile arrangements for which they require content-producing devices as the primary work outputs of knowledge workers are intangible, analytic, creative, and, often times today, digital (Jarrahi & Thomson, 2017; Nelson et al., 2017; Vishwanath, 2016).

As such, they require devices able to support the creation, analysis and consumption of digital content which narrows down the use of mobile devices to specialised artefacts that facilitate knowledge work. Further to this, Rieger and Majchrzak (2017) defined a mobile taxonomy based on two dimensions — degree of mobility and media richness of inputs — and identified the devices that provided the ideal capabilities required by mobile workers as those that contained input features such as peripherals (keyboard, pointing device, touch) and voice capabilities with a *moveable* degree of mobility (refer to Figure 2-8). Meeting that criteria, the mobile devices used for mobile work consequently comprise of the following: Smartphone, Tablet, Netbook and Notebook.

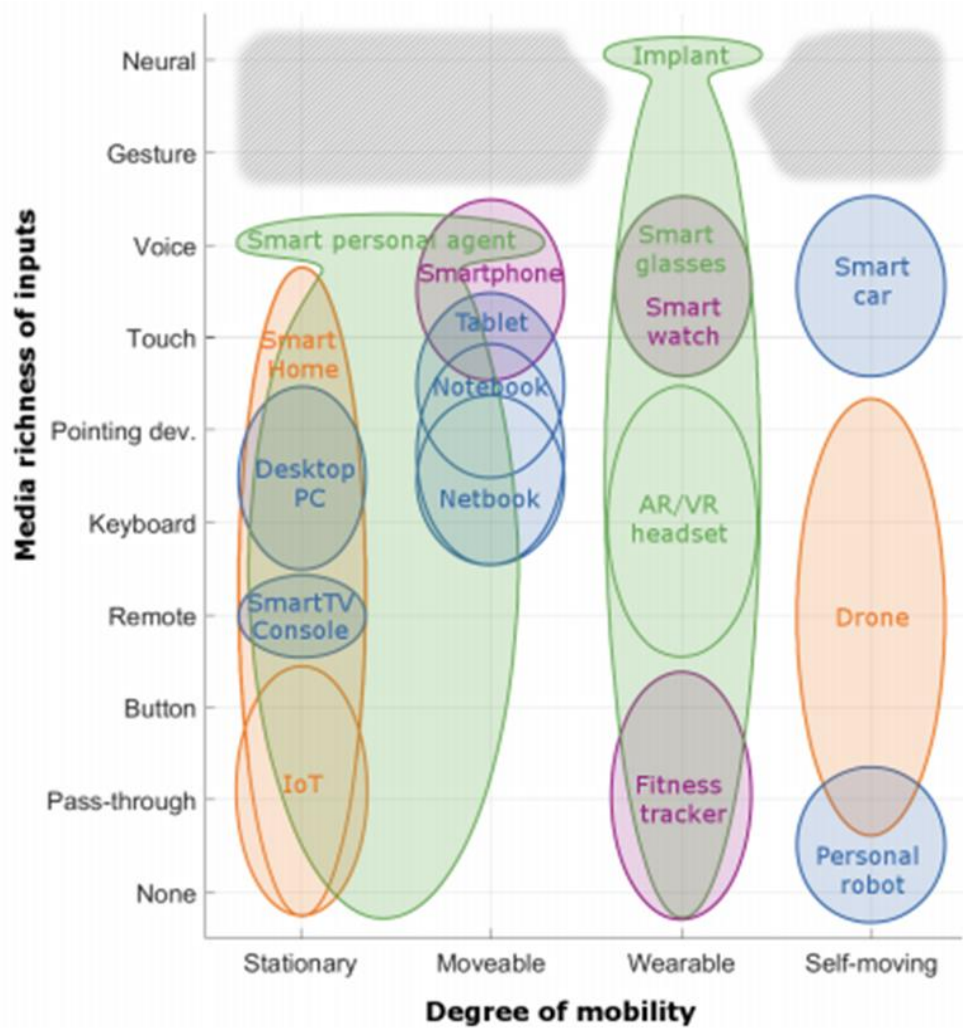

**Figure 2-8. Mobile Device Taxonomy - Adapted from Rieger (2017)**

Although the use of mobile devices (whether employee or organization owned) has shown to be convenient in the context of higher mobility, this convenience comes at a high security cost. The use of such devices by knowledge workers in knowledge-sharing activities poses a problem for confidentiality. Challenges in confidentiality occur as a result of employees' security (mis)behaviours.

While, as previously stated, there is abundant literature on leakage in terms of data and information (Abdul Molok et al., 2010b; T. Chen et al., 2014; D'Arcy et al., 2009; Gordon, 2007; Krishnamurthy & Wills, 2010; Morrow, 2012; Shabtai et al., 2012a; Yahav et al., 2014), knowledge leakage research, comparatively, continues to be underrepresented in the current knowledge management literature. However, most recent literature has emphasized the criticality of knowledge protection in organizations as a way to sustain competitive advantage, as well as the human dimension whilst dealing with knowledge assets (Ahmad et al., 2015; Ahmad, Bosua, et al., 2014; Bloodgood & Chen, 2021; Durst et al., 2015; Jiang et al., 2016; Kang & Lee, 2017; Mupepi, Mambo Governor Modak et al., 2017; Sumbal et al., 2017; Tan et al., 2016; Tsang et al., 2016; D. Zhang et al., 2017).

Therefore, knowledge leakage can be seen as a multi-dimensional problem and subsequently the focus should not only be on technological (e.g., firewall, antivirus, and compartmentalization) and formal (i.e., policies, standards and procedures) controls, but also on human factors and informal controls as well (Ahmad, Bosua, et al., 2014; Ahmad & Maynard, 2014; Bloodgood & Chen, 2021; D'Arcy et al., 2009; Foli & Durst, 2022; Jiang et al., 2016; Li et al., 2019; Mupepi, Mambo Governor Modak et al., 2017).

Whether deliberately, or inadvertently, workers are usually the main sources of information breaches rather than hackers. In fact, research has shown that the culture and people within an organization are just as likely to be the source of data leakage (Abdul Molok et al., 2011a; Ahmad et al., 2015; Colwill, 2009; Crossler et al., 2013; Ponemon Institute, 2021a). For example, confidential and sensitive information is sometimes shared inadvertently through social media and mobile devices exacerbate this issue further (Abdul Molok et al., 2011b; Yahav et al., 2014). In other cases, addressing human aspects through deterrence with education and awareness programs have proved to be effective (D'Arcy et al., 2009; Park et al., 2012; Tsang et al., 2016). Additionally, mobility and mobile technology have heightened the risk profile of organizations (Allam et al., 2014; Derballa & Pousttchi, 2006a, 2006b; Ghosh & Rai, 2013; Pratama & Scarlatos, 2020; D. Zhang et al., 2017).

## 2.6.1 Mobile contexts and Mobility

Although knowledge leakage is enabled by the employee in control of the mobile device, there are multiple environmental factors that affect the use of mobile devices for knowledge work. As previously stated, Nonaka and Toyama (2003) suggest that knowledge creation, sharing and distribution are achieved through the interactions between the individual, the organization and the environment. The environment influences the individual while, at the same time, individuals are continuously recreating their environment through their social interactions. This proposes that social factors in human interactions constantly change the environment in which knowledge is created. Further, Nonaka and Toyama (2003) developed a model of knowledge creation in order to explain the conversion of knowledge through interactions between individuals, groups of individuals, organizations and the environment. This model not only highlights the importance of the environmental and organizational circumstances around an individual, it also highlights the importance of the social environment where individuals interact within groups to obtain information (Ikujiro Nonaka & Konno, 1998; Ikujiro Nonaka & Toyama, 2003). The concept of mobility in knowledge management is not new and the substantial impact of mobile technology on knowledge management processes has been previously identified leading to the definition of *mobile knowledge management* and the importance of mobile environments (Derballa & Pousttchi, 2006a, 2006b; Franklin et al., 2020; Jarrahi & Thomson, 2017; Morrell, 2020; Nelson et al., 2017).

These environments are referred to in the literature, from a mobile device perspective, as the "context" of the mobile device usage. Context means the idea of *situated cognition*. In mobile computing, context defines the environment or situation surrounding a user or a device. (Abdoul Aziz Diallo, 2012; Al Sharoufi, 2021; L. Chen & Nath, 2008; B. N. Schilit & Theimer, 1994). Table 2-1 summarizes the mobile-usage context constructs and shows the taxonomy for mobile contexts from the literature. In understanding the different contexts of mobile device and mobility in these different settings (technological, environmental, organizational, social and personal), it is important to assess the overall security risk of the device as the potential enabler of, or medium through which knowledge leakage can occur, in conjunction with the user and the environment within which the device is used. The significance of mobile device contexts lies in the fact that knowledge leakage risk cannot be accurately assessed without considering its context within which such leakage occurs. (Benítez-Guerrero et al., 2012; Bradley & Dunlop, 2005; Xu, 2021). This concept also highlights the strong relationship between mobile and mobility, while the former refers to the device and technology, the latter refers to the behaviour of the mobile worker that changes with context. From there, a strategy to address the knowledge leakage risk through mobile devices should be built on an understanding of mobility (Derballa & Pousttchi, 2006b; Jiang et al., 2016; Povarnitsyna, 2020; Tavares, 2020).

**Table 2-2. Taxonomy of Relevant Mobile Usage Contexts derived from the literature[4]**

| Context Construct | Reference | Description |
|---|---|---|
| Environmental | (Agbo & Oyelere, 2019; Kofod-Petersen & Cassens, 2006; Luo et al., 2020; Nieto et al., 2006) | The environmental context is defined as the conjunction of the following contexts: temporal context, spatial context, social context, technological context, and business context |
| Personal | (Aliannejadi et al., 2021; Kofod-Petersen & Cassens, 2006) | The personal context provides the attributes of cognitive skills and draws on psychological and physiological contexts: psychological, goal, cognition, physiological, identity, actions |
| Social | (Jacob et al., 2020; Nieto et al., 2006; Xu, 2021) | Provides a social perspective of context, which captures the attributes of people (e.g. attitude, skills, and values) and the relationship of these people among one another and within the organization and collective structures (collective values and norms). |
| Technological | (Abdoul Aziz Diallo, 2012; Al Sharoufi, 2021; Jacob et al., 2020; Tamminen et al., 2004) | Provides the technological and technical attributes such as: network connections, infrastructure, equipment, devices and systems. It is an aggregate context which consists of other technical constituents such as spatial, user and location context. |
| Organizational | (Crossler et al., 2013; Furnell & Rajendran, 2012; Jacob et al., 2020; Whitman, Michael and Mattord, 2011) | Defines the social interactions within the workplace and security behaviour determined by Information Security Policies, Security Education Training and Awareness, Culture, Standards, organizational processes and procedures |
| Device | (Al Sharoufi, 2021; Belko Abdoul Aziz Diallo et al., 2014; Jacob et al., 2020; Kofod-Petersen & Cassens, 2006; Nieto et al., 2006) | Technological features such as device identifier, device type and processing capabilities that pertains exclusively to the communication device (i.e., laptop, tablet, smartphone) |

Mohamed et al. (2006) found that one of the key routes of knowledge leakage is *people* through social contexts of mobile usage. These routes include training courses, collaborations with universities, multi-disciplinary teams and temporary workers. Through social interactions in these different contexts, knowledge is shared or accessible to other users. Social context also includes the use of social networking platforms on mobile devices (Krishnamurthy & Wills, 2010; Xu, 2021).

---

[4] Published in: Agudelo-Serna, C. A., Bosua, R., Ahmad, A., & Maynard, S. (2017). Strategies to Mitigate Knowledge Leakage Risk caused by the use of mobile devices: A Preliminary Study.

Due to the nature of mobile device usage, the context of device usage transitions across many changes in technical, social and locational environments (technological, environmental, organizational and personal contexts). Through the interactions of these dynamic contexts with one another, the risk of knowledge leakage also becomes dynamic. Thus, knowledge can be leaked through the technological, organizational, personal, and social contexts (Aliannejadi et al., 2021; Dang-Pham & Pittayachawan, 2015; B A A Diallo et al., 2011; Belko Abdoul Aziz Diallo et al., 2014; Xu, 2021). As an illustration of this phenomenon, Astani et al (2013) found that a significant amount of employees from information sensitive industries such as banking, connected their mobile devices to unsecured public Wi-Fi networks (i.e., technological context, environmental context) which exposes the device to the security vulnerabilities of those networks and may be used as a vehicle for knowledge leakage. By simply changing the network connection to a public Wi-Fi network, these employees are drastically changing the technological and environmental contexts and, therefore, their "mobile device usage context" in which the device is operating, changing in this way the risk profile of their device, and drastically affecting the potential for knowledge leakage.

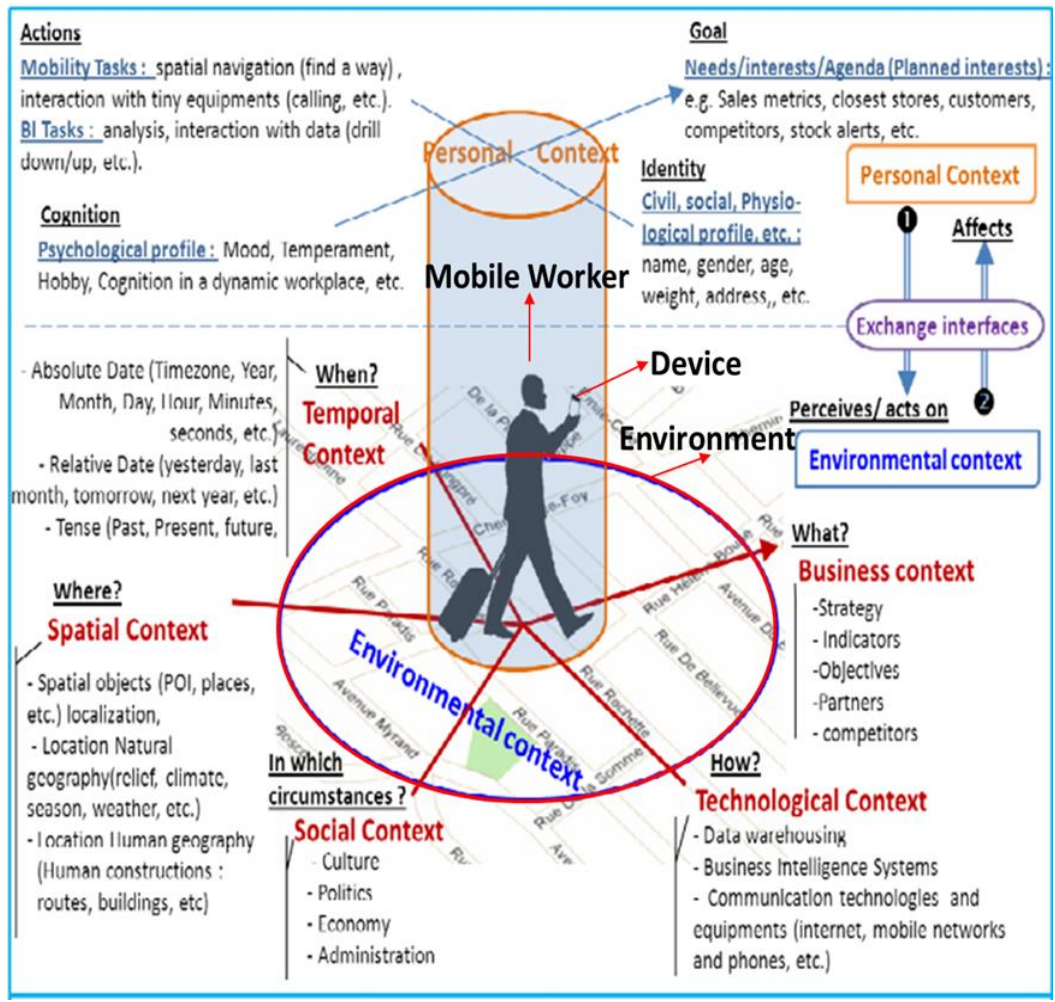

**Figure 2-9. Mobile Contexts and User Interaction with Mobile Device**

As shown in Figure 2-9, the interaction of the user with a mobile device and the multiple mobile contexts determines their knowledge leakage risk. These contexts are relevant to the use of the mobile device. If a user changes devices (device context, technological context), for example, then his/her overall context (user context) will change. The new device may not have the same functions as the previous one, resulting in a new number of contexts affecting the device. Since the old device is no longer used by the user, various contexts (e.g., social, user, and location contexts) no longer apply to it. This highlights the dynamic changes in knowledge leakage risks as the circumstances of how the knowledge worker uses their mobile device change.

Additionally, several researchers highlight how people and objects are constantly moving in and out of different context risks and the relevancy of these objects and people to the context are dynamic and hence the security threat of knowledge leaking is constantly changing (Al Sharoufi, 2021; Aliannejadi et al., 2021; Jarrahi & Thomson, 2017; Nelson et al., 2017; Rieger & Majchrzak, 2017; Tamminen et al., 2004). As an example, to understand this dynamic change of context, let us imagine that Gina is sitting in a coffee store reading corporate emails from a tablet before heading into work (environmental, personal and technological context) and a new customer sits down behind Gina (social context), Gina's risk context has changed as the customer may potentially read Gina's tablet screen (shoulder surfing). Gina then receives a phone call (personal and social context), which introduces a new person (caller) into the context, with whom the agenda of a morning meeting is discussed (organizational context). This change in context risk now involves the surrounding people within earshot drastically increasing the potential for knowledge leakage.

From the literature there have been many approaches to modelling the contextual information surrounding mobiles across many disciplines of Information Technology. However, most of the research into the *contextual information* and *context* of mobile devices has been focused on the technical and computing issues ignoring the human dimension and contexts associated to mobile usage (Agbo & Oyelere, 2019; Al Sharoufi, 2021; Aliannejadi et al., 2021; Benítez-Guerrero et al., 2012; Bradley & Dunlop, 2005; Belko Abdoul Aziz Diallo et al., 2012; Hofer et al., 2003; Kofod-Petersen & Cassens, 2006; Luo et al., 2020; B. N. Schilit & Theimer, 1994; Xu, 2021).

Similarly, Hofer et al. (2003) also extended and modelled these dimensions of context into device context (e.g. device identifier and device type) and network context (e.g. network connection types) which were included as the technical context, in more recent studies, by Abdoul Aziz Diallo (2012), Jarrahi and Thomson (2017), and Al-Sharoufi (2021). However, these studies failed to address the social context, neglecting the human perspective from the mobile contextual model, namely, user behaviour.

On the other hand, Chen & Nath (2008) asserted that the social context is not independent of the technical context; it is the "interaction and compatibility" between the two that determine the effectiveness of a work system. This interdependency of the social and technical context is further reflected by Bradley and Dunlop's, (2005) "Model of Context in Computer Science" which aims to illustrate the key components and characteristics of context which are present during user-computer interaction. The key idea derived from Chen and Nath's (2008) model of context is that there are multiple contexts that contribute to the mobile usage context of mobile devices.

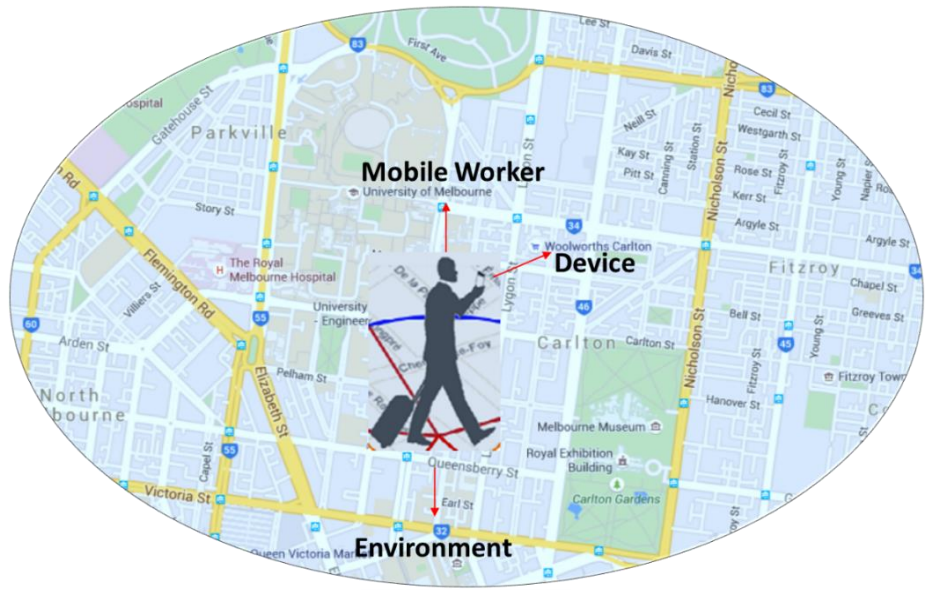

**Context**: Mobile Worker + Device + Environment

**Figure 2-10. Mobile Context Definition adopted from Chen and Nath , 2008**

Expanding on Chen & Nath's (2008) social context interaction framework and Bradley and Dunlop's (2005) model of context, Nelson et al (2017) defined the concept of mobile context as the interaction between the mobile worker, the device and the environment (see Figure 2-10). This definition of mobile context represents a fundamental shift away from the traditional and more technical definition established in the mobile computing literature and lays the foundation to include more social centric concepts from the human computer interaction domain, thus incorporating socio technical aspects into the usage of mobile devices and the interaction with people.

## 2.7 Strategies to Protect Organizational Knowledge

As discussed during this chapter so far, information and knowledge leakage has become a significant security risk to organizations, particularly in an increasingly networked and mobile society. This increased interconnectedness and mobility has resulted in a greater number of security incidents and breaches within organizations (Ahmad et al., 2020, 2021). A recent global research study on leakage conducted by the Ponemon Institute (2021a) found each security incident in business costs an average of US$4.24 million which is a 10% rise from the average cost in 2019 which was $3.86. Additionally, the aforementioned study also found that organisations spend on average US$2.2 million each on investigating and assessing information breaches (Ponemon Institute, 2021a).

In the same vein, recent literature, both professional and academic, show how organizations struggle with leakage of sensitive organizational resources across various avenues, such as espionage (i.e., industrial- and cyber-) , social media, cloud computing and portable data devices (Ahlfänger et al., 2021; Ahmad et al., 2015; Ahmad, Bosua, et al., 2014; Arias-Pérez et al., 2020; Bloodgood & Chen, 2021; Dtex, 2019, 2022; Inkpen et al., 2019; Jiang et al., 2013; Krishnamurthy & Wills, 2010; Mohamed et al., 2006; Nelson et al., 2017; Ponemon Institute, 2021b). As previously mentioned, although much of the literature has focused on technical security-related aspects of leakage (i.e., data and information), scant research has been conducted on knowledge leakage explicitly through mobile devices (Bouncken & Barwinski, 2021; Ghosh & Rai, 2013; Janssen & Spruit, 2019; Povarnitsyna, 2020; Tavares, 2020; Zahadat et al., 2015).

In this section, the author highlights and discusses the information security and knowledge protection strategies found in the current information security management and knowledge protection and security literature to protect and mitigate the aforementioned risk of organisational knowledge leakage.

## 2.7.1 Definition of Strategy

The term "strategy" has been used in military contexts for centuries, and it has been applied to business and other fields more recently. It is unclear who first coined the term or when it was first used in the sense of a plan of action designed to achieve a specific goal.

In business, strategy is defined as the plan of action an organization takes to achieve its goals and objectives and refers to the long-term plan of an organization to reach them. Strategy involves making choices about the types of products or services to offer, the target market to serve, and the resources to allocate to these efforts. (Hitt et al., 2017; Horne et al., 2017)

According to Mintzberg (1985), strategy is the pattern or plan that coordinates an organization's major goals and policies with its actions. In other words, it is the "*direction and scope of an organization over the long term: which achieves advantage for the organization through its configuration of resources within a challenging environment, to meet the needs of markets and to fulfill stakeholder expectations.*" (Johnson et al., 1999 P. 10)

Mintzberg (1985) further distinguishes between three types of strategies: deliberate, emergent, and realized. A deliberate strategy is one that is consciously planned and implemented by an organization. An emergent strategy arises through the day-to-day interactions and decision-making of an organization's members. A realized strategy is the actual strategy that an organization ends up following, which may be different from the deliberate or intended strategy.

When referring to *strategy* throughout this thesis, the researcher adopted the definition of *emergent strategy* as defined by Mintzberg. This type of strategy emerges over time as an organization responds to changes in its environment, it is not planned in advance, but rather evolves as an organization adapts to its surroundings.

Moreover, the emergent strategy represents a valid approach in situations where an organization operates in a rapidly changing environment and needs to be flexible and responsive in order to remain competitive. Unlike, deliberate (conventional) strategies that focus on long-term planning and are prescriptive in nature, emergent strategies are dynamic emphasize learning, agility and discovery, and make use of resources, *tactics* and *controls* which once reconfigured to address the environmental needs, constitute the new resulting emergent strategy.

In the context of information security, Park and Ruighaver (2008) define strategy as "*the art of deciding how to best utilize what appropriate defensive information security technologies and measures, and of deploying and applying them in a coordinated way to defense organization's information infrastructure(s) against internal and external threats by offering confidentiality, integrity and availability at the expense of least efforts and costs while to be effective*" (Park & Ruighaver, 2008, p. 27).

Park and Ruighaver (2008) further argue that the information security strategy should adopt a military rather than a business perspective whereby tactics, components and technologies of a strategy can also be considered strategies in their own right, as they contain sets of actions and objectives that are intended to achieve a specific outcome. For example, a component of an information security strategy may be to implement a firewall (network control), which is, in itself, a strategy for preventing unauthorized access to a network.

Furthermore, other strategist authors (David & David, 2011; Freeman, 2010) have also stated that integral components of strategy such as controls, technology and tactics can be considered strategies themselves insofar as they operate in tandem to facilitate the realization of strategic objectives.

Therefore, in this thesis the researcher uses the term *strategy* to also signify components, and elements of strategies and technologies that can be configured and deployed together to form dynamic *emergent strategies*.

## 2.7.2 Information Security Strategies

Within the information security management literature, Ahmad, Maynard and Park (2014) conducted a comprehensive literature review across different fields of information security, including the military, and defined and classified nine different security strategies in accordance with time, space and decision making dimensions that organisations may use to address the challenges of technological complexity and improve their security posture. The framework used for classifying the information security strategies comprises of three dimensions (Park & Ruighaver, 2008). **The time dimension** refers to strategies that can be implemented either proactively, before an attack occurs, or reactively, after an attack has taken place. **The space dimension** pertains to how the battlefield zone is arranged. For example, partitioning the space into different zones, in order to separate trusted and untrusted computer systems. This configuration can prevent a potential attack on an insecure system from spreading to more secure areas. **The decision making dimension** relates to decisions on specific, tactics (controls, measures, technologies) of attack and response that inform and configure the strategy to be applied. Figure 2-11 below depicts the framework for classifying information security strategies.

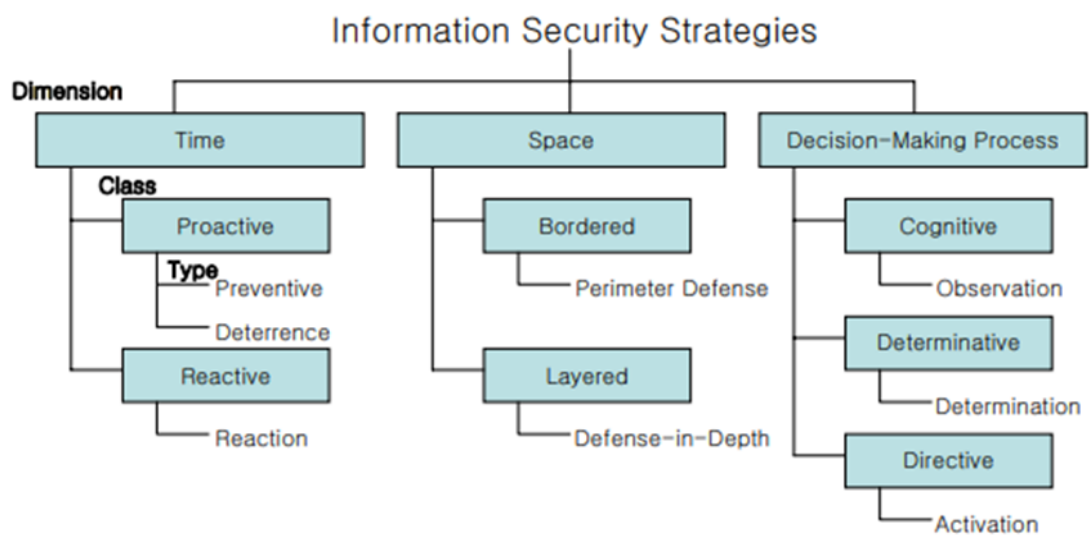

**Figure 2-11. Framework for Classifying Information Security adopted from Park & Ruighaver (2008)**

However, although some of these strategies can be applied to protect knowledge assets, their focus was mainly on information assets and technical risk rather than knowledge assets. see Table 2-3 for a summary of information security strategies.

**Table 2-3. Information Security Strategies adopted from in Ahmad et al. (2014)**

| Strategy | Definition |
|---|---|
| Prevention | Prevention aims to protect information assets prior to an attack by prohibiting unauthorized access, modification, destruction, or disclosure |
| Deterrence | Deterrence employs disciplinary action to influence human behaviour and attitude |
| Surveillance | Surveillance is the systematic monitoring of the security environment towards developing situational awareness to assist in adapting to fast-changing circumstances and threats |
| Detection | Detection is an operational-level strategy aimed at identifying specific security behaviour |
| Response | Response takes appropriate corrective actions against identified attacks |
| Deception | Deception distracts an attacker's attention from critical information assets using decoys thereby leading the attacker to waste time and resources |
| Perimeter Defence | Perimeter defence creates a boundary around information assets that is secured by regulating traffic at every incoming and outgoing information channel (choke points) |
| Compartmentalisation | Compartmentalization reduces an attacker's opportunities by dividing the intended area of attack into zones that are secured separately |
| Layering | Layering uses multiple countermeasures that function independently but increases the effectiveness of the defence when working together thereby posing a series of challenges to the attacker. The defensive system is designed to be resilient by overlapping the series of countermeasures, whereby each countermeasure complements the next so that if one fails another will back it up |

In contrast, other authors (Amara et al., 2008; Amoroso & Link, 2021; K. C. Desouza & Vanapalli, 2005; Grimaldi et al., 2021; Kitching & Blackburn, 1998; Manhart & Thalmann, 2015; Olander et al., 2014; Päällysaho & Kuusisto, 2011) have included strategies to protect knowledge from the perspective of additional dimensions such as: people, product, and process. The first dimension, people, refers to tacit knowledge, the second dimension, product, refers to knowledge as a product (knowledge asset - explicit knowledge), and the third dimension, process, refers to knowledge as an embodied organisational process. Each one of these dimensions contains different types of mitigation strategies from the organisational, legal, and technical views (see Table 2-4 Knowledge protection strategies found in the ).

**Table 2-4. Knowledge protection strategies found in the literature.**

| Strategy Type | People | Product | Process | Author(s) |
|---|---|---|---|---|
| Organisational | Awareness<br><br>Training<br><br>Indoctrination<br><br>Recruitment<br><br>Leadership<br><br>Education<br><br>Role creation<br><br>Counterintelligence<br><br>Human Resource Management | Lead time<br><br>Secrecy<br><br>Awareness<br><br>Training<br><br>Concealment<br><br>Standardisation<br><br>Annotation | Awareness<br><br>Training<br><br>Separation of duties<br><br>Accountability<br><br>Human Resource Management | (Amara et al., 2008; Boon et al., 2019; K. C. Desouza & Vanapalli, 2005; Figueiredo & de Matos Ferreira, 2020; Olander et al., 2014; Ritala et al., 2018) |
| Legal | Non-disclosure Agreements<br><br>Non-compete agreements<br><br>Ground rules | Patent<br><br>Trademark<br><br>Non-disclosure Agreements<br><br>Non-compete agreements<br><br>Ground rules<br><br>Intellectual Property Right | Copyright<br><br>Trademark<br><br>Non-disclosure Agreements<br><br>Non-compete agreements<br><br>Ground rules<br><br>Intellectual Property Right | (Amara et al., 2008; Amoroso & Link, 2021; Grimaldi et al., 2021; Link & van Hasselt, 2022; Manhart & Thalmann, 2015; Olander et al., 2014; Ritala et al., 2018) |
| Technical | | Standardisation<br><br>Annotation | Separation of duties | (Amara et al., 2008; Bolisani et al., 2013; K. C. |

| | Securing devices | Accountability | Desouza, 2011; |
|---|---|---|---|
| | Data Loss Prevention (DLP) | Securing comm. Channels | Dhillon, 2007; Shabtai et al., 2012b; Trkman et al., 2012; Yu et al., 2018) |
| | | (physical) access control | |
| | | Securing devices | |

For the first dimension, people, in order to protect knowledge held by people (tacit knowledge) organisational mechanisms such as indoctrination and recruitment are utilised to employ people meeting protection requirements, also counterintelligence may be used to monitor and identify potential leaks as evidenced and the defence and intelligence sector (K. C. Desouza & Vanapalli, 2005; Grimaldi et al., 2021). Legal strategies to protect tacit knowledge include contract clauses and ground rules contracts (Bolisani et al., 2013; Link & van Hasselt, 2022; Olander et al., 2014). No technical strategies to protect tacit knowledge were found in the analysed literature which highlights a gap in the current body of knowledge.

The second dimension, product, refers to the protection of knowledge as a product, that is, an object or artefact, explicit knowledge stored in documents, which may be addressed by common security strategies (K. C. Desouza & Vanapalli, 2005; Figueiredo & de Matos Ferreira, 2020). From the organisational perspective lead time may help organisations to enter the market first, preventing undesired knowledge appropriation, secrecy and concealment restricts knowledge sharing, and training and awareness also contribute to protect knowledge. Legal strategies refer to IPR (Intellectual property rights) and include copyright, trademarks, patents, and trade secrets (Amara et al., 2008; Amoroso & Link, 2021; M. Lee et al., 2018; Päällysaho & Kuusisto, 2011). And technical strategies to protect explicit knowledge refers to standardisation of documentation processes, segmentation, tagging of knowledge and clearance.

The third dimension, process, refers to the protection of knowledge as an organisational process, i.e., or strategic knowledge. Organisational strategies include training and awareness, organisational roles such as gatekeepers to monitor information flows, leadership, accountability, separation of duties and authorisation procedures(Ahmad, Bosua, et al., 2014; Boon et al., 2019; K. C. Desouza & Vanapalli, 2005; Grimaldi et al., 2021; Kitching & Blackburn, 1998; Zhao, 2006). Legal measures, similar to the previous dimension, product, contractual measures and IPR contribute to the protection of this type of knowledge. Technical strategies include access control, authentication, identification cards, biometrics, password, compartmentalisation, and encryption (Jaeger et al., 2017; Shabtai et al., 2012a, 2012b; Sonnenschein et al., 2017; Yu et al., 2018; Zahadat et al., 2015).

### 2.7.3 Informal Strategies to Protect Knowledge

In addition to the previously mentioned strategies, another important category that the researcher found in the existent knowledge management and knowledge protection literature referred to the use of formals vs. informal strategies to protect organizational knowledge, particularly, intellectual property (Amara et al., 2008; M. Lee et al., 2018; Päällysaho & Kuusisto, 2011). The consensus among these studies highlighted the importance and predominance of informal strategies across knowledge intensive firms due to their convenience, familiarity, economical value (cheaper), implementation efforts (easier), and as effective as compared to more formal legal strategies. Even when formal legal mechanism were adopted and implemented, most organization favoured the utilization of informal strategies(Figueiredo & de Matos Ferreira, 2020; Link & van Hasselt, 2022).

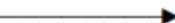

**Figure 2-12. A Continuum of Intellectual Property Protection Mechanisms. Adopted from Kitching and Blackburn (1998)**

These strategies were also classified as (informal vs formal) intellectual property protection mechanisms (i.e., IPPM) in the literature. A proposed continuum model by Kitching & Blackburn (1998) illustrates the degree of (in)formality of diferent IPPMs applied by knowledge intensive firms in the computer services, design, electronics and mechanical engineering industries (See Figure 2-12 above). Similarly, Päällysaho and Kuusisto (2011) also suggested a classification of intellectual property protection strategies adopted by knowledge intensive firms in Europe based on three levels of legal formality, that is, informal, semi-formal, and formal (Intellectual Property Rights — IPR) (See Figure 2-13 below). The semi-formal level was defined as the type of methods that relied on the legal system for their enforcement despite their lack of registration status. For instance, in the case of secrecy which may be dependant on physical security, it can also be enforeced through a legal definition on confidential information.

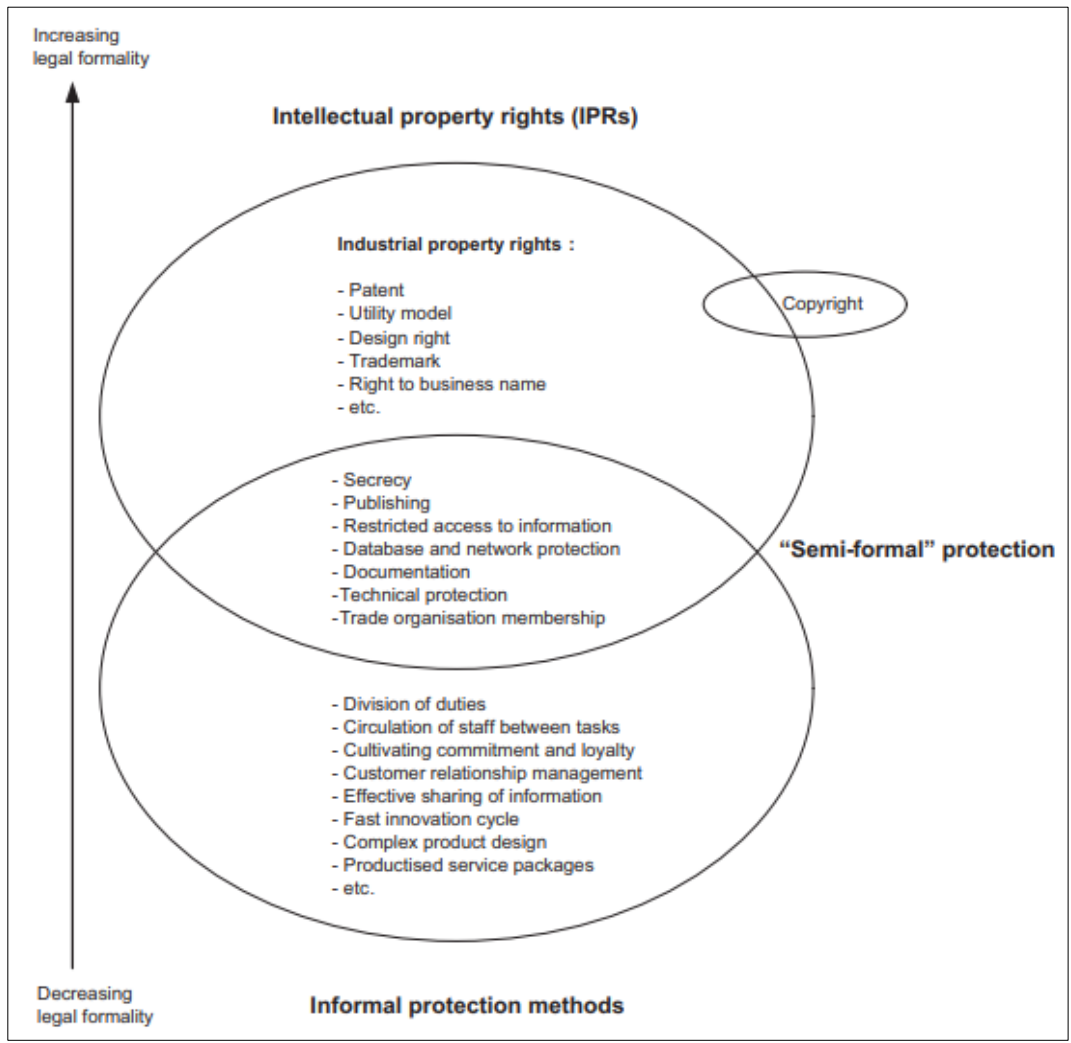

**Figure 2-13. Level of Legal Formality. Adopted from Päällysaho and Kuusisto (2011)**

Previous research (Kitching & Blackburn, 1998; M. Lee et al., 2018; Päällysaho & Kuusisto, 2011) has highlighted the importance of legal mechanisms to protect organizational knowledge such as intellectual property rights (e.g., patents, copyright, trademarks, trade secrets), clauses (e.g., non disclosure agreements -NDA, Non compete clauses - NCC. These previous studies found a relationship between industry type and level of formality whereby more innovative, and knowledge intensive organizations were more likely to have more informal and creative ways to protect their knowledge despite of also having in place formal mechanisms. For example, Päällysaho and Kuusisto (2011) found that business and management consultancy firms favored informal strategies, while software firms were the most active in using IPRs (See Figure 2-14 below). However, these aforementioned studies' findings showed no association between degree of strategy formality and other firm characteristics such as size, and risk profile. However, in contrast, a more recent study (Link & van Hasselt, 2022) showed a relationship between degree of formality and organizational size, that is, the bigger the organization' size, the more formal the knowledge protection strategies. Although organizational risk, determined by the level of exposure to competitor's absorption, was part of the study, no significant relationship was determined between risk and degree of strategy formality. Therefore, studying the relationship between organizational risk and degree of protection formality remains a future avenue for research.

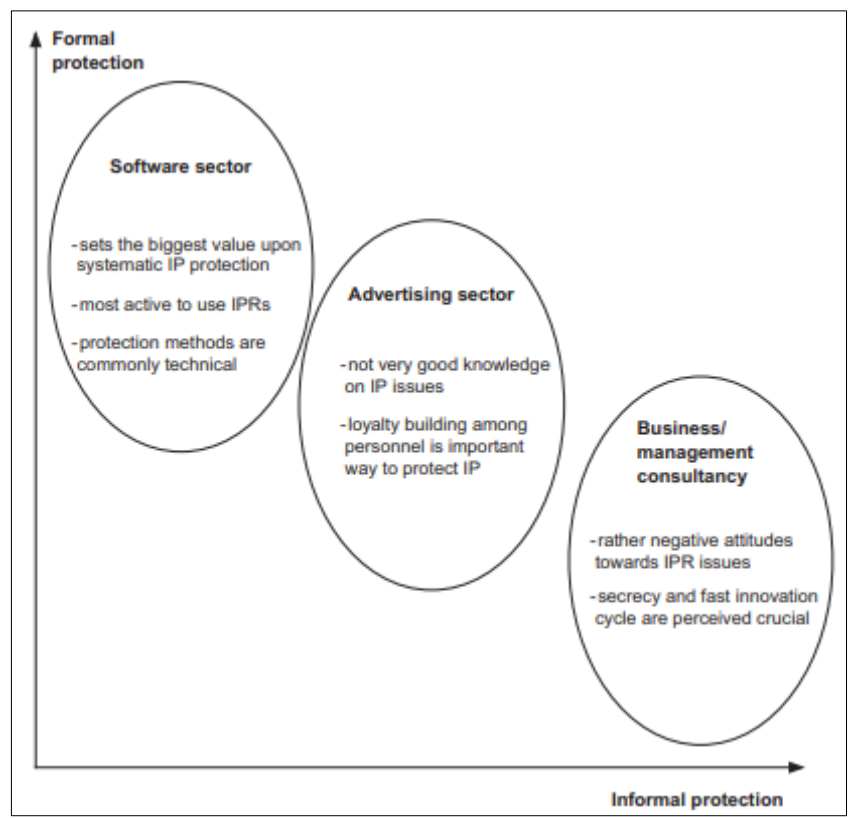

**Figure 2-14. IP Protection in three different knowledge intensive industries. Adopted from Päällysaho and Kuusisto (2011)**

### 2.7.4 Summary of Identified Gaps in Current Security Strategies

Finally, the researcher identified several gaps in the analysed information and knowledge management literature. First, lack of in-depth studies that provide specific strategies for knowledge intensive organizations and different type of industries based on different risk levels. Most of the analysed studies present high-level strategies that are generalised and fail to identify specific strategies for different organizations and risk levels. Second, mobile devices and mobile knowledge management are widely neglected, particularly for the protection of tacit knowledge. And third, the tacit dimension of knowledge remains largely understudied as the majority of the literature focus on formal and informal strategies to protect explicit knowledge. In the methodology chapter — chapter four — the writer proposes a research agenda to address these gaps.

## 2.8 Summary

This chapter provides the results of a comprehensive literature review across multiple domains of information systems, including knowledge, knowledge management, the knowledge-based view (KBV) theory of the firm, mobile computing, and knowledge protection and information security management. The literature review was conducted following the methodology proposed by Webster and Watson (2002), and the results were presented in this chapter, organized by guiding frameworks and concept matrices. The KBV theory of the firm was used as a conceptual integration point and theoretical lens for synthesizing the literature and understanding knowledge as an asset, a process, and its flow within and outside the organization. The chapter also explored topics related to knowledge leakage, such as leakage risk and vectors. In addition, mobile computing concepts and contexts were introduced to provide insight into the dynamic nature of knowledge in the context of mobility and mobile technology. The chapter also illustrated some of the current strategies organizations employ to better manage and protect their knowledge assets and resources. The subsequent chapter will describe the research model used in this study in more detail.

# Chapter 3. RESEARCH MODEL

The previous chapter presented the literature review and the different thematic areas explored during the research. This chapter[5] illustrates the methodology and steps taken to develop the research model based on the literature review and the synthesis of salient topics resulting from the review and analysis of different streams of literature such as information security management, knowledge management, and mobile/mobility literature.

## 3.1 Research Model Development Methodology

Following a methodology extensively used in information systems (IS) research to develop qualitative research models and theory in IS, the current study followed one of the five approaches described in the theory taxonomy in IS. The five categories – I)Analysis, II)Explanation, III)Prediction, IV)Explanation and Prediction, and V)Design and Action – illustrate the nature of research, theory building and model development in the IS discipline (Gregor & Jones, 2007) as shown in Table 3-1.

---

[5] Sections of this chapter have been published in the following publications:

- Agudelo, C. A., Bosua, R., Ahmad, A., & Maynard, S. B. (2016). Understanding knowledge leakage & BYOD (Bring Your Own Device): A mobile worker perspective. arXiv preprint arXiv:1606.01450.
- Agudelo-Serna, C. A., Bosua, R., Ahmad, A., & Maynard, S. (2017). Strategies to Mitigate Knowledge Leakage Risk caused by the use of mobile devices: A Preliminary Study.
- Agudelo-Serna, C. A., Bosua, R., Ahmad, A., & Maynard, S. B. (2018). Towards a knowledge leakage mitigation framework for mobile devices in knowledge-intensive organizations.

**Table 3-1. Theory Taxonomy in IS adopted from Gregor (2006)**

| Theory Taxonomy in IS | |
|---|---|
| **Type** | **Attributes** |
| I.   Analysis | Says what is without extending beyond analysis and description. No causal relationships among phenomena are specified and no predictions are made. |
| II.   Explanation | Says what is, how, why, when, and where. It provides explanations but without aiming to predict with any precision. There are no testable propositions. |
| III.   Prediction | Says what is and what will be. It provides predictions and has testable propositions but does not have well-developed justificatory causal explanations. |
| IV.   Explanation and Prediction | Says what is, how, why, when, where, and what will be. Provides predictions and has both testable propositions and causal explanations. |
| V.   Design and Action | Says how to do something. It gives explicit prescriptions (e.g., methods, techniques, principles of form and function) for constructing an artifact. |

This study approached the phenomenon of interest, through the *explanation* category (the second category in Table 3-1), part of the Theory Taxonomy in IS, in conjunction with the Knowledge based View of the firm (KBV) theory to develop and propose a research model that displays the main determinants in the Knowledge Leakage Risk (KLR) through mobile devices phenomenon.

Based on the Theory Taxonomy, this research placed in context the area of interest of KLR and considered the domain questions of what the core problems are, topics of interest, and what the boundaries of the phenomenon are.

In accordance with the theory taxonomy in IS, the explanation category describes *what is, how, when, and where, providing explanations but without aiming to predict with any precision. There are no testable propositions* (Gregor, 2006)

Table 3-2 provides an explanation of the different components of a theory and a model to allow IS researchers to identify the different parts of the proposed model.

**Table 3-2. Structural Components based on the Theory Taxonomy in IS adopted from Gregor (2006)**

| Table Model Structural Components based on the Theory Taxonomy in IS (Gregor, 2006) | |
|---|---|
| **Common Components** | **Definition** |
| Means of representation | Representation in some way: in words, mathematical terms, symbolic logic, diagrams, tables or graphically. Additional aids for representation could include pictures, models, or prototype systems. |
| Constructs | These refer to the phenomena of interest. Many different types of constructs are possible: for example, observational (real) terms, theoretical (nominal) terms and collective terms. |
| Statements of Relationship | These show relationships among the constructs. Again, these may be of many types: associative, compositional, unidirectional, bidirectional, conditional, or causal. Very simple relationships can be specified: for example, "x is a member of class A." |
| Scope | The scope is specified by the degree of generality of the statements of relationships (signified by modal qualifiers such as "some," "many," "all," and "never") and statements of boundaries showing the limits of generalizations. |
| **Contingent on Purpose** | **Definition** |
| Causable Explanations | Statements of relationships among phenomena that show causal reasoning (not covering law or probabilistic reasoning alone). |
| Testable propositions | Statements of relationships between constructs are stated in such a form that they can be tested empirically. |
| Prescriptive Statements | Statements in the theory specify how people can accomplish something in practice (e.g., construct an artifact or develop a strategy). |

The *means of representation* for the research model will use diagrams to depict the interaction of the different factors that determine the KLR through mobile devices phenomenon of interest. Similarly, the *constructs* used within the model rely on the main concepts and themes analysed during the literature review framed in the setting of mobility and mobile contexts. In a similar vein, the *statements of relationship* represent an association type of connection without an indication of causality, only correlation therefore propositions within the model are by no means, causable, testable or prescriptive in nature (Gregor & Jones, 2007).

Once the model conditions have been defined as well as the theory through which the researcher analysed the KLR phenomenon in the context of mobility, the next steps in developing the model include highlighting the salient topics of the literature review in the next section.

## 3.2 Mobility and Mobile Contexts approach to the Research Model Development

As discussed in the previous chapter - section 2.6, this study took a contextual approach to the KLR phenomenon, in particular considering the mobility and mobile properties, defined in the literature as *mobile contexts* and summarized in Table 3-3. IS researchers argue that modelling provide the tools to understand and analyse a problem due to the fact that this process identifies critical areas and help to represent systems from the real world into simplified versions that afford manipulation and testing (Gregor, 2006; Gregor & Hevner, 2013; Hevner et al., 2004). Modelling also enables further analysis of the phenomenon investigated and helps to stimulate theorising about the relationship of factors within the modelled system, and to enable predictions for future scenarios to be formulated (Neuman, 2006). In order to understand the knowledge leakage phenomenon, a conceptual research model is proposed to help to explain and to analyse the factors involved in the aforementioned research phenomenon and inform the research agenda. The development of the model was conducted in three stages in which the researcher:

1. Identified the mobile contexts, constructs and grouped them together based on their definition and the Integrative model of IT business value framework.

2. Indicated the relationship between the mobile contexts and the *knowledge leakage risk through mobile devices* construct.

3. Highlighted the relationship between *knowledge leakage risk through mobile devices* and *organizational knowledge mitigation capabilities* construct.

**Table 3-3. Mobile Computing Contexts Taxonomy from Literature Review**

| Context Construct | Reference | Description |
|---|---|---|
| Environmental | (Agbo & Oyelere, 2019; Kofod-Petersen & Cassens, 2006; Luo et al., 2020; Nieto et al., 2006) | The environmental context is defined as the conjunction of the following contexts: temporal context, spatial context, social context, technological context, and business context |
| Personal | (Aliannejadi et al., 2021; Kofod-Petersen & Cassens, 2006) | The personal context provides the attributes of cognitive skills and draws on psychological and physiological contexts: psychological, goal, cognition, physiological, identity, actions |
| Social | (Jacob et al., 2020; Nieto et al., 2006; Xu, 2021) | Provides a social perspective of context, which captures the attributes of people (e.g. attitude, skills, and values) and the relationship of these people among one another and within the organization and collective structures (collective values and norms). |
| Technological | (Abdoul Aziz Diallo, 2012; Al Sharoufi, 2021; Jacob et al., 2020; Tamminen et al., 2004) | Provides the technological and technical attributes such as: network connections, infrastructure, equipment, devices and systems. It is an aggregate context which consists of other technical constituents such as spatial, user and location context. |
| Organizational | (Crossler et al., 2013; Furnell & Rajendran, 2012; Jacob et al., 2020; Whitman, Michael and Mattord, 2011) | Defines the social interactions within the workplace and security behaviour determined by Information Security Policies, Security Education Training and Awareness, Culture, Standards, organizational processes and procedures |
| Device | (Al Sharoufi, 2021; Belko Abdoul Aziz Diallo et al., 2014; Jacob et al., 2020; Kofod-Petersen & Cassens, 2006; Nieto et al., 2006) | Technological features such as device identifier, device type and processing capabilities that pertains exclusively to the communication device (i.e., laptop, tablet, smartphone) |

For the first stage of the development of the model, the researcher started with the definitions of the mobile context taxonomy on the Table 3-3 and expanding on the *social context interaction framework* by Chen & Nath (2008) and Bradley and Dunlop's (2005) *model of context*, the author grouped the related mobile contexts into factors employing *the Integrative model of IT business value framework* developed by Melville et al (2004) in order to integrate and create group level constructs that are composed of individual but related contexts addressed in a similar way in the literature. This is an approach commonly used by IS researchers to combine and compose higher-level constructs that facilitate the analysis and simplify the relationship of multilevel variables and modelling(Preacher et al., 2010; Snijders & Bosker, 2011; Van Mierlo et al., 2009). In the case of mobile contexts, for example, the personal and social contexts, both refer to the human-centric aspect of mobile computing and interaction. Similarly, the device and technological contexts refer to the technical aspects of mobility that are related to the network, infrastructure, device type and processing capabilities. Comparatively, environmental and organizational contexts both relate to the focal firm's perspective both from the internal view — e.g., organizational processes, policies, culture — and from the external perspective — e.g., competitive environment, industry characteristics, regulations, political frameworks and macro environment like country and regional location.

As previously defined in the literature review chapter, these frameworks provide an organization-level perspective that explains how the three different elements of a firm's context influence adoption decisions and bring value to the firm. These three elements are the technological factors, the enterprise factors, and the human factors. All three are posited to influence technological adoption and innovation (Baker, 2012) .Table 3-4 lists the mapping between the factors and mobile contexts analysed in the literature.

**Table 3-4. Factors and Mobile Contexts mapping based on the literature**

| Factor | Definition | Mobile Context | Authors |
|---|---|---|---|
| Human | Refers to physical and cognitive properties of individuals and social behaviour that influence interaction amongst individuals and element of a system | Personal Context<br><br>Social Context | (Ajzen et al., 1991; Aliannejadi et al., 2021; Bandura, 1978; Johnston & Warkentin, 2010; Olander et al., 2014; Siponen, 2000; Whitman, Michael and Mattord, 2011; Xu, 2021) |
| Enterprise | Refers to internal and external entities, processes and situations that can influence a focal firm such as organizational structure, culture, internal capabilities, external market environment, competitors, industry, and regulations | Environmental Context<br><br>Organizational Context | (Agbo & Oyelere, 2019; Arias-Pérez et al., 2020; Leonard-barton, 1992; Liebeskind, 1996; Luo et al., 2020; Melville et al., 2004; Penrose, 1960; David J. Teece, 2007) |
| Technological | Refers to information and communication resources, tools and technologies used to improve a firm's capabilities and competitiveness. | Device Context<br><br>Technological Context | (Abdoul Aziz Diallo, 2012; Al Sharoufi, 2021; L. Chen & Nath, 2008; K. C. Desouza, 2011; K. C. Desouza & Vanapalli, 2005; Jarrahi & Thomson, 2017; Melville et al., 2004; Nelson et al., 2017) |

In the context of KLR through mobile devices, the framework and grouping of the mobile contexts provides the lens through which the mobile constructs were grouped. As the reader can see in the Figure 3-1, the high-level construct *mobile device usage contexts* — which is the first part of the conceptual research model — is comprised of the aforementioned factors, which in turn contain the respective contexts: 1) Personal context and 2) Social context are grouped together under Human factors which refer to motivations and cognitive processes, as well as social norms that are explicit and implicit from human behaviours and social interaction. 3) Environmental context and 4) Organizational context constitute the Enterprise factors and refer to the organizational culture and behaviour, operating environment (regulations) not only within the workplace but also outside (macro environment). Finally, the Technological factors are composed of 5) Device context and 6) Technological context and refer to the technology and information systems that enable and facilitate the adoption of technology and technical artefacts to perform knowledge-sharing activities such as the mobile devices.

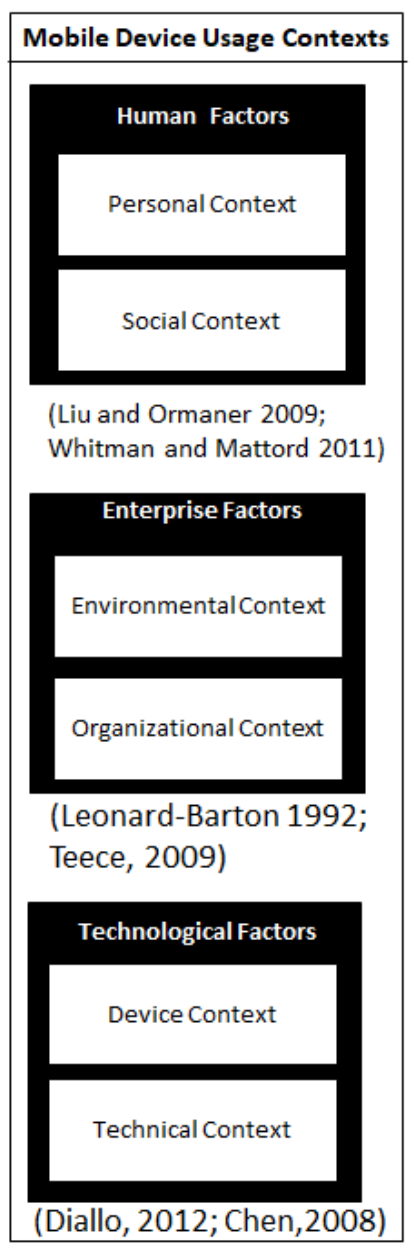

**Figure 3-1. Mobile contexts grouped using Chen and Nath's (2008) Social Context Interaction Framework and Bradley and Dunlop's (2005) Model of Context**

## 3.3 Knowledge based View Theory approach to the Research Model Development – Knowledge as an object and capability

For the second stage in the development of the conceptual model, the researcher used the knowledge based view of the firm, considering the point of view of knowledge as an object or as a process(Carlsson, S.A., El Sawy, O.A., Eriksson, I., and Raven, 1996; Carlsson, 2003; Kengatharan, 2019; Kianto et al., 2020; Melville et al., 2004; Parker, 2012; Tavares, 2020; Wernerfelt, 1984; Zack, 1999) which suggests that the concept of knowledge can be viewed in two ways, either as an object that is stored and manipulated, or as a process of knowing and acting with expertise. Organizational knowledge is therefore a process of developing and organizing knowledge to facilitate access and retrieval of content. This view is an extension of the view of knowledge as an object, with an emphasis on the accessibility of the knowledge objects. The goal of this perspective is to create an environment where knowledge is readily available and easily utilized to improve organizational performance (Alavi et al., 2005; Alavi & Leidner, 2001; Styhre, 2003).

Carlsson et al. (1998) proposed an additional perspective on knowledge, defining it as a capability that has the potential to influence future actions. These different views of knowledge lead to varied understandings of knowledge management. Regarding the view of knowledge as an object or information access, the focus of knowledge management lies in building and managing knowledge stocks. Alternatively, the view of knowledge as a process highlights the importance of knowledge flow and processes of knowledge creation, sharing, and distribution. In the case of the view of knowledge as a capability, knowledge management emphasizes building core competencies and understanding the strategic advantage of know-how, and creation of intellectual capital that must be protected in order to ensure competitive advantage and prevent knowledge erosion and loss that could result in the firm losing its position to competitors.

This perspective emphasizes the role of knowledge in shaping an organization's future and emphasizes its importance as a key resource that can provide sustainable competitive advantage. This consideration of knowledge as an object and/or capability bears importance for the following section and next step in the definition of the conceptual model.

## 3.4 Risk Perspective on Knowledge Leakage through Mobile Devices

In the previous section, the concept of knowledge as an object and a capability was introduced and as such knowledge may be subject to risks similar to other organizational assets such as loss, duplication spill over and leakage which warrants protection (Bloodgood & Chen, 2021; Parker, 2012; Ritala et al., 2015). As discussed in the previous chapter, a firm's competitive advantage lies in its ability to prevent such loss of knowledge across the organization's boundaries (Brown & Duguid, 2001; K. C. Desouza, 2006; Sveen et al., 2009). Knowledge leakage occurs when the ideas develop in the originating organization – where it is created and used — are leaked to its competitors (Annansingh, 2006, 2012; Nunes et al., 2006).

As previously discussed, knowledge can be tacit or explicit. Tacit knowledge, being uncodifiable and embedded in individuals, presents significant challenges in terms of transferability, while explicit knowledge, being easily codified and communicated, can be disseminated relatively effortlessly beyond the organizational boundaries (Alavi & Leidner, 2001; Ancori et al., 2000; Bouncken & Barwinski, 2021; K. C. Desouza, 2003b; Maravilhas & Martins, 2019; Polanyi & Sen, 1997). From the models of knowledge management, one of the key objectives of knowledge management is to promote the sharing, transfer, and dissemination of explicit knowledge. Nevertheless, the unstructured and ad hoc nature of knowledge sharing processes can lead to unintended knowledge leakage, which poses significant risks to the organization (Annansingh, 2012; Khatib, 2021; Trkman et al., 2012).

Organizations are increasingly reliant upon the use of mobile devices for inter-organizational communication and knowledge work (Jacob et al., 2020; Nelson et al., 2017; Rieger & Majchrzak, 2017). Moreover, most organizations employ mobile devices to facilitate off-site work which exacerbates the risk even further as these devices are usually outside the organizational perimeter, particularly in working off-site settings (Allam et al., 2014; Jarrahi & Thomson, 2017; Ortbach et al., 2015; Pratama & Scarlatos, 2020) where the security measures lack organizational rigour, and the aforementioned devices are outside the control of the centralized office, particularly in the current circumstances, at the time of writing, caused by the current COVID-19 pandemic where remote and mobile workers have become easy targets as well as their devices as highlighted by a number authors (Aliannejadi et al., 2021; Bloodgood & Chen, 2021; Lallie et al., 2021; Morrell, 2020; Weichbroth & Łysik, 2020; Xu, 2021).

The information and codified knowledge within these mobile devices, not just electronic files, but also images, voice, video, text and the knowledge elicited from the interaction of the worker with the device could be readily exposed and then used to infer or draw other knowledge insights from existing information or by putting pieces together (OSINT, Mosaic theory ,(Abdul Molok et al., 2010b)) resulting in knowledge leakage whether deliberately or unintentionally (Atwood, 2019; Dalton et al., 2020; Melnitzky, 2011; Promnick, 2017; Thorleuchter & Van Den Poel, 2013). This situation highlights the risk that organizations face, from the definition in the previous chapter, the author established that KLR is defined in the literature in terms of probability and consequence:

$$KLR\ rating = Probability\ Ocurrence\ x\ Consequence$$

However, this definition simplifies and fails to take into consideration other more complex and contextual factors that affect either the probability or consequence of the resulting risk, particularly, in the case of knowledge , where risk becomes a multi dimensional and multi disciplinary concept in which risk management and knowledge management fields are combined into one entity (Jiang et al., 2016; Khatib, 2021; Li et al., 2019; Vafaei-Zadeh et al., 2019; Wu et al., 2021). The mobile device usage contexts and its constituent factors in the mobile contexts will affect and determine a different level of KLR exposure of an organization; for example, human factors (employee behaviours and experience) could warrant a different risk when considering attitude, knowledge, background and culture of workers (Jarrahi & Thomson, 2017; Nelson et al., 2017; Vishwanath, 2016). Similarly, enterprise factors (such as type of industry, location and country) may influence the risk rating of leakage, for instance in military organizations versus small start-ups (Bosua & Scheepers, 2007; Grimaldi et al., 2021; Marsh & Stock, 2006; Van Wijk et al., 2008). Even more, technological factors (such as network, device configuration and technical controls) could have a significant effect on determining high-risk devices, for example devices with encryption and multi-factor authentication enabled are less likely to be compromised compared to unmanaged and unpatched devices without such controls (Armando et al., 2014; Dang-Pham & Pittayachawan, 2015; Dtex, 2022; Franklin et al., 2020; Toelle, 2021). This relationship is illustrated in Figure 3-2. The last step in the model development process is explained next.

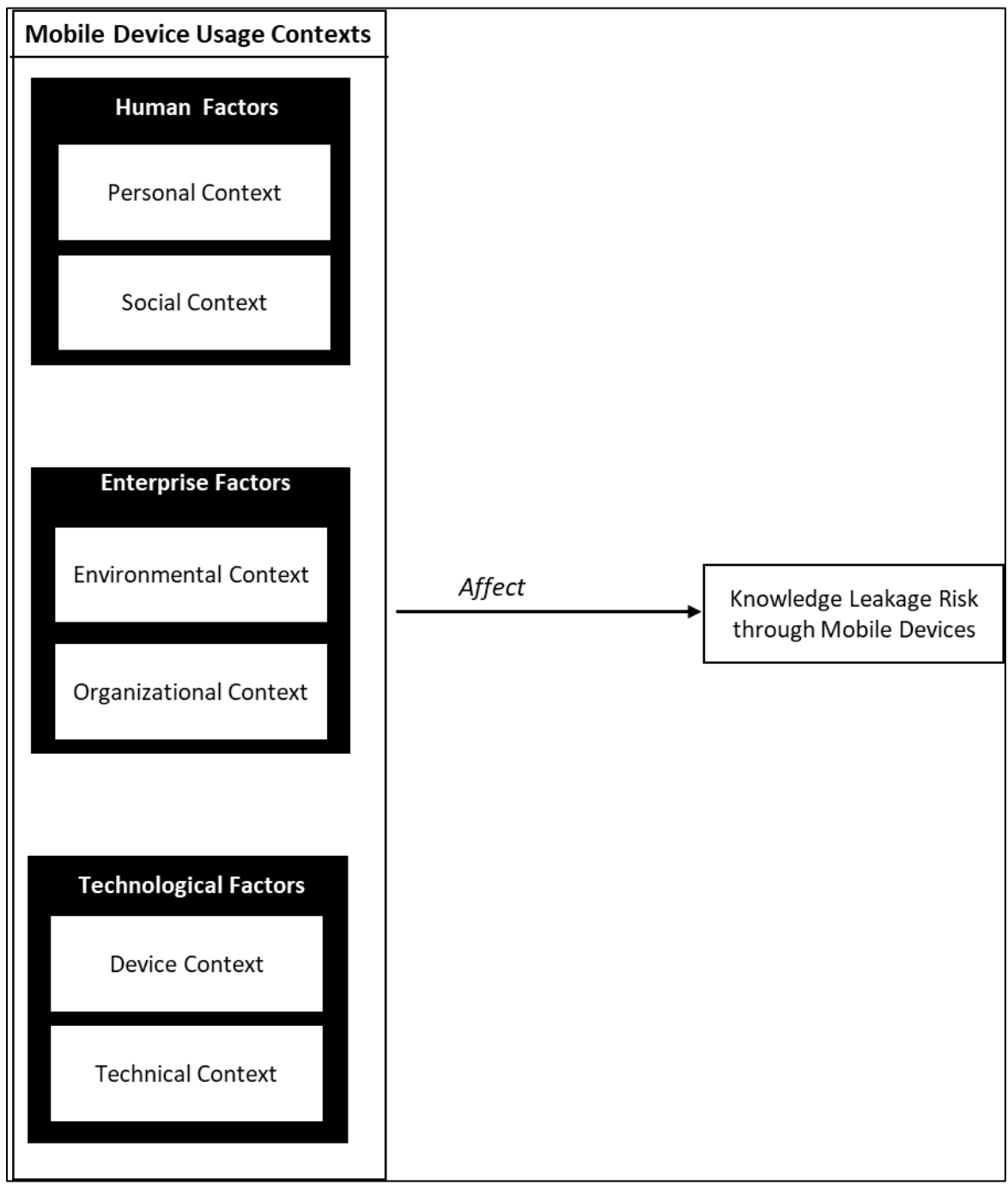

**Figure 3-2. Mobile Device Usage Contexts and Knowledge Leakage Risk relationship**

By understanding the different factors and consequently the different associated risks, organizations can develop and implement different levels of controls, prioritization and mitigation capabilities to protect knowledge in each instance. The literature review chapter presented different mitigation strategies for the previously discussed hypothetical scenarios. As a way to illustrate this, let us refer to the aforementioned human factor related risks in the previous example, in order to address the employee's lack of knowledge and haphazard attitude towards security, one of the strategies suggested by authors such as Ritala et al (2015) and Desouza and Vanapalli (2005) allude to indoctrination and training. To mitigate enterprise factor associated risks such as industry type where codified knowledge/information may not be properly managed, strategies such as compartmentalization, as illustrated by Ahmad, Maynard, et al. (2014), may be used to reduce the risk of leakage. And to combat technological factor linked risks such as lack of security Dhillon (2007) proposes securing devices and communication channels.

Therefore, for the third and final stage of the model development, the resulting risk profile dependent on the mobile device usage contexts and its aforementioned factors will inform the appropriate knowledge mitigation capabilities that organizations should develop in order to combat either the risk probability or the consequence of knowledge leakage. This relationship is illustrated in the Figure 3-3 Knowledge Leakage Risk and Organizational Mitigation strategies capabilities relationship.

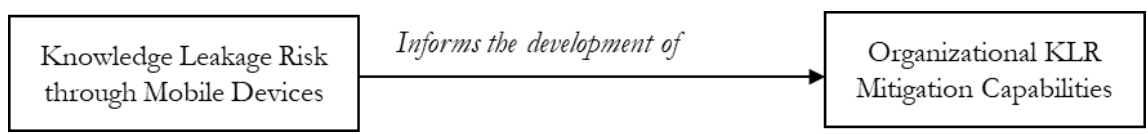

**Figure 3-3 Knowledge Leakage Risk and Organizational Mitigation strategies capabilities relationship**

Finally, the complete model is presented in Figure 3-4 integrating the different stages and indicating the relationships amongst constructs. This conceptual model provides a framework to understand how human, internal (organizational resources) and external (trading partners, competitive and macro environment), and technological factors impact the leakage risk caused mobile devices and how this risk, in turn, informs the organizational performance via mitigation capabilities which contributes to improvement of organizational information and knowledge security performance signified in the construct organizational KLR mitigation strategies.

 A summary of the constructs used in the research model is presented in Table 3-5 below

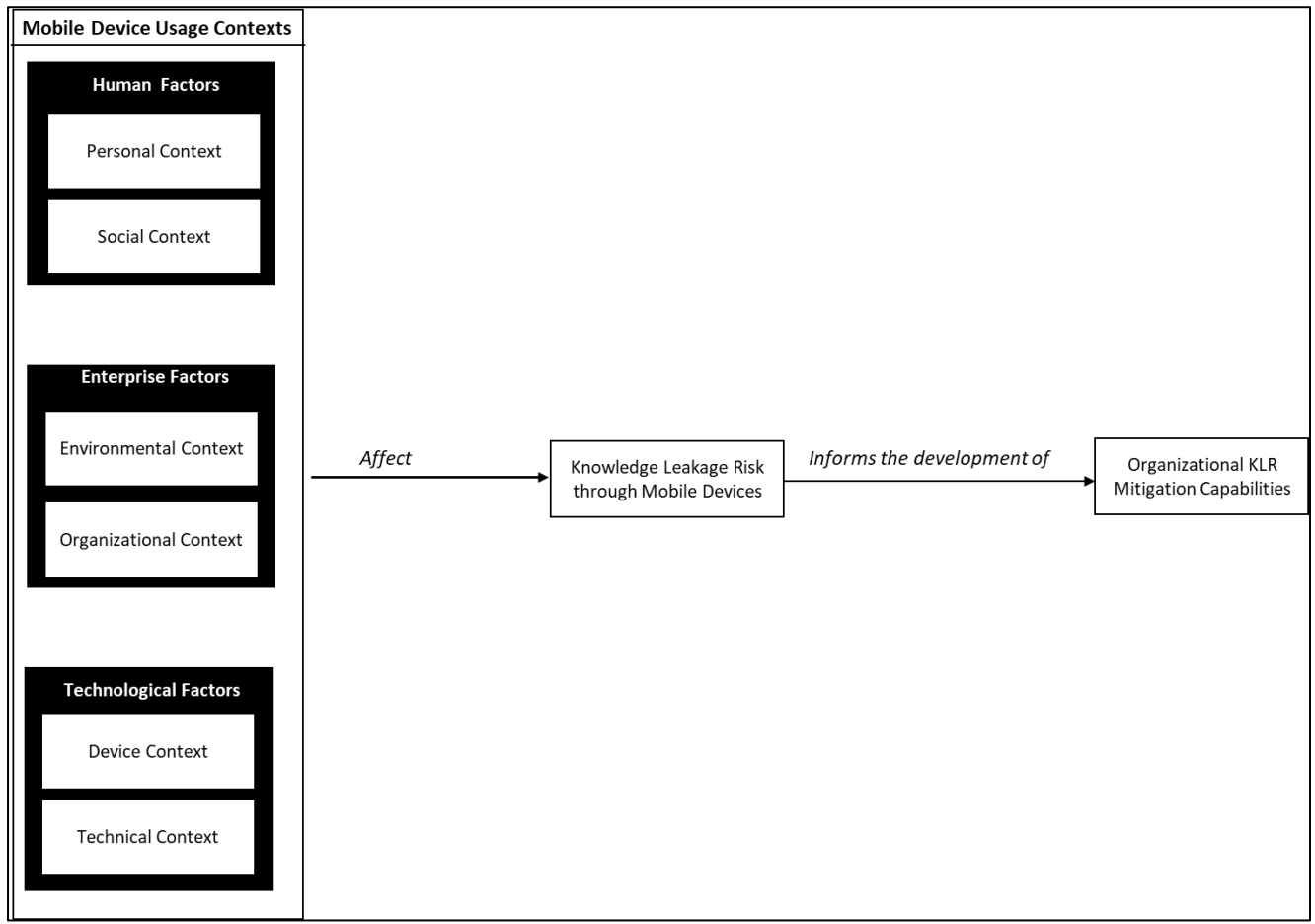

**Figure 3-4. Conceptual research model**

The next section elaborates on the propositions of the model.

**Table 3-5. Definition of Constructs in the Research Model**

| Construct | Definition | Reference |
|---|---|---|
| Organizational KLR Mitigation Strategies | Formal, informal and technical risk control strategies, used by organizations to safeguard knowledge assets at risk. Such strategies aim to reduce risk impact or probability (risk reduction), as well as share, avoid, transfer or accept any residual risk remaining after the risk treatment. | (27005:2011, 2011; Ahmad, Bosua, et al., 2014; Amara et al., 2008; Amoroso & Link, 2021; Arias-Pérez et al., 2020; Bouncken & Barwinski, 2021; Dhillon, 2007; Grimaldi et al., 2021; Link & van Hasselt, 2022; Mupepi, 2017) |
| Knowledge leakage Risk through Mobile Devices | Knowledge leakage risk caused by the use of mobile devices in organizations. This high-level construct is derived from the standard definition of risk and it is often operationalized using a qualitative scale, i.e., low, medium and high. However, it's important to note that this qualitative risk scale and its levels and impacts are highly dependent on the specific context of each organization, making it challenging to standardize or assign universal weightings. | (27005:2011, 2011; Ahmad et al., 2015; Ahmad, Bosua, et al., 2014; Bouncken & Barwinski, 2021; K. C. Desouza & Vanapalli, 2005; Mupepi, 2017; Scarfo, 2012; Shedden et al., 2010, 2011; Vafaei-Zadeh et al., 2019; Wu et al., 2021) |
| Human Factors | The combination of personal and social contexts referring to individual's behaviour, attitude, cognitive capabilities, motivations, experiences (personal context) as well as group's culture and values, social norms, peer's influence and superior's influence (social context). | (Ajzen et al., 1991; Bandura, 1978; Johnston & Warkentin, 2010; Olander et al., 2014; Siponen, 2000; Whitman, Michael and Mattord, 2011) (Aliannejadi et al., 2021; Kofod-Petersen & Cassens, 2006) (Jacob et al., 2020; Nieto et al., 2006; Xu, 2021) |
| Enterprise Factors | The combination of environmental and organizational contexts referring to external conditions (e.g., competitors, industry, external locations) as well as internal organizational resources and capabilities (e.g., policies, culture, processes, routines). | (Leonard-barton, 1992; Liebeskind, 1996; Melville et al., 2004; Penrose, 1960; David J. Teece, 2007) (Crossler et al., 2013; Furnell & Rajendran, 2012; Jacob et al., 2020; |

| Construct | Definition | Reference |
|---|---|---|
| | | Whitman, Michael and Mattord, 2011) (Agbo & Oyelere, 2019; Kofod-Petersen & Cassens, 2006; Luo et al., 2020; Nieto et al., 2006) |
| Technological Factors | The combination of device and technological contexts referring to the infrastructure and technological resources internal and external to the organization that enable and support knowledge-sharing activities. | (Abdoul Aziz Diallo, 2012; Al Sharoufi, 2021; Jacob et al., 2020; Tamminen et al., 2004) (Al Sharoufi, 2021; Belko Abdoul Aziz Diallo et al., 2014; Jacob et al., 2020; Kofod-Petersen & Cassens, 2006; Nieto et al., 2006) (Abdoul Aziz Diallo, 2012; L. Chen & Nath, 2008; K. C. Desouza, 2011; K. C. Desouza & Vanapalli, 2005; Melville et al., 2004; Nelson et al., 2017) |

## 3.5 Propositions and constructs within the model

A proposition is a tentative and conjectural relationship between constructs that is stated in a declarative form. Propositions are qualitative statements made to explore assumptions about social phenomena used in qualitative research and inductive studies, unlike hypothesis that are quantitative and deductive in nature. Propositions help to explore a phenomenon without prior empirical evidence, not to test measurable variables empirically (see Figure 3-5). Furthermore, propositions form the basis for scientific research. The validity of a research study is, to a large extent, evaluated on the criteria of its propositions. Propositions' main purpose are to suggest a link between two concepts in a situation where the link cannot be verified by experiment. Because of this, it relies on reasonable assumptions and existing correlative evidence. (Hevner et al., 2004). In the conceptual research model defined based on the literature review, the researcher postulated two propositions that are by no means measurable or testable, rather indicative of the relationship amongst the constructs.

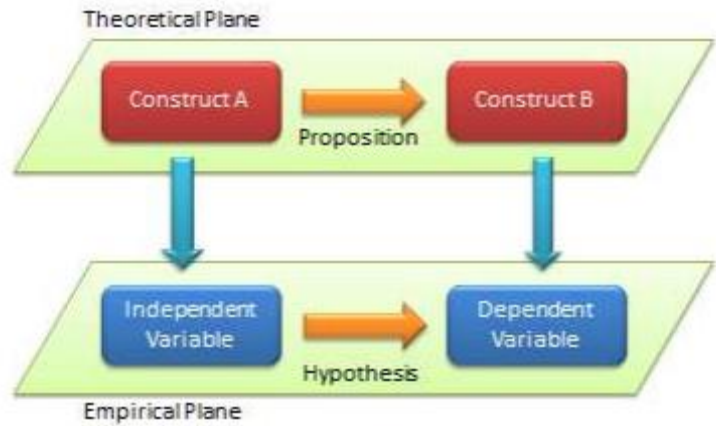

**Figure 3-5. Propositions vs Hypothesis**

### 3.5.1.1 Proposition 1. (P1)

*The knowledge leakage risk through mobile devices informs the development of organizational knowledge leakage risk mitigation capabilities.*

This proposition shows the relationship between risk and organizational response (strategic capability) to mitigate knowledge leakage caused by the use of mobile devices. Previous studies have highlighted the importance of risk assessment for operational performance and capabilities development within organizations (27005:2011, 2011; Ahmad et al., 2015; Ahmad, Bosua, et al., 2014; K. C. Desouza & Vanapalli, 2005; Nelson et al., 2017; Povarnitsyna, 2020; Scarfo, 2012; Shedden et al., 2010, 2011)

As discussed previously and expanding on KBV and the contextual frameworks by Chen and Nath (2008), previous studies have evaluated a considerable number of organizational characteristics as determinants of competitive advantage, which in turn have been classified within the broader category of basic competences or influencing factors, but critically knowledge represented as a capability remains the most important organizational asset to protect and develop (L. Chen & Nath, 2008; Grant, 1996b; Leonard-barton, 1992; J.-C. Spender & Grant, 1996a; D J Teece et al., 1991). The outcome of the risk assessment process allows the organization to understand the gaps to address, key competencies to develop and assets to protect that translates into the combination of internal competencies to form the required organizational capabilities to improve their competitive advantage leading to better security performance and mitigation capabilities (Kaplan et al., 2001; Melville et al., 2004; D J Teece et al., 1991; David J. Teece, 2007; Zaini et al., 2018)

### 3.5.1.2 Proposition 2. (P2)

*The mobile device usage contexts and its constituent human, enterprise and technological factors affect the likelihood and/or consequence of KLR through mobile devices.*

Human, enterprise, and technological factors characterize the distinctive but complementary contexts within which leakage occurs, that in turn, determines the organizational risk exposure. That is to say, these factors and associated contexts (personal, social, organizational, environmental, device and technical) have an impact on the level of the risk probability or repercussion as they provide the dynamic conditions and the changing circumstances that knowledge intensive organizations face when mobile workers conduct their knowledge work outside the organization's boundary. (Ahmad et al., 2015; Ahmad, Bousa, et al., 2014; Bouncken & Barwinski, 2021; Khatib, 2021; Li et al., 2019; Marjanovic, 2013). Figure 3-6 below shows the research model and propositions.

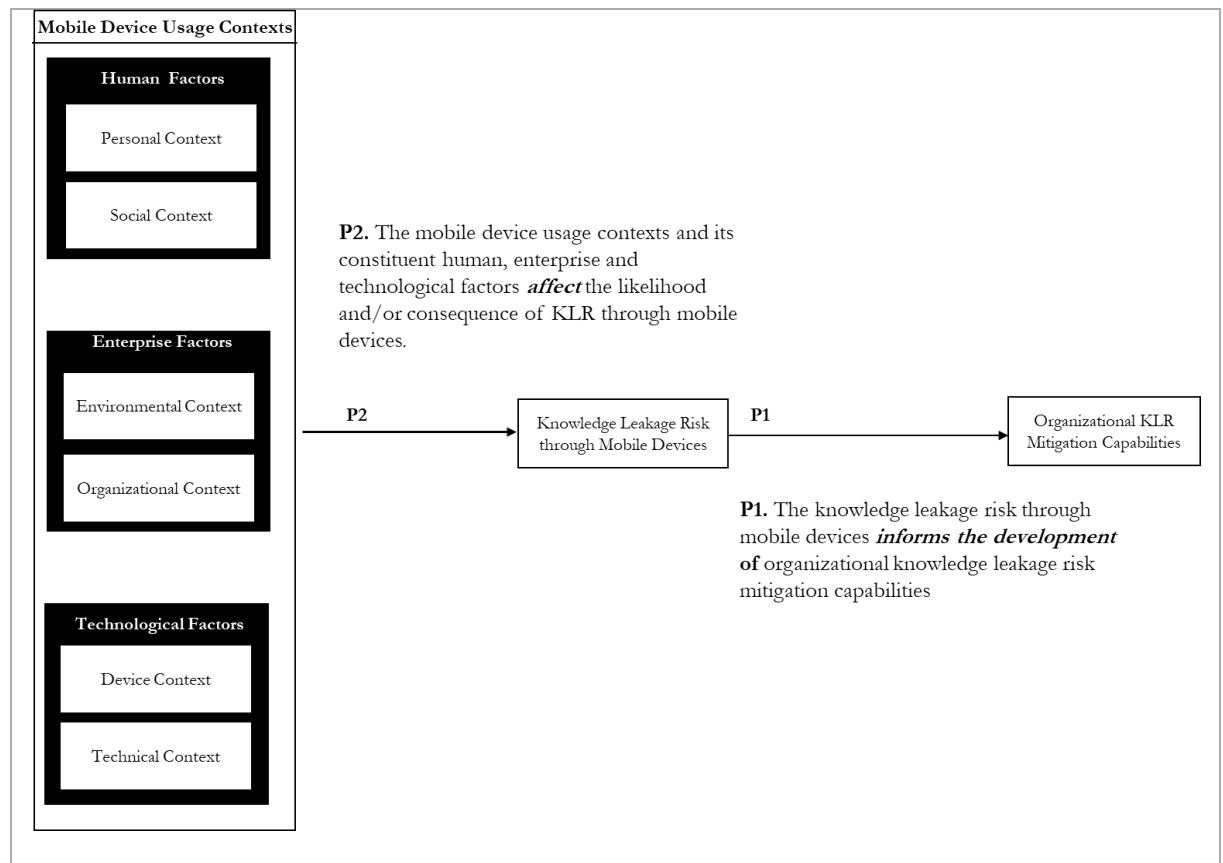

**Figure 3-6. Research Model and Propositions**

## 3.5.2 Human factors

Within the mobile context's literature, human factors refer to the combination of personal and social contexts with regard to individual's behaviour, attitude, cognitive capabilities, motivations, experiences (personal context) as well as group's culture and values, social norms, peer's influence and superior's influence (social context). These factors can play a critical role in the success or failure of mobile device practices and efforts, as they can be both a source of vulnerabilities and a potential solution to them

## 3.5.3 Enterprise Factors

Enterprise factors include environmental and organizational contexts with reference to external conditions (e.g., competitors, industry, external locations) as well as internal organizational resources and capabilities (e.g., policies, culture, processes, routines). Enterprise factors also refer to the various organizational and systemic aspects of an enterprise that can impact the risk and practices in the context of mobility and mobile devices. These factors may include governance, risk management, organizational structure and culture.

### 3.5.4 Technological Factors

Technological factors cover device and technological contexts in relation to the infrastructure and technological resources internal and external to the organization that enable and support knowledge-sharing activities. Technological factors refer to the various technological aspects of an enterprise that can impact the security of mobile devices and systems. These factors include devices, network architecture, security technologies, and controls (e.g., encryption, identity and access management).

## 3.6 Summary

This chapter presented the development of the research model based on the Theory Taxonomy in IS and the knowledge-based view of the firm theory. The conceptual model was developed in three stages and the relationships and constructs were defined. Additionally, the mobile computing concepts and contexts were also presented to understand the dynamic of knowledge in the context of mobility and mobile technology. The next chapter provides a detailed description of the methodology used in this research.

# Chapter 4. RESEARCH METHODOLOGY

The previous chapter presented the research model development, the steps and rationale behind the different salient constructs analysed and examined during the literature review of different streams of information security management, knowledge management, knowledge protection, and mobile/mobility literature. This chapter[6] presents and justifies the research paradigm, the research design, research method, as well as the data collection techniques, and the qualitative data analysis conducted based on the data gathered. Moreover, it also summarizes the steps and measures taken to ensure research rigour as well as data validity and reliability.

## 4.1 Research Paradigm

The research paradigm constitutes the theoretical framework that underpins the researcher's beliefs and perspectives and shapes their understanding of the world. It determines the researcher's approach to the research process, including the methods and techniques employed, and influences the inferences and conclusions that are

---

[6] Sections of this chapter have been published in the following publications:

- Agudelo, C. A., Bosua, R., Ahmad, A., & Maynard, S. B. (2016). Understanding knowledge leakage & BYOD (Bring Your Own Device): A mobile worker perspective. arXiv preprint arXiv:1606.01450.
- Agudelo-Serna, C. A., Bosua, R., Ahmad, A., & Maynard, S. (2017). Strategies to Mitigate Knowledge Leakage Risk caused by the use of mobile devices: A Preliminary Study.
- Agudelo-Serna, C. A., Bosua, R., Ahmad, A., & Maynard, S. B. (2018). Towards a knowledge leakage mitigation framework for mobile devices in knowledge-intensive organizations.

drawn from the findings.(Guba & Lincoln, 1994; Hevner et al., 2004; Wynn & Williams, 2012).

Information systems research employs three main philosophical paradigms (as illustrated in Table 4-1): positivist, interpretivist, and critical realism (Myers & Newman, 2007; Neuman, 2006; Orlikowski & Baroudi, 1991). However, information system researchers utilize positivist and interpretivist paradigms more commonly as compared to critical realism in IS research (Neuman, 2006; Orlikowski & Baroudi, 1991). Each of the three paradigms represents a different world view and is associated with contrasting ontological, epistemological, and methodological assumptions which are summarised in Table 4-1. Although these research paradigms are distinct, there is no such thing as a *better* or *stronger* paradigm, as there is no right or wrong way of interpreting the world around a particular individual (Guba, 1990; Guba & Lincoln, 1994; Shanks, 2002).

**Table 4-1. Research Paradigms. Adapted from Guba and Lincoln (1994) and Orlikowski and Baroudi (1991)**

|  | Positivist | Interpretivist | Critical Realism |
|---|---|---|---|
| Ontology<br><br>*What is real?* | There is a single reality or truth (Realist) | Person(researcher) and reality are inseparable (life-world).<br><br>Reality exists based on the individual (Relativism) | Reality is shaped by social, political, cultural, economic and gender values (Historical realism). Reality exists; contingent truth claims. |
| Epistemology<br><br>*What is true?* | Scientific knowledge is truth, reality is apprehensible. | Knowledge of the world is intentionally constituted through a person's lived experience | Findings are based on values, local examples of truth. It is possible to know context-sensitive reality by combining empirical observations and interpretations. |
| Methodology<br><br>*How to examine what is real?* | Mostly quantitative methods (Statistics, content analysis) | Interpretation. Mostly qualitative Methods. (Hermeneutics, phenomenology, etc) Purposive and multipurpose sampling. | Quantitative + qualitative methods |
| Validity | Certainty: data truly measure reality | Defensible knowledge claims | Claims supported by data. Reality is never fully apprehended. |
| Reliability | Replicability: research results can be reproduced | Interpretive awareness: researchers recognise and address implications of their subjectivity | Results can be interpreted, triangulation. |

As stated by Orlikowski and Baroudi (1991), in the interpretivist paradigm "*people create and associate their own subjective and intersubjective meanings as they interact with the world around them*" (Orlikowski & Baroudi, 1991, pp. 5–7). Interpretivists assert that a comprehensive understanding of a research phenomenon requires consideration of how it is subjectively experienced and constructed by the individuals involved.

Consequently, direct engagement with the individuals (research subjects) is critical to understand their perspectives and the surrounding environment (Dwivedi & Kuljis, 2008; Myers & Klein, 1999).

Interpretivist research is typically oriented towards developing a deep understanding of the research phenomenon and generating new theoretical insights based on the empirical findings. (Creswell, 2013; Myers & Newman, 2007).

The selection of a philosophical paradigm forms a critical aspect of research design, and should be guided by the nature and goals of the study as well as the researcher's orientation. Key criteria for determining the suitability of the interpretivist paradigm have been identified by leading scholars in the field, including Darke et al. (1998) and Orlikowski and Baroudi (1991). These criteria suggest that the researcher should recognize that reality is socially constructed and subjective, and that the research process is not free from values (*value free*) and assumptions.

In addition, the interpretivist paradigm is appropriate when the aim is not to identify generalizable patterns, but rather to achieve a deep and contextually situated understanding of the phenomenon under investigation. By adopting an interpretivist approach, the researcher can appreciate the complexity and diversity of human experiences and meanings, and contribute to the development of new theoretical insights and practical applications.

Therefore, this research study adopted the interpretivist paradigm as it represents the most appropriate paradigm given that the study seeks to provide an in-depth understanding of how the phenomenon of knowledge leakage risk occurs within knowledge intensive organisations.

For this reason, the researcher should engage and interact with people who participate in the management and protection of organizational knowledge and information assets inside organisations. Input from research participants (i.e., information security and knowledge management professionals/ practitioners) represented the primary source of information for collecting, informing and analysing the research phenomena. The input was obtained by conducting qualitative interviews which provided rich data and an in-depth understanding, and some cases supported by the analysis of complementary organizational documentation such as policies, procedures and redacted accounts of security incidents relating to organizational knowledge leakage.

## 4.2 Research Nature

According to Neuman (2006), there are three types of research: descriptive, explanatory, and exploratory. Descriptive research seeks to collect a comprehensive range of data about an existing situation or issue (Creswell, 2013; De Moya & Pallud, 2017). Explanatory research aims to clarify the reasons behind a particular phenomenon and is employed when the issue has already been described (Neuman, 2006). On the other hand, exploratory research is utilized to gain a deeper understanding of a phenomenon (Yin, 2003) and is typically employed in theory-building research (Shanks et al., 1993). For this reason, this study project adopts the exploratory type of research, as it aims to better understand and identify the managerial practices to mitigate the knowledge leakage risk caused by mobile devices inside knowledge-intensive organizations.

## 4.3 Research Methods

This study adopts a qualitative approach due to its interpretivist, theory building and explorative nature. This approach allows for an in-depth understanding of the research phenomenon and the context by investigating the perceptions, beliefs, and attitudes of the participants that influence their behaviors. Therefore, from an interpretivist philosophical perspective, a qualitative approach is most suitable (Myers & Newman, 2007; Shanks et al., 1993).

The data collection process involves interviews, document analysis, and observations conducted in the organizations under study. Qualitative data analysis techniques are employed to deeply investigate organizational behaviors and practices. However, the results are context-specific, limiting their generalizability to other contexts (Creswell, 2013; Neuman, 2006).

A qualitative research approach is of great importance in academic research as it provides the opportunity for researchers to investigate the specific context in which the research phenomenon and research questions are being examined (Stiles, 1993). This approach also allows for a better understanding of the perceptions, beliefs, and attitudes of the participants that influence their behaviours, thereby enabling a more comprehensive analysis (Myers & Newman, 2007; Neuman, 2006).

Hence, this study adopted the qualitative research methodology because of the need to understand the specific phenomenon of knowledge leakage risk caused by the use of mobile devices from the participants (information security and knowledge management professionals and practitioners) in a natural setting. Moreover, the qualitative approach demonstrates adaptability as the research design can be customized to meet the unique demands of the research situation. In addition, data collection and analysis instruments in qualitative research remain typically tailored to the specific context rather than following standardized procedures. (T. W. Lee et al., 1999).

While a quantitative approach, for example a survey, can be used to investigate knowledge leakage in knowledge-intensive organisations, it will fail to capture the participants' perspectives on how the leakage phenomenon takes place and why a particular approach was taken for a mitigation strategy. Studying the knowledge leakage occurrence using the qualitative approach will enable a better understanding of the phenomena. The development of a proper understanding from a qualitative perspective will allow a quantitative approach to be used in the future to provide a generalisation and testing of the research findings (Creswell, 2013; Myers & Newman, 2007; Neuman, 2006).

# 4.4 Research Design

The research design for this study was staged in three main phases: a contextual stage, a model development stage, and an empirical study in which the conceptual model was validated and refined (see Figure 4-1).

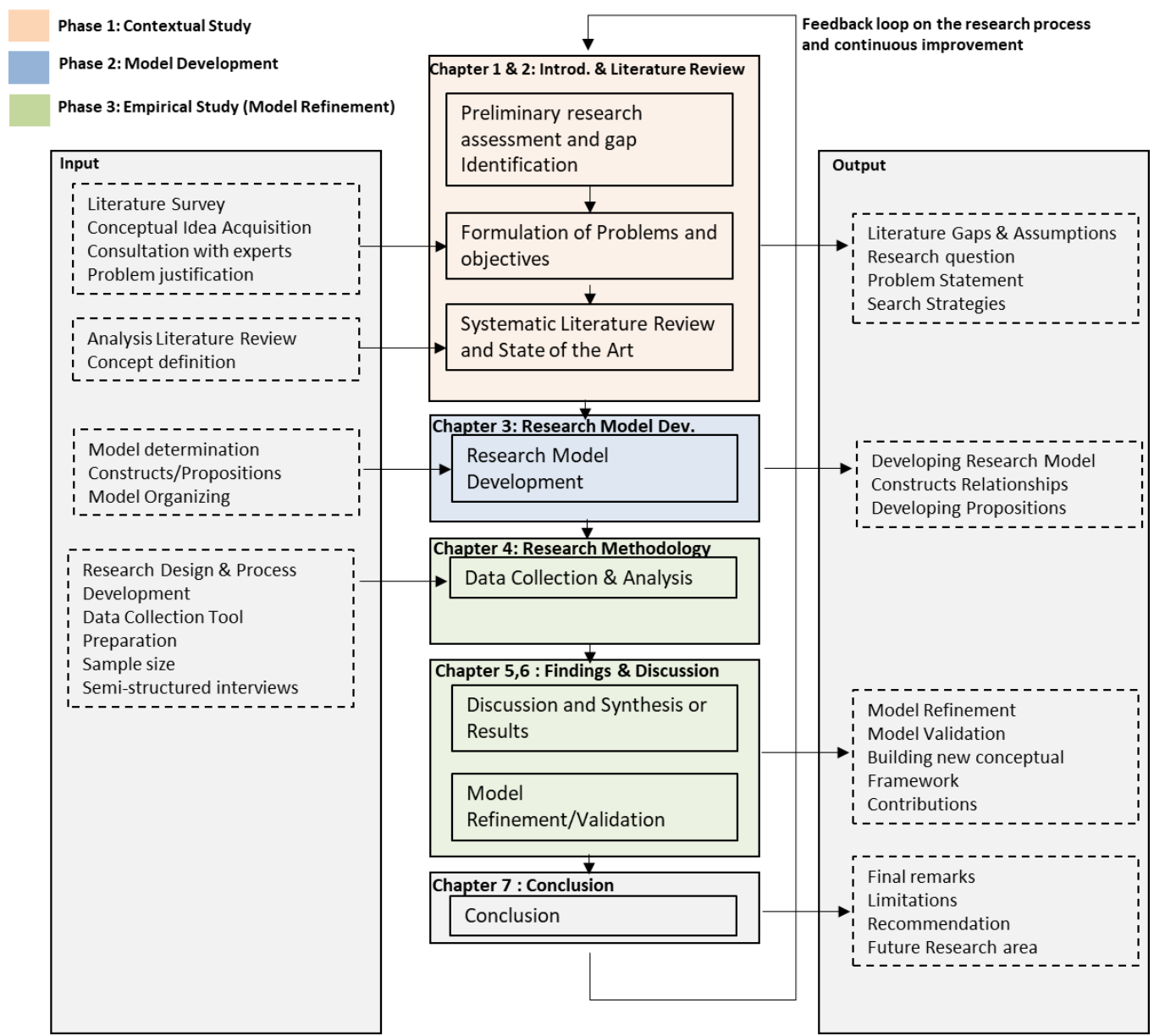

**Figure 4-1. Summary of Research design**

## 4.4.1 Phase 1: Contextual Study

As illustrated in Figure 4-1, the first stage of the research design, the contextual study, refers to the work described in chapters one and two —Introduction and Literature Review — in which the researcher introduced the research topic and question for this study, and conducted a systematic literature review to develop an understanding of the knowledge leakage phenomenon framed within the information security management, knowledge management and mobile computing literatures. Moreover, within each key section a number of limitations and gaps were identified in regard to mitigation strategies and knowledge protection in the context of mobile device usage.

## 4.4.2 Phase 2: Model Development

The second stage of the research design, the model development, refers to the work contained in chapter three — Research Model Development — in which the researcher elaborated on the process followed to build the research model. The three steps the researcher conducted to develop the model can be summarised as: 1) identify the mobile contexts, constructs and group them together based on their definition and the *Integrative model of IT business value* framework; 2) establish the relationship between the mobile contexts and the *knowledge leakage risk through mobile devices* construct; and 3) determine the relationship between *knowledge leakage risk through mobile devices* and *organizational knowledge mitigation capabilities* construct. Further to this, the propositions between the constructs were developed and referenced to the literature sources (for more detail, please refer to chapter three).

## 4.4.3 Phase 3: Empirical Study

The study aimed to validate and refine the proposed conceptual model through empirical investigation. The researcher conducted semi-structured interviews with experts in security and knowledge management in the industry and analysed the obtained data. Based on the analysed data, the researcher conducted several iterations as part of the coding process until reaching theoretical saturation which was constantly contrasted against the literature review and the research model. This process allowed the researcher to further develop a framework and classification of the leakage mitigation strategies used by organizations.

# 4.4.4 Interviews

In this study, semi-structured interviews with Information Security and Knowledge managers were the primary source of data collection that were employed to validate the preliminary model that was developed in chapter three from the conceptual phase based on the literature review. However, although interviews are frequently employed as a qualitative data collection technique (Ezzy, 2013), data may also be collected using other means such as personal observations, and internal or external documents.

One of the advantages of this research method is the opportunity to explore various social, cultural, and political factors related to the research phenomenon that may not have been anticipated beforehand, through interactions with participants (Myers & Newman, 2007). However, the method has limitations. Data analysis is typically qualitative and complex due to contextual factors. Findings depend on the researcher's ability to observe and integrate data, which can lead to subjective interpretations. Establishing causality is also challenging as control over variables is limited. Furthermore, the results may not be generalizable to other contexts due to the contextualized and nuanced nature of the data.

To enhance the generalizability of research findings, the researcher can adopt a triangulation approach by incorporating various data collection methods such as observation and documentation. This approach enables the researcher to verify the data by comparing and contrasting the results from multiple sources, which increases the credibility and dependability of the findings (Myers & Klein, 1999).

Triangulation also minimizes the risk of bias and subjectivity in data collection and analysis, and can help to ensure the validity and reliability of research results strengthening the generalizability of research outcomes (Neuman, 2006).

Myers and Newman (2007) distinguish three types of interview structures: unstructured, semi-structured, and structured. Unstructured interviews have no predetermined questions and are guided by the participant's responses. Semi-structured interviews begin with primary questions, followed by further questioning to elicit rich information. Structured interviews employ the same pre-planned questions for all participants, with the next question being asked once the previous one has been answered (McLellan et al., 2003; Neuman, 2006). These interview structures differ in the way the interview is conducted, particularly in the type of questions presented, allowing for different types and depths of information to be obtained.

For this research project, the researcher selected a semi-structured interview research method due to the fact that it allows flexibility to ask questions dependent on the particular circumstances of the interview such as research model, further queries, additional clarifications and to gain in-depth information based on the interviewee's responses.

## 4.4.5 Interview Protocol

To commence the interview process, the researcher started by developing the interview questions. The formulation of the interview questions was based on the research aim, questions and preliminary research model framework, as outlined in the preceding chapter, and with reference to Myers and Newman's (2007) guidance and framework, which aligns with the unique features and objectives of this research study. The interviews were categorized into four primary parts (refer to *Appendix D - Interview Protocol).*

### 4.4.5.1 Background questions and Basic Information Section

In this section, the researcher asked general questions about the organization, position, role, experience, and professional background of the participant to set the scene and establish the initial context.

### 4.4.5.2 Opening Questions

Subsequently, after the previous section, the researcher proceeded to introduce questions about the specific phenomenon of knowledge leakage, perceptions and general understanding about these types of incidents.

### 4.4.5.3 Scenario Questions

Due to the fact that managers may be reluctant to share information about security incidents that have occurred within their organizations, the researcher employed hypothetical scenarios to elicit responses via proxy and enquire about organizational strategies used to mitigate such incidents. The scenarios used a combination of different situational circumstances that mixed tacit and explicit knowledge in combination with the use of mobile devices such as smartphones, tablets and laptops (For examples on the scenarios used, refer to appendices: *Appendix B – Examples of Scenarios used in the interviews and discussions* and *Appendix C – Example of a Mobile Contexts Table (Transitions)*).

### 4.4.5.4 Knowledge and Information Distinction Information given to Participants

The next section after this, involved a series of questions with respect to knowledge vs information strategies in the context of mobile device usage and in different settings such as partnerships, competitors, suppliers and vendors in line with the different knowledge leakage vectors identified during the literature review.

### 4.4.5.5 Closing Questions

Finally, in the closing section, the researcher asked the participant if there were any questions or final remarks on their part and finalized the interview.

## 4.4.6 Selection and Recruitment of Participants

The selection of study participants was based on their expertise in the domains of information security and knowledge management. The recruitment process involved several approaches, including personal contacts of the researcher, snowball sampling, perusal of organizational websites, and LinkedIn. The researcher initially identified and contacted prospective participants through participation in various workshops and development sessions organized by security groups such as the Australian Information Security Association (AISA). These events provided opportunities for the researcher to introduce the study and invite participation from a diverse pool of security and knowledge professionals.

An application for research ethics was submitted to the Melbourne School of Engineering and IT Human Ethics Advisory Group at the University of Melbourne as part of the preparation for the data collection phase. (**ID 1646440**) – See *Appendix F – Ethics Approval*. Ethics approval was mandatory in order to ensure that the ethics requirements of the research project were in accordance with the university regulations and privacy conditions to preserve the confidentiality and the privacy of the participants. As part of the research ethics application process, a plain language statement and consent form were developed as well (See the following appendices: *Appendix F – Ethics Approval*, *Appendix G – Plain Language Statement*, and *Appendix E – Consent Form*).

These documents were sent to the participants via email with in-depth information about the aims and objectives of the research project, the privacy requirements of individuals and organizations and their information, as well as the reasonable steps the researcher took to make sure the confidentiality and privacy were preserved at all times, the importance of their participation, and their right to withdraw their participation at any time without penalty (see *Appendix E – Consent Form*).

Once the prospective participants were identified, the researcher sent invitation emails with an outline of the research project (research topic, research problem, methodology, and expected contributions). Additionally, participants received a short description about the expected role in the research. Upon reception of emails with positive response to the invitation, the researcher contacted the prospective participant once more to organize an appropriate time and location to conduct the interview. For the participants who failed to answer the first invitation email, the researcher emailed them to enquire about their participation. In the initial stage, 23 emails were sent, of which, 11 positive responses were received, 6 negative responses were obtained and 6 failed to respond to the invitation. Subsequently, by ways of snowballing sampling, a second round of emails were sent to 21 new prospective participants who were referred by the 11 previously confirmed participants. Out of these 21 potential participants, only 9 replied positively to the email invitation, which amounted to 20 total participants altogether (11 from the first round of emails, plus, 9 from the second round of emails). Participants met the following criteria: held a managerial position in either information security or knowledge management areas, within a knowledge-intensive organization where they were allowed to use mobile devices (mobile device policy in place) such as smartphones, tablets or laptops to conduct their work (see Table 4-2). Specific information concerning demographic information of research participants are detailed in chapter five — findings.

**Table 4-2. Background Information about the research participants**

| ID | Role | Industry | Knowledge Asset(s) | Experience (Years) | Self-Reported Leakage Risk | Size (Headcount) |
|---|---|---|---|---|---|---|
| CIO1* | Chief Information Officer | Government | People, Policy, Processes, Strategy | 10+ | Medium | Enterprise |
| SM1 | Security Manager | Banking | Strategy, Processes, | 15+ | Medium | Large |
| CISO1*# | Chief Information Security Officer | Consultancy | People, Intellectual Property, Processes, Product, Methodology | 20+ | High | Large |
| SM2* | Security Manager | IT Provider | Product, Methodology | 10+ | Low | Medium |
| CTO1# | Chief Technical Officer | IT Services | People, Intellectual Property, Product, Methodology | 15+ | High | Medium |
| SM3# | Security Manager | Insurance | People, Processes, Methodology | 10+ | High | Enterprise |
| CISO4 | Chief Information Security Officer | Health Care | Product, Methodology | 10+ | Medium | Medium |
| CSM# | Cyber Security Manager | Consultancy | People, Intellectual Property, Processes, Methodology | 15+ | High | Large |

| ID | Role | Industry | Knowledge Asset(s) | Experience (Years) | Self-Reported Leakage Risk | Size (Headcount) |
|---|---|---|---|---|---|---|
| CISO2 | Chief Information Security Officer | Telecommunications | Product, Methodology, Intellectual Property, People | 15+ | High | Enterprise |
| CIKO# | Chief Information and Knowledge Officer | Government | Policy, Strategies, Intellectual Property | 10+ | Medium | Enterprise |
| CKO1 | Chief Knowledge Officer | Food | People, Intellectual Property, Process, Product | 15+ | High | Medium |
| KM1 | Knowledge Manager | Health Care | Client, Product, Processes | 10+ | Medium | Medium |
| KM2# | Knowledge Manager | Government | Policy, Strategy | 10+ | Medium | Enterprise |
| CKO2 | Chief Knowledge Officer | Consultancy | Intellectual Property, Product, Methodology, People | 15+ | High | Large |
| KM3 | Knowledge Manager | Consultancy | Intellectual Property, Processes, Methodology, People | 10+ | High | Large |
| CKO3 | Chief Knowledge Officer | Government | Processes, Strategy, Intellectual Property | 15+ | Medium | Enterprise |
| KM4* | Knowledge Manager | Health Care | Product, Processes | 5+ | Low | Medium |
| KM5 | Knowledge Manager | Education | Methodology, Research findings | 10+ | Medium | Large |
| KM* | Knowledge Manager | Not-for-Profit | Client relationship, processes | 15+ | Low | Medium |
| CISO3 | Chief Information Security Officer | Government | Processes, Product, Methodology, Intellectual Property, Strategy, people | 15+ | High | Enterprise |

## 4.4.7 Data collection

As described previously, given the explorative nature of this study, the research followed a qualitative research design using different participants. Data collection comprised of 20 interviews of information security and knowledge managers of medium to large knowledge-intensive organizations in Australia (see Table 4-2).

In some cases, supplementary documentation (policies, procedures, reports and organizational standards) was also provided by the organizations, and hence it was also examined for triangulation purposes as a way to confirm that the data reported in interviews matched organizational documented processes and procedures (Refer to

The selection of participants was a critical step in ensuring the validity and reliability of the study's findings. The participants were chosen based on their expertise in the domains of information security and knowledge management, specifically within knowledge-intensive organizations. This was crucial as these individuals are at the forefront of managing leakage associated risks in their respective organizations.

The qualifying criteria for potential interviewees included their role within the organization, their experience in dealing with knowledge management and information security, and their experience with the organization's sanctioned mobile device policies such as BYOD (Bring Your Own Device), CYOD (Choose Your Own Device), COPE (Company Owned/Personally Enabled), COBO (Company Owned/Business Only), and COSU (Company Owned/Single Use). These criteria ensured that the participants had a deep understanding of the phenomenon under study and could provide rich, detailed, and contextually relevant insights.

The selection process involved initial contact to gauge interest and availability, followed by a screening process to ensure they met the qualifying criteria. This was followed by scheduling and conducting the interviews. The recruitment process was facilitated by various approaches, including personal contacts of the researcher, snowball sampling, perusal of organizational websites, and LinkedIn.

The recruitment process was also guided by ethical considerations. An application for research ethics was submitted to the Melbourne School of Engineering and IT Human Ethics Advisory Group at the University of Melbourne. As part of the research ethics application process, a plain language statement and consent form were developed and sent to the participants, ensuring that the confidentiality and privacy of the participants were preserved at all times.

The selection and recruitment process was rigorous and systematic, ensuring that the participants were well-suited to contribute valuable insights to the study. The added explanation in section 4.4.6 provides a more detailed account of this process, which should help to further clarify the rigorous approach taken in selecting interviewees for this study.

Appendix E – Example of Supplementary Documentation provided by Organizations). In order to analyze the supplementary data, a qualitative document analysis on the provided documentation was conducted looking for supporting evidence on the secondary sources that corroborated the interviews statements. In many cases, information provided by interviewees failed to match the organizational documents, highlighting the issue of either outdated or incomplete documentation. In other cases, the practices reported by participants were not formally institutionalized but rather a cultural practice (informal mechanisms). By conducting an organizational document review, the researcher identified the strategies that were formally conducted by organizations and practices, that although informal, were rooted in the organization's culture. The content review process as well as further information on the data analysis is provided in the next section.

## 4.4.8 Interview and Supplementary Data

The objective of every interview was to ascertain the methods employed by each organization to facilitate the exchange, flow and dissemination of knowledge, particularly through mobile devices, while also preventing knowledge leakage from occurring. During the interviews the researcher was able to identify what strategies organizations had in place to prevent knowledge leakage (proposition 1 in the conceptual model).

Interviews were conducted over a period of one year and a half. Each interview lasted on average 1 hour, within a range of approximately 45 minutes (shortest interview) to 1 hour and a half (longest interview). The interviews, when allowed by the participant, were audio-recorded with the consent of the interviewee and transcribed verbatim and shared with each interviewee to check validity and verify the content.

During the interviews different context scenarios were presented to illustrate different levels of risk exposure. The scenarios illustrated the relationship between contexts and risk (proposition 2 in the conceptual model).

Additionally, as previously indicated, documents such as policies and procedures were also analysed in order to gain a comprehensive understanding of the knowledge protection mechanisms employed by the organizations. Following transcription, the data was subjected to a qualitative analysis methodology consisting of selective, axial, and thematic content analysis (Krippendorff, 1980; Miles & Huberman, 1994) This methodology was applied by adopting the framework provided by the Gioia methodology (Gioia et al., 2012) and drawing on the different mobile contexts outlined in the research model to classify collected evidence (as discussed in the preceding chapter).

Further details of the analysis process and the findings will be presented in the next section: Data Analysis, and further explained in the next chapter: Research Findings.

## 4.4.9 Overview of Data Analysis Procedure Using Gioia Methodology

In order to achieve qualitative rigor and systematically transform raw data into theoretical interpretations, this study adheres to the guidelines proposed in Gioia et al. (2012).

The data analysis process was iterative, aimed at improving insights and generalizability, despite the linear structure of the study. The initial phase involved reviewing the background materials, documentation, observation during interviews, interview transcripts, and field notes, which recorded impressions at the time of each interview (Langley, 1999; Locke et al., 2008; Yin, 2003, 2015, 2017). The objective was to identify indicators of how knowledge leakage risk incidents caused by the use of mobile devices occurred in organizations.

The analysis of interview data in this study employed constant comparative techniques and a combination of open, axial, and selective coding following the guidelines of Corbin and Strauss (1990). The process was iterative, commencing with coding the raw data and advancing to the development of second-order themes and aggregate dimensions. This method allowed for a more nuanced and comprehensive analysis of the collected data, enhancing the validity and reliability of the study's findings. By using constant comparative techniques and different types of coding, the researcher was able to systematically organize the data, draw connections between emerging themes, and generate insights that contributed to the conceptual framework of this research study (see *Figure 4-2*).

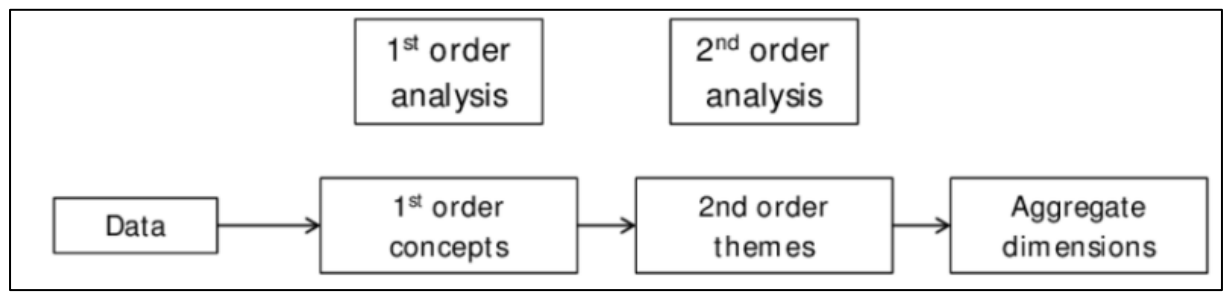

**Figure 4-2. Gioia Methodology Process based on Grounded Theory. Adapted from Gioia et al (2012)**

## 4.4.9.1 Stage 1. Identifying First Order Concepts

Following Corbin and Strauss (1990), the researcher employed constant comparative techniques and the combination of open, axial and selective coding for analysing the interview data and supplementary documentation in the cases that it was provided by participants. The initial codes that resulted from the interview and supplementary documentation included first order concepts such as mobile device usage based on the type of work conducted, risk management, mitigation strategies, formal, informal controls, as well as strategic, tactical and operational levels. These concepts were assigned to words, sentences and paragraphs as well, and used in the margin of transcriptions, documents, notes and memos, and subsequently organized in Microsoft Excel tables (see Figure 4-1 and Figure 4-4). Overall, more than 150 codes were identified that related to mobile device usage, knowledge risk and mitigation strategies. This process was conducted iteratively until no more new concepts were found, that is, the first order concepts were applied contrasted against other codes to ensure no new ideas emerged from the raw data, reaching, in this manner, theoretical saturation (J. M. Corbin & Strauss, 1990; J. Corbin & Strauss, 2014; Glaser & Strauss, 1968; Suddaby, 2006) (See Figure 4-5). Next, the researcher focused on the development of second order themes, described in the following section.

| Influencing Factors | Contexts | | Formal | Informal | Technical | Risk Framework / Policies/Guide |
|---|---|---|---|---|---|---|
| Human Factors | Personal Context | Strategic | Corporate Governance | Security Culture strategy | Score card System to measure performance and KM results | ISO 28000; Governance; Protective Security Policy Framework (ASD); Information Security Manual (ISM) - ASD; Resilia Framework; NIST Framework |
| | | Tactical | Closed-looped processes for knowledge-sharing and replication in mission critical areas- not left to choice or chance; Identifying People of security concern; Managing People of High Risk Positions; Application of "Need to Know" Principle; Insider Threat Approach - NIST , ISM (Military, R&D) | Culture & Change Management; KM Initiatives (such as knowledge-bases, communities of experts, and collaboration) are centered around predefined "mission-critical" areas; Security Education, Training & Awareness; Organizational Strategy; SETA targeted to user profiles.; Establish Whistleblowing channels to report non-compliant groups/users; Decoy shilling targeted to employees and senior managers); Behavioral based security training | Monitoring; Evaluation; Personnel Security Clearance | Personnel Security Handbook (PSPF) - Managing Insider Threat; Tele-working Handbook (PSPF); Mobile Computing Checklist (PSPF); Resilia Framework; Maslow's Motivation Model (Human Resources); Psychological theories |
| | | Operational | On-going HR Screening Processes; Background Checks; on-going Personnel checks; Personnel Risk Assessment; Psycological Profiling (Personality Traits); Formal Punishments /Sanctions (Military) analyze personality; non-competition clause; NDA, ; Staff Rotation | Computer Based training (Simulation-Cyber Siege); Security Awareness Assessment; Security Reminders; Analysis Change of Circumstances; Social Engineering Assessments (Decoy fish emails; Address Insider Threat training (Dobbing); Relaxation Techniques such as Yoga, Tai chi to engage awareness and motivation leading to less disgruntlement; Foster Innovation and Creativity to improve org. climate | Personnel Information comming from Online Social Networks Big data - Machine Learning (IBM Watson) - Military- Behaviour Analytics Predictive Analysis (Machine Learning); Gamification; Simulation | Personnel Security Handbook; Personnel Security checklist; NIST Guidelines |
| | Social Context | Strategic | Corporate Governance | Security Culture strategy | Score card System to measure performance and KM results | |
| | | Tactical | Closed-looped processes for knowledge-sharing and replication in mission critical areas- not left to choice or chance; Business Unit/ Group Policies tailored according to group risk profile; KM roles and Structure; Name a leader /owner in charge of process (Knowledge); Division of Duties; Organizational Hierarchy / reporting chain to management | KM initiatives (such as knowledge-bases, communities of experts, and collaboration) are centered around predefined "mission-critical" areas; Senior Management Involvement; Staff rotation; Rewards system to compliant users (scoring system - Highest scores receive prizes); Establish Whistleblowing channels to report non-compliant groups/users | Virtual Communities to prevent tacit knowledge leakage (Employees leaving the organization); Taxonomies - Organized content to enable search and retrieval - Knowledge Retention | |
| | | Operational | Gamification; Roles; Separation of duties; Shared administrative Access; Knowledge / Information Mirroring (Shadowing); Knowledge Portals (Wiki) | Security Culture Development; Gamification; Lead by example; Peer's influence (Managers to staff members); Social Engineering assessments; Inform on non-compliance staff (whistleblowing); Sharing experiences; set up meet ups with employees and senior managers; Sharing tips; Foster Discussion; Group Laboratories/workshops for mobile technologies (Devices/ Applications) | Yellow Pages to enable explicit knowledge management; Knowledge Gateways to prevent | Maslow's Motivation Model (Human Resources); Psychological theories; Emotional Intelligence; Telecommuting Handbook (working from home and out of the office) - PSPF - Physical Security Management Guidelines |
| | Organizational Context | Strategic | Corporate Governance; KM Vision; KM Deliberate Strategy; KM Emergent Strategy - to enable KM changes | Security Culture strategy | Score card System to measure performance and KM results | |
| | | Tactical | Risk Management Framework | Rewards system to compliant users (scoring system - Highest scores receive prizes) | | |
| | | | Organizational Init. Security Policy, Processes Information (Knowledge Governance: Owners, custodians and users); Classification of Information/knowledge (Confidential/public) | | | AS/NZS 4360 Risk Management; ISO 28000 - Risk Management; ISO 38500 - Governance; COBIT; ISO 1799 Risk Analysis; Resilia Framework |

Figure 4-3. Examples of First Order Concepts

| | A | B | C | D | E | F | G |
|---|---|---|---|---|---|---|---|
| 9 | Enterprise Factors | Organizational Context | Strategic | Corporate Governance KM Vision KM Deliberate Strategy KM Emergent Strategy - to enable KM changes (Militarg) | | Score card System to measure performance and KM results | |
| 10 | | | Tactical | Risk Management Framework | Security Culture strategy Rewards system to compliant users (scoring system - Highest scores receive prizes) | | AS(NZS) 4360 Risk Management ISO 31000 - Risk Management ISO 38500 - Governance COBIT ISO 1799 Risk Analysis |
| 11 | | | Operational | Organizational Inf. Security Policy, Processes Information (Knowledge Governance: Owners, custodians and users) Classification of Information/Knowledge (Confidential/public) Information Segmentation Tagging documents (RFID) Legal Mechanisms (NDA), confidentiality Agreements, Patents,Copyright, Trade Secrets, rights, trademark, non competition clause Compartmentalization Knowledge documentation policy (Wiki, Portals) | Informal Protection mechanisms: Secrecy, Noise, Need to know Secrecy Misinformation | | Resilia Framework. OCTAVE OCTAVE-S CRAMM PCIS - DSS National Institute of Standards and Technology (NIST) – Federal Information Processing Standard (FIPS) Publications |
| 12 | | Environmental Context | Strategic | Intra KM Strategic liason with external organizations such as agencies to have response mechanisms in case of leakage Fast Innovation cycle | | | |
| 13 | | | Tactical | Knowledge Communities KM Processess - Standard Procedures to share Knowledge/Information outside | Virtual communities | | |
| 14 | | | Operational | Inter (External) organizational Information Sharing Agreements (Government Agencies) Partnerships Guidelines and Procedures to perform work externally other locations, e.g, PSPF, ISM | Security Culture Securing of Knowledge/ Information channels when working from home, out of the office, from a client or in-transit Secrecy Misinformation Complexity | Secure Knowledge Sharing Platforms to be used when offsite Knowledge Portal and Gateways | Protective Security Policy Framework - guides Telecommuting Handbook (working from home and out of the office) - PSPF - Physical Security Management Guidelines |
| 15 | Technical Factors | Device Context | Strategic | Enterprise Mobility Management Strategy | | | |
| 16 | | | Tactical | Policy notification system | | | |
| 17 | | | Operational | End Point Security Compartmentalization Virtualization Mobile Device Tracking Device compliance check Corporate App Store Device Messaging System Look down/ Wipe out Mechanisms Mobile Device RFID (tagging) Standard Operating Environment Audit logs Mobile Data Analytics (Big data ) Predictive Analysis / Forecasting (Machine Learning) | Security Behaviour - use of mobile devices in external environments (tele-commuting) Mobile computing and communications Training | | Enterprise Mobility Management (EMM) - Mobile Device Management (MDM) OVASP (The Open Web Application Security Project) Telecommuting Handbook (working from home and out of the office) - PSPF - Physical Security Management Guidelines |
| 18 | | Technological Context | Strategic | KM Strategy (Sharing, retention, replication) | | | |
| 19 | | | Tactical | Development of taxonomy/Ontology to capture organizational Knowledge Knowledge Graphs (NASA) | Security reminder system to raise awareness and influence behaviour | | Common Vulnerability Scoring System (CVSS) OVASP |
| 20 | | | Operational | Knowledge Portals (Wiki, SharePoint) Information Rights Management (IRM) Corporate Firewalls, IDS, IPS, Antivirus, Standard Operating Environment Information Security Policy Corporate Patching Policy Policy compliance check Audit logs Security Operation Center (SOC) Incident Response Teams - Information Leakage (Militarg) | Security Behaviour - use of mobile technology and communications in external environments Guideline | | Information Security Manual (ISM) by ASD Protective Policy Framework (PSPF) Top 4 Mitigation Strategies - by ASD Institute of Standards and Technology (NIST) – Special Publications (SP) Telecommuting Handbook (working from home and out of the office) - |

**Figure 4-4. Examples of First Order Concepts. Second Iteration**

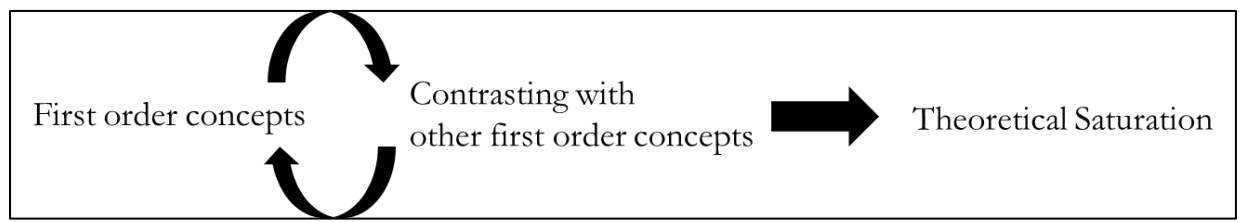

**Figure 4-5. Iterative Process to reach Theoretical Saturation**

## 4.4.9.2 Stage 2. Second Order Themes

In the next phase of the data analysis, the researcher grouped similar first order concepts into second order themes, using the terms used by the participants whenever possible (J. M. Corbin & Strauss, 1990; Gioia et al., 2012). The second order themes emerged as the linkages amongst the concepts, forming categories that became evident and that were sorted further and analysed in the context of the study main overarching research question:

- *How can knowledge-intensive organizations mitigate the knowledge leakage risk caused by the use of mobile devices?;*

and supporting secondary questions:

1. *What strategies are used by knowledge-intensive organizations to mitigate the risk of knowledge leakage (KLR) caused by the use of mobile devices?;*

2. *How does the perceived KLR level inform the strategies used by KI organizations?;*

3. *What knowledge assets do knowledge intensive organizations protect from KL?;*

As well as the knowledge based view theory as the underlying lens to guide the process and several iterations, during these iterations some first order concepts were merged, refined and others discarded in order to accomplish a higher level of abstraction and obtain the second order themes as indicated by Gioia et al (2012). This iterative process was also conducted in Microsoft Excel (see Figure 4-6)

| | | Controls | | | |
|---|---|---|---|---|---|
| Influencing Factors | Contexts | Formal | Informal | Technical | Risk Framework / Policies/Guides |
| Human Factors / Personal Context | Personal Context | HR Screening Processes<br>Background Checks<br>on-going Personnel checks<br>Monitoring<br>Evaluation<br>Personnel Security Clearance<br>Personnel Risk Assessment<br>Psycological Profiling (Personality Traits)<br>Identifying People of security concern<br>Formal Punishments /Sanctions (Military)<br>Managing People of High Risk Positions<br>Behaviour Analytics<br>Predictive Analysis (Machine Learning)<br>Personnel Information comming from Online Social Networks Big data - Machine Learning (IBM Watson) to analyse personality<br>Gamification<br>Managing Insider Threat | Security Education, Training & Awareness Campaigns<br>Computer Based training (Simulation-Cyber Siege)<br>Security Awareness Assessment<br>Security Reminders<br>Change of Circumstances<br>Social Engineering Assessments (Decoy fish emails and whailing targeted to employees and senior managers)<br>Behavioral based security training<br>Address Insider Threat<br>Relaxation Techniques such as Yoga, Tai chi to engage awareness and motivation<br>Foster Innovation and Creativity | | Personnel Security Handbook (PSPF) - Managing Insider Threat<br>Tele-working Handbook (PSPF)<br>Mobile Computing Checklist (PSPF)<br>Resilia Framework<br>Maslow's Motivation Model (Human Resources)<br>Psychological theories<br>Emotional Intelligence |
| | Social Context | Gamification<br>Roles<br>Organizational Hierarchies<br>Separation of duties<br>Shared administrative Access<br>Knowledge / Information Mirroring (Shadowing)<br>Knowledge Portals (Wiki) | Security Culture Development<br>Gamification<br>Lead by example<br>Peer's influence<br>Social Engineering assessments<br>Inform on non-compliance staff (whistleblowing)<br>Sharing experiences<br>set up meet ups with employees and senior managers<br>Sharing tips<br>Foster Discussion<br>Group Laboratories/workshops for mobile technologies (Devices/Applications)<br>organizational App store<br>Conversation Security (Handbook) | | Maslow's Motivation Model (Human Resources)<br>Psychological theories<br>Emotional Intelligence<br>Telecommuting Handbook (working from home and out of the office) - PSPF - Physical Security Management Guidelines |
| Enterprise Factors / Organizational Context | Organizational Context | Organizational Inf. Security Policy, Processes<br>Information (Knowledge Governance: Owners, custodians and users)<br>Classification of Information/Knowledge (Confidential/public)<br>Information Segmentation<br>Tagging documents (RFID)<br>Legal Mechanisms (NDA), confidentiality Agreements, Patents<br>Compartmentalization<br>Knowledge documentation policy (Wiki, Portals) | Informal Protection mechanisms: Secrecy, Noise, Need to know | | AS/NZS 4360 Risk Management<br>ISO 31000 - Risk Management<br>ISO 38500 - Governance<br>COBIT<br>ISO 1799 Risk Analysis<br>Resilia Framework<br>OCTAVE<br>OCTAVE-S<br>CRAMM<br>PCIS - DSS<br>National Institute of Standards and Information Processing Standard (FIPS) Publications<br>ISO/IEC 27000 |
| | Environmental Context | Inter (External) organizational Information Sharing Agreements (Government Agencies)<br>Partnerships<br>Guidelines and Procedures to perform work externally/ other locations, e.g., PSPF, ISM | Security Culture<br>Securing of Knowledge/ Information channels when working from home, out of the office, from a client or in-transit | | Protective Security Policy Framework - guides<br>Telecommuting Handbook (working from home and out of the office) - PSPF - Physical Security Management Guidelines |
| | | End Point Security<br>Compartmentalization<br>Virtualization<br>Mobile Device Tracking<br>Device compliance check | | | |

**Figure 4-6. Second Order Theme Iteration**

### 4.4.9.3 Stage 3. Aggregate Dimensions

In the final phase, the previously developed data was further analysed by aggregating concepts identified in the second order themes to generate even higher level of abstraction, i.e., dimensions, which contain the overarching basic concepts relevant to the research aim of this study of understanding what mitigation strategies knowledge intensive organizations use in order to mitigate the knowledge leakage risk caused by the use of mobile devices in knowledge intensive organizations (see Figure 4-7).

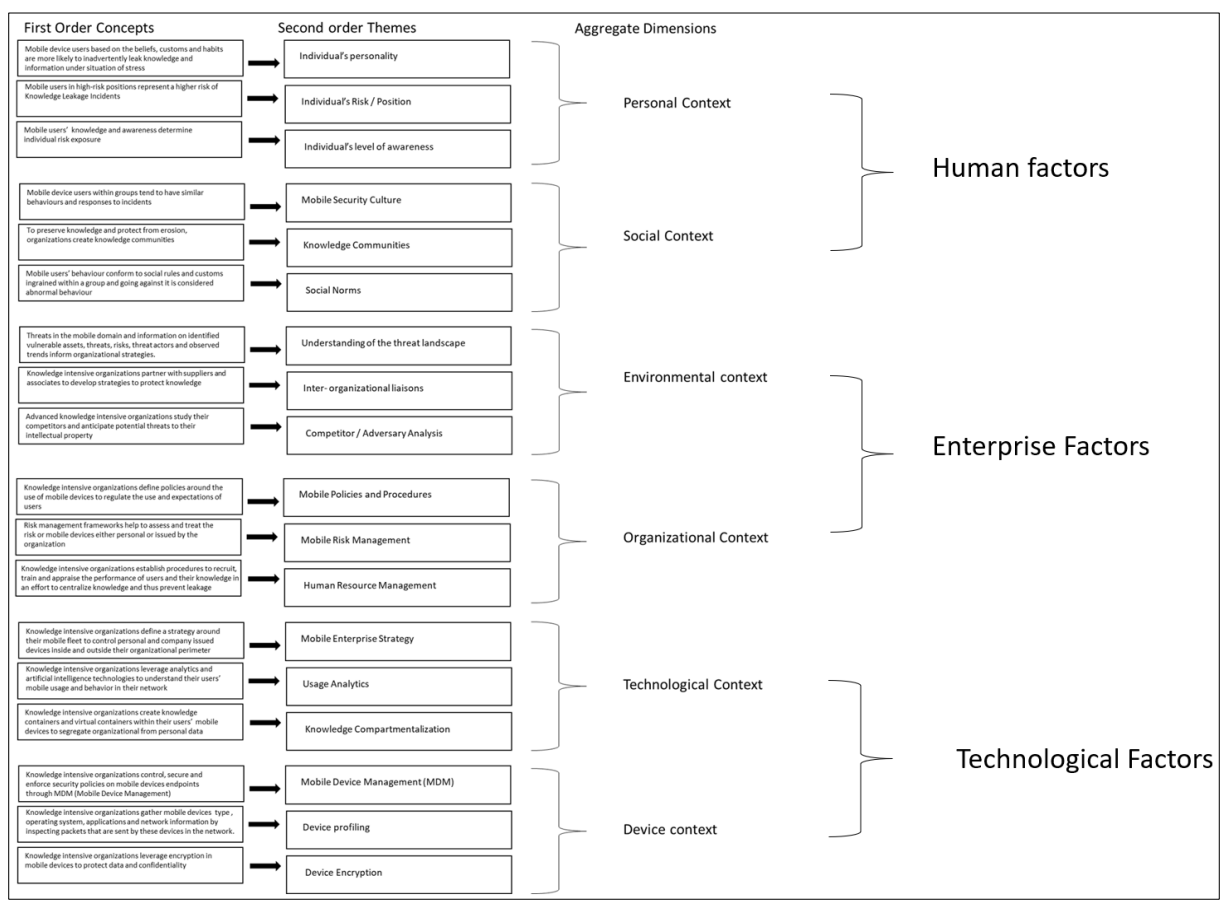

**Figure 4-7. Complete Data Structure with dimensions based on the research model depicted in Chapter 3.**

## 4.5 Data Analysis

In order to ensure that the findings are presented in a concise, logical and easy to understand manner, the researcher engaged a systematic approach to structure the presentation of the data analysis. Due to the complex nature of reporting the research findings in a way that remains within the context of the research agenda, the researcher shows the process in a linear fashion, that in no way, reflects the iterative and repetitive cyclical process undertaken by this study. See Figure 4-8. Flowchart of Systematic Process followed in the empirical study leading to the Findings for a visual understanding of the process engaged during this stage of the study.

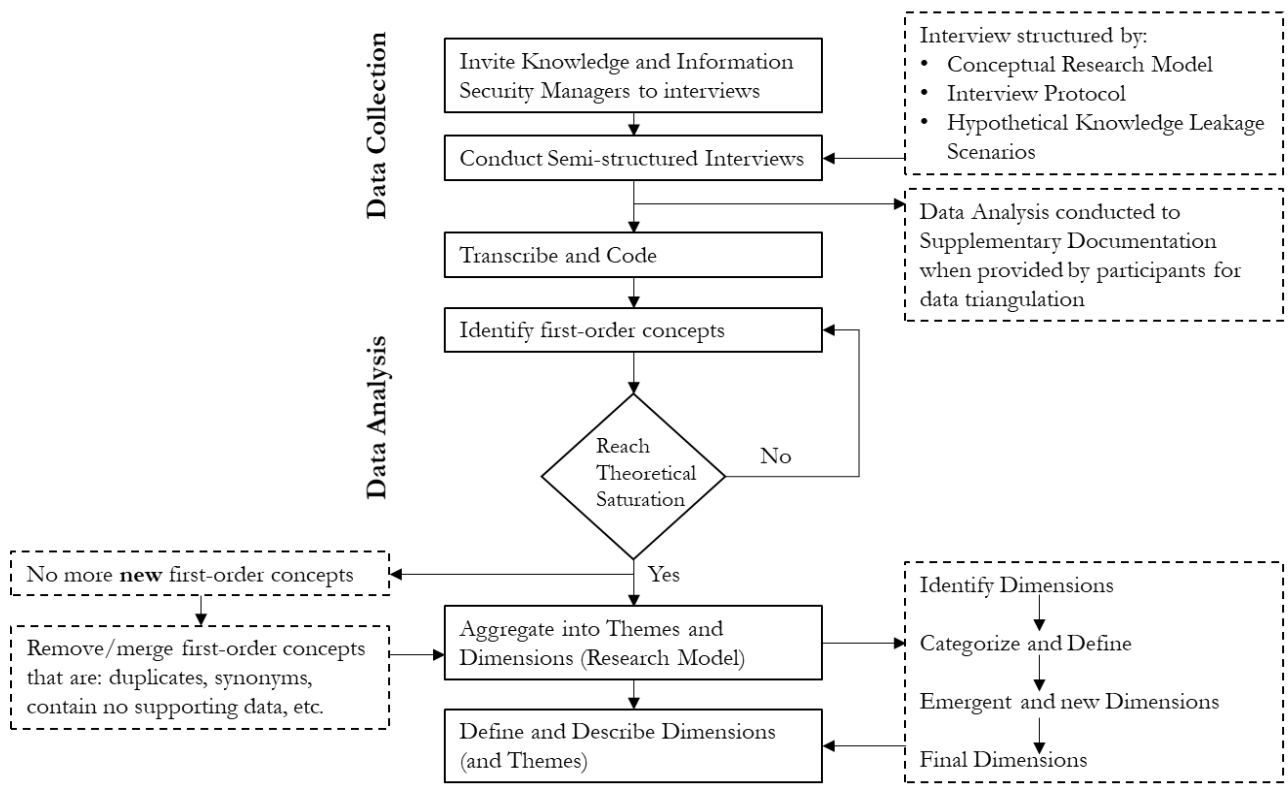

**Figure 4-8. Flowchart of Systematic Process followed in the empirical study leading to the Findings**

As described in the previous chapter, semi-structured interviews were conducted to refine and validate the preliminary research model developed in chapter three (Conceptual Research Model) which was based on the literature review (chapter two). A total of twenty interviews with information security and knowledge managers and supplementary document analysis (when provided) were performed and analysed according to the parameters defined in the methodology section (chapter four). It is important to remind the reader that the conducted empirical study was qualitative in nature, and **no** quantitative results, i.e., hypotheses testing, were conducted in this research project. Rather, the researcher developed and evaluated theoretical constructs and conceptual propositions describing the relationships amongst such constructs, and subsequently validated the aforementioned conceptual research model with its constructs and propositions through the data collection process in the empirical study.

Due to the fact that the research focus of this study is exploratory and even explanatory to a certain extent (as an attempt to explain a social phenomenon), the author focused on theory building and developing approaches established by other qualitative researchers (Creswell, 2013; Myers & Klein, 1999; Myers & Newman, 2007; Neuman, 2006) in order to extract meaningful abstractions from the raw data and, recurrent and emerging patterns, via the identification of themes and sub themes (J. M. Corbin & Strauss, 1990) leveraging grounded theory tools such as open coding, axial coding and selective coding, as well as adopting a related methodology that expands on these techniques to facilitate the abstraction of aggregated concepts and higher level dimensions as stated by Gioia (Gioia et al., 2012), thereby, improving the rigour in this qualitative and inductive study. As a result, this qualitative study seeks to build a profound understanding of a phenomenon from the rich data analyzed (Zmud, 1998).

## 4.5.1 Research Aim and Questions

As mentioned during the introduction chapter, the researcher identified gaps in the current body of knowledge that showed how the knowledge leakage phenomenon caused by the use of mobile devices specifically within knowledge intensive organizations, has been under studied and underrepresented in the current body of literature, despite the abundant literature on knowledge management, and information security addressing related topics such as knowledge sharing, information and data leakage, as well as mobile computing literature addressing mobile device usage and mobility.

However, in proportion, scant studies have focused on knowledge leakage, and, no significant studies, to date, have considered both knowledge leakage and mobile device usage together and their relationship.

Therefore, this study aims to explore and analyse the social phenomenon of knowledge leakage inside knowledge intensive organizations and understand what strategies can be used by such organizations to prevent leakage from happening particularly in the context of mobile device usage.

In order to address these research gaps, this research study seeks to answer the primary and overarching question:

- *How can knowledge intensive (KI) organizations mitigate the knowledge leakage risk (KLR) caused by the use of mobile devices?*

To this end, the following secondary supporting questions aim to assist the main research question by providing the necessary foundation and information required to answer the primary question:

1. *What strategies are used by knowledge-intensive organizations to mitigate the risk of knowledge leakage (KLR) caused by the use of mobile devices?*
2. *How does the perceived KLR level inform the strategies used by KI organizations?*
3. *What knowledge assets do knowledge intensive organizations protect from KL?*

In the next subsections, the researcher shows how these primary and secondary questions informed the research design, agenda, and data collection process in order to answer them.

## 4.5.2 Unit of Analysis vs. Unit of Observation

During this research project, the researcher considered the unit of analysis and unit of observation that were applied to this study, as these units may usually differ in research, depending on factors such as whether the study involves quantitative or qualitative methods, and also according to the approach and data methods employed to study a phenomenon (Sedgwick, 2014).

A **unit of analysis** refers to the entity that the researcher wishes to investigate and report about at the end of a study. On the other hand, a **unit of observation** pertains to the item or items that the researcher actually observes, measures or collects while attempting to learn and understand about the unit of analysis (Hopkins, 1982; Kumar, 2018). In some cases, the unit of observation and unit of analysis may coincide, in others, however, these units may differ. Moreover, the unit of analysis is determined by the research question of the study, while the unit of observation, on the other hand, will be determined largely by the method of data collection. In IS research, the most common units of analysis and units of observation include the following (Hopkins, 1982; Kumar, 2018):

- Individuals
- Groups
- Organizations
- Documents

In this research study, based on the research question and sub-questions, methodology and data collection methods as defined in chapter four, the unit of analysis will be represented by the knowledge intensive organization, whereas the unit of observation will be represented by individuals, particularly, knowledge managers and information security managers, as well as organizational documents such as policies, procedures and documented processes.

As stated earlier, in this study, the primary and secondary research questions informed the unit of analysis and influenced the data collection method and process to reflect the research aim. (See Table 4-3 below)

**Table 4-3. Unit of Analysis vs. Unit of Observation**

| Research (sub)-Questions | Unit of Analysis | Data Collection | Unit of Observation | Type of Findings |
|---|---|---|---|---|
| How can knowledge-intensive organizations mitigate the knowledge leakage risk (KLR) caused by the use of mobile devices? | Organization | Interviews Documents | Knowledge Manager Information Security Manager | Framework |
| 1. What strategies are used by knowledge-intensive organizations to mitigate the risk of knowledge leakage (KLR) caused by the use of mobile devices? | Organization | Interviews Documents | Manager | Strategies Categorization |
| 2. How does the perceived KLR level inform the strategies used by KI organizations? | Organization | Interviews Documents | Manager | Strategies |
| 3. what knowledge assets do knowledge intensive organizations protect from KL ? | Organization | Interviews Documents | Manager | Knowledge Asset |

A common mistake researchers make with respect to both causality and units of analysis and observation relates to **ecological fallacy** (Sedgwick, 2014). Ecological fallacy occurs when claims about one lower level unit of analysis are made based upon data resulting from one higher level unit of observation. For example, claims about individuals are made built from data gathered at the organizational level or group level. Conversely, when claims about a higher level unit of analysis are made determined from one lower level unit of observation, these claims may result in an error of **reductionism** (Hopkins, 1982)**.** In this research study, the researcher makes claims concerning knowledge intensive organizations based on individual data gathered from knowledge managers, information security managers, and organizational documents. Therefore, a possibility exists that causal explanation and reasoning could derive from the error of reductionism.

To safeguard against the reductionism error and increase the trustworthiness of the findings, the researcher has employed the matrix rigor from Carson et al (2001) as explained and developed in the previous chapter — Methodology — in order to avoid deviation from the scientific method and draw well reasoned conclusions from the data collected. Nonetheless, the error of reductionism still remains to a certain degree and represents a limitation for this type of studies as denoted by other researchers in the social sciences and the IS research field (Hasan & Banna, 2012; Hopkins, 1982; Kumar, 2018; Sedgwick, 2014).

### 4.5.3 Gioia Methodology

This research aims to answer the research question of*: How knowledge intensive organizations can mitigate the knowledge leakage risk caused by the use of mobile devices* and, in doing so, explores the mitigation strategies used by the aforementioned organizations to combat such risk. The data captured through the interviews and document analysis describes the use of different approaches that assisted organizations to develop higher-order knowledge leakage preventing capabilities and mitigation strategies. Moreover, the data also illustrates the impact of using such strategies on the overall enterprise security performance in terms of its knowledge protection capabilities. In addition to the rich description provided by the data, a further data structure, based on the Gioia methodology, displays the relationships amongst concepts, themes and dimensions (See Figure 5-15 Complete data structure for all Contexts and Figure 5-16. mind map of first order concepts, second order themes and dimensions). In addition to this, multiple extracts and quotes from the interviews are provided which complements the structure by providing further evidence of the salient and emergent constructs. Finally, the findings from the data analysis are integrated with the existing IS literature to complement and refine the overall model to address the knowledge leakage risk caused by the use of mobile devices in knowledge intensive organizations. As previously described, although the researcher presents the data coding process in a linear manner (see Figure 4-9), in reality, the aforementioned process was conducted in a non linear, multi stage, iterative fashion.

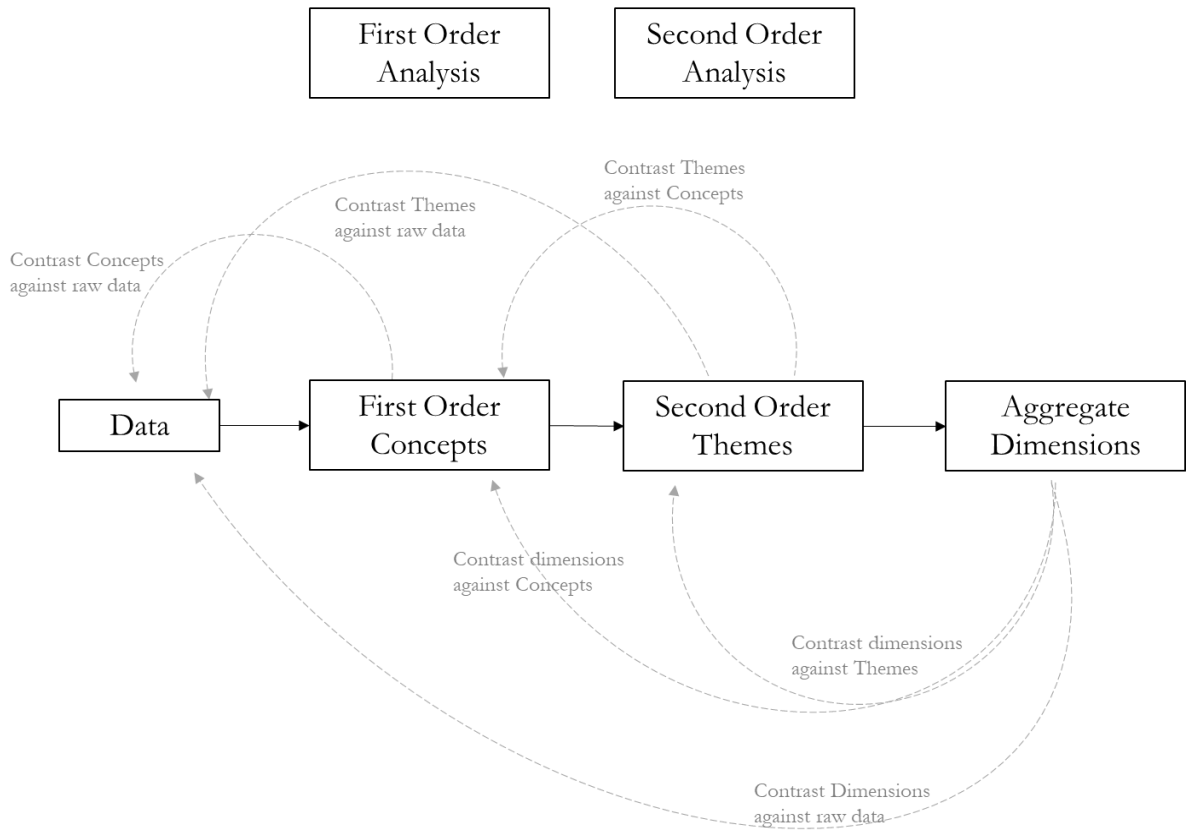

**Figure 4-9. Gioia Methodology Process applied in this empirical study. Adapted from Gioia et al (2012)**

As mentioned in the methodology chapter (chapter four), the Gioia methodology is a qualitative data analysis method with an inductive approach, that is, a bottom up method whereby the raw data is synthesized, analyzed and aggregated in such a way that produces a general principle, i.e., the building block of theory. The beginning of the analysis consists of developing the first-order concepts, followed by developing the second-order themes to finally aggregate the second order themes into the aggregate dimensions (see Figure 4-9).

# 4.6 Interview Process

Collecting sensitive data from organizations such as information security related data via interviews poses several challenges as sometimes said organizations may be reluctant to share incidents or reveal enough information that may tarnish its reputation, which in turn, affect the research outcomes due to limited and restricted accounts around the phenomenon being investigated. In order to overcome these challenges, Myers and Klein (1999) advise to develop a strong relationship and trust with interviewers by reassuring participants that the research design is conducted in a sensitive manner and confidentiality and anonymity are always preserved at all times. Due to the fact that the research topic of this study involved confidential topics such as leakage incidents and discussion of internal organizational processes and intellectual property, and sensitive projects, the researcher had to resort to the use of scenarios to elicit responses that otherwise would have been problematic to obtain, if not, impracticable.

The interview data was gathered during a period of approximately two years where the researcher constantly compared the conducted interviews, as part of the coding process, and in an effort to refine the interview protocol in the early stages, the researcher regularly reworded problematic sections and questions that proved confusing to the research participants in order to improve interviewees' engagement and elicit responses and information about the relevant topics pertaining to the research objective of this study. As Corbin and Strauss (2014) advise, the constant interaction with the data leads to the emergence of the relevant questions. This process, in turn, implied that interview questions were modified to allow for changes resulting in a more structured and refined protocol that was conducted in an iterative fashion during the course of the data collection process.

The data analysis process was initiated after the first interview, and the process of constant comparison was employed to refine the interview protocol and questions to enhance coherence and more accurately investigate the key concepts emerging from the data. This iterative process allowed for the vetting of ideas and the enhancement of the validity and reliability of interpretations. In total twenty interviews were conducted with information security and knowledge managers who belonged to knowledge intensive organizations in which mobile policies were in place and the use of mobile devices was predominant as part of their knowledge work.

Most interviews were conducted face to face and in-situ (inside the participant's organization), when possible, to improve validity and enhance the context-rich research interaction between the participant and the researcher, this situation also facilitated the note-taking process and the development of journal and memo notes that assisted in the triangulation strategy. By writing memos and field notes about the interviews, participant's attitudes and natural environment, the researcher developed a better understanding of the phenomena being researched and improved validity through the convergence of information from different sources, which, in turn, contributed to the verification of findings and results (Patton, 1999).

The interviews lasted on average one hour, with individual variability in the duration, spanning from 45 minutes (shortest interview) to one and a half hours (longest interview). Although the researcher conducted the interviews as consistently as practicable, the semi structured nature of the interview meant that they remained flexible and on occasions, particular topics were more emphasized to the detriment of others. Occasionally, some participants explicitly requested not to be recorded due to confidentiality concerns or fear of repercussions, in those specific cases, the researcher wrote notes during the interview with the key points for each question and revisited them with the interviewee afterwards to confirm accuracy. In circumstances like the one just described, some authors suggest avoiding recording altogether as the best approach (Mbonye et al., 2013; Rutakumwa et al., 2020; Schulkind et al., 2016), instead using interview scripts from field notes becomes the preferred option. Particularly, when participants' concerns and fears in relation to being recorded may affect their engagement, their answers to the questions and the outcome of the interview. Hence, as previously mentioned, in these cases, the researcher opted for field notes in lieu of audio recordings.

The interview protocol consisted of semi-structured, open-ended, discovery-oriented questions (Myers & Newman, 2007). The data collection process from the interviews focused on information security and knowledge management topics, including organizational knowledge and information, different types of policies, risk management strategic documentation, different types of controls, use of mobile devices for knowledge work, contextual factors, and organizational capabilities (Refer to *Appendix D - Interview Protocol* for the interview protocol). In the cases where a relevant organizational document existed and supplied by the participant to the researcher, the aforementioned document constituted a secondary source of primary data collection, improving internal construct validity through triangulation(Creswell, 2013; Neuman, 2006). As previously mentioned, most interviews, when authorized by participants, were audio-recorded with permission and subsequently these interviews were intelligently transcribed instead of verbatim, that is, transcribing research subjects' words, but making an interpretation to exclude pauses, status, and filler words and potentially cleaning up the grammar (Halcomb & Davidson, 2006; McLellan et al., 2003; McMullin, 2021). Furthermore, the intelligent transcription approach allowed the researcher to eliminate irrelevant details from the interview transcription including slang, stammers, stutters, non-standard language (*gonna, ain't, 'cause*), and any other form or detail describing disruptions in speech, such as throat clearing, and coughing. In other words, the goal was to eliminate phrases that add no value to the discussion's research topic. The transcriptions and the provided supplementary documentation that participants supplied were copied into Microsoft Word and Microsoft Excel for a qualitative analysis process, and the content manually coded into concepts, themes and dimensions as guided by the Gioia (2012) methodology , which expands on the constant comparative method as outlined within the grounded theory methodology defined in the seminal book *The Discovery of Grounded Theory* authored by Barney G. Glaser and Anselm L. Strauss (1968).

## 4.6.1 Supplementary document analysis

Gathering and analyzing documents and artefacts from the field represents an invaluable source of data, particularly for understanding complex and abstract topics such as an organization's culture, strategy, information security and knowledge management practices, policies, processes and procedures, which , more often than not, are contained in organizational repositories and documents (Creswell, 2013; Yin, 2003). Even though more demanding and time consuming than other collection methods, document collection and examination will often result in additional primary data and secondary data available to verify and validate other forms of data collection and observation such as interviews (Neuman, 2006; Yin, 2011). Furthermore, combining the collection and examination of documents with interviews also provides an alternative way to expedite and simplify the interview process, as the researcher can focus on continuing the interview without interrupting the participant to ask for clarification on different pieces of organizational related information such as processes , locations, titles and names, for example (Myers & Newman, 2007).

The process of supplementary document analysis followed a similar approach as that of the interview analysis, i.e., open coding, axial coding, and selective coding which contributed to validating and verifying the data collected from the research subjects. While the coding process for the interviews usually took more than one iteration, the supplementary documentation coding process, on the other hand, often resulted in one or two iterations at most in some rare cases, likely due to the refined final state of the documentation provided such as policies, processes and procedures which unlike the text of the interviews, lacked most of the subjective nature and contextual approach present in the interview transcripts.

# 4.7 Evaluation of the Research Method

The current qualitative body of knowledge in Information Systems research recommends different approaches to evaluate the rigor in qualitative studies both for the positivist (Neuman, 2006; Yin, 2003, 2017) and interpretivist (Gioia et al., 2012; Myers & Klein, 1999) paradigms.

To ensure the methodological rigor of this exploratory qualitative research, the framework proposed by Carson et al (2001) was employed. This comprehensive set of techniques incorporates various suggestions from the existing literature, with a specific focus on establishing the trustworthiness of the study's findings. By utilizing this framework, the study aimed to enhance the credibility, dependability, confirmability, and transferability of the research results. Table 4-4 lists the thirteen techniques for improving trustworthiness (Carson et al., 2001) and the approaches used in this research are correlated with the techniques discussed below.

**Table 4-4. Trustworthiness of the findings. Adapted from Carson et al (2001)**

| Technique | Application in this study | Status |
|---|---|---|
| Researching in the field | Interviews were conducted in participant's organizations | ✓ |
| Purposive sampling | As indicated in section 4.4.6 Selection and Recruitment of Participants, snowballing sampling was used which is a type of purposive sampling. | ✓ |
| Cross-context comparison of results | As indicated in section 4.4.8 Interview and Supplementary Data, interview data insights were compared with the results of supplementary documentation analysis data insights, when this was supplied. | ✓ |
| Depth/intimacy of interviewing | The flexible nature of the semi-structure interviews lent itself to develop rapport with participants and encourage the discussion and reflexion of the research topics | ✓ |
| Negative case analysis | Participants were actively encouraged to disprove statements and contrasting opinions as a way to elicit discussion and explore alternative views. | ✓ |

| Technique | Application in this study | Status |
|---|---|---|
| Debriefing by peers | Debriefing sessions were held with fellow academics right after interviews and with former supervisor to reflect on the process and refine the interview technique | ✓ |
| Maintaining a journal | The researcher recorded notes and memos throughout the study to document the research findings at various stages. | ✓ |
| Multiple interviewers | Although interviews were conducted by the researcher only, the interview notes were discussed with the supervisors right after the same day, or the day after the interview to improve inter-coder reliability | ✗ |
| Present the findings to respondents | Preliminary findings and insights from the interviews were emailed to participants to assess comprehensiveness and reliability. Some key participants replied and their comments were taken into consideration | ✓ |
| Data triangulation | As mentioned in section 4.4.8 Interview and Supplementary Data, supplementary documentation (policies, procedures, reports and organizational standards) was also provided by the organizations and therefore examined for triangulation purposes | ✓ |
| Draft review by respondents | This was not conducted due to time constraints | ✗ |
| Independent audits | The execution of this task was not possible due to constraints such as confidentiality, privacy, and the unavailability of an external independent auditor. | ✗ |
| Prolonged and persistent observation | As participant observation was not incorporated in the research design, this technique is not relevant to the study. | N/A |

As illustrated in Table 4-4, this study is compliant with the relevant techniques except for the techniques associated to multiple interviewers, draft review by informants, and independent audits.

Further to this, this research project also implemented recommended methods for data collection and analysis in order to further improve the trustworthiness of the findings, including:

1. The use of other sources of data to triangulate perspectives (Eisenhardt, 2016; Neuman, 2006);

2. The development of thick descriptions, as part of the Gioia methodology, including participant feedback, when possible, to enhance the richness of the context and ensure the quality and validity of the interpretations, as well as a thorough data collection and analysis process. (Gioia et al., 2012; Myers & Klein, 1999; Myers & Newman, 2007);

3. The review of the emergent models and constructs by other fellow student researchers, not involved in the study, to scrutinize ideas, and to reinforce the validity and reliability of interpretations (Guba, 1990; Guba & Lincoln, 1994).

## 4.8 Summary

This chapter presented and elaborated on the research methods and design of the research. This study is exploratory in nature and seeks to stimulate future theory building and develop a framework to guide organisations in implementing comprehensive and effective knowledge leakage mitigation strategies for mobile devices. By adopting an interpretive qualitative approach, this study aims to answer the research question addressed in this research project. For that reason, the research was conducted in multiple stages, first, a conceptual model was developed based on the current literature and state of the art (chapter 1 and 2). Second, a model was developed based on the salient concepts of information security management, knowledge management, and mobile computing literatures applying the *Integrative model of IT business value* framework (chapter 3). Third, an empirical study where twenty semi-structured interviews with information security and knowledge managers, as well as documents were analysed in order to aim in the validation and refinement of the conceptual model. The following chapter discusses the findings of the study.

# Chapter 5. FINDINGS

T he previous chapter outlined the research design and methodology, in that chapter, the researcher discussed and justified the methodological approach to address the overarching research question of how knowledge intensive organizations can mitigate the risk of knowledge leakage caused by the use of mobile devices. Following on the previous section, this chapter[7] presents the results and key findings based on the qualitative data analysis conducted, which as previously stated, contains the results of the empirical study and validates the conceptual research model.

## 5.1 Research Participants' Demographics

Twenty subjects were recruited for this research project. As discussed in the methodology chapter (chapter four), the purposive sample, also known as judgmental, selective or subjective sample, is a form of non probability sampling, in which the researcher relies on their judgement and convenience when choosing members of the general population to participate in a study (Neuman, 2006). This sampling approach is widely used in qualitative research for the identification and selection of information rich cases related to the phenomenon of interest. The demographics of the twenty research participants is summarized in Table 5-1. Demographic Information about the Research Participants below. In Table 5-1, the purposive sample contains multiple attributes (table columns) such as role, type of industry, knowledge assets, experience in years, self-reported knowledge leakage risk and organizational size (using headcount as a proxy). These attributes are defined and further explained in the following sub-sections.

---

[7] Sections of this chapter have been published in the following publications:
- Agudelo-Serna, C. A., Bosua, R., Ahmad, A., & Maynard, S. (2017). Strategies to Mitigate Knowledge Leakage Risk caused by the use of mobile devices: A Preliminary Study.
- Agudelo-Serna, C. A., Bosua, R., Ahmad, A., & Maynard, S. B. (2018). Towards a knowledge leakage mitigation framework for mobile devices in knowledge-intensive organizations.

**Table 5-1. Demographic Information about the Research Participants**

| ID | Role | Industry | Knowledge Asset(s) | Experience (Years) | Self-Reported Leakage Risk | Size (Headcount) |
|---|---|---|---|---|---|---|
| CIO1* | Chief Information Officer | Government | People, Policy, Processes, Strategy | 10+ | Medium | Enterprise |
| SM1 | Security Manager | Banking | Strategy, Processes, | 15+ | Medium | Large |
| CISO1*# | Chief Information Security Officer | Consultancy | People, Intellectual Property, Processes, Product, Methodology | 20+ | High | Large |
| SM2* | Security Manager | IT Provider | Product, Methodology | 10+ | Low | Medium |
| CTO1# | Chief Technical Officer | IT Services | People, Intellectual Property, Product, Methodology | 15+ | High | Medium |
| SM3# | Security Manager | Insurance | People, Processes, Methodology | 10+ | High | Enterprise |
| CISO4 | Chief Information Security Officer | Health Care | Product, Methodology | 10+ | Medium | Medium |
| CSM# | Cyber Security Manager | Consultancy | People, Intellectual Property, Processes, Methodology | 15+ | High | Large |

| ID | Role | Industry | Knowledge Asset(s) | Experience (Years) | Self-Reported Leakage Risk | Size (Headcount) |
|---|---|---|---|---|---|---|
| CISO2 | Chief Information Security Officer | Telecommunications | Product, Methodology, Intellectual Property, People | 15+ | High | Enterprise |
| CIKO# | Chief Information and Knowledge Officer | Government | Policy, Strategies, Intellectual Property | 10+ | Medium | Enterprise |
| CKO1 | Chief Knowledge Officer | Food | People, Intellectual Property, Process, Product | 15+ | High | Medium |
| KM1 | Knowledge Manager | Health Care | Client, Product, Processes | 10+ | Medium | Medium |
| KM2# | Knowledge Manager | Government | Policy, Strategy | 10+ | Medium | Enterprise |
| CKO2 | Chief Knowledge Officer | Consultancy | Intellectual Property, Product, Methodology, People | 15+ | High | Large |
| KM3 | Knowledge Manager | Consultancy | Intellectual Property, Processes, Methodology, People | 10+ | High | Large |
| CKO3 | Chief Knowledge Officer | Government | Processes, Strategy, Intellectual Property | 15+ | Medium | Enterprise |
| KM4* | Knowledge Manager | Health Care | Product, Processes | 5+ | Low | Medium |
| KM5 | Knowledge Manager | Education | Methodology, Research findings | 10+ | Medium | Large |
| KM* | Knowledge Manager | Not-for-Profit | Client relationship, processes | 15+ | Low | Medium |
| CISO3 | Chief Information Security Officer | Government | Processes, Product, Methodology, Intellectual Property, Strategy, people | 15+ | High | Enterprise |

* Participant provided supplementary documentation.

# Participant requested not to be audio recorded.

## 5.1.1 Research Participants' Role within the Organization

The characteristics of the purposive sample for this research project are summarized in terms of different attributes and criteria. The first attribute relates to the participant role within the organization, one of the selection criteria stated that participants must have held at least a managerial position, as one of the study's aims required the understanding of mitigation strategies, that is, processes and policies at the strategic level as opposed to tactical or operational level roles. Out of the twenty participants, ten held managerial positions in the knowledge management area, and ten in the information security space. In the knowledge management category, 20% (four) belonged to executive positions and 30% (six) to managerial positions. As for the information security category, 30% (six) held executive positions, and 20% (four) managerial positions (See Table 5-2 and Figure 5-1).

**Table 5-2. Research Participants' Role**

| Role Level | Area | Number | Percentage |
|------------|------|--------|------------|
| Executive | Knowledge Management | 4 | 20% |
| Managerial | Knowledge Management | 6 | 30% |
| Executive | Information Security | 6 | 30% |
| Managerial | Information Security | 4 | 20% |

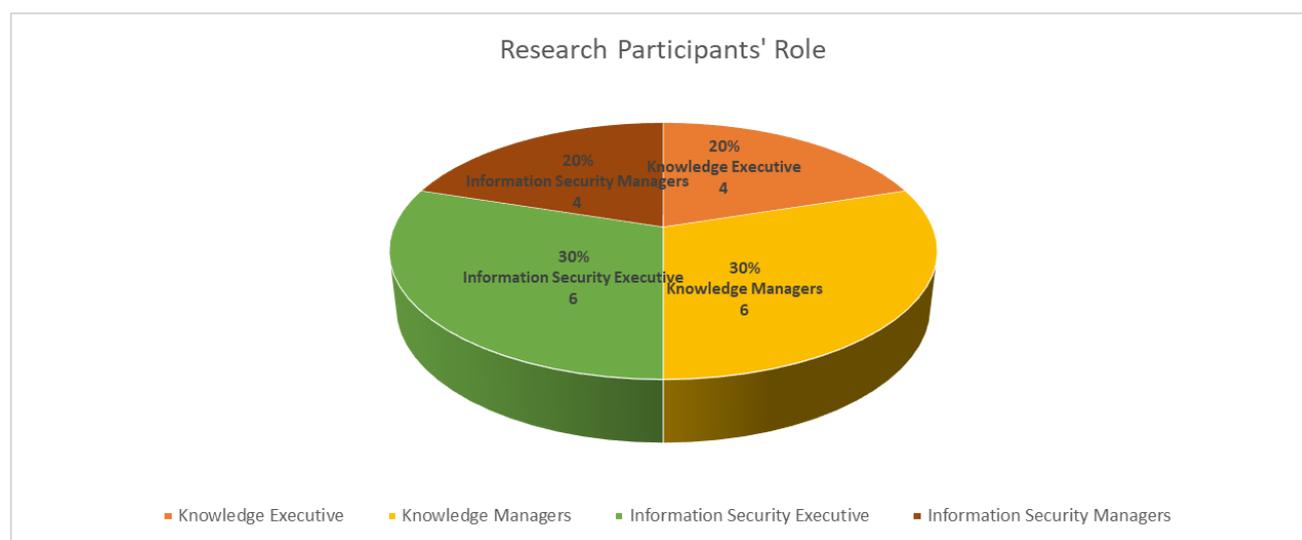

**Figure 5-1. Research Participants Role distribution**

## 5.1.2 Research Participants' Type of Industry

As stated earlier in section *4.4.6 Selection and Recruitment of Participants*, due to the nature of the purposive sample employed in this study and the research questions, only some industry sectors were included. The industry sectors that were included in the study were chosen chiefly because of three reasons: 1) Knowledge Intensive reliant organizations 2) Higher adoption of mobility and 3) Accessibility to interviewees. So, for example, sectors such as agriculture, construction and manufacturing were not part of the cohort of participants due to the fact that, first, the researcher was unable to contact representatives of such industries. And second, some organizations are more likely to adopt mobile policies than others, for example, consulting and technology organizations compared to agriculture organizations. The researcher contacted only knowledge intensive organizations that implemented and displayed sanctioned mobility and mobile device policies in place. Therefore, for the second demographic attribute that relates to type of industry, only **8** types of industry sectors were included in this study divided as follows: out of the twenty research subjects, 25% (five participants) belonged to government, 20% (four participants) worked in the consultancy services organizations, 15% (three participants) in technology services, 15% (three participants) in health care organizations, 10% (two participants) in banking, 5% (one participant) in the food industry, 5% (one participant) in education and 5% (one participant) in the not for profit sector (See Table 5-3 and Figure 5-2).

**Table 5-3. Research Participants' Type of Industry**

| Type of Industry | Number | Percentage |
|---|---|---|
| Government | 5 | 25% |
| Consultancy Services | 4 | 20% |
| Technology Services | 3 | 15% |
| Health Care | 3 | 15% |
| Banking | 2 | 10% |
| Food | 1 | 5% |
| Education | 1 | 5% |
| Not For Profit | 1 | 5% |

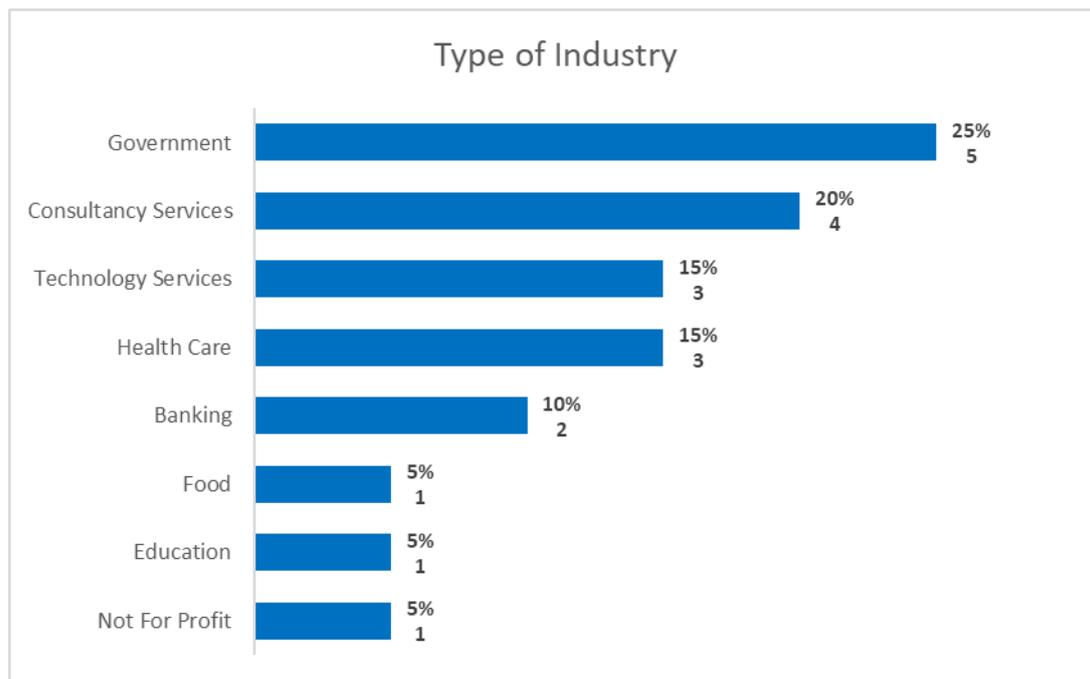

**Figure 5-2. Research Participants' Type of Industry distribution**

## 5.1.3 Reported Organizational Knowledge Assets

During the interviews research participants were asked to name the most critical organizational knowledge assets that warranted protection from competitors, customers, partners, and suppliers within their firms. Overall, organizational knowledge assets mentioned by participants varied from organization to organization, but ultimately fell into the following categories ordered from most common to least common: people – 60% (12 participants), process - 55% (11 participants), methodology – 55% (11 participants), product – 50% (10 participants), intellectual property – 50% (10 participants), strategy -25% (5 participants), policy – 15% (3 participants), client information – 10% (2 participants), and research findings – 5% (1 participant). See Table 5-4 and Figure 5-3 for summarized information on the reported organizational knowledge assets.

**Table 5-4. Reported Organizational Knowledge Asset**

| Identified Organizational Knowledge Asset | Number | Percentage |
|---|---|---|
| People | 12 | 60% |
| Process | 11 | 55% |
| Methodology | 11 | 55% |
| Product | 10 | 50% |
| Intellectual Property | 10 | 50% |
| Strategy | 5 | 25% |
| Policy | 3 | 15% |
| Client Information | 2 | 10% |
| Research Findings | 1 | 5% |

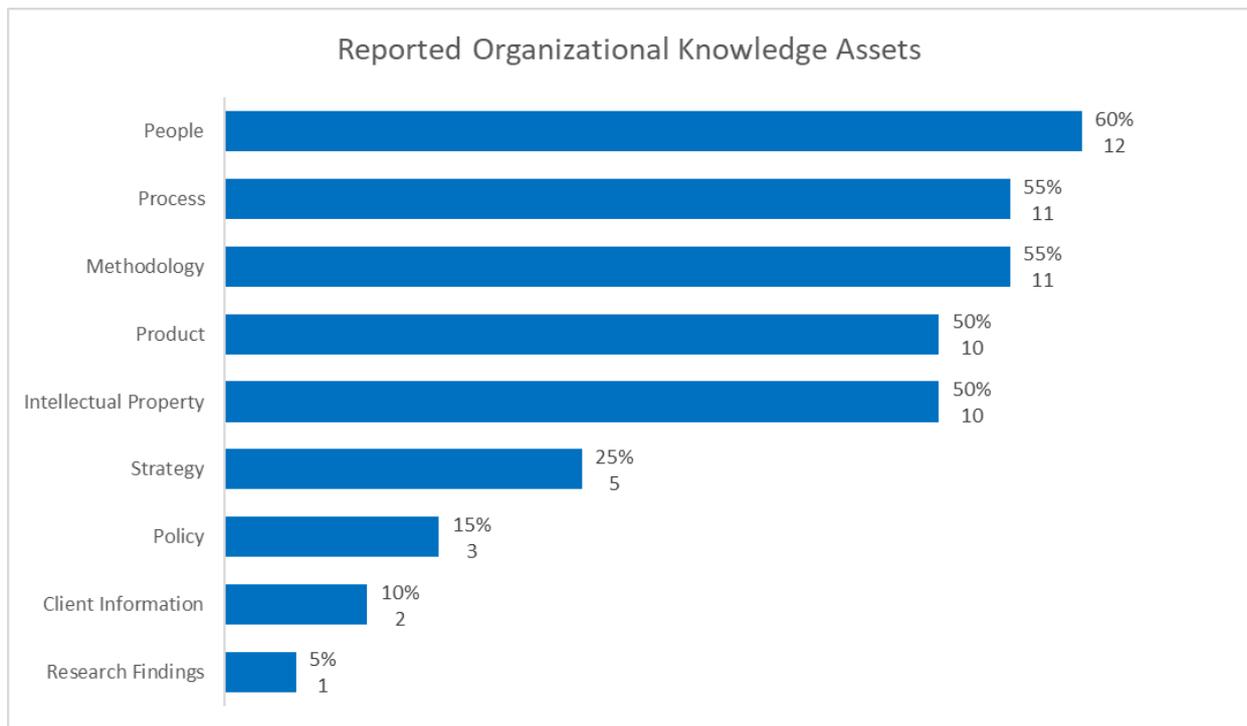

**Figure 5-3. Reported Organizational Knowledge Asset Distribution**

## 5.1.4 Research Participants' experience

As previously described, the selection criteria for participants dictated that they must have had at least five years of experience in the relevant field which meant that research subjects were considered experts in their respective professional areas by industry standards, which coupled with the managerial position requirement signified that these candidates were suitable for the parameters of this research study.

Overall, experience spanned from five years to more than twenty years of experience, with most of the research subjects (90%) with experience ranging between ten and nineteen years of experience. Out of the twenty participants, 45% (9 participants) had between ten and fourteen years of experience, 45% (9 participants) had between fifteen and nineteen years of experience, 5% (1 participant) had more than twenty years of experience, and 5% (1 participant had between 5 and nine years of experience. See Table 5-5 and Figure 5-4 for a summarized information on the participants' years of experience.

**Table 5-5. Participants' Experience in Years**

| Participants' range of Experience in years | Number | Percentage |
|---|---|---|
| 5 - 9 | 1 | 5% |
| 10 - 14 | 9 | 45% |
| 15 - 19 | 9 | 45% |
| 20 + | 1 | 5% |

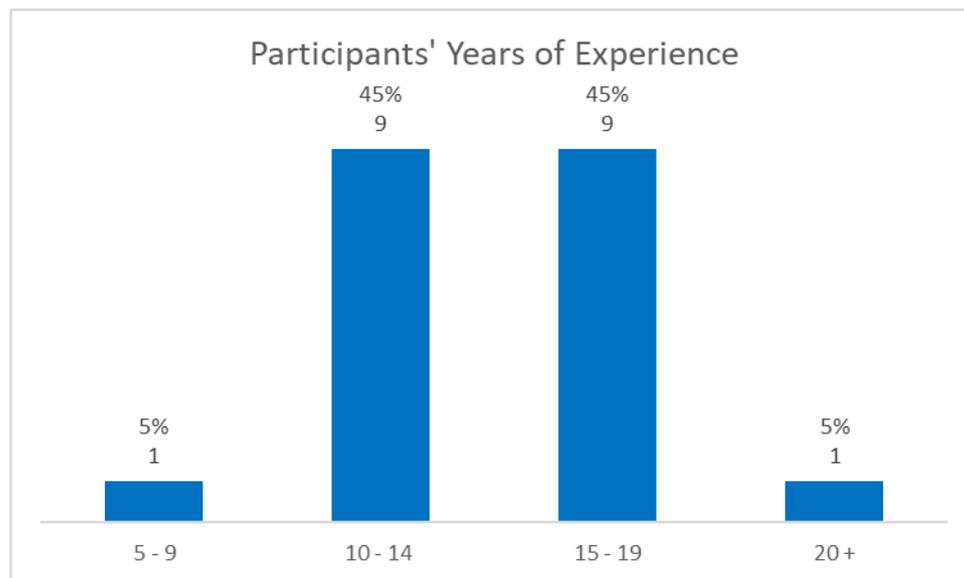

**Figure 5-4. Participants' Years of Experience Distribution**

## 5.1.5 Participants' Self-Reported Leakage Risk

During the interviews research participants were asked to qualify the knowledge leakage risk within their organizations based on their perceptions of exposure amongst the following risk levels: high, medium, and low. The definition of knowledge risk and risk level was based on the standard definition of risk in the ISO 27001 standard within the risk management section (27005:2011, 2011) which, in turn, extends from the definition of the ISO 31000 standard for Risk Management – Principles and Guidelines. In this ISO 31000 standard, risk is defined as the *effect of uncertainty on objectives* (Australia & New Zealand, 2009). ISO 27001 standard leverages this definition to use a more substantial and specific meaning in the context of information security: *Information security risk is associated with the potential that threats will exploit vulnerabilities of an information asset or group of information assets and thereby cause harm to an organization* (27005:2011, 2011, sec. Risk). Furthermore, risk — and by extension knowledge leakage risk — is defined as a function of two dimensions, i.e., likelihood (or probability) and impact (or consequence), as was indicated in the literature review chapter (chapter two). Therefore, participants were asked to select their perceived knowledge leakage risk in accordance with the likelihood of a knowledge leakage incident and the impact should such an incident occur within their organizations as indicated in Figure 5-5. It is worth noting that this was a subjective appraisal of the participant's perceived leakage risk and was, by no means, measured or verified by the researcher during the research study, this subjective assessment was only used as a proxy to determine the organization's risk profile in the context of the conceptual research model defined in chapter three.

It is important to clarify that the risk matrix depicted in *Figure* 5-5 below, is by no means a research contribution. Rather, it was utilized as a tool to standardize the perceived risk of participants, providing a common language for discussing and understanding the potential impact and likelihood of different risk ratings. It served as a means to facilitate the conversation around risk in the context of knowledge leakage through mobile devices. Due to the self-reported nature of the risks, the researcher acknowledges the potential inaccuracies as a result of participant bias.

To mitigate this, the study employed rigorous interview protocols and triangulation methods to ensure the reliability and validity of the data. The participants were also selected based on their expertise and experience in the field, which further enhances the credibility of their responses.

The study claims no means to provide an objective measure of risk, but rather seeks to understand the perceptions and experiences of individuals within their specific organizational contexts. Therefore, the potential bias does not undermine the value of the insights gained, but rather adds to the richness and depth of the data.

In general, out of the twenty participants, 50% (10 participants) deemed their knowledge leakage risk level as high, 35% (7 participants) considered their risk level as medium and the remaining 15% (3 participants) regarded their risk level as low (See Table 5-6 and Figure 5-6).

| | | Negligible | Minor | Moderate | Significant | Severe |
|---|---|---|---|---|---|---|
| **Likelihood** | Very Likely | Low | Medium | High | High | High |
| | Likely | Low | Medium | Medium | High | High |
| | Possible | Low | Low | Medium | Medium | High |
| | Unlikely | Low | Low | Medium | Medium | Medium |
| | Very Unlikely | Low | Low | Low | Medium | Medium |

*(Column header group: Impact)*

**Figure 5-5. Knowledge Leakage Risk as a function of Likelihood and Impact**

**Table 5-6. Participants' Self-Reported Leakage Risk Level**

| Self-Reported Risk Leakage Level | Number | Percentage |
|---|---|---|
| High | 10 | 50% |
| Medium | 7 | 35% |
| Low | 3 | 15% |

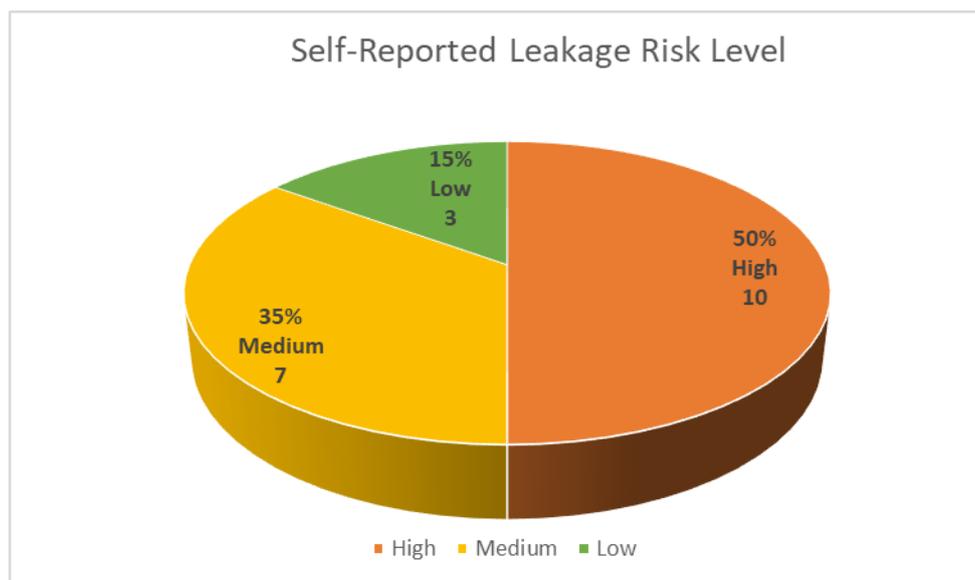

**Figure 5-6. Participants' Self-Reported Leakage Risk Level**

## 5.1.6 Research Participants' Organization Size

Throughout the interviews, participants were requested to supply the estimated number of employees within their organizations. In this analysis of the impact of the organization size, the researcher examined the employee headcount supplied by the research subjects as a proxy for the organization's size. Subsequently, In order to use a standardized approach to determine the right classification size, the researcher employed the standard definition as stated by the Australian Bureau of Statistics (Statistics, 2020) that uses the classification displayed in Table 5-7 to define an organization's size based on the number of employees.

**Table 5-7. Organization Size based on the Australian Bureau of Statistics**

| Classification | Number of Employees |
|---|---|
| Micro | 0-4 |
| Small | 5-19 |
| Medium | 20-199 |
| Large | 200 -1000 |
| Enterprise | 1000+ |

According to the classification in Table 5-7, the researcher assigned the right classification size to the participant's organization based on the employee headcount answer provided by the research subjects. Out of the twenty interviewed subjects, 35% (7 participants) reported that their organizations employed between 20 and 199 people, this category of organizations were assigned the medium classification. Another 35% (7 participants) stated that their organizations contained more than 1000 employees, such organizations were allocated a classification of enterprise. Finally, the remaining 30% (6 participants) informed that their organizations hired between 200 and 1000 employees, these organizations were designated as a large organization. Micro and small size organizations were not represented in this purposive sample. See Table 5-8 and Figure 5-7 for the summarized information on the participants organizations' sizes.

**Table 5-8. Participants' Organization Size**

| Organization Size | Number of Organizations per Size | Percentage |
|---|---|---|
| Micro | 0 | 0% |
| Small | 0 | 0% |
| Medium | 7 | 35% |
| Large | 6 | 30% |
| Enterprise | 7 | 35% |

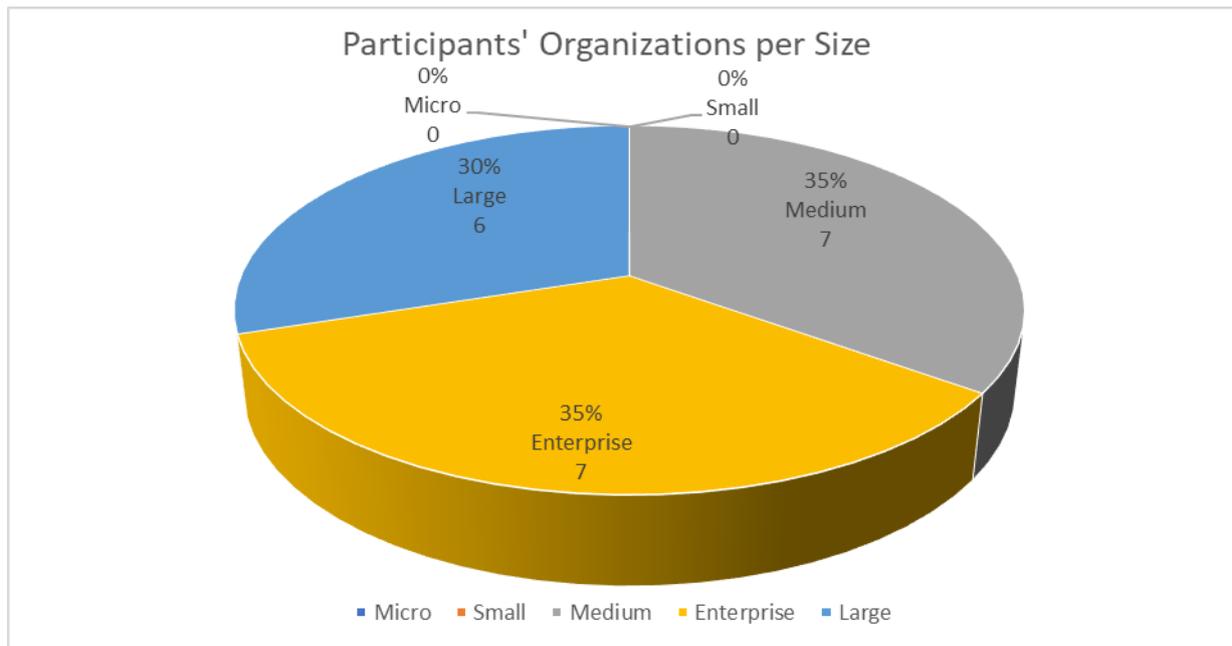

**Figure 5-7. Participants' Organization Size Distribution**

## 5.1.7 Data Structure

For the data structure, the researcher developed the structure based on the first order concepts (initial open codes) as a result of the interviews and document reviews leveraging the qualitative data that indicated the purpose of different mitigation strategies used by organizations to combat knowledge risk caused by mobile devices. Initially, **269** codes were identified following the procedures described in the methodology chapter (chapter four). After reviewing and contrasting the first order concepts and codes, and as interviews progressed, constant comparison amongst different interviewees' terms were conducted in subsequent iterations leading to the reduction of the number of codes due to similarity, relevance to the research aim, duplicates and merging of concepts (J. M. Corbin & Strauss, 1990; Gioia et al., 2012). This process resulted in **106** as a final number of concepts. For the first order concepts, the researcher used, as much as possible, in-vivo codes — participants terms or informant centric codes when these terms described the phenomenon — otherwise, literature terms (chapter two) were employed based on the research aim and questions (J. Corbin & Strauss, 2014). The list of first order concepts and their frequency, organized by frequency in descending order — from most common concepts to least common ones — is illustrated in Figure 5-8, Figure 5-9 and Figure 5-10.

In the following sections of the data structure, the researcher explains how the first order concepts (informant centric terms) were grouped into second order themes (theory centric constructs) using extracts from the interviews and supplementary documentation data, and guided by the research model defined in chapter three as the underlying framework to unify the themes. Subsequently, the second order themes were distilled into theoretical dimensions based on the literature and finally, additional consultation with the literature to refine the articulation of the emergent concepts and relationships was conducted as indicated by Gioia (2012) in his methodology.

| Number | First Order Concepts | Frequency | Visual |
|--------:|------------------------|----------:|:------:|
| 1 | Risk Management | 16 | |
| 2 | Policies and Procedures (Knowledge management and Information Security) | 14 | |
| 3 | Endpoint Security for Mobile Devices | 13 | |
| 4 | Information Systems (Databases, Datawarehouses, Repositories) | 11 | |
| 5 | Mobile Device Management (MDM) | 11 | |
| 6 | Limiting/blocking access | 11 | |
| 7 | Mobile Security Policies (BYOD, CYOD, COPE, COBO) | 11 | |
| 8 | Meeting Compliance requirements | 11 | |
| 9 | Organizational Mobility Policy | 11 | |
| 10 | Firewall | 10 | |
| 11 | Multi-Factor Authentication | 10 | |
| 12 | Mobile Detection and Response | 10 | |
| 13 | Employee motivation and rewards | 10 | |
| 14 | Virtualization | 10 | |
| 15 | Identity and Access Management (IAM) | 10 | |
| 16 | User and Entity Behavioural analytics - UEBA (User profiling) | 10 | |
| 17 | User Activity Monitoring | 10 | |
| 18 | Anti-Virus and Anti-Malware | 10 | |
| 19 | Training, Awareness education | 9 | |
| 20 | IPR (Patents, Copyrights, Trademarks, Tradesecrets) | 9 | |
| 21 | Legal strategies (clauses and contracts, non-compete agreements, NDA) | 9 | |
| 22 | Mobile Application Management (MAM) | 9 | |
| 23 | Security Information and Event Management (SIEM) | 9 | |
| 24 | Perimeter security | 9 | |
| 25 | Separation of Duties | 9 | |
| 26 | Industry standards (ISO, NIST, Risk) | 9 | |
| 27 | Simulations in Organizations | 9 | |
| 28 | Policy Compliance check | 9 | |
| 29 | Human Resource Management | 8 | |
| 30 | Artificial Intelligence tools (Machine Learning and Deep Learning) | 8 | |
| 31 | HR Screening process | 8 | |
| 32 | VPN | 8 | |
| 33 | Device Profiling | 8 | |
| 34 | Employee rotation | 8 | |
| 35 | Role based access | 8 | |
| 36 | Trust | 8 | |

**Figure 5-8. List of First Order Concepts and Frequency (1/3)**

| Number | First Order Concepts | Frequency | Visual |
|---|---|---|---|
| 37 | mentoring / Coaching | 8 | 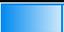 |
| 38 | Data Loss/Leak Prevention (DLP) platforms | 8 | 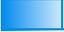 |
| 39 | Device, OS and Application Whitelisting | 8 | 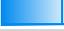 |
| 40 | Knowledge Sharing | 8 | 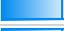 |
| 41 | Machine Learning Threat Protection | 8 | 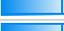 |
| 42 | Deterrance (Formal Punishment) | 8 | 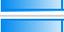 |
| 43 | Decoy employs | 8 | 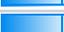 |
| 44 | Mobile Analytics | 8 | 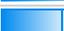 |
| 45 | Information Exfiltration Protection | 8 | 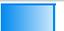 |
| 46 | Informal Rules (ad hoc processes) | 7 | 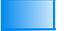 |
| 47 | Encryption | 7 | 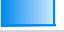 |
| 48 | Information Rights Management (IRM) | 7 | 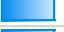 |
| 49 | Knowledge Transfer | 7 | 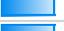 |
| 50 | Contextual access | 7 | 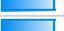 |
| 51 | Adopting a least-privilege strategy | 7 | 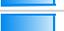 |
| 52 | Micro-perimeters | 7 | 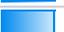 |
| 53 | Gamification | 7 | 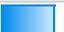 |
| 54 | Knoledge Culture | 7 | 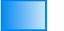 |
| 55 | Need-to-know principle | 7 | 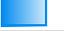 |
| 56 | Knowledge Management | 6 | 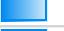 |
| 57 | Recruitment process | 6 | 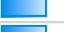 |
| 58 | Knowledge Codification (Explicit Knowledge) | 6 | 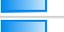 |
| 59 | Threat Intelligence | 6 | 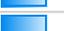 |
| 60 | Retention Management | 6 | 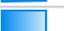 |
| 61 | Contextual authentication | 6 | 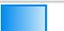 |
| 62 | Microsegmentation | 6 | 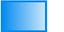 |
| 63 | Noise generation | 6 | 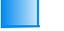 |
| 64 | Motivation Rewards | 6 | 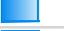 |
| 65 | Risk Scoring system | 6 | 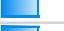 |
| 66 | Community of practice (virtual community) | 5 | 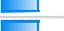 |
| 67 | Sandboxing (environments, devices, documentation) | 5 | 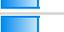 |
| 68 | Conditional Access | 5 | 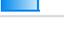 |
| 69 | Defense-in-depth | 5 | 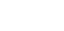 |
| 70 | Biometric authentication | 5 | 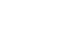 |
| 71 | Identifying people of high security concern | 5 | 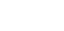 |

**Figure 5-9. List of First Order Concepts and Frequency (2/3)**

| Number | First Order Concepts | Frequency | Visual |
|--------|----------------------|-----------|--------|
| 72 | Knowledge Systems (Groupware, knowledge portals) | 4 |  |
| 73 | Peer mentoring Policy | 4 |  |
| 74 | Layering (device, network, access) | 4 |  |
| 75 | Supply Chain Risk Management (clients, competitors, partners , suppliers) | 4 |  |
| 76 | Identification of critical knowledge assets | 4 |  |
| 77 | Secrecy | 4 |  |
| 78 | User Authentication | 4 |  |
| 79 | Zero Trust | 4 |  |
| 80 | Storytelling | 4 |  |
| 81 | containerization | 4 |  |
| 82 | Knowledge workflow approval | 4 |  |
| 83 | Social media and Darkweb Intelligence gathering | 4 |  |
| 84 | Compartmentalization | 3 |  |
| 85 | Knowledge Governance | 3 |  |
| 86 | Knowledge Classification (Information Classification) | 3 |  |
| 87 | Environmental Analysis | 3 |  |
| 88 | Indoctrination | 3 |  |
| 89 | lead-time on competitors | 3 |  |
| 90 | Fast Innovation Cycle | 3 |  |
| 91 | Keeping knowledge tacit intentionally | 3 |  |
| 92 | Risk assessment based on employee position | 3 |  |
| 93 | Alliance and partnerships with other organizations | 3 |  |
| 94 | Knowledge decontextualization | 3 |  |
| 95 | Tagging | 2 |  |
| 96 | Market Analysis | 2 |  |
| 97 | Competitor Analysis | 2 |  |
| 98 | Informal monitoring of knowledge flows | 2 |  |
| 99 | Complexity | 2 |  |
| 100 | Knowledge Protection Roles | 2 |  |
| 101 | Security Clearances | 2 |  |
| 102 | Deception | 2 |  |
| 103 | Sensitive Compartmentalized Information Facility (SCIF) | 1 |  |
| 104 | counter Intelligence (gathering intelligence) | 1 |  |
| 105 | Misinformation / Disinformation | 1 |  |
| 106 | Offensive Strategy (reconnaissance,espionage, gather information on competitors) | 1 |  |

**Figure 5-10. List of First Order Concepts and Frequency (3/3)**

## 5.1.7.1 The Conceptual Model

As previously discussed in chapter three, the conceptual model has been derived from the literature and it helps to organize the first order concepts in line with the theory centric themes relevant to the research aim and questions of this project (See Figure 5-11).

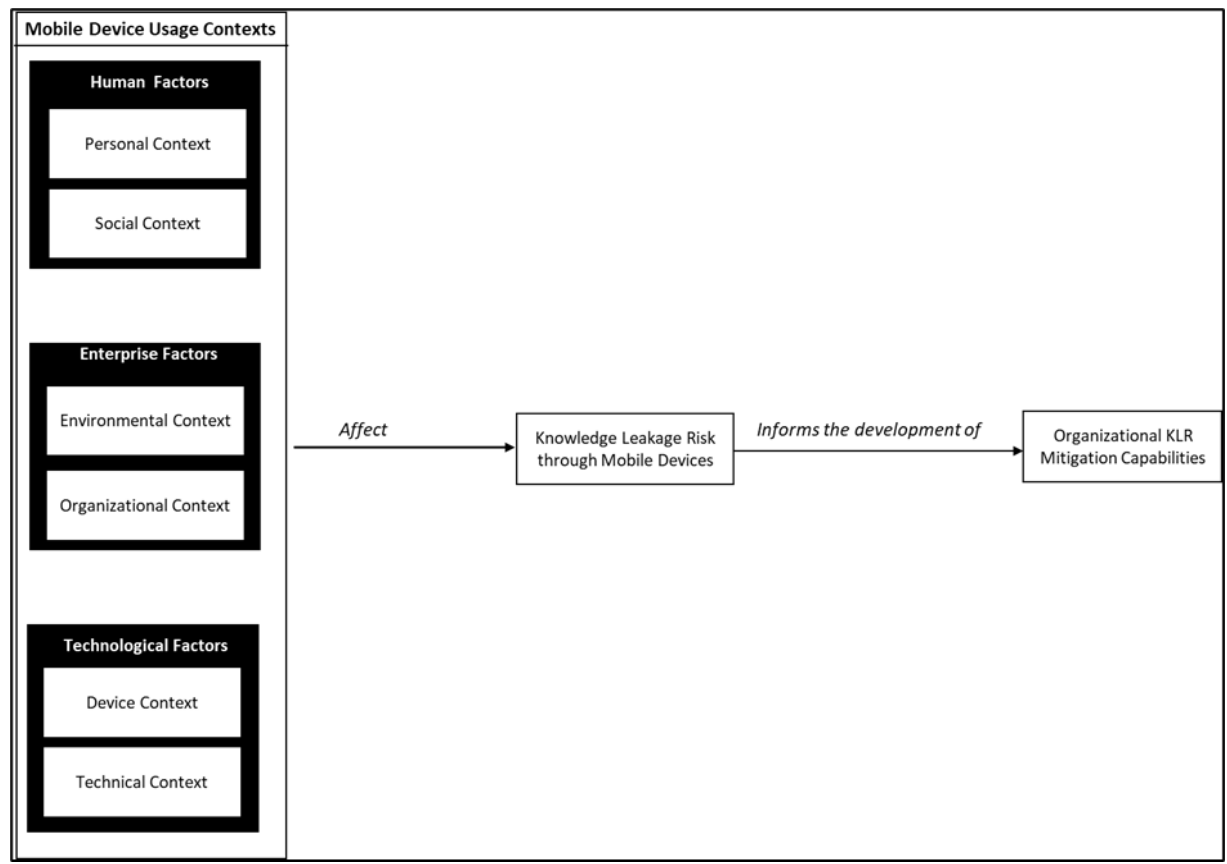

**Figure 5-11. Conceptual Research Model developed in Chapter three.**

In the following sections, the researcher shows how the different constructs in the research model which are grounded in the theory, were used to group the first order concepts into themes and dimensions in line with the Gioia (2012) methodology.

### 5.1.7.2 Personal Context

As mentioned in chapter 3, in the research model, the personal context provides the attributes of cognitive skills and draws on psychological and physiological features such as attitudes, beliefs, experiences, goals, cognition skills, physiological identity, intrinsic motivations amongst others (Abdoul Aziz Diallo, 2012; Ajzen, 1991; L. Chen & Nath, 2008). Overall, during the interviews, knowledge and information security experts mentioned how the human aspect of knowledge and information security is often neglected and barely represented in solutions, products and processes, as one participant noted:

*"…in our area, we always try to focus and address first our people issues,*

*before even looking at the technology side of things, I know that in other*

*departments, the first port of call is to first call ICT and ask for changes or*

*new applications without even asking employees prior to beginning a new project*

*to fix something that could have been easily addressed with just counselling and*

*maybe a simple change in a process.." (SM3)*

In addition to this, interviewees also reported how employees, and more specifically, employee's behaviour, had become a risk to their organizations not only in terms of leakage, but also in terms of sabotage and fraud, and therefore highlighted the need to apply a risk management approach to insiders, i.e., *insider risk management*, as a strategy to deal with insider threats and understand the underlying behavioural indicators that increase the overall risk of malicious and non malicious workers. By understanding that every employee represents an insider risk, but only a few of them will actually turn into insider threats either intentionally or unintentionally — by action, intent or inaction, negligence and mistake, respectively— a risk based model provides the best way to deal with such a liability. In this respect, an information security participant added:

*"…We have been using a workforce cyber intelligence platform called DTex,*

*that has helped us detect and respond to potential internal risk incidents before*

*a breach even occurs thanks to its risk score system that predictively has*

*indicated which employees are being negligent, taking unsafe actions, so for*

*example, we can talk to them before they even email sensitive information to the*

*wrong person. In other cases, it has alerted us of users intentionally trying to*

*bypass security policies, so we can block their network access immediately before*

*they even try to exfiltrate intellectual property …" (CTO1)*

Participants also mentioned how the threat landscape has evolved throughout the years and how difficult it is to secure mobile workers, particularly when working offsite and away from the centralized offices due to the fact that people working away from the office are more likely to fall for scams and leak sensitive intellectual property (Nelson et al., 2017; Vishwanath, 2016). This issue highlights how workers' behaviours change in the context of mobility. As one of the knowledge experts stated:

> *"…we've noticed that staff constantly working outside [ORGNAME*] tend to fall for phishing scams and social engineering attempts more easily than our internal staff. Recently we've seen an uptake in the amount of phishing emails and malware attacks coming specifically from our mobile staff, these statistics are coming from our internal security team and our board is really concerned with the figures" (KM5)*

*Note: *[ORGNAME]* all organization names have been redacted hereafter.

Subjects also illustrated how an organizational culture around security and increase of awareness of the current threats, as well as a motivation and rewards program can positively influence individual's behaviour, and help to foster proper security attitudes and culture in mobile workers, increase compliance and therefore improve the overall security posture of an organization:

> *"… at [ORGNAME] there has been so much traction with a new series of workshops the Learning and Development team have set up for our different business areas. In there [workshops], our employees learn how to be more aware of different things like phishing emails, dodgy links, how to report suspicious websites or documents. The [Learning and Development] team have also designed games and rewards for staff that do really well and that seems to have improved our employees' awareness and make them more proactive. They even have a hall of fame with the names of the staff who are leading in terms of rewards… it's pretty funny, actually" (CIKO)*

Another participant added about the importance of peer mentoring and community for mobile workers (or mobile workforce as they stated):

> *"… for us it's all about our people and making sure they're prepared as we have so much to lose if things go wrong, so to that end, we have created kind of mentoring and coaching programs and communities around different areas and types of staff, so the ones that are usually on the go, we have a mobile workforce community and set of guides, tips and documentation to ensure they are comfortable doing their work in a safe and secure way as they don't have the same support as the staff working at head office and they're mostly working in the field or customer's site" (CKO2)*

Risk management was a recurrent theme throughout the interviews and was also reflected in the people centric comments. As CKO3 recounted:

> *"…onboarding new staff is always challenging and background checking is just one of the steps we conduct along with a risk assessment process we have in place, where we identify possible high risk candidates based on their positions and past history, assess their risk using a risk scoring system, remediate any possible risks usually based on their score and monitor the risk score to check is not increasing…we do similar assessments with existing staff based on their positions…" (CKO3)*

Similarly, training and awareness was used as a way to influence behaviour but also as a way to motivate employees to stay within the firm, in an effort to prevent tacit knowledge leakage via workers leaving the organization. Some of the participants noted:

> *"…our CEO always says: if people are the weakest link then education is the strongest link, this is why we are investing in a comprehensive education program for our staff…so they [people] remain up-to-date and that shows that the company cares for you" (SM2)*

> *"…education for us is very important and we take every opportunity to train our workforce to make them aware of new trends, security wise and also to improve productivity, every month you have access to new courses and training* *options in our LMS based on your job function and your goals in the organization…" (CIO1)*

*"… I like the word indoctrination, I have a background in intelligence and we use it as a way to make sure employees understand our code of ethics and instil the right attitudes and strategies to remain in this job, in our division, we have access to classified information and we need to know the right people are handling it appropriately …" (CISO3)*

### 5.1.7.3 Social Context

The social context provides a social perspective of the interaction of the user with the mobile artefact in the context of a group setting, this context captures the attributes of people (e.g. attitude, skills, and values) and the relationship of these people among one another and within the organization and collective structures such as informal groups and formal work divisions (collective values and norms). In other words, the social context refers to the social environment or socio cultural milieu (social setting) in which the people interact and it includes culture and other people and institutions with whom they interact (Ajzen, 1991; L. Chen & Nath, 2008).

During the interviews, participants reported the importance of culture and the sense of belonging to communities to promote appropriate behaviours aimed to influence mobile workers as a way to mitigate knowledge leakage risk. As one participant stated:

*"… we pride ourselves in having an excellent knowledge culture, we have communities of practice both face to face as well as virtual where we talk about work issues but sometimes other topics are discussed too… I find it really beneficial particularly for the employees that are unable to have physical interactions as they're working at customer sites, sometimes in other states or overseas…" (KM3)*

Research subjects also mentioned the social aspect of peer mentoring programs and storytelling initiatives as a way to foster a secure knowledge culture. As KM4 narrated:

> *"…the good thing about the program that we started a few years ago is that now the employees teach the new staff our values and principles and often I see how they report any suspicious activity before actioning on requests coming from strangers or emails" (KM4)*

Gamification was another strategy used by organizations to create social connection amongst employees and foster a culture where knowledge leakage is minimized through the use of games to show their workers current trends and maintain a high level of awareness. Gamification refers to the use of game mechanics and psychology to drive specific behaviours within a group of people. This technique is widely used in knowledge management (Friedrich et al., 2020) and more recently has been adopted to target mobile knowledge workers as well as described in Abedi et al (2018) and also in Sampaio et al (2019):

> *"… it is exciting to see our employees engaged in competitions across different areas within [ORGNAME] each month you see in SharePoint the list of people who are at the top and the best areas in terms of their [awareness] score, also it's always fun to solve the puzzles the security team create to see how you compare with the rest of the organization…" (CIKO)*

In addition to gamification, other participants referred to the use of knowledge management portals within their organizations' intranets which contained libraries of knowledge in the form of videos, audio, documentation, models, and pictures. These portals acted as knowledge gateways and as knowledge directories, where workers could query knowledge experts or *gurus* within their organization, therefore promoting a way to keep tacit knowledge available via knowledge management systems and at the same time allowing tacit knowledge holders to codify their knowledge (K. Desouza, 2009). In this CKO1 regard narrated:

*"…last year we rolled out our enterprise contentment management system with SharePoint. Our guys created a portal where we've placed sites, subsites and document libraries for different areas that have enabled role based access control, meaning only authorized users can see the content, this is because we have tender information, and product designs we can't afford to lose, we had a data breach the year before and it was a disaster…" (CKO1)*

The data structure representing the personal and social contexts is displayed in Figure 5-12.

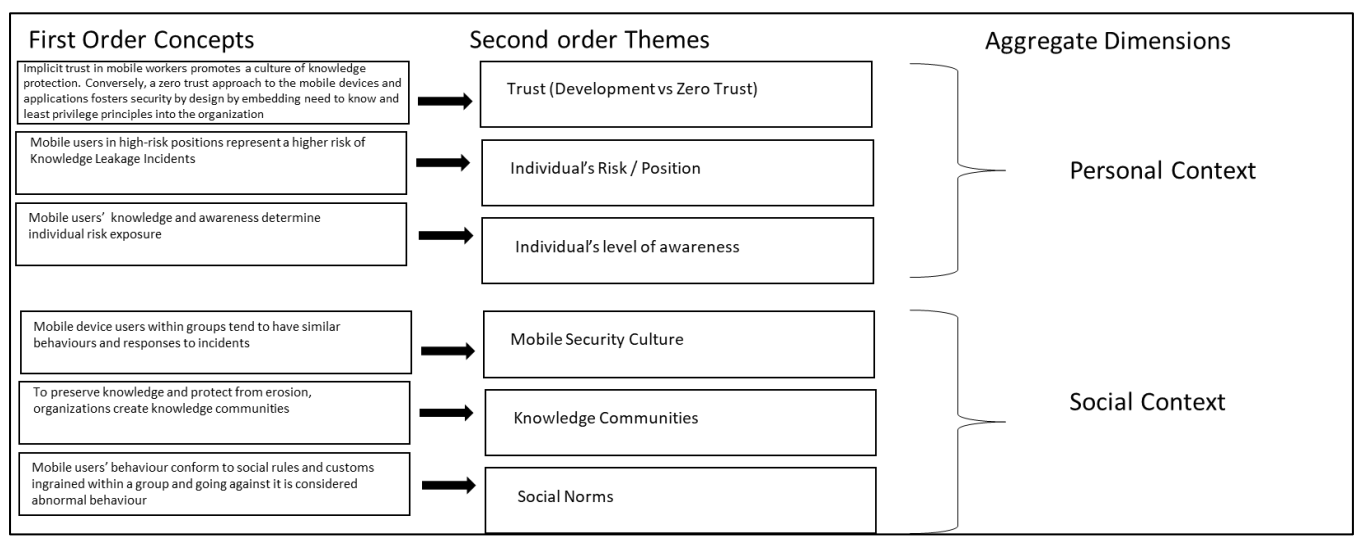

**Figure 5-12 Data Structure for Personal and Social Contexts**

## 5.1.7.4 Environmental Context

The environmental context, defined as the conjunction of the different contexts in the mobile computing literature, includes temporal context, spatial context, social context, technological context, and business context (L. Chen & Nath, 2008; Belko Abdoul Aziz Diallo et al., 2012). From the organizational perspective, it refers to the entities that do not have a direct effect on the operations of the firm but affect the way the organization performs its work (Melville et al., 2004). Examples of such entities in the management literature include competitors, regulators, customers, partners and suppliers.

Research subjects remarked how regulatory environment and legal requirements remained one the main drivers to adopt specific mitigation measures as sometimes "being compliant" seemed more important than actually implementing capabilities that truly contributed to the overall performance of the organization. In this concern, one subject stated:

> *"…in our industry we have to be compliant with the GDPR because we have partnerships with European organizations and this has required a lot of effort and taken up so much time and resources, which has detracted from other projects we were working on prior to this. I can see the benefit long term, but in the short term it has changed our priorities and focus as failure to comply puts us in a bad position open to potential lawsuits and financial liability …"*
>
> *(CISO2)*

Other aspects addressed in the interviews from the environmental context perspective included the understanding of the current threat landscape, specifically, regarding competitors or adversaries and any type of supply chain risk derived from customers, competitors, partners, suppliers and regulators which aligns with the current literature on knowledge leakage risk and different knowledge leakage vectors formerly discussed in the literature review (chapter two). As expressed by

> *"…we have a third-party risk management program with CyberGRX that enables us to assess our clients, vendors and business partners, which for us is a priority given the supply chain attacks we've seen in the last few years. The process is comprehensive, their [CyberGRX] global risk information exchange is always updated, you can compare your risk against benchmarks and baselines per industry …" (CISO1)*

Another theme identified during the interviews was the use of alliances or partnerships with other organizations as a way to protect organizational knowledge and extend competitive advantage against competitors. This finding also concurs with previous literature that highlights that knowledge leakage in the context of alliances may prove beneficial as long as, the focal firm defines formal contracts and a proper level of trust exists between the partners (Jiang et al., 2013, 2016; Moein et al., 2015)

> *"So far we have different partnerships with universities in the UK and the US to help us with research and development of technologies, however we only give them the bare minimum just to make sure there's no chance of a breach and they usually work in another location isolated from us " (CISO2)*

## 5.1.7.5 Organizational Context

The organizational context defines the social interactions within the workplace as well as the security behaviours of mobile workers in the context of a group of people working together and their interactions. The organizational context is also influenced by the culture and affected by organizational documentation and processes such as knowledge management policies, information security policies, security education training and awareness, security culture, standards, organizational processes and procedures (Alshaikh et al., 2021; S. Maynard et al., 2018; S. B. Maynard et al., 2011).

Within the organizational context theme, a number of formal, informal and legal protection mechanisms were identified. Examples of informal organizational mechanisms mentioned during the interview included misinformation and disinformation, secrecy, lead time on competitors, adding extra complexity to processes, noise generation and fast innovation cycle. These mitigation strategies were not documented in any of the processes or procedures, instead these ad-hoc practices were embedded in the culture of the organizations. The researcher found that organizations whose perceived risk (self reported risk) was deemed as high, were more likely to report any of these informal mechanisms as part of their risk mitigation strategies. As CISO3 expressed:

> *"… we protect our IP [intellectual property] from different kinds of actors, not just regular criminals, but also organized groups and even governments that* *have enough resources to keep trying until they break in, that's why on top of*

*regular controls and classification structures, we often resort to unorthodox ways*

*like creating deceptive IP and misleading documentation to give ourselves some*

*advantage, some cushion in case something goes wrong …" (CISO3)*

The most common type of organizational strategy cited during the interviews referred to risk management procedures, as mentioned earlier, in the personal context, risk management was a recurrent theme at different levels —individual, group and organizational. The use of risk management processes and strategies— in the context of mobile workers and mobile work — in conjunction with mobile policies and procedures (such as BYOD — *Bring Your Own Device*, CYOD — *Choose Your Own Device*, COPE — *Company Owned/Personally Enabled*, COBO — *Company Owned Business Only*), and human resource management mechanisms (Knowledge sharing, employee retention, employee rotation, HR screening and onboarding/offboarding processes) was a common combination and approach followed by most organizations to protect organizational knowledge in mobile work flows. This confirms what previous researchers have reported in the knowledge management literature (Olander et al., 2011, 2014; Ritala et al., 2018). On human resource management mechanisms, one participant added:

*"We have a really strict screening policy, once a knowledgeable person leaves the*

*organization, in fact, the policy states that screening is on-going. So when you*

*join the company you have to undertake a long screening process and after that*

*every year HR reminds us the process and even when you leave you need to*

*follow an exit policy to make sure there is no liability for the company" (KM1)*

In addition, the researcher observed that the ownership of the mobile device, whether employee or organizationally owned, seemed to have no bearing on the knowledge leakage risk perception by research subjects. This finding coincides with recent research that shows how device ownership barely influences user behaviour (Pratama & Scarlatos, 2020). Research participant KM, in this regard commented:

> "… we make sure that our staff are aware of our BYOD [Bring Your Own
> Device] policy before they start using their smartphones and tablets to do their
> work, [ORGNAME] pay for the phone and tablet and the organizational
> plan, but first you have to enrol them into the MDM [Mobile Device
> Management] platform with Airwatch …" (KM)

The data structure representing the environmental and organizational contexts is illustrated in Figure 5-13.

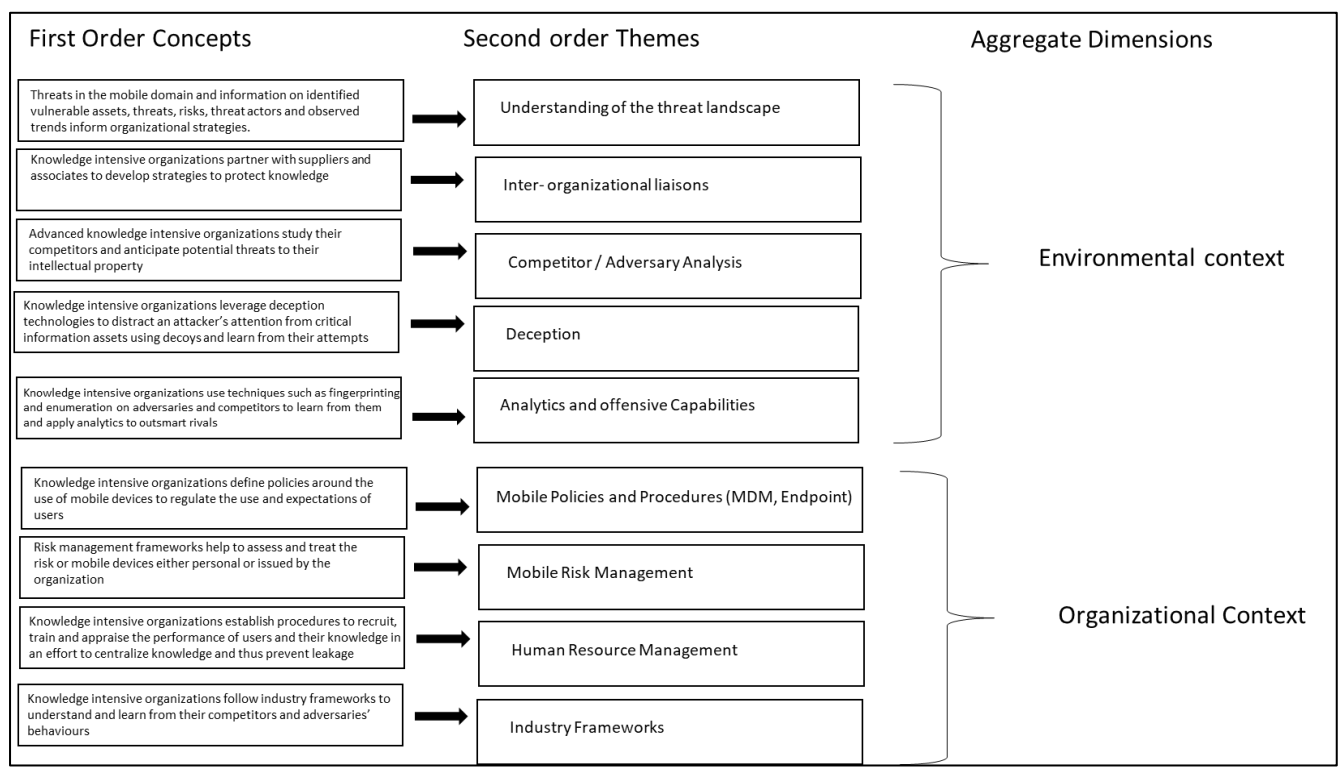

**Figure 5-13 Data Structure for Environmental and Organizational Contexts**

## 5.1.7.6 Technical Context

This context provides the technological and technical attributes for the device and its computing capabilities such as network connections, infrastructure, equipment, devices and systems. It represents an aggregate context that consists of other technical constituents such as spatial, user and location context (Belko Abdoul Aziz

Diallo et al., 2012, 2014). In the mobile computing literature, the technological context also refers to the technical mechanisms employed by the mobile users during the interaction with the device (Hofer et al., 2003; B. N. Schilit & Theimer, 1994). During the interviews, participants detailed technical controls and mechanisms utilized in their organizations mostly in three main categories, mobile analytics (analytics feeds, analytics platforms and artificial intelligence powered platforms such as user and entity behaviour analytics— UEBA), mobile security strategies (detection, prevention, response, deception, compartmentalization, perimeter defence, layering, deterrence), mobile tactical and operational controls (endpoint security for mobile devices, EDR — *Endpoint Detection and Response*, MDM — *Mobile Device Management*, encryption, MFA — *multi-factor authentication*, SIEM — *Security Incident and Event Management*, DLP — *Data Loss/Leak Prevention*, VPN — *Virtual Private Network*, IAM — *Identity and Access Management*, IRM —*Information Right Management,* conditional and contextual access, Firewall, Anti virus, Anti malware, RBAC — *Role Based Access Control*, Biometrics, layering and tagging). Overall these findings suggest that technical controls are widely used to prevent leakage of knowledge that have been reduced to codified information (explicit knowledge) and corroborates what multiple researchers have previously reported in the knowledge protection literature (Ahmad, Bosua, et al., 2014; Ahmad, Maynard, et al., 2014; K. C. Desouza & Vanapalli, 2005; Manhart & Thalmann, 2015). In reference to mobile analytics controls, one participant outlined:

> *"…what I love about our AlienVault SIEM [Security Incident and Event Management] is that it displays a nice dashboard showing the patterns for a particular individual, so we know more about their usage and their profiles and sometimes it even notifies us when a possible person may be at risk of leaving the organization as their behaviour changes and, for instance, starts sending a lot of company information to other accounts outside our authorized domains…"(SM1)*

Similarly, with respect to mobile tactical and operational controls, participant *CSM* mentioned:

> *"…our zero trust architecture always verifies the access to our organizational resources using contextual access which means that it only grants access to users once the originating IP address has been verified, the device has been recognized by our endpoint using the MAC address and device id, and the geolocation of the device matches our records, if any of those parameters are different, then access is denied, only when all parameters are green access is granted …"*
>
> *(CSM)*

Participant *SM3* also described the use of mobile security strategies in his organization:

> *"…as most of our users work outside the perimeter, in the cloud, we implemented micro segmentation of our offices and networks to isolate sensitive resources and users based on the business impacts so every user is within their own virtual segment and each segment becomes a micro core and a perimeter, …the current trust model is broken and this model considers all network traffic untrusted…" (SM3)*

## 5.1.7.7 Device Context

The device context describes the technological features pertaining to the device itself such as device identifier, device type, location, device profile and processing capabilities (i.e., laptop, tablet, smartphone). Unlike the technological context, the device context encompasses attributes and conditions that only alter the state of a single device instance as opposed to a network of devices(Benítez-Guerrero et al., 2012; Hofer et al., 2003; Jarrahi & Thomson, 2017). Throughout the interviews, participants highlighted tactical and operational controls that targeted the mobile devices focusing on three categories, device profiling, MDM — *Mobile Device Management*, and encryption and compartmentalization of codified knowledge. These findings seem to indicate that technical controls focusing on the device itself only address codified knowledge — that is explicit knowledge—that has been reduced to information and therefore treated as an object and even as data at this level, neglecting tacit knowledge. In this regard, taken together the findings from the technological and device contexts show that there is a research gap in the mobile knowledge management literature concerning the tacit knowledge management and the use of technological tools that represent an important future research direction. These findings corroborate previous deficiencies raised by a body of research in the knowledge management literature and calls for concerted efforts to investigate and better understand organizational strategies in the context of tacit knowledge and mobility (Ahmad, Bosua, et al., 2014; Jarrahi & Thomson, 2017; Manhart & Thalmann, 2015; Nelson et al., 2017).

Examples of remarks of device leakage control mechanisms are presented below. One knowledge management expert, in relation to the use of MDM, outlined

*"…we use a feature within Airwatch MDM [Mobile Device Management]*
*that is called Secure Content Management that allows our mobile workforce to*
*access documents on the go via their laptops or iPads but the physical location*
*of the document is stored back on our servers so if anything happens we just*
*revoke access to the content without messing with their actual equipment …"*
*(KM5)*

With respect to compartmentalization in mobile devices, one information security participant remarked:

*"…the Dtex platform provides insights on staff behaviour using UEBA*
*[User and Entity Behaviour Analytics ] and creates securely isolated*
*containers that authorized staff access from their devices, no one else has access*
*to …" (CTO1)*

The data structure representing the technological  and device contexts is presented in Figure 5-14.

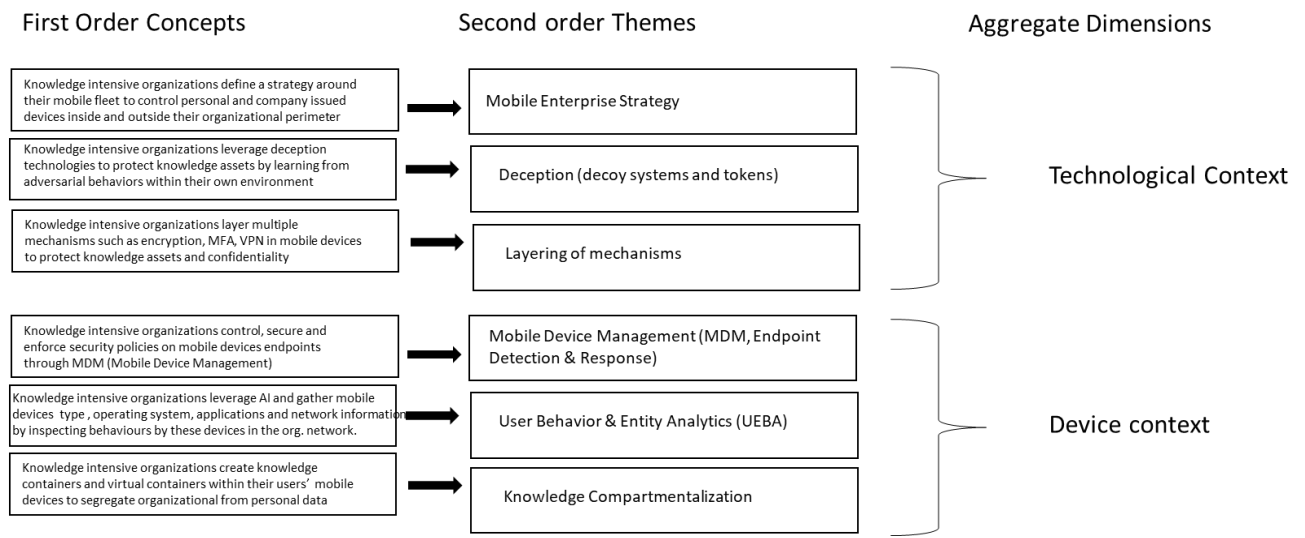

Figure 5-14 Data Structure for Technological and Device Contexts

## 5.1.7.8 Knowledge Leakage Risk Through Mobile Devices

As defined in the conceptual research model in chapter three, knowledge leakage risk caused by the use of mobile devices in organizations refers to a high-level construct derived from the standard definition of risk and it is often operationalized using a qualitative and subjective scale, i.e., low, medium and high (27005:2011, 2011; Amara et al., 2008).

During the interviews, the researcher asked the participants to provide their perceived risk, i.e., self-reported knowledge leakage risk. The process of how the perceived risk was assessed was explained in section 5.1.5 Participants' Self-Reported Leakage Risk.

Overall, research subjects with a high level or perceived knowledge leakage risk were more likely to include within their mitigation strategies more unconventional and active defense approaches such as gathering intelligence about competitors, misinformation/disinformation, and deception. This approach was evident in government and consultancy organizations with a focus on defense and intelligence. In this regard, one participant elaborated:

*"… to protect against attacks like [cyber] espionage our team of analysts employ an offensive strategy gathering insights from OSINT [Open Source*

*Intelligence] feeds and threat intelligence sources including dark web monitoring services, the idea is to gain an understanding of our threat environment and adversaries…" (CISO3)*

Another participant reiterated a similar approach:

*"… as part of our blue and red team exercises, we run simulations to execute targeted attacks and emulate post-exploitation tactics and techniques of advanced threat actors… we set up Cobalt Strike or Armitage as exploit framework and keep an updated database from CTI [Cyber Threat Intelligence] feeds..." (CSM)*

Participant *CKO2* also elaborated on how they use deception to their advantage and as a learning tool that informs their leakage mitigation capabilities:

*"… our security analysts collect information from our honeypot and learn from the information gathered so we can harness this to improve our defenses, we've been doing it for a few years now and it's valuable information that gives us a lot of insights on the latest trends …" (CKO2)*

On the other hand, organizations whose perceived risk level was low and medium, had a tendency to employ similar approaches in managing their knowledge leakage risk. The only difference the researcher found was that self-reported low risk organizations had more informal controls, however, participants seemed to be confident with their mitigation measures and thought such measures were enough to protect their knowledge assets in case of an incident. In similar cases, these participants failed to identify people in their risk assessments and only identified explicit knowledge, i.e., codified knowledge and information.

Research subject, *KM*, whose organization leakage risk was reported as low, stated:

*" … we audit our mobile workforce devices periodically, and make sure they're compliant with our policies, every time there is an indication of malicious activity we contact them and run scans on their devices, so far we haven't had*

> *any breach and our ICT team is always proactive with our compliance*
> *requirements, we just recently implemented the essential eight controls mandated*
> *by the government, so we try to manage our risk levels and keep the residual*
> *risk low…" (KM)*

Similarly, participant *KM5*, who indicated a medium level leakage risk, expressed:

> *" … our recent risk management framework review showed that our*
> *information repositories stored internal non-public data information and that in*
> *the case of a compromise, even though the probability was likely, the impact*
> *was minor which resulted in a medium risk …" (KM5)*

## 5.1.7.9 Organizational Knowledge Leakage Risk Mitigation Strategies

As defined in the conceptual research model (chapter three), formal and informal, organizational risk control strategies are used by organizations to safeguard knowledge assets at risk. Such strategies aim to reduce risk impact or probability (risk reduction), as well as share, avoid, transfer or accept any residual risk remaining after the risk treatment. These strategies target different contexts within higher order dimensions (factors) such as human, enterprise and technological.

Throughout the interviews, participants identified different organizational mitigation strategies in the formal and informal categories, and most importantly different types of knowledge — tacit and explicit knowledge. As identified in the literature tacit knowledge protection remains disproportionately underrepresented in comparison to explicit knowledge protection (Bosua & Scheepers, 2007; Manhart & Thalmann, 2015).

In this respect, the researcher identified and categorized the leakage mitigation strategies for both — tacit and explicit — knowledge types based on the level of formality — formal vs. informal — and proposed a leakage mitigation framework for mobile knowledge workers that is described in section 5.2 below Proposed Knowledge Leakage Mitigation Framework for Mobile Knowledge workers.

Research participants elaborated on formal and informal knowledge leakage mitigation strategies as expressed by *CISO4* on informal mechanisms:

> *"… at [ORGNAME] we work with customers from hospitals designing new clinical devices that are trialled and tested in laboratories and we have to ensure the designs are protected at all times, sometimes we speed up our testing procedures and cut down the time to market so we can be the first ones in releasing a product which gives us a huge advantage…" (CISO4)*

With respect to formal mitigation strategies, a knowledge management expert commented on human resource management strategies:

> *"… our retention management policy has decreased the high employee turnover rate we had a few years ago, we offered perks to our subject matter experts and provided them with training and different professional development options so they have the freedom to choose and that keeps them engaged and …" (KM2)*

As illustrated so far, the first order concepts as extracted from the data collected and reported by participants, supported the definition of the constructs, based on the literature, within the conceptual model, as depicted in section 5.1.7.1. In line with the model, the higher order dimensions — human, enterprise and technological factors — are used to group the different contexts as described in the following sub sections.

## 5.1.7.10 Human Factors

Human factors refer to the overall motivations and cognitive processes, as well as social norms that are explicit and implicit from human behaviours inherent to individuals, and social interactions in group settings. Human factors are comprised of personal and social context.

## 5.1.7.11 Enterprise Factors

The enterprise factors refer to the organizational scope and its internal components such as culture and behaviour. Enterprise factors also include the external operating environment within which the organization operates (competitors, customers, regulators, suppliers and partners) as well as the political, demographic, economic and socio-cultural elements (macro environment). Enterprise factors are comprised of organizational context and environmental context.

## 5.1.7.12 Technological Factors

The technological factors are composed of the device context and technological context and refer to the overarching technology, infrastructure, tools, and information systems that enable and facilitate the interaction of users with technology and technical artefacts in the context of knowledge-sharing activities and mobile knowledge work facilitated by the use of mobile devices.

The data structure based on the different contexts grouped by the higher order dimensions is illustrated in Figure 5-15.

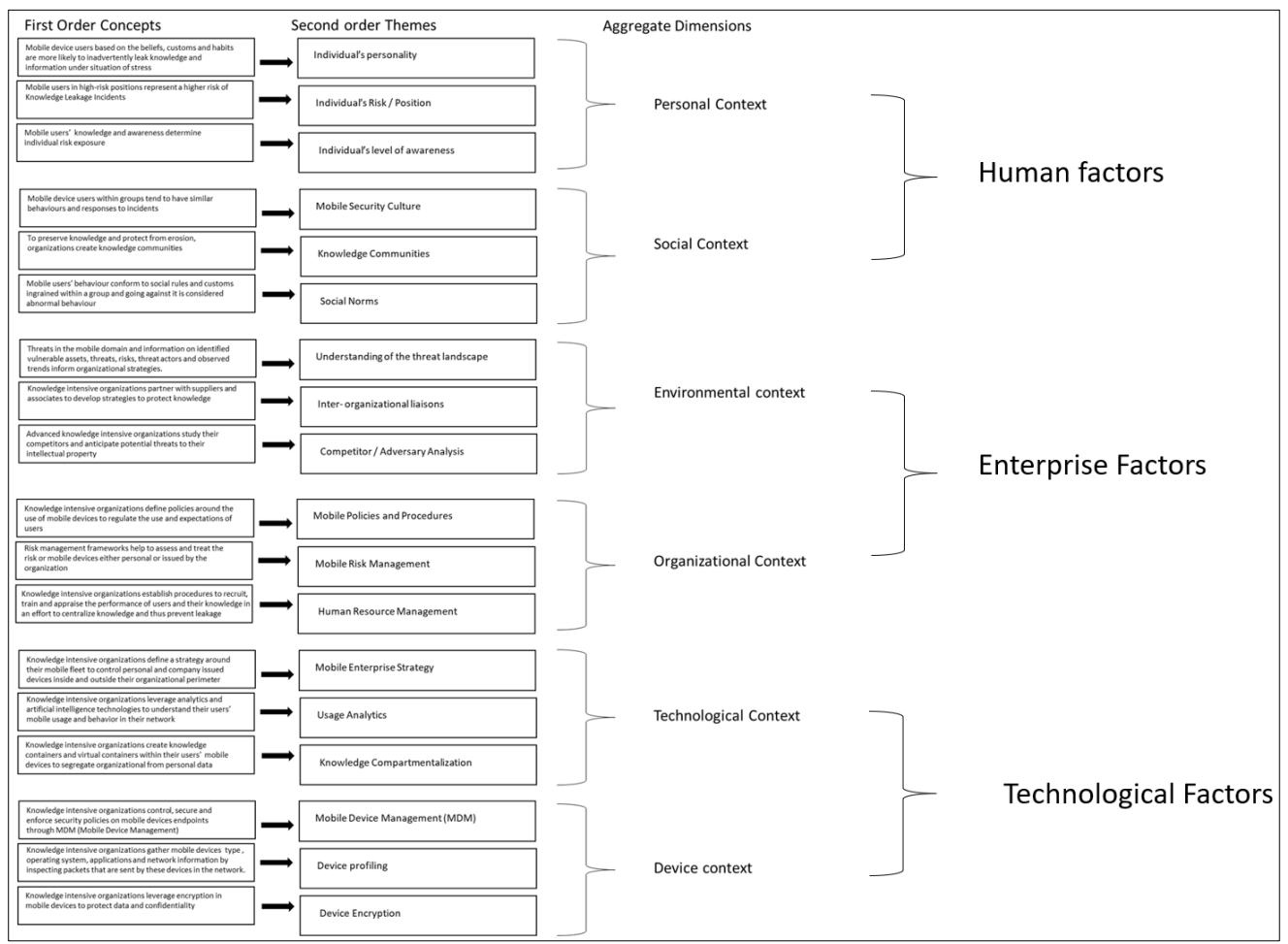

**Figure 5-15 Complete data structure for all Contexts**

The mind map of the first order concepts, second order themes, and dimensions is illustrated in Figure 5-16. This mind map presents a first iteration, found by the researcher, of the mitigation strategies employed by different knowledge intensive organizations to ameliorate their security posture in the context of mobile devices.

Further to this mind map, and extending on the first order concepts, second order themes, and dimensions, the researcher further refined and in a subsequent iteration, categorized the leakage mitigation strategies as reported by research participants into a classification scheme. The collected information was then grounded by the research literature and the research model in order to categorize and structure the various strategies under the different research model factors and mobile contexts (as previously explained in chapter three, model development). The classification scheme is shown in Figure 5-17 below (page 173). The researcher will further detail the classification scheme in the discussion chapter linking each one of the listed strategies back to literature.

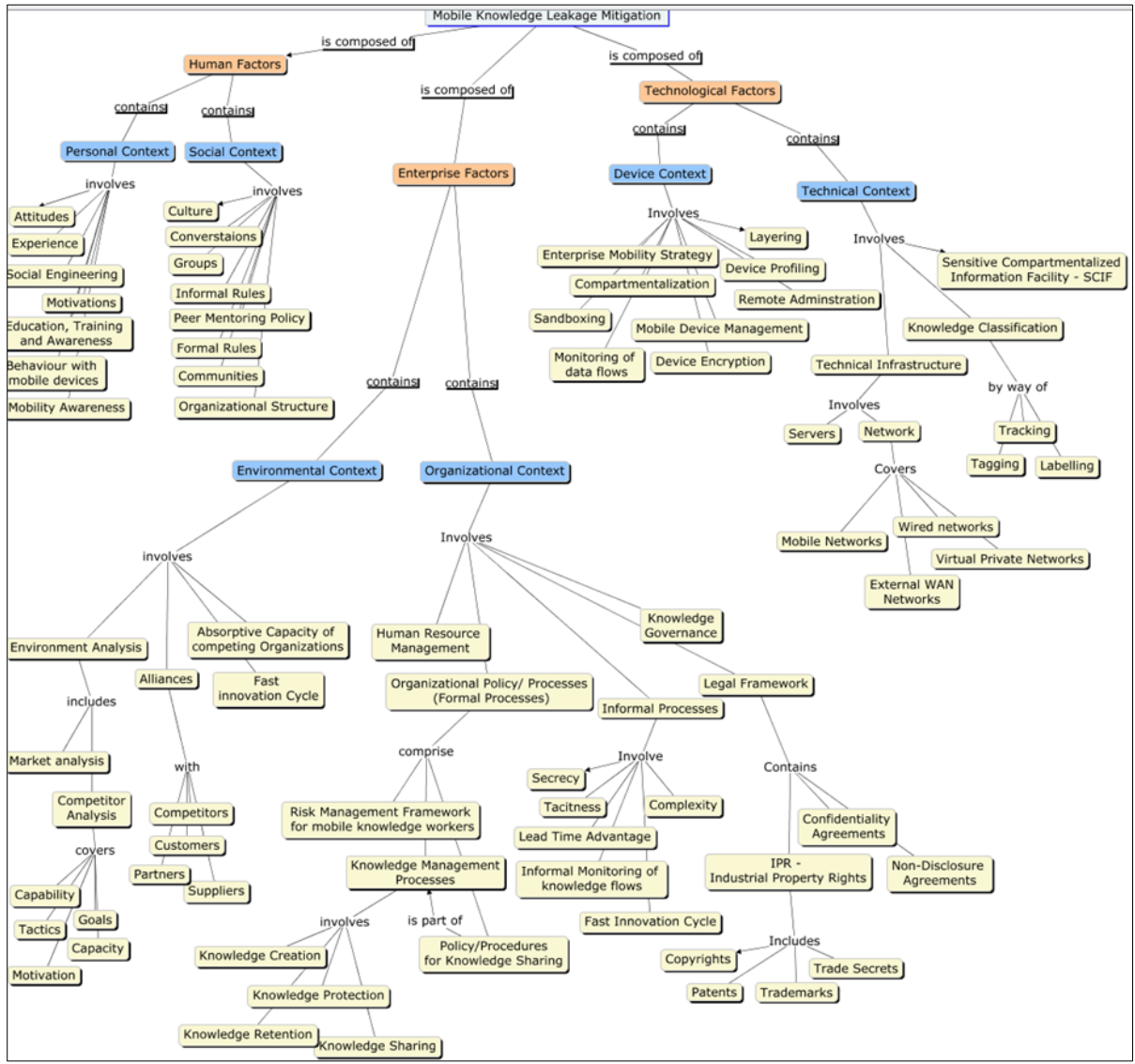

**Figure 5-16. mind map of first order concepts, second order themes and dimensions**

During the course of the interviews and as a result of the analysed documents, the researcher observed several recurring leakage mitigation strategies employed by organizations that rank under the enterprise, human and technological dimensions. As previously presented, upon the classification of these strategies under the aforementioned dimensions and their constituent contexts, i.e., Enterprise — comprised of organizational and environmental; Human — comprised of personal and social; and Technological — comprised of device and technological contexts— the investigator created a classification scheme of strategies used by organizations to address the knowledge leakage risk caused by the use of mobile devices.

Each strategy outlined within the boxes below in the classification represents a different relevant approach and summarizes a particular measure to address the leakage risk.

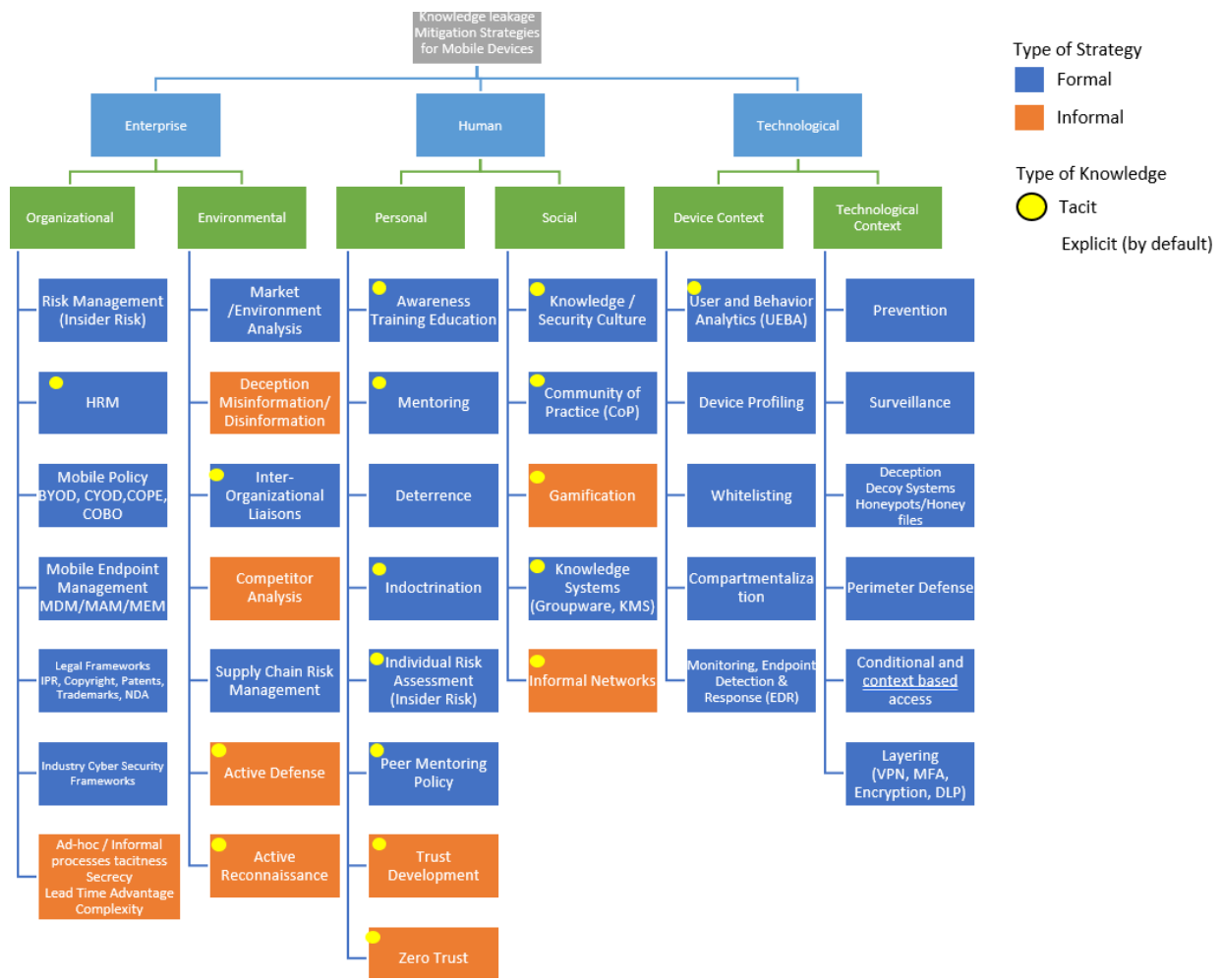

**Figure 5-17. Classification scheme of Knowledge Leakage Mitigation Strategies for Mobile Devices**

## 5.2 Proposed Knowledge Leakage Mitigation Framework for Mobile Knowledge workers (Initial approach)

Based on the interviews, supplementary documents and the conceptual model developed in chapter three and depicted in the present chapter, the researcher, in an effort to show the application of the conceptual model and address the gaps in the current literature, initially arranged the framework in a 2 x 2 matrix and categorized the reported mitigation strategies drawing on two dimensions grounded in the knowledge management literature, that is, degree of formality of leakage mitigation strategies — formal vs informal — (Amara et al., 2008; Bolisani et al., 2013; Dhillon, 2007; Sveen et al., 2009) and type of knowledge —Tacit vs Explicit — (Ikujiro Nonaka, 1991, 1994; Polanyi & Sen, 1997). This framework was subsequently rearranged in a hierarchical manner in a later iteration (see Figure 5-17). The initial framework is depicted in Figure 5-18, in there, it can be seen that the first quadrant, 'Formal Tacit', applies to organizations that have processes in place to protect knowledge that have not been articulated yet. Examples of typical organizations in this quadrant are intelligence and military organizations. The second quadrant 'Formal Explicit' best suits organizations that need to protect codified knowledge from competitors, for example, software and service companies. The 'Informal Tacit' quadrant is well suited for organizations that need to protect intellectual property but do not have knowledge processes in place (for example., small organizations, start-ups). The last quadrant 'Informal Explicit' can be utilized by organizations that rely on codified knowledge but do not have proper secure knowledge processes in place. It is worthwhile noting that, these strategies are not mutually exclusive; rather they can be applied in conjunction in a layered manner.

The research findings suggest that informal protection methods are more commonly adopted than formal methods. Formal methods are based on legal measures and organizational processes. These methods comprise policies and procedures that manage the access to and use of knowledge such as human resource management (Alshaikh et al., 2015, 2019). Formal methods include policies and procedures that establish and ensure the effective use of technical controls. Examples of such mechanisms include system audits, update mechanisms, risk assessments, identification of security roles, segregation of responsibilities and implementation of indicators (Ahmad, Bosua, et al., 2014). In contrast, informal protection methods are based on relationship, trust and organizational ad hoc arrangements. Typically, these methods involve actions related to deploying security in organizations by creating a security culture (Lim et al., 2010, 2009). Examples of informal controls include training employees, implementing security incentives, increasing the commitment to security and motivating users.

Among informal methods, some are more popular than others (for example, secrecy, trust, and fast innovation cycles). Less common techniques reported during the interviews, include offensive strategies (active intelligence gathering regarding the environment and competitors), competitor analysis and zero trust (Ainslie et al., 2023; Kotsias et al., 2022). Moreover, organizations tend to use different types of strategy (formal, informal) for achieving a better overall protection. These strategies as well as the leakage mitigation framework are further discussed and explored in the following discussion chapter.

| Type of Knowledge | | |
|---|---|---|
| | Tacit | Explicit |
| **Formal** | **Quadrant 1: Formal Tacit** | **Quadrant 2: Formal Explicit** |
| | **Human** | **Human** |
| | 1. Peer mentoring | 1. Security Gamification |
| | 2. Education Training Awareness | 2. Mobile Behaviour Profiling/Analysis |
| | 3. Security Culture Development | 3. On-going Screening / monitoring |
| | 4. Communities of Practice | **Enterprise** |
| | 5. Conversation Security (Phone) | 4. Enterprise Mobile Strategy |
| | **Enterprise** | 5. Mobile Risk Management |
| | 6. HR Management | 6. Legal Framework (Contracts) |
| | 7. Knowledge Sharing Policy | 7. Industrial property Rights |
| | 8. Agreements with other organizations | 8. Patents/Trademarks/Copyrights/Designs |
| | 9. Patents/Trademarks/Copyrights/Designs | **Technological** |
| | **Technological** | 9. Securing Communication channels and Devices |
| | 10. Expert Directories | 10. Compartmentalization |
| | 11. Case conferences | 11. Device Monitoring / Detection |
| | 12. Content/ People separation | 12. Perimeter Defence |
| **Informal** | **Quadrant 3: Informal Tacit** | **Quadrant 4: Informal Explicit** |
| | **Human** | **Human** |
| | 1. Trust development | 1. Discretionary Access to knowledge base |
| | 2. Secrecy | 2. Competitor Analysis |
| | 3. Staff rotation | **Enterprise** |
| | 4. Staff Shadowing | 1. Knowledge-sharing with business partners |
| | 5. Offensive Strategy (Competitors) | 2. Deception/Misinformation |
| | **Enterprise** | 3. Complexity of Design |
| | 1. Secrecy | **Technological** |
| | 2. Fast Innovation Cycle | 1. Restricted access |
| | 3. Complexity in Knowledge processes | |
| | 4. Lead Time Advantage | |
| | **Technological** | |
| | 1. Deception (Decoy campaigns) | |

*(Left axis label: Knowledge Leakage Mitigation Strategy Type)*

**Figure 5-18. Knowledge Mitigation Framework for Mobile Knowledge Workers**

The initial iteration of the framework in Figure 5-18 above serves only as a preliminary version of the final model based on the findings of this chapter. It is designed to provide a structured approach to guide knowledge leakage mitigation efforts in knowledge-intensive organizations. However, it is important to clarify that this framework is not claimed to be exhaustive or definitive.

The framework represents a synthesis of the insights gained from the study and is intended to serve as a starting point for further exploration and refinement. Furthermore, the framework lacks rigorous testing and validation in various contexts to ascertain its comprehensiveness and applicability. Therefore, while the initial framework offered a representative view of the key factors and strategies identified in this chapter, the initial version as well as the final version are open to further development and enhancement as more empirical evidence becomes available and as the field evolves.

The framework has not been tested on organizations for its effectiveness. The reason for this is twofold. Firstly, the time constraints associated with the completion of the research study prevented the subsequent phase of empirical testing within the organizations involved in the study.

Secondly, and perhaps more importantly, the research methodology adopted in this study, which is based on the Gioia methodology, is primarily concerned with generating rich, in-depth insights from the data, rather than testing hypotheses or validating models. The framework derived from this study is intended to provide a conceptual understanding of the phenomenon of knowledge leakage and to guide mitigation strategies. It represents an initial iteration that is open to further refinement and testing in future research.

The testing of the framework for its effectiveness in real-world settings would indeed be a valuable next step. However, this would involve a different research approach, possibly involving experimental or quasi-experimental designs, and is beyond the scope of the current study. This is an area that future research could potentially explore to further validate and enhance the framework.

## 5.3 Summary

This chapter presented the results and key findings of the data analysis as a result of the interviews, and supplementary documentation. A detailed description of the process followed by the researcher was presented. First, the data analysis using Gioia methodology was addressed. Second the interview process was described along with the supplementary document analysis. Third, a detailed report on the demographics of research participants was informed. Four, the data structure process was displayed, and the conceptual research model applied to establish the relationships amongst the first order concepts. Finally, a knowledge leakage mitigation framework is proposed based on the conceptual model constructs and the dimensions of formality (formal vs informal) and knowledge type (tacit vs explicit).

The subsequent chapter presents an in-depth discussion of the results and findings obtained from the data analysis and interviews, contextualized within the existing information systems (IS) literature. The researcher highlights the key themes and implications of the results, along with integrating the findings into the literature. Furthermore, this study's implications for IS research and practice are also thoroughly examined and discussed.

*"All warfare is based on deception. Hence, when we are able to attack, we must seem unable; when using our forces, we must appear inactive; when we are near, we must make the enemy believe we are far away; when far away, we must make him believe we are near."*

**—Sun tzu, The Art of War**

# Chapter 6. DISCUSSION

The previous chapter presented the overall findings of this research study. These findings explain the role that mobile devices play in the knowledge leakage risk phenomenon within organizations. Furthermore, the findings chapter also illustrated how the mitigation framework was developed based on the mitigation strategies presented by interviewees and identified during the interviews and supplementary document analysis. The mitigation framework represents one of the main contributions of this research project along with the identification of main strategies implemented by organizations. Additionally, the details of the framework explain how different controls and strategies can be applied to different industries and risk profiles. This chapter[8] highlights the contributions that these findings bring to IS research and practice in general.

The discussion chapter starts by explaining the findings and key insights of the study in relation to the existing knowledge leakage and mobile computing and knowledge management literature. The chapter also highlights the connection of this research to broader topics in the IS literature. The final part of the chapter outlines the implications and conclusions of this study for both IS research and practice.

---

[8]Sections of this chapter have been published in the following publications:

- Agudelo-Serna, C. A., Bosua, R., Ahmad, A., & Maynard, S. (2017). Strategies to Mitigate Knowledge Leakage Risk caused by the use of mobile devices: A Preliminary Study.
- Agudelo-Serna, C. A., Bosua, R., Ahmad, A., & Maynard, S. B. (2018). Towards a knowledge leakage mitigation framework for mobile devices in knowledge-intensive organizations.

# 6.1 Chapter Overview

As was mentioned before, knowledge intensive organizations operate in dynamic and complex environments that, coupled with environmental and technological threats from insiders, competitors, and partners, amongst other risks, form a hostile and changing landscape that forces organizations to adapt and evolve.

In addition to this, as evidenced in the previous chapter, the use of mobile devices and the mobility of (knowledge) workers exacerbate the risk of leakage and loss of competitive advantage such as organizational knowledge represented in the form of people, intellectual property and processes, just to name a few.

During the interviews and analysed documents, the researcher observed a number of recurrent leakage mitigation strategies utilized by organizations that fall under the enterprise, human and technological dimensions. As displayed in the findings chapter, upon the classification of these strategies under the aforementioned dimensions and their constituent contexts, i.e., Enterprise — comprised of organizational and environmental; Human — comprised of personal and social; and Technological — comprised of device and technological contexts— the investigator developed a classification scheme of strategies employed by organizations to address the knowledge leakage risk caused by the use of mobile devices based on the findings and framework described in the previous chapter (see Figure 6-3).

Each strategy outlined within the boxes below in the classification represents a different salient approach and summarizes a specific measure to address the leakage risk. The next sections describe each one of these strategies in more detail, grounded in the current body of IS, knowledge management, knowledge protection, information security and mobile computing literature as well as contrast the findings with the research questions this study aimed to address.

# 6.2 Research Questions, Propositions and Model

This study aimed to answer the following overarching research question:

- *How can knowledge intensive (KI) organizations mitigate the knowledge leakage risk (KLR) caused by the use of mobile devices?*

In order to address this main question, the following supporting sub-questions assisted by providing the necessary foundation and context required to answer the primary question:

4. *What strategies are used by knowledge-intensive organizations to mitigate the risk of knowledge leakage (KLR) caused by the use of mobile devices?*

5. *How does the perceived KLR level inform the strategies used by KI organizations?*

6. *What knowledge assets do knowledge intensive organizations protect from KL?*

In relation to the research questions, the researcher developed a research model based on the literature review and derived two theoretical propositions, as explained in chapter 3, to inform the methodology and design of the study (see summary in Table 6-1 below):

1. **P1:** *The knowledge leakage risk through mobile devices informs the development of organizational knowledge leakage risk mitigation capabilities.*

2. **P2:** *The mobile device usage contexts and its constituent human, enterprise and technological factors affect the likelihood and/or consequence of KLR through mobile devices.*

In the next section, the key findings and insights related to the research questions are presented and discussed.

**Table 6-1. Research Model, Questions and Propositions**

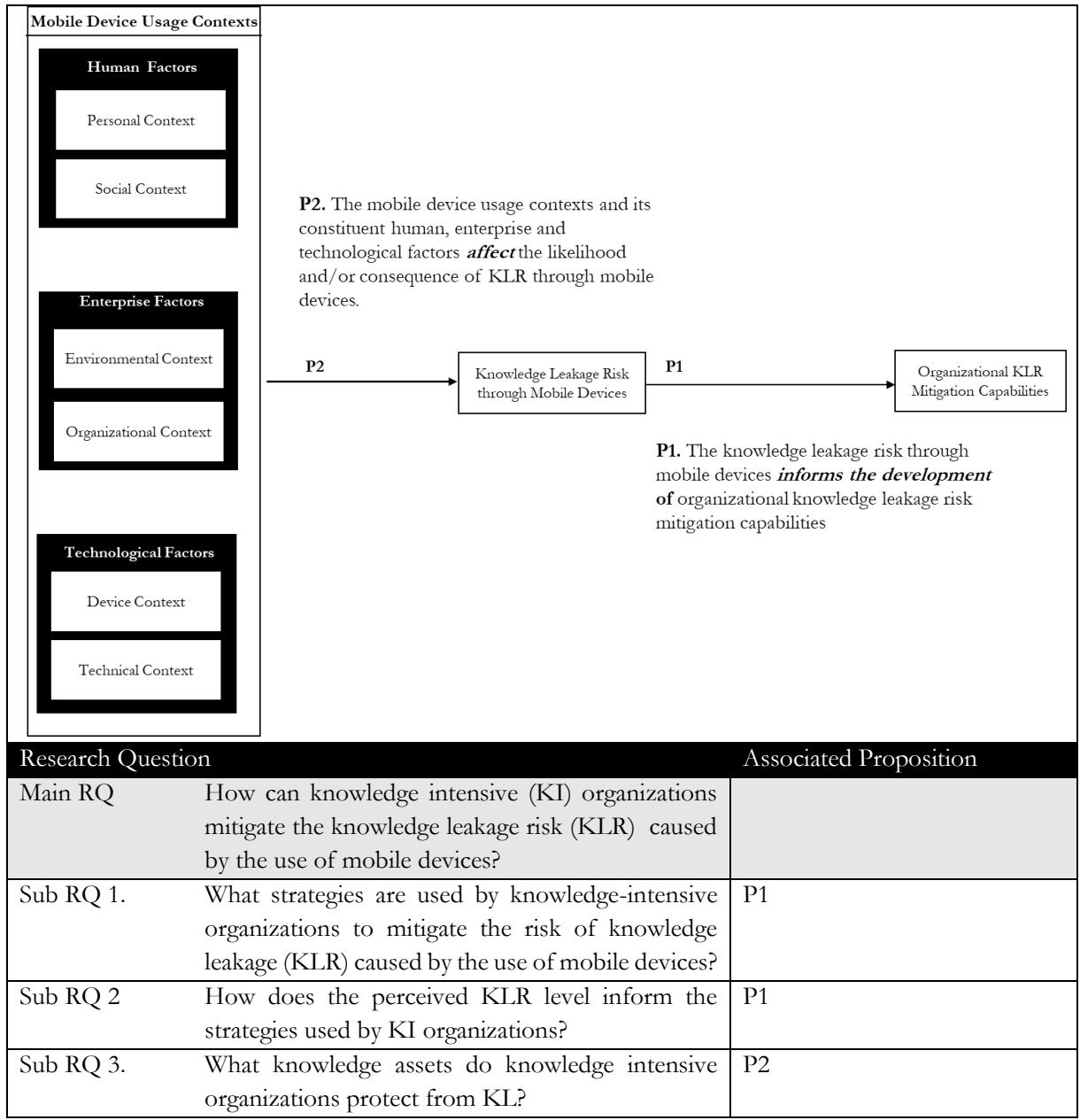

| Research Question | | Associated Proposition |
|---|---|---|
| Main RQ | How can knowledge intensive (KI) organizations mitigate the knowledge leakage risk (KLR) caused by the use of mobile devices? | |
| Sub RQ 1. | What strategies are used by knowledge-intensive organizations to mitigate the risk of knowledge leakage (KLR) caused by the use of mobile devices? | P1 |
| Sub RQ 2 | How does the perceived KLR level inform the strategies used by KI organizations? | P1 |
| Sub RQ 3. | What knowledge assets do knowledge intensive organizations protect from KL? | P2 |

# 6.3 Discussing the Propositions and Research Questions

In this study, the researcher presented two theoretical propositions that highlighted the relationship between 1) KLR and Organizational mitigation capabilities; and 2) Mobile device contexts and KLR (refer to Table 6-1 above).

## 6.3.1 Proposition 1

**P1**: *The knowledge leakage risk through mobile devices informs the development of organizational knowledge leakage risk mitigation capabilities.* (Figure 6-1)

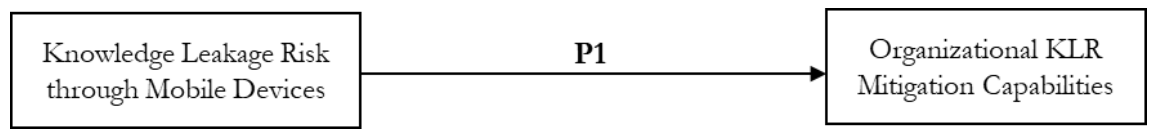

**Figure 6-1. Proposition 1**

In line with ***P1***, the findings suggest and support the idea that knowledge intensive organizations employ different strategies based on their perceived organizational risk (see Figure 6-1). Moreover, these organizational strategies constitute a way to configure or reconfigure organizational capabilities based on knowledge assets, internal competencies and resources to achieve their knowledge protection goals, offset their risk exposure, and sustain competitive advantage in a dynamic and complex environment (Grant, 1996b). In some cases, this process requires improving some capabilities, developing or acquiring completely new ones, or even removing some non-core, non-strategic capabilities (Kaplan et al., 2001; Leonard-barton, 1992). In this regard, this proposition helps to understand the capabilities, configured in the form of strategies, used by knowledge intensive organizations to combat the risk of leakage through mobile devices. Therefore, providing support to the proposition and also an answer to the first research sub-question: *What strategies are used by knowledge-intensive organizations to mitigate the risk of knowledge leakage (KLR)* *caused by the use of mobile devices?.* The answer to this sub-question and the others will be provided in the next chapter (See Chapter 7).

The findings also seem to indicate that organizations that assess their self perceived risk level as high tend to employ more proactive and innovative strategies that, for instance, leverage new and emergent technologies such as artificial intelligence to protect organizational knowledge. Examples of such strategies are Deception, Zero Trust, Behaviour Analytics, Active defence, and Insider Risk Management.

On the other hand, the findings appear to suggest that organizations that rank their perceived risk as medium favour more conventional knowledge protection mechanisms, both formal and informal. E.g., IPRs, Secrecy, Tacitness. Whereas self perceived low risk level organizations are more inclined to prefer informal strategies over formal ones as these measures are considered more cost effective and faster to implement. Therefore, the risk level seems to have an impact on the strategies employed by knowledge intensive organizations. This relationship offers an answer to the second research sub-question: *How does the perceived KLR level inform the strategies used by KI organizations?* (See Chapter 7 - section *7.1.2.2 Secondary Research Question 2: Knowledge Leakage Risk Level*).

This interesting association between perceived risk and strategy selection may appear somewhat surprising compared to previous literature which purports that organizations adopt controls for a variety of reasons, not just because of perceived risk. While perceived risk is certainly a factor in the decision-making process, other factors such as compliance requirements, industry standards, best practices, perceived value of controls, and organizational culture play a more substantial role (Cram et al., 2017, 2019; Palanisamy et al., 2020; Puhakainen & Siponen, 2010; Von Solms, 1999). For this reason, these findings must be interpreted with caution as this may be the result of a small sample size and therefore might not be transferable to other type of organizations, geographies and/or circumstances.

Overall, these findings collectively support **Proposition 1** as the evidence from the study indicates an association between knowledge leakage risk and mitigation capabilities, and that knowledge leakage risk may have an effect on and is likely to inform the mitigation capabilities in knowledge intensive organizations.

The relationship between perceived risk and strategy represents a novel finding of this study and contributes to the existing knowledge protection and knowledge management literature. This correlation between these two constructs may be explained by the knowledge based view and resource based view theory of the firm literature that state that organizational level actions are a response to, and informed by, the market threats and constitute the core of business strategy and contribute to competitive positioning. Such organizational level actions can be protective and competitive in nature. Competitive actions aim to defend and improve a firm's relative competitive position, while a protective action aims to protect organizational assets in the market (Grant, 1996b; Leonard-barton, 1992; David J. Teece, 2007).

However, caution should be exercised when extrapolating and generalizing these findings to other organizations due to limitations of the study such as the sample size, specific demographics, and geography (Australian based organizations). Thus, further research to substantiate these findings is warranted.

## 6.3.2 Proposition 2

**P2:** *The mobile device usage contexts and its constituent human, enterprise and technological factors affect the likelihood and/or consequence of KLR through mobile devices.*

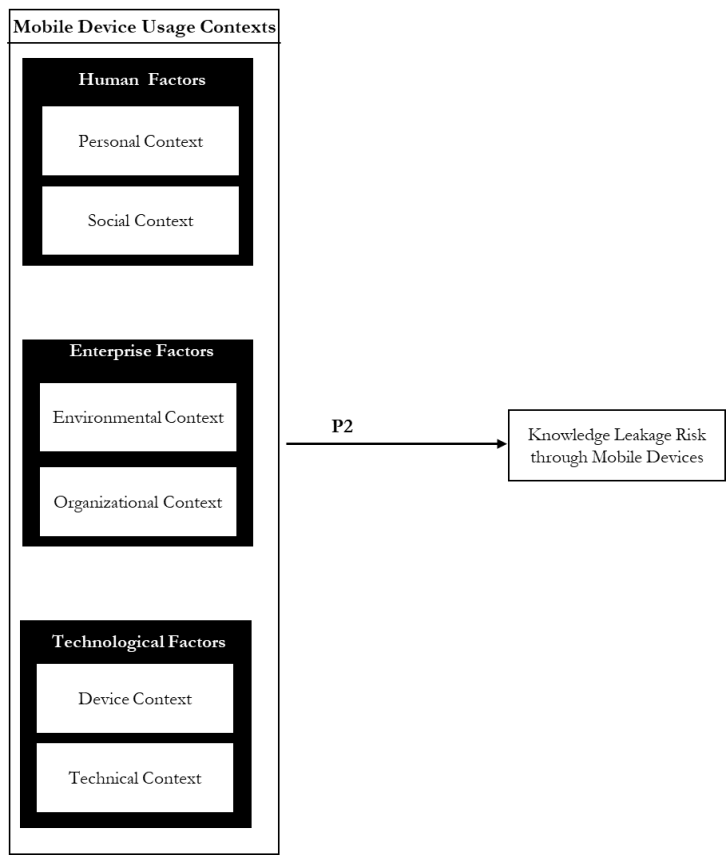

**Figure 6-2. Proposition 2**

With regard to P2, the findings indicate an association between the constructs: *mobile usage contexts* and *knowledge leakage risk through mobile devices* (as shown in Figure 6-2). The results show that the different mobile contexts (grouped into factors) are correlated to the level of risk that knowledge intensive organizations perceive when assessing their risk exposure. For example, mobile workers conducting knowledge work from overseas in complex environments are ranked as a higher risk level employee than someone working from the office protected by the organizational perimeter, controls, and policies (Organizational / Environmental context). The likelihood of a leakage risk eventuating escalates in conditions of complexity and uncertainty, and in the possible presence of competitors or adversaries the consequence also increases. Similarly, a worker using their personal device to access corporate resources would require more conditional access, and validation controls to confirm identity due to the higher perceived risk, as opposed to the same worker accessing resources from their sanctioned, company issued mobile device (Device /

Technical Context). The same applies for workers in positions where due to their job function, they have access to sensitive and confidential information and are privy to critical knowledge assets (Personal Context) or workers socializing with colleagues and competitors at a conference or other social events (Social Context). These workers pose a greater risk due to the fluctuating nature of the contexts they navigate through.

The results of this study also show that the correlation between mobile contexts and leakage risk has an association with the knowledge assets that organizations tend to protect, this seems to be particularly the case for high and medium risk level organizations. For example, with regard to the personal and social context (Human factors), knowledge intensive organizations are concerned with the protection of their people and intellectual capital (i.e., tacit knowledge). Evidence of this, is reflected on the emphasis on the insider risk and individual risk management processes as well as mentoring, indoctrination, and community strategies used by organizations and illustrated in the classification scheme of strategies (see Figure 6-3). Equally important, in relation to the organizational and environmental context (Enterprise factors), participants indicated their interest in safeguarding knowledge assets such as organizational processes and methodologies as presented in Chapter 5 Findings – Section 5.3.3. With respect to the device and technological context (Technological factors), participants stated their concern with organizational assets and resources physically stored on the devices such as product, client, and research related information. Therefore, this association assist in providing an answer to the third research sub-question: *What knowledge assets do knowledge intensive organizations protect from KL?*

This relationship between mobile contexts, knowledge leakage risk and knowledge assets should be interpreted with caution and may not be extrapolated to other organizations as this may be the result of the small sample size as well as the distribution of the sample that exhibits a predominance of high and medium risk compared to low risk level organizations (85% vs 15% - See section *5.1.5 Participants' Self-Reported Leakage Risk*).

Overall, these findings collectively support **Proposition 2** as the data suggests a relationship between *mobile device usage contexts* and *knowledge leakage risk through mobile devices* and likely the dynamic nature of contexts informs the risk.

Interestingly, the relationship between mobile contexts and risk represents a novel contribution to the existing body of knowledge protection and knowledge management literature for mobile knowledge workers. This relationship highlights the dynamic nature of risk associated with the complexities of the environment and interactions of mobile users and contexts. Borrowing from the information security management literature, the *Information Warfare Paradigm* proposed by Richard Baskerville (2005) and supported by other IS authors (Baskerville et al., 2014, 2022; Spagnoletti & Resca, 2007, 2008) with its fundamental assumptions and premises may assist in explaining this correlation (see Table 6-2).

"*The warfare paradigm assumes that risks are unpredictable, not measurable, and transient. It assumes a dynamic relationship with safeguards and a causal structure based on process. It draws its principles from possibility theory, its strategy from agility theory, and its organizational learning from exploration. The shifting context of many organizations promises to increase the presence of the warfare paradigm as balanced against the business paradigm. This shift means that assumptions about the transience of risk, unpredictability of risks, and the consequential emergence of safeguards will grow. An increasing belief that the essential causal structure of security is based on process will lead to a greater perception that security events are more important than static threats; and security failures are a process failure rather than a simple failure of a security safeguard. This shift may lead to increasing use of possibility theory, agility strategies, and exploitative learning strategies.*" (Baskerville, 2005, p. 23)

Similar to what the Information Warfare Paradigm purports in terms of assumptions, logical structure and organizing principles, the mobile contexts are dynamic in nature, and therefore the *context - risk* relationship is also dynamic, highlighting the *transience* and *unpredictability* of the risk and the *emergence*, *agility*, and *explorative* nature of the control strategies as they need to be swiftly reconfigured (*on the fly*) to form mitigation capabilities that adapt to the changing environment and shifting mobile contexts within which the mobile worker moves through. That is to say, the dynamic strategies are *consequential* to risk, the risk level is never *static*, rather it is *dynamic*, it changes as the mobile worker changes contexts. Due to variability of the mobile contexts informed by the threat landscape, the risk and threats become *unpredictable*, difficult to *measure* and warrant an *explorative* approach that calls for the variation of, and the quick deployment of the strategies applied to knowledge workers upon discovery of the *transient* risks as result of the fluctuating contexts. In other words, different strategies can be applied variably to the same mobile worker based on the changing risk level in conjunction with conditions (contexts) highlighting the importance of *events* (as a result of changing contexts) rather than static risks.

**Table 6-2. Features distinguishing two security paradigms. Adopted from Richard Baskerville (2005)**

| | Features | Business Security Paradigm | Information Warfare Paradigm |
|---|---|---|---|
| **ASSUMPTIONS** | Risk Tempo | Premise 1: Information systems security risks are persistent. | Premise 1: Information warfare risks are transient. |
| | Safeguards Tempo | Corollary 1.1: Effective information systems security safeguards must be persistent | Corollary 1.1: Effective information Warfare security safeguards must be emergent |
| | Risk-safeguarding timing | Corollary 1.2: Risks safeguards share a static relationship. | Corollary 1.2: Risks safeguards share a dynamic relationship. |
| | Risk forecasting | Premise 2: Risks are predictable | Premise 2: Risks are unpredictable |
| | Risk measurement | Corollary 2.1: Risks are measurable | Corollary 2.1: Risks are not measurable |
| | Risk safeguard logical form | Premise 3: The relationship of safeguards to threats is determinate. | Premise 3: The relationship of safeguards to threats is consequential. |
| **LOGICAL STRUCTURE** | Causal Structure | Variance | Process |
| | Unit of Analysis | Variables | Events |
| | Safeguard-Risk Reduction Definition | Safeguard is necessary and sufficient to reduce risk | Safeguards are part of a necessary sequence of conditions to reduce risk |
| | Relationship to time | Static | Dynamic |
| | Logical Form | "if X then Y" | "if not X then not Y" |
| **ORGANIZING PRINCIPLES** | Operating Mathematics | Probability | Possibility |
| | Strategic goal | Quality | Agility |
| | Learning strategy | Exploitation | Exploration |

# 6.4 Key themes and Insights

The researcher identified two key themes within the findings of this study that assist in addressing the overarching research question (i.e., *How can knowledge intensive (KI) organizations mitigate the knowledge leakage risk (KLR) caused by the use of mobile devices?*) and represent a novel contribution to the knowledge management and protection literature for mobile knowledge workers.

These themes are:

1. The dichotomy of Trust in the context of mobility
2. The use active defence approaches to protect knowledge assets

The next section elaborates on each one of these themes

## 6.4.1 The dichotomy of Trust in the context of Mobility

Within this theme the researcher identified a stark contrast between the development of trust in some knowledge intensive organizations versus the zero trust approach of others.

### 6.4.1.1 Trust Development

The concept of trust remains critical within the knowledge management literature, as trust leads to cooperative behaviour and it is used as a coordinated mechanism in facilitating knowledge creation and transfer among knowledge workers within a firm, therefore trust contributes to the protection and capture of **tacit** knowledge (Becerra et al., 2008).

An interesting finding established during the interviews referred to how trust lays the foundations for the institution of alliances and partnerships with other organizations as a way to protect organizational knowledge and extend competitive advantage against competitors. This finding match those observed in previous studies that highlights that knowledge leakage in the context of alliances may prove a positive and beneficial outcome as long as the leakage is properly managed, the focal firm defines formal contracts and a proper level of trust exists between the partners (Jiang et al., 2013, 2016; Moein et al., 2015).

Of particular interest to the researcher was the fact that this result commonly occurred amongst knowledge intensive organizations whose risk graded as low and medium, where it was stated that trust development fostered the knowledge sharing and protection at the group level within organizations and facilitates collaboration at the inter and intra levels. In this regard, these findings remain consistent with those of Bertino et al. (2006) and Ford (2004) who have extensively studied the concept of trust management in the knowledge management literature. This concept is further developed in the next section in 6.5.2.2.1.1 as part of the explanation of the different strategies within the framework.

## 6.4.1.2 Zero Trust

Conversely, research subjects, mostly from organizations that reported their perceived risk as medium and high, expressed a contrasting but interesting view, whereby a zero trust approach formed a preferred way to protect knowledge within organizations stemming from the fact that in the current threat landscape trust is considered a liability, even a vulnerability.

Furthermore, research participants pointed out how in multiple leakage incidents the main culprit often involves an insider in a position of trust — sometimes even a privileged user and therefore a trusted employee — rather than a malicious outsider.

This finding corroborates the ideas of several authors in information security (Campbell, 2020; Kindervag et al., 2010; Zaheer et al., 2019) who have suggested that the current information security model is broken and trust is no longer feasible.. Changing the thinking about trust models and becoming aware of the misuse of the word *trust* in relation to knowledge security, becomes paramount to overcome the challenges posed by a mobile workforce outside the control of the organizational perimeter (Campbell, 2020; Kindervag et al., 2010). Zero trust is a novel approach that addresses many of the gaps highlighted by previous knowledge management and protection studies (Manhart & Thalmann, 2015; Olander et al., 2011, 2014; Ritala et al., 2015) particularly in the context of mobile devices, and more broadly, mobility. The concept of zero trust is non-existent in the current knowledge management literature and it even contravenes the conventional notion of knowledge sharing which presupposes the prevalence of trust. Therefore, one of the contributions of this study, is to challenge the status quo in the knowledge management and protection literature in relation to the use of *trust* and propose the adoption of the *zero trust* model from the information security realm and application to mobile knowledge management and protection processes and workflows within *knowledge intensive* organizations. This concept is further developed in sections 6.5.1.3.2.4 and 6.5.2.2.1.2.

While the zero trust model is designed to enhance security and protect against internal and external attacks, it also poses a significant challenge to the process of knowledge sharing. The strict controls and authentication requirements of this model may inhibit the free flow of knowledge and ideas, which are essential for innovation. In this regard, it may be argued that the zero trust model, if implemented without proper consideration of the need for knowledge sharing and innovation, may lead to the stifle of creativity and impede the organizations ability to innovate.

To balance the need for knowledge protection with the need for knowledge sharing and innovation, this study suggests that organizations adopt a "hybrid trust" model. This hybrid model proposes a middle way between total trust and zero trust, whereby organizations adopt a risk-based approach to trust and security. This approach would involve assessing the potential risks associated with different types of knowledge sharing and implementing appropriate strategies and information security requirements accordingly. For example, based on the level of risk, type of asset, knowledge and sensitivity thereof, different controls and strategies may be applied (i.e, explicit vs tacit; high- vs. medium- vs. low-risk levels).

## 6.4.2 The use active defence approaches to protect knowledge assets

During the interview, participants from high risk level knowledge intensive organizations highlighted the use of active defence approaches such as active reconnaissance, deception, environmental analysis, Open source intelligence (OSINT), insider risk management, behaviour analytics and market/ environmental analysis and intelligence as a way to protect knowledge assets from leakage.

### 6.4.2.1 Open-Source Intelligence (OSINT) and Market/Environmental analysis

Participants reported that in order to understand their environment where their organizations operate and prepare for leakage scenarios and to safeguard their knowledge assets, they would gather available intelligence (open source and active reconnaissance) on the market and competitors, as well as conduct risk analysis on business associates and affiliates. These findings are consistent with previous research by authors in the field of the military and intelligence literature (Fredrich et al., 2019; Peyrot et al., 2002; Yarhi-Milo, 2014) who have studied the competitor as a mechanism to understand and *know* adversaries and their behaviour in the operating environment including legislation, competition and governing bodies. This strategy is detailed in section 6.5.1.1.2.1.

### 6.4.2.2 Deception

Another interesting finding in this regard relates to the sophisticated use of deception technologies, to protect their knowledge assets from leakage incidents. The use of different types of decoy entities such as honeypots, honeynets, honeyfiles or -tokens, even honey personas and honey user accounts were used to misdirect, deceive, and even interact with adversaries and malicious insiders to learn from their behaviours. These findings match those observed in previous studies by Libicki et al. (2015) where the authors suggest that this strategy is used to increase the complexity, cost, uncertainty, and time for the attacker, and slightly tip the balance in favour of the defender This finding is neither mentioned in the knowledge management nor knowledge protection literature and hence represents another contribution to the existing literature form the information security literature. More detailed information on these techniques is further explained in sections 6.5.1.3.2.2 and 6.5.2.1.2.1.

### 6.4.2.3 Behaviour Analytics and Insider Risk for mobile devices

Organizations outlined the use of profiling tools to identify malicious insiders and assess internal staff when using mobile devices that otherwise may compromise their organizations and expose them to leakage incidents either intentionally or unintentionally. The use of behaviour analytics, and insider risk tools were listed to baseline the organization and benchmark against future behaviour and to detect abnormal activities. Examples of abnormal behaviours relate to logging in outside of usual hours, emailing sensitive data to outside parties, an insider worker or a privileged user sending emails with high volumes of organizational data to external or personal email accounts, employees copying a high number of files from a network to an external drive or unmanaged device. These findings seem to be consistent with more recent research (Arohan et al., 2020; Martín et al., 2021) that found that user entity behaviour analytics (UEBA) associated with device usage as part of an insider risk management program can be used to effectively analyse patterns of human behaviour, enforce security policy, and pre-empt incidents by identifying intent and potential high risk users and malicious insiders. Given that UEBA and Insider risk management tools assist in determining possible intent (i.e., inadvertent vs. malicious) it may assist in preventing accidental tacit knowledge leakage when this incident results from (non-malicious) unintentional or negligent behaviourbehaviour that can be addressed pre-emptively.

Therefore, this finding addresses one of the gaps previously mentioned in the literature review regarding the lack of technological measures to specifically address tacit knowledge protection which was also supported and raised by previous research (Ahmad, Bosua, et al., 2014; Bolisani et al., 2013; Manhart & Thalmann, 2015; Trkman et al., 2012). Therefore, the application of this particular strategy adds to the current body of knowledge and contributes to the knowledge protection and mobile knowledge management literature. Further information on these topics is elaborated in sections 6.5.1.3.1.1 and 6.5.1.2.1.1.

For a complete summary and list of all the strategies identified in the findings of these study please refer to Table 6-3.

The next section details each one of the mitigation strategies following the structure of the classification scheme developed during based on the findings of these study.

**Table 6-3. Classification Scheme of Knowledge Leakage Mitigation Strategies for Mobile Devices**

| Factor | Context | Strategy | Type of Strategy | | Type of Knowledge | |
|---|---|---|---|---|---|---|
| | | | Formal (29) | Informal (9) | Tacit (17) | Explicit (21) |
| **Human (13)** | Personal (8) | 1. Awareness Training and Education | ✓ | | ✓ | |
| | | 2. Mentoring | ✓ | | ✓ | |
| | | 3. Deterrence | ✓ | | | ✓ |
| | | 4. Indoctrination | ✓ | | ✓ | |
| | | 5. Individual Risk Assessment (Insider Risk) | ✓ | | ✓ | |
| | | 6. Peer Mentoring Policy | ✓ | | ✓ | |
| | | 7. Trust Development | | ✓ | ✓ | |
| | | 8. Zero Trust | | ✓ | ✓ | |
| | Social (5) | 9. Knowledge Security Culture | ✓ | | ✓ | |
| | | 10. Community of Practice | ✓ | | ✓ | |
| | | 11. Gamification | | ✓ | ✓ | |
| | | 12. Knowledge Systems (Groupware, KMS) | ✓ | | ✓ | |
| | | 13. Informal Networks | | ✓ | ✓ | |
| **Enterprise (14)** | Organizational (7) | 14. Risk Management (Insider Risk) | ✓ | | | ✓ |
| | | 15. Human Resource Management (HRM) | ✓ | | ✓ | |
| | | 16. Mobile Policy (BYOD, CYOD, COPE, COBO) | ✓ | | | ✓ |
| | | 17. Mobile Endpoint Management (MDM, MAM, MEM) | ✓ | | | ✓ |
| | | 18. Legal Frameworks (IPR, Copyright, Patents, Trademark, NDA) | ✓ | | | ✓ |
| | | 19. Industry Cyber Security Frameworks | ✓ | | | ✓ |
| | | 20. Ad-hoc Informal Processes (Tacitness, Secrecy, Lead time advantage, Complexity) | | ✓ | ✓ | |
| | Environmental (7) | 21. Market/ Environment Analysis | ✓ | | | ✓ |
| | | 22. Deception/ Misinformation/Disinformation | | ✓ | | ✓ |
| | | 23. Inter Organizational Liaisons | ✓ | | ✓ | |
| | | 24. Competitor Analysis | | ✓ | | ✓ |
| | | 25. Supply Chain Risk Management | ✓ | | | ✓ |
| | | 26. Active Defence | | ✓ | ✓ | |
| | | 27. Active Reconnaissance | | ✓ | ✓ | |
| **Technological (11)** | Device (5) | 28. User and Behaviour Analytics (UEBA) | ✓ | | ✓ | |
| | | 29. Device Profiling | ✓ | | | ✓ |
| | | 30. Whitelisting | ✓ | | | ✓ |
| | | 31. Compartmentalization | ✓ | | | ✓ |
| | | 32. Monitoring and Endpoint Detection & Response (EDR) | ✓ | | | ✓ |
| | Technological (6) | 33. Prevention | ✓ | | | ✓ |
| | | 34. Surveillance | ✓ | | | ✓ |
| | | 35. Deception (Decoy systems, honeypot, honey files) | ✓ | | | ✓ |
| | | 36. Perimeter Defence | ✓ | | | ✓ |
| | | 37. Conditional and Context based access | ✓ | | | ✓ |
| | | 38. Layering (Defence in Depth, MFA, Encryption, DLP) | ✓ | | | ✓ |

# 6.5 Classification scheme of Knowledge Leakage Mitigation Strategies for Mobile Devices

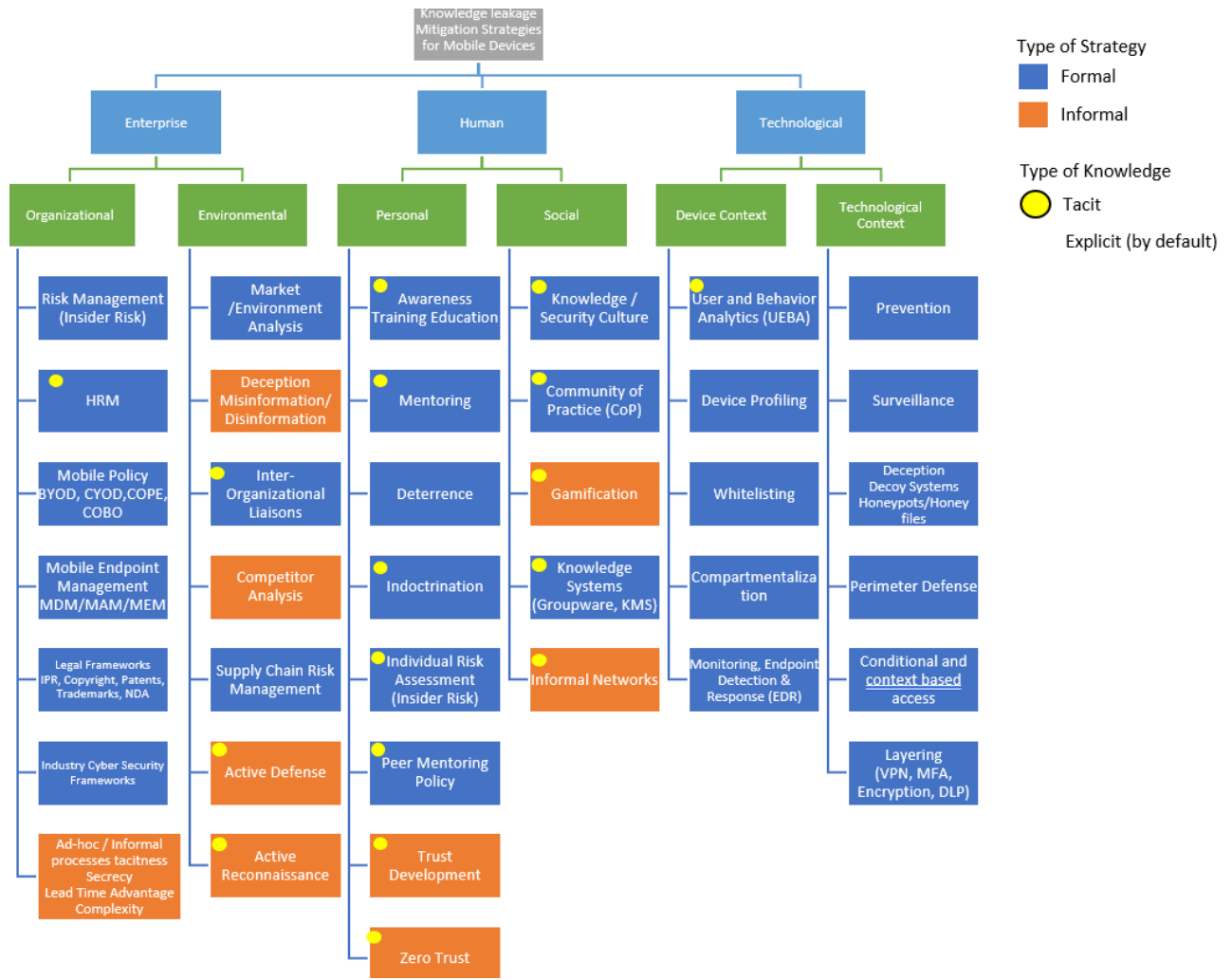

**Figure 6-3. Classification scheme of Knowledge Leakage Mitigation Strategies for Mobile Devices**

As mentioned in the findings chapter, the researcher developed the classification scheme (see Figure 6-3 above) derived from the strategies mentioned by research participants during the interviews, organizational documents provided by some research participants as well as field notes taken during the interactions with interviewees. The collected information was then grounded by the research literature and the research model in order to categorize and structure the different strategies.

The developed classification scheme contributes to the knowledge management and knowledge protection literature by identifying specific strategies used by knowledge intensive organizations. Moreover, the classification also develops a deeper understanding of the different measures used by organizations according to different aspects such as their perceived risk profile (i.e., high, medium, low), level of strategy formality (i.e., formal, informal), and type of knowledge being protected (i.e., tacit, explicit).

The following sections detail the salient strategies based on these aspects and especially, two broad categories of strategies evident from the data collection, i.e., formal and informal strategies.

Formal strategies refer to the specific policies, procedures, and guidelines that an organization implement to protect its sensitive knowledge assets, processes, people and systems. These strategies are usually developed and implemented by the organization are often based on industry standards, culture and best practices. Examples of formal strategies include the implementation of a knowledge management System (KMS). This KMS would outline the procedures for capturing, storing, and disposing of knowledge in a way that complies with legal and regulatory requirements. Such system would also facilitate the retrieval and analysis of historical knowledge, which can be useful for decision-making and planning.

On the other hand, informal strategies refer to the informal, everyday actions that employees take to protect the organization's knowledge assets, processes, people and systems. These strategies are less formal and may not be specifically outlined in the organization's policies procedures and documentation, but they still play a crucial role in keeping the organization operating and competitive. Examples of informal strategies include employees' behaviours, actions or practices to promoting a culture of knowledge-sharing among employees, through informal meetings, mentoring programs, or communities of practice that allow employees to share their expertise.

## 6.5.1 Formal Strategies

A number of formal strategies have been documented in the knowledge management, knowledge protection and security literature. As highlighted throughout this study, a lack of knowledge security may result in the replication of ideas by competitors and impede the exploitation of innovations (Bolisani et al., 2013). Similarly, knowledge leakage, as mentioned previously, may cause reputational damage, productivity and loss of revenue (Ahmad, Bosua, et al., 2014; Bloodgood & Chen, 2021). Therefore, striking the perfect balance between knowledge security and knowledge sharing becomes paramount to addressing the challenges organizations presently face in the current threat landscape (Friedrich et al., 2020; Inkpen et al., 2019; Manhart & Thalmann, 2015).

This problem becomes further exacerbated by the use of mobile devices and mobile technologies in general which favour knowledge sharing to the detriment of security (Jarrahi & Thomson, 2017; Nelson et al., 2017). In order to overcome this challenge, organizations should employ knowledge protection strategies to mitigate the risk of leakage. However, the prevalent knowledge management literature suggests that organizations struggle to define and implement a clear knowledge protection and security strategy (Ahmad, Bosua, et al., 2014; Arias-Pérez et al., 2020; Olander et al., 2014; Ritala et al., 2015).

The interviews and organizational documents provided insights regarding the formal knowledge protection strategies organizations employ to mitigate leakage in the context of mobile devices and worker mobility. Table 6-4 below illustrates the list of formal strategies gathered from the data collection and the supported body of IS in the different fields of knowledge management, information security and mobile computing literature that correlates to the mentioned mitigation strategies and documentation provided by the research participants. Additionally, constructs and literature references have been added to connect and position the research findings within the IS field, particularly in the knowledge management, information security and mobile computing areas.

**Table 6-4. Formal strategies highlighted collected during the interviews and organizational documents**

| Factor | Context | Formal Strategy from data collection | Constructs from Literature | Literature references |
|---|---|---|---|---|
| Enterprise | Organizational | 1. Risk Management<br>2. Human Resource Management (HRM)<br>3. Mobile Policies (MDM, MAM, MTD, MEM)<br>4. Legal Frameworks<br>5. Industry cyber security Frameworks | 1. Risk Management<br>2. HRM<br>3. Mobile Strategy<br>4. IPR<br>5. Mitre ATT&CK, Mitre D3fend, Zero Trust Architecture | (Ahmad, Bosua, et al., 2014; Allam et al., 2014; Amara et al., 2008; Janssen & Spruit, 2019; Kaloroumakis & Smith, 2021; Kindervag, 2010; Morrell, 2020; Olander et al., 2011; Pratama & Scarlatos, 2020; Rose et al., 2020; Strom et al., 2018; Thorleuchter & Van Den Poel, 2013; Weichbroth & Łysik, 2020) |
|  | Environmental | 1. Market/Environment analysis<br>2. Inter-organizational Liaisons<br>3. Supply Chain Risk Management | 1. Environment Analysis<br>2. collaboration/coopetition<br>3. SCRM | (Fredrich et al., 2019; Lascaux, 2020; Maravilhas & Martins, 2019; Pournader et al., 2020; Schniederjans et al., 2020; Zahoor & Al-Tabbaa, 2020) |
| Human | Personal | 1. Awareness, Training and Education<br>2. Mentoring<br>3. Deterrence<br>4. Indoctrination<br>5. Individual Risk Assessment<br>6. Peer Mentoring Policy | 1. Awareness, Training Education<br>2. Mentoring<br>3. Deterrence<br>4. Indoctrination<br>5. Insider Risk/Insider Threat<br>6. Peer Mentoring (HRM) | (D'Arcy & Hovav, 2009; K. C. Desouza et al., 2003; K. C. Desouza & Vanapalli, 2005; Dhillon, 2007; Olander et al., 2011; Ritala et al., 2018; Trkman et al., 2012) |
|  | Social | 1. Knowledge Security Culture<br>2. Community of Practice<br>3. Knowledge Management Systems | 1. Knowledge Management/protection culture<br>2. CoP<br>3. KMS | (Alavi & Leidner, 2001; Bolisani et al., 2013; K. C. Desouza & Vanapalli, 2005; Di Vaio et al., 2021; Intezari et al., 2017; Olander et al., 2011; Ritala et al., 2018) |
| Technological | Device | 1. User behaviour and Entity Analytics<br>2. Device Profiling<br>3. Whitelisting<br>4. Compartmentalization | 1. UEBA<br>2. Mobile Endpoint management | (Ahmad, Maynard, et al., 2014; Arohan et al., 2020; Franklin et al., 2020; Martín et al., 2021) |

| Factor | Context | Formal Strategy from data collection | Constructs from Literature | Literature references |
|--------|---------|--------------------------------------|----------------------------|------------------------|
| | | 5. Monitoring, Endpoint Detection and Response (EDR) | 3. Information Security Strategies | |
| | Technical | 1. Prevention<br>2. Surveillance<br>3. Deception (Honeypots, honey files)<br>4. Perimeter Defence<br>5. Conditional and context based access<br>6. Layering (VPN, MFA, Encryption, DLP) | 1. Information Security Strategies<br>2. Zero Trust Architecture<br>3. Mitre ATT&CK and Mitre D3fend | (Ahmad, Maynard, et al., 2014; Campbell, 2020; Kaloroumakis & Smith, 2021; Karabacak & Whittaker, 2022; Kindervag, 2010; Kindervag et al., 2010; Rose et al., 2020; Strom et al., 2018; Zaheer et al., 2019) |

### 6.5.1.1 Enterprise Factor - Formal Strategies

As mentioned earlier, the enterprise factors refer to the organizational scope and its overarching internal and external components such as culture, behaviour, processes, policies as well as its external stakeholders and relations. Enterprise factors also include the external operating environment within which the organization operates (competitors, customers, regulators, suppliers and partners) as well as the political, demographic, economic and socio-cultural elements (macro environment). Enterprise factors are comprised of organizational context and environmental context. In Table 6-4 above, the researcher collected the different strategies that fall under the enterprise factor umbrella and explored further in the sections below.

#### 6.5.1.1.1 Organizational Context

Under organizational context, the researcher focused on strategies that influenced the internal areas of the firm.

#### 6.5.1.1.1.1    Legal Frameworks

Based on Table 6-4 above, within the enterprise formal strategies which in turn are composed of organizational and environmental contexts, the most salient mitigation strategies reported by participants that pertain to the organizational context, from an (inwardly) internal perspective include risk management, insider risk management, human resource management and legal frameworks clauses (e.g., Non-disclosure agreement, Non-compete, IPR – Copyright, patent, trademark, industrial designs, trade secrets). The findings observed in this study mirror those of the previous extensively studied in the IS field (Amara et al., 2008; Colwill, 2009; Galati et al., 2019; Olander et al., 2011). These formal strategies have already been reviewed in the literature review, but overall organizations that reported the extensive use of legal frameworks tended to be large and enterprise sized organizations. Medium sized organizations, on the other hand, preferred less formal legal mechanisms, which is also in accordance with previous research in the field (Amara et al., 2008; Bolisani et al., 2013; Olander et al., 2009)

#### 6.5.1.1.1.2    Mobile Policies

Interestingly, also within the organizational context, as reported by the research participants, mobile management policies, which collectively refer to organizational measures and rules to manage mobile devices endpoints within organizations — and include mobile device management (MDM), mobile application management (MAM), mobile threat defense (MTD) and mobile enterprise management (MEM) protocols — was not found within the knowledge management literature in the context of knowledge protection for mobile workers and mobile work, but instead within the mobile computing field. This highlights the existing gap in the knowledge management literature in reference to the management and protection of knowledge created as a result of the interaction with mobile devices and therefore represents a contribution to the mobile knowledge management literature. As stated by interviewees, the significance of mobile policies lies in the standardization of procedures. These observations agree with the findings of previous studies in the mobile computing and information security fields that argue that mobile policies ensure proper knowledge management and protection as well as information security measures which  are employed to reduce the risk of explicit and codified knowledge present in mobile devices, i.e., data and information (Song & Lee, 2014; Weichbroth & Łysik, 2020; Zahadat et al., 2015).

### 6.5.1.1.1.3    Cyber Security Frameworks

Equally important, within the enterprise context, research participants also reported the implementation and application of several industry cyber security frameworks used to protect (explicit) knowledge (intellectual property, trade secrets) within organizations, predominantly frameworks such as Mitre ATT&CK, Mitre D3fend and Zero Trust Architecture. These findings seem to be consistent with previous research in related areas such as industry security frameworks (Strom et al., 2018), best practices in information security management (Kaloroumakis & Smith, 2021), and new shifts in paradigms(Campbell, 2020; Kindervag, 2010; Kindervag et al., 2010; Rose et al., 2020) that state that these cyber security frameworks assist organizations in developing capabilities and understanding of specific adversaries' behaviour and how they operate, provide adversary emulation, apply behaviour analytics to identify potentially malicious activity within the organization, and evaluate defensive gap assessments to determine what parts of the organization lack defences and / or visibility, and improve the organization's cyber threat intelligence and understanding of current threats and actors that impact its operating environment. Additionally, these frameworks provide specific components tailored exclusively to address mobile worker, devices, and mobility behaviours. These organizational capabilities contribute to predict and control potential leakage circumstances and preventively limit the negative impact if or when they materialize.

### 6.5.1.1.2 Environmental Context

Within the environmental context, the researcher observed strategies that impact the surrounding of the organization and that unlike the organizational context, such strategies have an outwardly perspective (externally facing).

### 6.5.1.1.2.1    Market/Environment analysis

Similarly, correlated to the previous formal strategy, within the environmental context, a novel strategy in the literature, as reported by research participants, market or environment analysis, although not a traditional measure to mitigate leakage, provides the elements to understand the surrounding context within which the organization operates and confers the tools to pre-emptively manage leakage situations and design knowledge protection mechanisms to safeguard innovation. This finding match those observed in an earlier study (Fredrich et al., 2019) that refer to this strategy as competitor intelligence and it is grounded on the military and intelligence literature. This finding also corroborates what others authors have found and suggest that this strategy is used as a mechanism to understand and *know* adversaries and their behaviour in the operating environment including legislation, competition and governing bodies (Peyrot et al., 2002; Yarhi-Milo, 2014).

### 6.5.1.1.2.2 Inter/Intra Organizational liaisons

In line with the previous measure, another innovative strategy, informed by research participants, the inter / intra organizational liaisons strategy, as gathered from the data collected, corresponds to the collaboration and *coopetition* construct found in the knowledge management literature. This finding is consistent with previous research and more recent research that found that **Coopetition** pertained the cooperation between competing organizations that gain an advantage from integrating with suppliers, customers and other firms that provide complementary products and services and serves to protect organizational knowledge among the participating organizations (Bouncken et al., 2015; Fredrich et al., 2019; Lascaux, 2020). Although paradoxical, the protection of knowledge through sharing seems to be rooted in a more fundamental concept — trust. This research finding is also supported by extensive research in the area of organizational trust and contracts that highlights the importance of trustworthy behaviours and attitudes that reinforce or impede trust level amongst organizations (Jiang et al., 2013; Moein et al., 2015; Ortbach et al., 2015)

### 6.5.1.1.2.3 Supply Chain Risk Management

Finally, another important emerging strategy highlighted by participants within the environmental context is the **supply chain risk management (SCRM)** that although part of the overall risk management strategy mentioned in the organizational context, warrants, according to the researcher, explicit mention in the environmental context as well, as it was overtly addressed on multiple occasions by the interviewees, which emphasizes the importance of knowing the operating environment and the risk posed by third parties such as customers, providers, partners and competitors and their relationships with organizations. In the context of mobile devices, participants alluded to the risk associated with supplier and vendors involved in the production, distribution and maintenance of devices.

Overall, these findings seem to be consistent with more recent research by Pournader and colleagues (2020) which found that Supply chain risk management, also referred to as **third party risk management** (**TPRM**) in the literature, in a more general way, describes the process organizations implement to identify, assess, and mitigate the risk in their supply chain. Furthermore, as emphasized by participants, the study findings show that SCRM / TPRM apply risk management processes, tools, techniques independently or in conjunction with supply chain partners to minimize risks, uncertainties and the adverse impact caused by supply chain or third party entities, which reflects the findings of multiple studies in supply chain risk management (K. C. Desouza et al., 2003; Pournader et al., 2020; Tan et al., 2015; Wong et al., 2021).

## 6.5.1.2 Human Factor - Formal Strategies

Human factors refer to the overall motivations, intents, behaviours, cognitive capabilities and processes, as well as social norms that are explicit and implicit from human behaviours inherent to individuals, and social interactions in group settings. Human factors are comprised of the personal and the social contexts. In Table 6-4 above, the researcher gathered the different strategies that fall under the human factor category and elaborated further in the next sections below.

### 6.5.1.2.1 Personal Context

Following on from Table 6-4 above, the next row in the table (Human) refers to the human factor comprised of the personal and social contexts. Similar to the previous factor, some strategies have been extensively studied and addressed in the knowledge management and information security literature. Strategies mentioned by participants included **awareness, training and education** aimed to prevent leakage. This finding supports previous research in this area (D'Arcy et al., 2009; K. C. Desouza, 2003a; Dhillon, 2007). Similarly, another finding, the use of **deterrence** to minimize leakage as reported in the interviews, it is also supported by the literature (D'Arcy & Hovav, 2007, 2009). In the same vein, the use of **peer mentoring,** as reported by participants, as a way to address the leakage risk remains consistent with previous studies.(Bandura, 1978; Herath & Rao, 2009). Likewise, participants also stated the employment of **indoctrination** as a means to lessen the risk of leakage which also accords with earlier observations (K. C. Desouza, 2011; K. C. Desouza & Vanapalli, 2005; Trkman et al., 2012), that found that by developing the individual's awareness, improving adherence to organizational policies, and influencing their behaviours and attitudes towards knowledge protection and information security measures resulted in a reduction of improper behaviours and decrease in the risk of leakage incidents.

#### 6.5.1.2.1.1    Individual Risk Assessments (Insider Risk/Insider Threat)

Interestingly, the researcher noted that research subjects, from organizations whose reported self perceived risk level was deemed as high, outlined that **individual risk assessments** were often conducted in order to safeguard against knowledge leakage caused by mobile workers and mobile devices. This was rarely the case in medium risk organizations and no evidence was found in low level organizations. This finding is in accordance with previous and recent literature and research in the area of *insider risk* and *insider threat* that have found that insiders who have legitimate access to an organization's resources are often the cause of leakage incidents and warrant the monitoring of employee activities on devices to identify unusual patterns of data access or exfiltration. (Bishop et al., 2010; Colwill, 2009; Crossler et al., 2013; Dtex, 2022; Homoliak et al., 2019; Ponemon Institute, 2021b; Toelle, 2021).

Another interesting finding in relation to the individual risk assessment that the researcher found shows that these types of  risk assessments in high risk organizations constituted a multi-dimensional approach that included different risk factors such as job function, exposure to past and present incidents, access level to sensitive knowledge (and information) within the organization, behaviour and level of compliance to organization policies, which was determined based on behaviour analytics, and compliance in organizational security training. An example of the risk factors and what a personal risk assessment looks like taken from provided documentation from one of the high level organizations is shown below in Figure 6-4 and Figure 6-5 respectively.

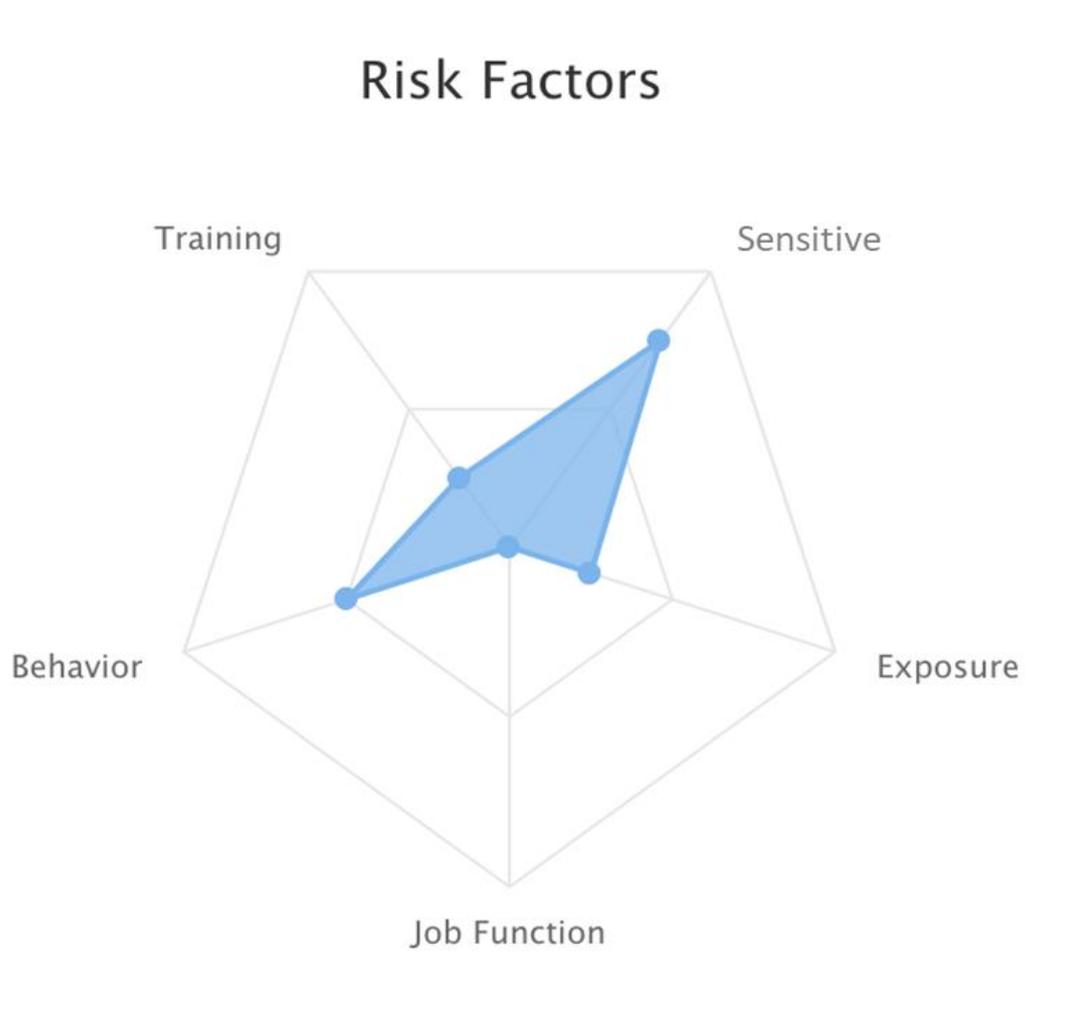

**Figure 6-4. Risk Factors for Individual risk assessment**

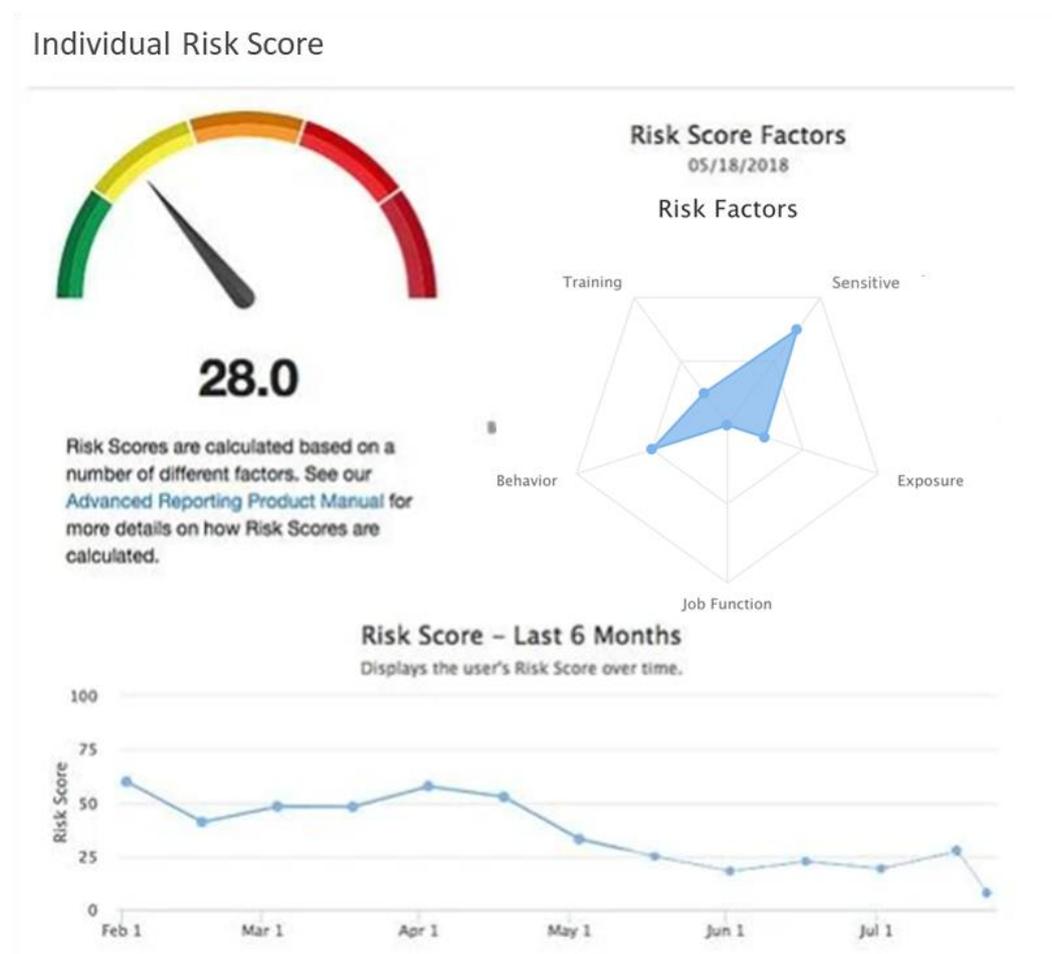

**Figure 6-5. Example of Individual Risk Assessment - Personal Risk Score**

These findings are consistent and mirror those of previous studies such as Olander (2014), Ritala (2018), Mupepi (2017) and more recently Galati (2019) and Bloodgood (2021) that have all examined the effect of accidental leakage and risk management in knowledge intensive organizations, in particular, assessing individual risk as part of the human resource management (HRM) process to emphasize on the human factors and human aspect of knowledge creation and knowledge sharing and as an avenue to proactively protect tacit knowledge that may be inadvertently leaked due to worker's (mis)actions. These findings also illustrate how human factors such as motivations, previous experiences with leakage incidents, previous training and behaviours along with exposure based on job function could determine the level of risk of a particular individual in an organization. These findings corroborate the ideas of Toelle (2021) and Bishop (2010) who suggested that factors such as behaviours attitudes, cognitive capabilities, motivations, and experiences play a crucial role in defining the risk level of people.

However, these previous studies have neglected the analysis of mobile workers and the use of mobile devices as part of the knowledge leakage phenomenon which represents a gap in the current knowledge management literature. In this regard, the researcher found from the participants' interviews that the use of mobile policies in combination with mobile protection technologies to protect knowledge flows as supported by previous research in other fields (Jarrahi & Thomson, 2017; Nelson et al., 2017) can be used in conjunction with the aforementioned individual strategies using a layered approach which is a common method as part of information security strategies (Ahmad, Maynard, et al., 2014) to combine multiple independent countermeasures to further increase the protection strategy effectiveness and reduce the risk of leakage. The mobile policies and technological strategies found by the researcher are further explained under the technological factors in section 6.5.1.3.

### 6.5.1.2.2 Social Context

Within the social context, the researcher grouped strategies that address and affect behaviours and actions at the group level within organizations.

#### 6.5.1.2.2.1 Knowledge Security Culture

With respect to the social context, as displayed in Table 6-4, one of the recurring themes reported by the research participants referred to **knowledge security culture** as a strategy to foster proper behaviours and management of organizational knowledge to prevent unintended disclosure. This research finding further support the notion from authors in the knowledge management literature, who have stated that culture in organizations can be defined in terms of ideologies, sets of beliefs, basic assumptions, shared sets of core values, important understandings and collective will (Al Saifi, 2015; Alavi et al., 2005). Furthermore, in line with the observations and comments from research participants in relation to organizational mobile policies, procedures and practices, these findings reflect what other knowledge researchers have suggested that knowledge management culture should also include other more explicit cultural artifacts such as norms, practices, symbols and ideologies (Alavi & Leidner, 2001; Intezari et al., 2017). During the study interviews, the researcher also found that these constructs lay the foundation to understanding social group behaviour within organizations. More specifically, these findings further support the idea that the knowledge protection culture, which as defined in the knowledge protection literature, alludes to the organizational practices and behaviour shaping activities that addresses group behaviours and attitudes towards the security of knowledge within organizations (K. C. Desouza & Vanapalli, 2005; Olander et al., 2011). Although as it was discovered during the interviews, organizational behaviour focuses on three levels, namely the individual, group and organizational levels; the organizational culture, however, emphasizes mostly on the group and organizational levels over the individual one, which corroborates the ideas of researchers (Al Saifi, 2015; J. B. Barney, 1986; Intezari et al., 2017) who suggest that organizational culture seeks to influence the collective actions, practices and habits of groups to improve organizational performance, which emphasizes the significance of knowledge security culture as a strategy approach.

### 6.5.1.2.2.2    Communities of Practice (CoP)

Similar to knowledge protection culture, the next recurrent strategy theme in the interviews pertained to **communities of practice** (**CoP**) within organizations as a way to ensure tacit knowledge remained as part of the organizational knowledge repository to safeguard against the risk of leakage caused by mobile devices and also when employees leave the firm. This strategy also reinforces the previous measure, i.e., culture, as it promotes collaboration, sharing, community management and coordination. This finding matches those of Alavi et al. (2005) who suggested that CoP help organizations foster a culture of continuous learning and improvement, leading to the development of internal capabilities and resources aimed to protect internal knowledge. Moreover, as observed and mentioned during the interviews, communities of practice also encourage the organic growth of local cultures within the organization which in turn fosters the transfer and protection of tacit knowledge which confirms what Nonaka and colleagues (1995) found in their earlier study.

### 6.5.1.2.2.3    Knowledge Management Systems (KMS)

Not surprisingly, **Knowledge Management Systems** (i.e., KMS), as reported by research participants, related to yet another key strategy, within the social context, aimed towards storing and retrieving of codified knowledge and to improving the understanding of tacit knowledge as well as the encouraging of collaboration amongst knowledge workers. Participants informed that KMS served as a way to centralize the knowledge repository in their organizations, and also provided structure to the knowledge creation and sharing process by facilitating the templates and frameworks to different knowledge assets such as intellectual property, processes, methodologies, software, research, proposal templates and product designs. Examples of KMS, mentioned during the interviews included SharePoint, Google Collaboration Suite, Amazon and Office365 groupware platforms. The use of messaging systems such as Yammer, Jabber, Microsoft Teams among others were also highlighted as a way to access tacit knowledge via subject matter experts within organizations. Incidentally, along with collaboration suites, email also remained a common and preferred communication channel to share and seek organizational (tacit) knowledge from knowledge experts in organizations whose reported risk ranked as low and medium.

In this regard, the findings of this study under the social context with respect to knowledge protection culture, communities of practice, and knowledge management systems corroborate the results of previous research in the knowledge management field (Al Saifi, 2015; Alavi et al., 2005; Alavi & Leidner, 2001; D G Schwartz, 2006; Trkman et al., 2012) which highlights the importance of developing each one of these aspects to improve the overall social environment of the firm and to promote a safe knowledge sharing culture amongst knowledge workers. Moreover, as mentioned by interviewees, the combination of these (social) strategies can be synergistic and produce better results than using only one or the other, due to the fact that the advantages of one measure will usually complement the disadvantages of another.

## 6.5.1.3 Technological Factor – Formal Strategies

The next item in Table 6-4 above relates to the technological factor. The technological factors are composed of the device context and technological context and refer to the overarching technology, infrastructure, tools, and information systems that enable and facilitate the interaction of users with technology and technical artefacts in the context of knowledge-sharing activities and mobile knowledge work facilitated by the use of mobile devices.

Under this factor, the researcher has collected and listed the technical strategies recounted by research informants as the salient mechanisms to protect against the risk produced by mobile devices.

### 6.5.1.3.1 Device Context

First, in the device context, which pertains to the specific measures that address the communication device itself (i.e., mobile, tablet, laptop), the most significant informed strategies include: user behaviour and entity analytics (UEBA), Device Profiling, Whitelisting, compartmentalization, Monitoring, Endpoint Detection and Response. Each one of these strategies corresponds to different information security measures in the mobile computing and Information security management literature.

#### 6.5.1.3.1.1    User Entity Behaviour Analytics (UEBA)

Research participants reported how **UEBA** served as a security approach that leverages the use of artificial intelligence and machine learning to monitor and analyze user's behaviour within their organizations to establish a behavioural baseline to help detect insider threats, and abnormal behaviour that may indicate attempts to exfiltrate intellectual property and other anomalies that may lead to potential leakage incidents, particularly through the use of mobile devices. Further, UEBA was also reported as a way of analyses patterns of human behaviours linked to the device usage and individual characteristics such as the device ID, location, time, regular resource access and usage, as well as conditional access and contextual authentication (i.e., based on user and device context, need to know and least privilege principles, and strictly enforced by information security policy) to determine abnormal behaviour. This finding supports what recent information security researchers (Arohan et al., 2020; Martín et al., 2021) have found in the application of behaviour analytics within the context of mobile devices.

Interestingly, UEBA constitutes the only technological measure derived from this study's findings that may be used to protect tacit knowledge due to the predictive capabilities and the use of artificial intelligence and machine learning to pre-emptively determine and detect a potential tacit knowledge leakage incident based on the (abnormal) behaviour of a particular user that while may not be stopped, it may well assist organizations in applying preventive measures to contain and limit the extent of the leakage that otherwise would have been undetectable and potentially greater in extent.

Similarly, UEBA also helps to determine intent (i.e., inadvertent vs. malicious) which means it may also assist in preventing accidental tacit knowledge leakage when this incident results from (non malicious) unintentional or negligent mis-behaviours and mis-habits that can be addressed pre-emptively. Examples of abnormal behaviours relate to logging in outside of usual hours, emailing sensitive data to outside parties, an insider worker or a privileged user sending emails with high volume of organizational data to external or personal email accounts, employee copying a high number of files from a network to an external drive or unmanaged device. Each one of these types of behaviours would trigger UEBA alerts, inform preventive actions and determine risk scores for a particular user and device.

Furthermore, this particular finding constitutes a novel and significant contribution to the knowledge management protection literature as it addresses one of the gaps previously mentioned in the literature review regarding the lack of technological measures to address tacit knowledge protection which was also supported and raised by previous research (Ahmad, Bosua, et al., 2014; Bolisani et al., 2013; Manhart &

Thalmann, 2015; Trkman et al., 2012). Therefore, the application of this particular strategy adds to the current body of knowledge and contributes to the knowledge protection and mobile knowledge management literature.

#### 6.5.1.3.1.2 Device Profiling and Whitelisting

In line with the previous strategy, other measures such as **device profiling** and **whitelisting** as expressed by the research informants, constitute a part of the overall and more encompassing mobile endpoint management process as referred to in the mobile computing literature. As stated by participants, mobile endpoint management indicates how to manage mobile devices within the enterprise and relates to the access control and identification of mobile devices, applications and contextual mobile features (e.g., context as determined by user id, digital fingerprint, location, device signature, time, user and IP address) that the organizational mobile policy has explicitly approved and allowed to access the enterprise resources. This finding supports what other authors (Alexandrou & Chen, 2022; Franklin et al., 2020) have found in the mobile computing and the information security domains which showed that whitelisting represents a much stricter measure than blacklisting as by default, the former denies access to resources and only explicitly grants access to those resources proven to be safe, reducing in this way, the potential risk of leakage as compared to the latter approach.

#### 6.5.1.3.1.3 Compartmentalization, monitoring, Endpoint Detection and Response (EDR)

With regard to the **compartmentalization**, **monitoring**, **detection** and **response** as informed by interviewees, the researcher found that these information security strategies as reported earlier in the literature review, have been previously studied by IS researchers (Ahmad, Maynard, et al., 2014; Herath & Rao, 2009; Ifinedo, 2012; Sveen et al., 2009), however the focus has been predominantly on information rather than knowledge using the more conventional approach of a perimeter based model security that disregards the challenges posed by mobile workers, therefore the mobility component and mobile device usage element have, for the most part, been neglected in those studies.

The findings in this study show that organizations are starting to include solutions that address the mobile workflow as well as devices adopting a more novel approach by securing data and devices wherever they are located, regardless of their location or whether they are inside or outside the corporate network.

This finding confirms what more recent studies in the information security domain have also found, such as those by Karabacak & Whittaker (2022), Rose et al. (2020), and Zaheer et al. (2019). These studies have considered the mobility and pervasiveness of mobile devices and have offered a perimeter-less security architecture perspective and less conventional approaches, which differ from the traditional and extensive approaches found in the knowledge management literature.

Intriguingly, based on the interviews, the researcher observed that, where once the perimeter was restricted to the organization itself, this is no longer the case, organizations now exist in an age where the mobile user themself has become the new perimeter. Therefore, a new approach must evolve in order to fit this new reality. In the following section, these novel approaches are further explored.

### 6.5.1.3.2 Technical Context

In accordance with the previous measures related to the device (context), in the technical context, the researcher focuses on the strategies that surround the device (context) and target the operating environment within which the mobile worker interacts with the device and the organization's infrastructure.

As stated in the preceding section, the device is subject to different measures that, by themselves, produce incomplete results unless used in combination with the overarching framework of the organizational measures. As per Table 6-4 above, such organizational strategies include prevention, surveillance, deception, perimeter defence, layering, conditional and contextual based access. Most of these strategies found by the researcher during the data collection have been already described in the literature review and support the previous research associated to knowledge protection (Ahmad, Maynard, et al., 2014; Herath & Rao, 2009; Ifinedo, 2012; Sveen et al., 2009), however, as indicated earlier in the preceding section, the same caveat applies — i.e., the emphasis on these studies referred to information rather than knowledge, failure to address mobile workers and mobility, and the perspective of analysis considered the organizational perimeter as one of the main criteria to define what can be trusted , that is, inside vs outside of the organizational boundaries.

### 6.5.1.3.2.1    Layering (Defence in Depth, VPN, MFA, Encryption, DLP)

Furthermore, the layering of strategies as reported by research participants focused on the combination of independent technical information security measures that, combined together, increase the effectiveness of the organizational defence. Salient measures referenced by interviewees, in this regard, included multi-factor authentication (MFA), virtual private networks (VPN), encryption and data loss prevention or data leakage prevention (DLP). Incidentally, with respect to DLP, participants commented how this measure was extensively used to detect and prevent data breaches and leakage of organizational knowledge (e.g., intellectual property) as DLP focuses on detecting illicit transfer of data outside the enterprise perimeter. However, interestingly, because of this same reason, DLP failed to work properly in the context of mobile worker and mobile devices due to the fact that the organizational perimeter was not as clearly delimited, and in some cases, it was non-existent. Additionally, given that DLP uses a strict, even rigid, rule based system, in some cases, it may fail to adapt to more complex and dynamic situations like the ones presented by mobile workers and mobile devices. This highlights the issue that solutions like DLP that may work effectively under normal circumstances when the user is behind and protected by the organizational perimeter, will become less effective in cases where mobile workers and mobile devices interact outside the aforementioned perimeter. This finding corroborates the ideas of Yu and colleagues (2018) who suggested that current DLP solutions may be ineffective and inefficient for mobile devices and require a novel contextual model approach to address these deficiencies and dynamically adjust to the changing conditions. Therefore, it remains critical to reinforce the premise of using layering of multiple strategies to overcome and counteract the deficiencies of one measure with the strengths of other strategies.

### 6.5.1.3.2.2    Deception

Interestingly, the researcher found another remarkable recurring theme that pertains to the sophisticated use of **deception** strategies, particularly in organizations that ranked their self reported risk of leakage as high as a means to protect their knowledge assets from knowledge leakage incidents. In these organizations, the use of different types of decoy entities such as honeypots, honeynets, honeyfiles or -tokens, even honey personas and honey user accounts were employed to misdirect, deceive, and even interact with adversaries in an effort to learn from their operating and complex environment as well as from these cyber actors' behaviours. For mobile devices, in particular, the use of MDM (Mobile Device Management) software to create a fake mobile device, was reported to be used to deceive attackers and insiders by mimicking the device's behaviour during phishing attacks and simulations. The main objective of this strategy, as reported by participants, is to increase the complexity, cost, uncertainty, and time for the attacker, and slightly tip the balance in favour of the defender, for a change, given that defenders are usually at a disadvantage. This finding corroborates the ideas of previous security authors such as Ross Anderson (2001), Libicki and colleagues (2015) who state that as has often been the case in information security, the attacker has the advantage and can strike at any time. Further, to the concerns expressed by participants in relation to defenders, the research findings are consistent with those of security author Libicki and colleagues who suggested that this concept is known as the attacker's advantage and the defender's dilemma (Libicki et al., 2015) and summarized it in four principles:

1. Principle one: the defender must defend all points while the attacker can choose the (one) weakest point.

2. Principle two: the defender can defend only against known attacks while the attacker can probe for unknown vulnerabilities.

3. Principle three: the defender must be constantly vigilant while the attacker can strike at will.

4. Principle four: the defender must play by the rules while the attacker can play dirty.

In addition to this, and also in line with the findings, Ross Anderson (2001) stated that: "*Attack is simply easier than defence, defenders have to be right 100% of the time while attackers only need to be right once*" (Anderson, 2001, p. 7). The findings from the research interviews show that by leveraging deception strategies, defenders can afford to spend more time analysing and have the opportunity to learn from the attack, assess their own response capabilities, and level the playing field by allowing the defender to actively observe the attacker potentially causing the adversary to waste resources and reveal tools, tactics, techniques and procedures (TTTPs), intent, and targeting. In doing so, defenders (organizations) will be turning the attacker's advantage into the defender's advantage. This finding corroborates the ideas of security researchers (Fai & Goh, 2021; Matre et al., 2021) who state that such an advantage alludes to the concept that organizations defend against attackers in their own environment. This presents a major advantage due to the fact that organizations have control over the landscape where they will face their attackers.

The application of deception strategies from the information security domain constitutes a novel approach and significant contribution within the knowledge protection literature as it provides innovative means to prevent knowledge leakage through mobile devices when used in combination with MDM.

### 6.5.1.3.2.3    Active defence

In line with the previous finding, participants reported the use of proactive strategies to detect, prevent, and respond to leakage incidents through mobile devices. Particularly, participants that deemed their self reported risk of leakage as high and medium. These approaches involved actively seeking out potential threats and vulnerabilities in their systems, devices and operating environments, as well as using various techniques to disrupt, degrade, or neutralize cyber threats. This finding further validates the idea of using **active defence** mechanisms and seems to be consistent with other research (Denning, 2014; Stech et al., 2016; Strom et al., 2018) that found that organizations use deception tools, tactics, techniques and procedures (TTTPs) into adversary modelling systems to improve organizational leakage mitigation capabilities, learn from competitors and adversaries, and assess internal response capabilities.

Such modelling systems as mentioned by research participants, included the Mitre ATT&CK and Mitre D3fend cyber security industry frameworks (these strategies were also mentioned as part of the organizational context strategies). The Mitre ATT&CK framework (stands for Adversarial Tactics, Techniques and Common Knowledge), created in 2013 by the Mitre corporation. As explained by research participants, the framework is a curated knowledge base and model for adversary behaviour in the different phases of the attack life cycle (i.e., Reconnaissance, Resource Development, Initial Access, Execution, Persistence, Privilege Escalation, Defense Evasion, Credential Access, Discovery, Lateral Movement, Collection, Command and Control, Exfiltration and Impact) and provides an understanding on attacker models methodologies and mitigation. This observation is in line with previous research (Heckman, Stech, Schmoker, et al., 2015; Heckman, Stech, Thomas, et al., 2015) in the field of threat intelligence within the information security domain.

The framework features matrices (TTTPs) for enterprise, mobile and industrial control systems (See Figure 6-6 below for the Mitre ATT&CK framework for mobile devices).

The Mitre D3fend framework, on the other hand, as reported by participants, complements the Mitre ATT&CK framework as it provides defensive countermeasures to assist their organizations in planning and tailoring their defences for common Mitre ATT&CK techniques. The findings also suggest that the D3fend framework in combination with the ATT&CK framework clearly identifies and precisely specifies strategies and capabilities used by organizations to improve their defence and proactively reduce the leakage risk through devices (see Figure 6-7 below). This observation seem to be consistent with more recent findings by Kaloroumakis & Smith (2021) who found that by using both frameworks together, organizations can gain a more complete view of their security posture and develop more effective defensive strategies that address the specific risks and threats posed by mobile devices.

## Device Access

| Initial Access | Execution | Persistence | Privilege Escalation | Defense Evasion | Credential Access | Discovery | Lateral Movement | Collection | Command and Control | Exfiltration | Impact |
|---|---|---|---|---|---|---|---|---|---|---|---|
| 9 techniques | 4 techniques | 9 techniques | 4 techniques | 21 techniques | 11 techniques | 9 techniques | 2 techniques | 18 techniques | 9 techniques | 4 techniques | 11 techniques |
| Deliver Malicious App via Authorized App Store | Broadcast Receivers | Broadcast Receivers | Code Injection | Application Discovery | Access Notifications | Application Discovery | Attack PC via USB Connection | Access Calendar Entries | Alternate Network Mediums | Alternate Network Mediums | Call Control |
| Deliver Malicious App via Other Means | Command-Line Interface | Code Injection | Device Administrator Permissions | Access Sensitive Data in Device Logs | Evade Analysis Environment | Exploit Enterprise Resources | Access Call Log | Call Control | Commonly Used Port | Carrier Billing Fraud |
| Drive-by Compromise | Native Code | Compromise Application Executable | Exploit OS Vulnerability | Delete Device Data | Access Stored Application Data | File and Directory Discovery | | Access Contact List | Commonly Used Port | Data Encrypted | Clipboard Modification |
| Exploit via Charging Station or PC | Scheduled Task/Job | Foreground Persistence | Exploit TEE Vulnerability | Device Lockout | Capture Clipboard Data | Location Tracking | | Access Notifications | Domain Generation Algorithms | Standard Application Layer Protocol | Data Encrypted for Impact |
| Exploit via Radio Interfaces | | Modify Cached Executable Code | | Disguise Root/Jailbreak Indicators | Capture SMS Messages | Network Service Scanning | | Access Sensitive Data in Device Logs | Remote File Copy | | Delete Device Data |
| Install Insecure or Malicious Configuration | | Modify OS Kernel or Boot Partition | | Download New Code at Runtime | Exploit TEE Vulnerability | Process Discovery | | Access Stored Application Data | Standard Application Layer Protocol | | Device Lockout |
| Lockscreen Bypass | | Modify System Partition | | Evade Analysis Environment | Input Capture | System Information Discovery | | Call Control | Standard Cryptographic Protocol | | Generate Fraudulent Advertising Revenue |
| Masquerade as Legitimate Application | | Modify Trusted Execution Environment | | Geofencing | Input Prompt | System Network Configuration Discovery | | Capture Audio | Uncommonly Used Port | | Input Injection |
| Supply Chain Compromise | | Scheduled Task/Job | | Hooking | Keychain | System Network Connections Discovery | | Capture Camera | Web Service | | Manipulate App Store Rankings or Ratings |
| | | | | Input Injection | Network Traffic Capture or Redirection | | | Capture Clipboard Data | | | Modify System Partition |
| | | | | Install Insecure or Malicious Configuration | URI Hijacking | | | Capture SMS Messages | | | SMS Control |
| | | | | Masquerade as Legitimate Application | | | | Data from Local System | | | |
| | | | | Modify OS Kernel or Boot Partition | | | | Foreground Persistence | | | |
| | | | | Modify System Partition | | | | Input Capture | | | |
| | | | | Modify Trusted Execution Environment | | | | Location Tracking | | | |
| | | | | Native Code | | | | Network Information Discovery | | | |
| | | | | Obfuscated Files or Information | | | | Network Traffic Capture or Redirection | | | |
| | | | | Proxy Through Victim | | | | Screen Capture | | | |
| | | | | Suppress Application Icon | | | | | | | |
| | | | | Uninstall Malicious Application | | | | | | | |
| | | | | User Evasion | | | | | | | |

**Figure 6-6. Mitre ATT&CK for Mobile Devices.**

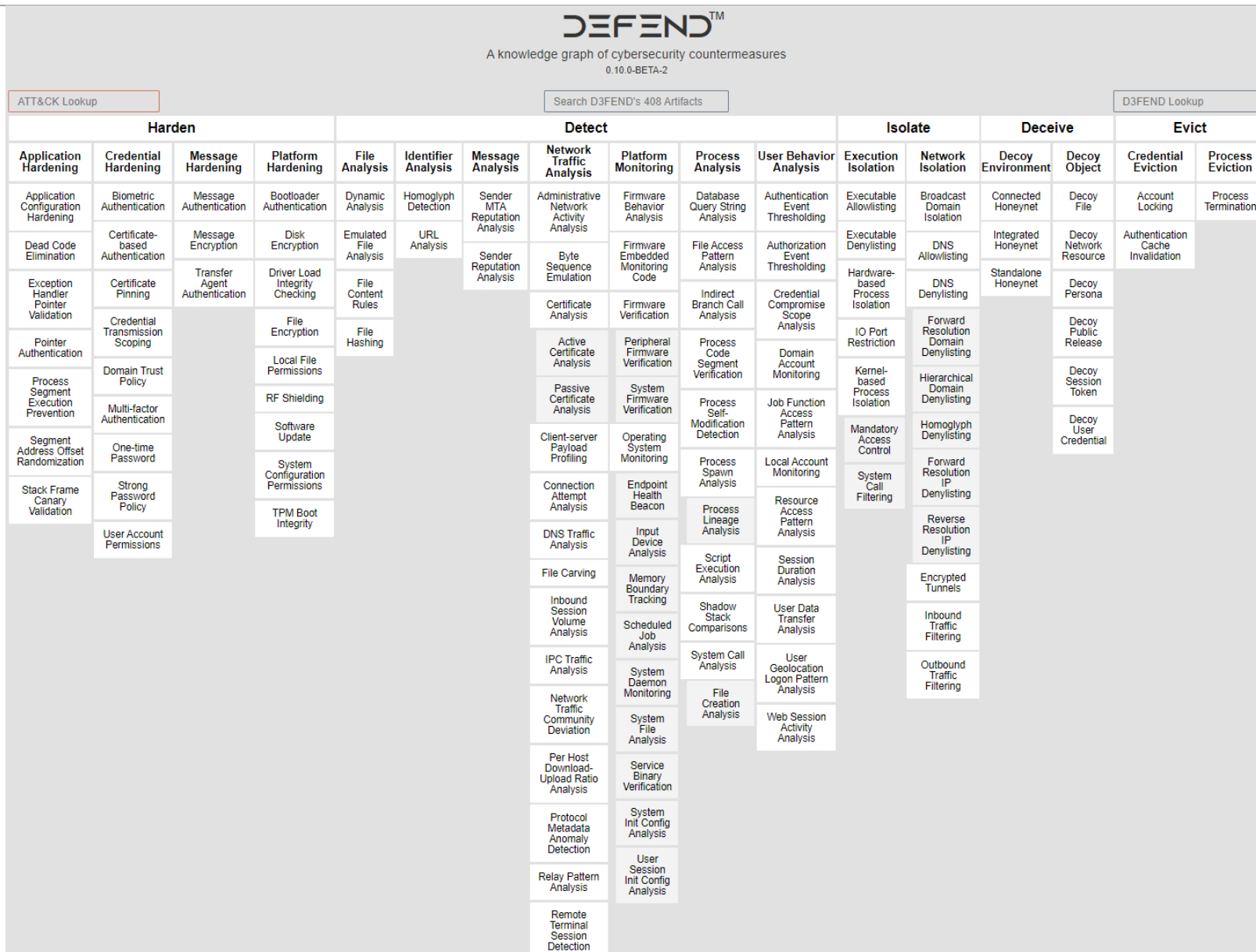

**Figure 6-7. Mitre D3fend Framework**

#### 6.5.1.3.2.4    Zero Trust

Another key finding derived from the data collection process relates to the use of a **zero trust** concept as a strategy to protect organizational explicit knowledge (i.e., codified) leakage. Surprisingly enough, in the information security literature, the concept of **zero trust architecture** is not new, and already existed from 2010 when the term was coined by John Kindervag (Kindervag, 2010). Research participants indicated that due to mobile devices and the mobility of the workforce zero trust was based on the initial concept of moving from a perimeter based security model to one that focused on the security of individual transactions, i.e., de-perimeterization. The de-perimeterization concept evolved and improved into the larger concept of zero trust. Zero trust architecture (ZTA) challenges the idea of the current trust paradigm and trust model conventionally used in information security and questions the organizational perimeter based model whereby organizations contain trusted and untrusted zones and networks (i.e., inside organizational perimeter vs outside organizational perimeter), trusted and untrusted users (i.e., authenticated vs non authenticated), trusted and untrusted applications and data (authorized vs non authorized) and instead  ZTA proposes that everything should be untrusted and always validated and verified by default, before access is granted to resources, thereby embedding security by design, and not as an afterthought as has historically been the case.

This finding matches those observed by other researchers (Colwill, 2009; Kindervag, 2010; Zaheer et al., 2019) who showed that this change in paradigm arises from the fact that in the current paradigm, malicious insiders are trusted by default. This situation demands a new trust model.

The researcher also observed that participants mentioned strategies for ZTA that reuse several information security strategies such as compartmentalization which is referred to in ZTA as micro segmentation and it is applied holistically to devices, network, users and applications to create what ZTA defines as micro perimeters, that is, small secured boundaries defined by a device, an application or data as compared to the regular network perimeter that includes multiple network nodes and resources. This observation supports what Rose and colleagues (2020) showed in their research concerning the idea that micro perimeters embed prevention, detection, and perimeter defence mechanisms as part of a single management unit.

Additionally, as its name suggests, zero trust assumes that no user or asset is trustworthy based on location, status, profile, or asset ownership (enterprise owned vs personal devices like BYOD) unlike the regular trust model. This characteristic makes it ideal for the management of mobile workers and mobile devices given that as stated previously, and emphasized by participants, the organizational perimeter, as we know it, no longer exists.

This strategy in combination with mobile endpoint management explained in the previous section (device context) ensures that mobile devices are contained and better controlled in their own micro perimeter, reducing the risk of leakage.

This finding represents a novel contribution from the information security domain into the knowledge protection literature and addresses the gap highlighted by previous authors in relation to knowledge and mobile devices (Ahmad, Bosua, et al., 2014; Ilvonen et al., 2018; Manhart & Thalmann, 2015).

### 6.5.1.3.2.5    Conditional and Contextual Access

Participants also highlighted that ZTA also stresses the importance of conditional and contextual access to resources which is particularly relevant for mobile devices. During the interviews, interviewees mentioned that conditional access alluded to the different conditions that a user must meet in order to access an organizational resource (e.g., knowledge asset), the most common access conditions include identity, location, time, device, application, verification, authorization level, risk level and exceptions. Conversely, contextual access makes use of context-based policies which leverage on and extend access conditions. This observation reflects those ideas of Karabacak, Whittaker (2022) as well as Rose (2020) who also highlight the importance of taking into account previous rich historical information related to the specific entity (i.e., context). For example, before granting access to a knowledge asset, a contextual access policy will first gather rich information about the user's context (i.e., who is the user?; is the user in a risky group? Has the user been assigned a high risk score?), application context (which application or resource is the user trying to access?; do they usually access it or is it the first time?), device context (is the user using a recognized device? ; is it a new device? Is it patched?; is it a corporate issued or personal owned device?; what is its security posture?), location context ( is the user accessing from a new geo-location or their regular location?), network context (is the user using a VPN? ; Is it a new IP address?). Once these contexts have been evaluated, the contextual policy could be dynamically set to permit seamless access to the resource if the device is a corporate owned device within the organization, but otherwise if the device is in a different country and personally owned, then the contextual policy could prompt the user for authentication and MFA, for example.

Overall, this important finding represents a novel contribution to the knowledge protection literature and supports more recent research from the information security domain (Buck et al., 2021; Campbell, 2020; Collier & Sarkis, 2021) and corroborates the idea that trust remains a misused concept, particularly in the realm of information security, and thus having trusted actors and zones (within the organization) vs untrusted actors and zones (external to the organization) represents a flawed model and should be revised, and therefore assuming that the untrusted users (whether intentionally or unintentionally) are already within the organization, i.e., insider threats, should become the new norm.

## 6.5.2 Informal Strategies

During the interviews, research participants talked about several informal strategies that although were not documented or formally established within organizational processes and procedures warrant mentioning as they were deemed critical to define knowledge protection strategies to mitigate leakage in the context of mobile devices and worker mobility. Also, important to note is the fact that participants only reported informal strategies for the enterprise and human factors, no informal strategies were mentioned for the technological factors. Table 6-5 below illustrates the list of informal strategies gathered from the data collection and the supported body of knowledge management, information security and mobile computing literature that correlates to the mentioned mitigation strategies. Additionally, constructs and literature references have been added to connect and position the research findings within the IS field, particularly in the knowledge management, knowledge protection, information security and mobile computing areas.

**Table 6-5. Informal Strategies collected during the interviews**

| Factor | Context | Informal Strategy from data collection | Constructs from Literature | Literature references |
|---|---|---|---|---|
| Enterprise | Organizational | 1. Ad-hoc Processes (Secrecy, Tacitness, Lead time advantage, complexity, Fast Innovation Cycle) | 1. Informal Intellectual property protection mechanisms (Informal IPPMs) | (Amara et al., 2008; Grimaldi et al., 2021; Manhart & Thalmann, 2015; Miric et al., 2019; Päällysaho & Kuusisto, 2011; Pedraza-fariña, 2017; Ritala et al., 2015) |
| | Environmental | 1. Deception / Misinformation/ Disinformation 2. Competitor Analysis 3. Active defence 4. Active Reconnaissance | 1. Misinformation & Disinformation 2. Competitive Intelligence Analysis 3. Active Defence 4. Active Reconnaissance | (Atwood, 2019; Dalton et al., 2020; Denning, 2014; Fleisher & Bensoussan, 2003; Guess & Lyons, 2020; Hodges, 2005; Melnitzky, 2011; Promnick, 2017) |
| Human | Personal | 1. Trust development 2. Zero Trust | 1. Trust Management 2. Zero Trust Architecture | (Bertino et al., 2006; Ford, 2004; Kindervag, 2010; Kindervag et al., 2010; Rose et al., 2020) |
| | Social | 1. Gamification 2. Informal networks | 1. Knowledge Management/protection culture 2. CoP 3. KMS | |

### 6.5.2.1 Enterprise Factors – Informal Strategies

Based on Table 6-5 above, under the enterprise factor, the researcher compiled the informal strategies reported by research informants that belong to the organizational and environmental context.

#### 6.5.2.1.1 Organizational Context

Under the organizational context, the researcher listed the informal strategies that address internal resources and capabilities of the firm.

##### 6.5.2.1.1.1    Ad-hoc processes

Within the organizational context, **ad-hoc processes** reported by research participants  such as secrecy, tacitness, lead time advantage, and complexity have been extensively studied in the knowledge management literature by IS researchers (Amara et al., 2008; Grimaldi et al., 2021; Manhart & Thalmann, 2015; Miric et al., 2019; Päällysaho & Kuusisto, 2011; Pedraza-fariña, 2017; Ritala et al., 2015) and already discussed in the literature review.

Research informants from organizations whose self perceived risk ranked as low and medium, stated that these ad-hoc routines were a preferred way of protection of organizational knowledge over more formal or legal frameworks stemming from the fact that these practices are usually simple and economical to use. However, although organizations with a reported high risk level, for the most part, favoured formal practices, they employed these informal practices in combination with formal and legal strategies to improve their mitigation capabilities. These research findings match those observed in earlier studies (Amara et al., 2008; Olander et al., 2014; Päällysaho & Kuusisto, 2011) and suggest that an association between risk level and degree of protection formality (formal vs informal) may exist.

However, contrary to expectations and previous research (Kitching & Blackburn, 1998; M. Lee et al., 2018; Zhao, 2006), this study did not find a significant association between size and degree of protection formality, which may likely be due to the nature of the sample size. This represents a limitation of the research study and provides a further avenue for future research.

### 6.5.2.1.2 Environmental Context

Under the environmental context, the researcher grouped the informal strategies that target external components of the firm that affect its surrounding and operating environment.

#### 6.5.2.1.2.1    Misinformation/Disinformation (Deception)

Interviewees mentioned a few important informal strategies within the environmental context. Interestingly, **deception** remained one of the most common reported strategies amongst high risk level organizations, particularly the use of misinformation/disinformation to mislead competitors and adversaries and as a means to protect organizational knowledge (please refer to section *6.5.1.3.2.2 Deception*).

As participants indicated, misinformation relates to misleading information presented as facts that may be unintentional in nature, disinformation, on the other hand, as a subset of misinformation, refers to deliberately deceptive information. This observation echoes that of Guess and Lyons (2020) who suggested that the difference between misinformation and disinformation lies in the intentionality of the claim.

According to research participants, this practice remained undocumented with the intention of concealing it from potential rivals. By fabricating information, organizations often gain more time to work on their research and development products and services, and also serves as a disguise for manipulating their environment and adversarial stakeholders. Research informants stated that the use of message boards and dark web forums were often used to spread false information about priorities, technologies and fictional projects as well as gathering insights which provided them with time and intelligence about the market and adversaries. Aligned with what was discussed in the formal strategies within the technical context (see section 6.5.1.3.2 , page 216), deception represents an emerging mitigation and proactive strategy that allows organizations to lead the knowledge protection process and gather intelligence about the market and competitors which translates into better organizational leakage protection and mitigation capabilities. Although other deception strategies have been previously studied in the technical information security literature (Denning, 2014; Stech et al., 2016; Strom et al., 2018), the specific research findings from this study in relation to misinformation and disinformation within the environmental context as a knowledge protection strategy have not been previously described in the knowledge management and protection literature. The researcher considers that this constitutes a novel approach and warrants future research.

### 6.5.2.1.2.2    Competitor Analysis

In line with this, the next strategy from the environmental context alludes to **competitor analysis** which is an informal strategy, although related to the formal market & environment analysis strategy discussed in section 6.5.1.1.2 (page 204), differs from the latter in that, first, according to the research participants, it focuses on identifying only relevant competitors in the field with their weaknesses and strengths, second, it remains embedded in the organizational routine with no supporting documentation, and third, it leverages misinformation and disinformation to gather intelligence about competitor's behaviours and competitive advantage. Although most research informants commented on the use of competitor analysis/intelligence strategies, the researcher found that predominantly medium and high level risk organizations were more likely to apply such an approach.

Earlier research has studied the use case and impact of competitive analysis in business strategy (Fleisher & Bensoussan, 2003; Hodges, 2005) which matches and further supports the research findings in this regard.

### 6.5.2.1.2.3    Active defence

Another interesting strategy from the data collection refers to the use of an **active defence** approach to pre-emptively protect tacit knowledge, conforming with the previous strategies, this mitigation approach leverages other aforementioned measures such as deception to increase the cost of conducting knowledge appropriation for competitors or adversaries and reduce the cost of knowledge protection for organizations. Research subjects often mentioned the use of misdirection usually via misinformation and disinformation as well as decoys used as bait to waste rivals' time and efforts all while organizations gather counterintelligence on their behaviours and motives. Moreover, complementing active defense, organizations also informed on the use of artificial intelligence and machine learning powered platforms to proactively analyze their internal employees' behaviour (behavioural analytics) and pre-emptively establish behavioural and risk baselines and profiles (insider risk) that assist them in determining potential knowledge leakage incidents caused either by worker's inadvertent negligence or purposeful intent. By focusing on individuals' behavioural and risk profiles, organizations actively prevent or at least limit the impact of tacit knowledge leakage incidents, arising from the fact that the behaviour drives the action, particularly when such incidents are the result of accidental conduct. Conversely, in the case of malicious or intentional leakage behaviour (insider threat), such systems uncover suspicious activities that may lead to a breach and alert the organizations before a leakage incident transpires. While these artificial intelligence technologies may not be able to completely prevent an *intentional* tacit knowledge leakage incident from occurring, they, however, will limit the impact that such incidents have upon organizations.

As specified by participants, active defence is a concept rooted in the military arena that migrated into the information security space and defines the use of actions to outmanoeuvre an adversary and cause an attack to be more difficult to undertake by impeding them from completing or advancing their attack which in turn increases the probability of failure. This observation further support the ideas of Denning (2014) who showed how these approaches increase the cost of an attack in terms of an adversary's time and resources and how organizations use active defence to gather intelligence and learn from their competitor's behaviours and strategies.

Interestingly, as observed during the interviews, organizations are using active defence not only *outwardly* to combat external threats, but also *inwardly* to protect against internal threats. In this regard, the research study findings associated to active defence seem to be consistent with other research in the information security field which found that organizations use active defence to protect from and combat against espionage and information theft (Atwood, 2019; Melnitzky, 2011; Promnick, 2017). However, the researcher observed that the use of active defence in the knowledge management and knowledge protection literature to guard against knowledge leakage remains unexplored and represents a novel contribution in this space. This finding presents an opportunity for further research on this front.

### 6.5.2.1.2.4    Active reconnaissance

Following on from the previous finding, the next informal strategy described by research subjects alludes to **active reconnaissance** which illustrates how organizations actively gather information from competitors using open source intelligence (OSINT) to collect and analyse insights from diverse sources including the dark web via monitoring and intelligence services. Although active reconnaissance may appear similar to the competitor analysis strategy mentioned earlier in section *6.5.2.1.2.2 Competitor Analysis* , the difference lies in that the latter can be thought of as a more passive scouting process, whilst the former strategy uses more active means to gather information and therefore requires some level of engagement with the competitor or adversary party in order to obtain the insights. One such active means, for example, may require contacting members of the competing or adversarial organization to elicit intelligence that may be used to gain an advantage over the rival firm. Examples of interactions or engagements with competing firms mentioned by research informants included the use of intelligence sharing platforms such as message boards and online forums to exchange intelligence about new technologies, products or services.

Participants noted that similar to active defence, active reconnaissance is also a military grounded concept adopted by the information security field to describe exploration of an enemy area to obtain intelligence about an adversary. This observation is consistent with those of Dalton an colleagues (2020) as well as Denning (2014) who showed that organizations use active reconnaissance as part of a proactive defense strategy to identify and address potential vulnerabilities before they are exploited.

The research study findings in this respect suggest that medium and high level risk organizations conduct active reconnaissance on their competitors to gather intelligence and shift accordingly the priority and focus to areas of their knowledge leakage mitigation and protection capabilities in order to improve their competitive advantage and overall risk posture. By conducting active reconnaissance and collecting intelligence, firms stand a better chance to prepare for tacit knowledge leakage situations. To illustrate this point, one of the research subjects commented how by conducting active reconnaissance, their organization learned that one of its subject matter experts engaged in conversations with a competitor, the organization swiftly reacted and addressed the subject matter expert concerns preventing in this manner, the departure of a knowledge worker thereby averting tacit knowledge leakage.

Overall, the research findings under the enterprise factor as described until now suggest that a relationship exists between risk and degree of application of these informal strategies. The higher the reported risk level (self perceived), the greater the degree of implementation for ad-hoc processes, deception, competitor analysis, active defence and active reconnaissance strategies. This study findings regarding the association between risk and mitigation strategies agree with the results of previous research (Amara et al., 2008; M. Lee et al., 2018; Päällysaho & Kuusisto, 2011), which showed that higher risk level organizations tend to hold in place a combination of more advanced formal and informal strategies.

Surprisingly, however, in contrast to earlier research (Heckman, Stech, Schmoker, et al., 2015; Stech et al., 2016; F. Zhang et al., 2003), the researcher observed no association between organization size, industry type and more advanced strategies. This discrepancy could be attributed to the sample size as indicated earlier or could also arise from the fact that there might exist a confounding variable that was not taken into consideration. Nevertheless, further work is required to establish this difference and future research is warranted in this regard.

## 6.5.2.2 Human Factor – Informal Strategies

Based on Table 6-5 above, under human factor, the researcher compiled the informal strategies reported by research informants that belong to the personal and social context.

### 6.5.2.2.1 Personal Context

Under personal context, the researcher categorized the informal strategies that focus on (human) individuals' aspects.

Within the informal strategies, under the personal context, the emerging salient concept referred to **trust**, particular to two interesting although contrasting views, that is, **trust development** vs. **zero trust**.

#### 6.5.2.2.1.1 Trust Development

As expressed by research participants and explained before in the previous chapter, trust plays a crucial role in the preservation and protection of knowledge as it serves as a prerequisite for sharing and collaboration. These observations are in agreement with those of Becerra and colleagues (2008) who found that the concept of trust remains critical within the knowledge management literature, as trust leads to cooperative behaviour and it is used as a coordinated mechanism in facilitating knowledge creation and transfer among knowledge workers within a firm, therefore trust contributes to the protection and capture of **tacit** knowledge.

Furthermore, as established during the interviews, a fascinating finding that the researcher observed referred to how trust lays the foundations for the institution of alliances and partnerships with other organizations as a way to protect organizational knowledge and extend competitive advantage against competitors. This finding also concurs with previous literature that highlights that knowledge leakage in the context of alliances may prove a positive and beneficial outcome as long as the leakage is properly managed, the focal firm defines formal contracts and a proper level of trust exists between the partners (Jiang et al., 2013, 2016; Moein et al., 2015).

Research informants, mostly from organizations whose perceived risk graded as low and medium, described how the trust development fostered the knowledge sharing and protection at the group level within organizations. This finding mirrors that of Moein and colleagues (2015) who showed that trust represents a fundamental social construct and it is defined as *a belief in the character, strength, or truth of something or someone; to rely on the truthfulness or accuracy of something or someone* (Moein et al., 2015, pp. 9–11). As such, trust constitutes a subjective idea or sense that is difficult to manage. According to the research findings, the importance of trust is determined by the degree to which trust facilitates the knowledge sharing, creation and transfer process as knowledge workers need to first trust the individuals with whom they intend to interact. Likewise, organizations may have to negotiate contracts with other parties with whom they collaborate, which means trust has to be embedded in such alliances and partnerships. In this regard, this finding regarding the organizational alliances agrees with and corresponds to the concept of trust management in the knowledge management literature (Bertino et al., 2006; Ford, 2004) which has been extensively studied.

### 6.5.2.2.1.2 Zero Trust

Conversely, research subjects, mostly from organizations that reported their perceived risk as medium and high, expressed a contrasting but interesting view, whereby a zero trust approach formed a preferred way to protect knowledge within organizations stemming from the fact that in the current threat landscape trust is considered a liability, even a vulnerability. This observation is consistent with that of Campbell (2020) who proposed that trust constitute a weakness and should be eliminated. Moreover, research participants pointed out how in multiple leakage incidents the main culprit often involves an insider in a position of trust — sometimes even a privileged user and therefore a trusted employee — rather than a malicious outsider. In line with this view and with what was described in the technical context in the zero trust for mobile devices in section 6.5.1.3.2.4 (page 224), participants commented how the current trust paradigm within information security is broken and fails to address the present situation within firms, particularly in the context of remote work and mobile users who predominantly conduct their knowledge work outside the control of their organizations. In such an environment, trust becomes a secondary notion, and verification, access control, and a least privilege approach become the primary strategies. Changing the thinking about trust models and becoming aware of the misuse of the word *trust* in relation to knowledge security, becomes paramount to overcome the challenges posed by a mobile workforce outside the control of the organizational perimeter. These views from the research participants further support the ideas of Campbell (2020) and Kindervag and colleagues (2010) who presented zero trust as a novel approach that addresses many of the gaps highlighted by previous knowledge management and protection studies (Manhart & Thalmann, 2015; Olander et al., 2011, 2014; Ritala et al., 2015) particularly in the context of mobile devices, and more broadly, mobility. Therefore, one of the contributions of this study, is to challenge the status quo in the knowledge management and protection literature in relation to the use and reliance of *trust* and propose the adoption of the *zero trust* model from the information security realm and its application to the knowledge protection literature and mobile knowledge management processes and workflows within *knowledge intensive* organizations.

In this regard, one research participant reported that their organization followed the NIST (National Institute of Standards and Technology) Special Publication SP 800-207 guideline, which contains a set of security principles for implementing a zero trust architecture. This guideline as developed by Rose and colleagues (2020) outlines the principles and steps for an agnostic zero trust model in conjunction with risk management practices that can be adapted into knowledge management systems and comprises of six steps:

1. Identification of knowledge assets (Protect surface)
2. Conduct risk assessment on knowledge assets
3. Treat risks (apply mitigation strategies)
4. Map knowledge flows
5. Create knowledge risk management policies
6. Validate, monitor and review

This finding makes a unique contribution by proposing the adoption of zero trust models from the information security domain into the knowledge protection literature, with the aim of mitigating knowledge leakage within knowledge-intensive organizations caused by mobile devices and remote workers operating outside the organizational perimeter.

### 6.5.2.2.2 Social Context

Within the social context, the researcher classified the informal strategies reported by research subjects that concern the social setting in which workers interact. Such strategies influence the organizational culture and group behaviour.

#### 6.5.2.2.2.1 Gamification

An important finding reported by research participants concerned the use of gamification as a means to encourage knowledge sharing, capture and protection, particularly in high performing teams within their organizations. The practice, although informal, seemed to appear common amongst all types of organizations irrespective of size, type, and risk. Aimed at improving the culture and behaviour of groups, as well as fostering social connections, these practices contributed to maintaining a positive work environment that translated into a more productive workforce that was more likely to transfer tacit knowledge amongst peers and knowledge repositories thereby protecting it from leakage. These findings support previous research in this area (Friedrich et al., 2020; Sampaio et al., 2019; Swacha, 2015) which demonstrated that the use of gamification embedded in knowledge management processes resulted in a positive impact on workers' motivation and behaviours that promoted tacit knowledge sharing, creation, transfer and capture.

### 6.5.2.2.2.2    Informal networks

Surprisingly, research subjects mentioned the use of informal networks, as opposed to formal hierarchies and communities of practice as described in section *6.5.1.2.2.2 Communities of Practice (CoP) (page 211)*, as preferred ways of communicating amongst employees, particularly the ones who belonged to different areas. Informal face to face meetings as well as the use of collaboration suites and enterprise messaging systems in these organizations such as Microsoft Teams, Slack, Google, Cisco amongst others, meant that currently any subject matter expert within these knowledge intensive firms remained readily available to the whole organization.

Moreover, informants reported on how this capability proved to be practical and convenient particular to mobile (external/remote) workers. Intriguingly, the researcher also noted that this informal practice was more likely to be adopted and widespread amongst organizations whose self perceived risk scored as low and medium, by contrast, high level risk firms were far less likely to encourage the use of informal networks for organizational knowledge flows, although allowed, these communication channels remained, for the most part, limited and monitored.

The results in this regard, confirm earlier research (Bosua & Scheepers, 2007; Jewels et al., 2003; M. Schwartz & Hornych, 2011) on the impact of informal networks within knowledge management processes which establishes that informal communication networks operate independently of any formal structure or hierarchy and promote knowledge sharing, transfer, and capture activities, thereby enhancing the protection of knowledge and mitigating leakage. However, the explicit association of organizational leakage risk level with the use of informal networks as reported by research subjects in this research study, has remained understudied in the current knowledge management and protection literature. It is important to note that these findings cannot be extrapolated to all knowledge intensive organizations and must be interpreted with caution as this association could be a result of the small sample size. Scant literature focuses on the impact of informal networks to other risks such as knowledge waste, spillover and hoarding (Awazu, 2004; Durst et al., 2018) but has failed to address the explicit knowledge leakage risk in the context of mobility. Therefore, this constitutes an important avenue for future research.

## 6.6 Implications of the Research Study

This section of the research study addresses the theoretical and practical implications of the findings discussed throughout this chapter. The present study was designed to answer the main overarching research question of:

- *How can knowledge intensive (KI) organizations mitigate the knowledge leakage risk (KLR) caused by the use of mobile devices?*

In order to answer the main overarching research question, the following secondary supporting questions aim to assist the main research question by providing the necessary foundation and context required to answer the primary question:

1. *What strategies are used by knowledge-intensive organizations to mitigate the risk of knowledge leakage (KLR) caused by the use of mobile devices?*
2. *How does the perceived KLR level inform the strategies used by KI organizations?*
3. *What knowledge assets do knowledge intensive organizations protect from KL?*

In answering this main research question and supporting questions, this study discovered a number of findings that hold theoretical and practical significance, as well as utility and usefulness for the fields of knowledge management, knowledge protection and information security.

The major implications of this research study lie in its contributions to the literature and advancement of research of knowledge management, knowledge protection and information security. Further, the findings from this research carry both practical and theoretical implications and benefits to organizations, academia, and practitioners. Next, the researcher will expand on the significance of the research study and discuss its implications.

## 6.6.1 Implications for Research

This research study offers a number of important implications for the information systems literature, particularly on the knowledge management, knowledge protection and information security literature.

This thesis adopted the theoretical lens of the knowledge-based view of the firm that posits that organizations' competitiveness and performance is dependent on their ability to create, acquire, and utilize knowledge effectively. However, the role of trust is neglected in this theory. Given that trust is a critical enabler of knowledge sharing and collaboration, the theory can be extended to include trust. However, the use of mobile devices can erode trust and increase the risk of knowledge leakage in organizations. Therefore, this study proposes that organizations can maintain trust and prevent knowledge leakage in the context of mobile device use by adopting a "hybrid trust" model whereby a middle way between total trust and zero trust can be achieved, through the adoption of a risk-based approach to trust and security, that can balance the need for knowledge protection with the need for knowledge sharing and innovation.

First, the research study identifies specific knowledge leakage mitigation strategies employed by knowledge intensive organizations that in conjunction contribute to improving organizational knowledge protection capabilities and add to the protection literature. These strategies came as a result of the data collection process conducted in this study and are positioned within the knowledge management, knowledge protection, information security and mobile computing literature. Several of these knowledge leakage mitigation strategies are novel approaches or adaptations of existing strategies from other fields into the knowledge management and knowledge protection domains.

Second, the study also provides a novel classification scheme of different leakage mitigation strategies that contribute to the knowledge management and protection literature. The classification scheme addresses gaps in the current knowledge protection literature highlighted by previous research, in particular with regard to both areas tacit knowledge protection and knowledge worker mobility. To the best of the researcher's knowledge, to date, no other similar knowledge leakage classification scheme exists.

Third, leveraging on the classification scheme, the study assists in developing a deeper understanding of the different measures used by organizations. Such measures are further categorized by the human, enterprise, and technological factors which address different facets within organizations and their environments and are aimed at focusing on diverse organizational issues collectively.

Fourth, this research study also examines the relationship of different organizational aspects such as perceived leakage risk profile (i.e., high, medium, low), level of strategy formality (i.e., formal, informal), and type of knowledge being protected (i.e., tacit, explicit) with the leakage mitigation strategies and the impact of these aspects on the organization protection capabilities.

Fifth, the research study also contributes to the mobile knowledge management field which in contrast to the knowledge management discipline, has remained disproportionally understudied. Specifically, by providing guidelines and strategies that can be applied to mobile workers and mobile devices.

## 6.6.2 Implications for Practice

This research study holds practical significance for knowledge management and information security practitioners as well as organizations.

The key contribution to practice is the development of a strategic approach to mitigation predicated on the concept of dynamic risk. This approach posits that the mobile contexts are dynamic in nature, and therefore the context - risk relationship is also dynamic, which highlights the transience and unpredictability of the risk and the emergence, agility, and explorative nature of the strategies that organizations select and (re)configure to form mitigation capabilities to adapt to the changing environment and fluctuating mobile contexts.

First, one of the findings of this research study, the classification scheme provides support to both knowledge management and information security practitioners in identifying, explicitly naming, characterising, and classifying different mitigation strategies that can be applied to firms based on specific organizational attributes.

Second, the findings propose that organizations can take advantage of the classification scheme as a framework to improve protection capabilities and determine what strategies are best suited for their specific needs and knowledge assets based on their risk profile.

Third, the classification scheme proposes that organizations may use novel strategies such as active defence, active reconnaissance and deception supported by innovative technologies and new paradigms such as user behaviour analytics, cyber security frameworks and zero trust, that in combination will contribute to improving organizations' protection capabilities and overall performance over time.

Fourth, the findings from this research study also suggest that knowledge management professionals can leverage the classification scheme as well as the framework described in the previous chapter (findings) to focus on tacit knowledge protection by selecting strategies that best meet their requirements and build upon these mitigation strategies to create new capabilities.

Fifth, the results of this study posit that information security professionals concerned with the risk posed by mobile workers outside the organization, can harness the potential of artificial intelligence and machine learning platforms such as insider risk and behavioural analytics solutions to monitor and create dynamic least privilege policies that change, adapt and evolve based on the mobile context and conditional access of mobile workers irrespective of their title, location, profile or authorization level thereby ensuring that mobile workers are always verified before granting access to knowledge repositories, improving in this way, the organizational knowledge protection capabilities.

Finally, the findings from this research study also suggest that organizations can employ the classification scheme as a flexible way to configure specific strategy bundles (e.g., select specific strategies from the different contexts, for example: environmental, organizational, personal, social, etc) through different periods of time addressing different knowledge protection requirements according to the specific knowledge assets, risk profile and required degree of strategy formality (i.e., formal vs informal).

## 6.7 Summary

This chapter provided a comprehensive analysis and synthesis of the key findings presented in the preceding chapter, summarizing the major concepts of interest and their relationships. Moreover, this section situated the study's results within the existing IS research literature and highlighted the study's significant contributions to both theory and practice. The chapter concluded by outlining the implications of the findings for IS research and practice. The subsequent chapter, Conclusions, summarizes the research methodology, the research background, and the significant contributions of this study. Additionally, it presents avenues for future research and limitations of this study.

*"Know thy self, know thy enemy. A thousand battles, a thousand victories."*

**—Sun Tzu, The Art of War.**

# Chapter 7. CONCLUSIONS & FUTURE DIRECTIONS

The previous chapter elaborated on the discussions and key implications of this study. The discussion chapter described the different knowledge leakage mitigation strategies employed within knowledge intensive organizations. Furthermore, the discussion chapter also illustrated how the classification scheme was derived from the data collection process. The classification scheme represents one of the main contributions of this research. Additionally, the details of the classification scheme explain how different controls and strategies can be applied to different contexts and risk profiles across organizations. This chapter[9] will conclude the study by summarizing the key research findings in relation to the research aims and research questions, as well as the value and contribution thereof. The conclusion chapter will also review the limitations of the study and propose opportunities for further research.

The conclusion chapter is structured as follows, first, it starts by outlining the research questions and answering them based on the findings. Second, the chapter

---

[9] Sections of this chapter have been published in the following publications:

- Agudelo, C. A., Bosua, R., Ahmad, A., & Maynard, S. B. (2016). Understanding knowledge leakage & BYOD (Bring Your Own Device): A mobile worker perspective. arXiv preprint arXiv:1606.01450.
- Agudelo-Serna, C. A., Bosua, R., Ahmad, A., & Maynard, S. (2017). Strategies to Mitigate Knowledge Leakage Risk caused by the use of mobile devices: A Preliminary Study.
- Agudelo-Serna, C. A., Bosua, R., Ahmad, A., & Maynard, S. B. (2018). Towards a knowledge leakage mitigation framework for mobile devices in knowledge-intensive organizations.

highlights both the theoretical and practical contributions of the research study in relation to the research questions and research aim. Third, the next section after that will present the limitations of the study. The final part of the chapter will describe the future avenues for research based on the work developed in this study and its limitations.

# 7.1 Research Questions

The main purpose of this research study was to gain an understanding of how knowledge intensive organizations can address the phenomenon of knowledge leakage caused by the use of mobile devices.

To that end, the study posed the following main overarching research question:

- *How can knowledge intensive (KI) organizations mitigate the knowledge leakage risk (KLR) caused by the use of mobile devices?*

Answering the main question remains important for both practice and academia. For practice, understanding how in the context of the knowledge economy and the current threat landscape, the pervasiveness and convenience of mobility and mobile devices pose a leakage risk to organizational knowledge assets — represented in the form of people, intellectual property, processes, and products amongst others. Therefore, it remains paramount to provide knowledge intensive organizations with the right tools and strategies to combat the aforementioned risk. By offering organizations protection strategies to mitigate the leakage risk, firms can build upon these strategies to improve protection capabilities that will lead to superior knowledge management, security posture and overall performance.

For academia, addressing the lack of knowledge leakage and tacit knowledge strategies in IS literature (particularly, in the knowledge management literature with respect to mobile technologies and more specifically mobile devices) helps to advance the field and provides new insights and novel approaches in mobile computing and information security. Additionally, by providing a synthesis of different strategies, new insights and approaches to knowledge protection grounded in the empirical evidence collected in this study, this research also contributes to addressing the gap with respect to the lack of specific knowledge leakage mitigation strategies for knowledge intensive organizations. Further, this research offers organizations with a valuable tool to protect and improve their leakage mitigation capabilities.

### 7.1.1 Answer to the Main Research Question

In order to answer the main research question:

- *How can knowledge intensive (KI) organizations mitigate the knowledge leakage risk (KLR) caused by the use of mobile devices?*

The review of both academic and professional literature, as explained in the literature review chapter, uncovered a series of gaps in the current body of knowledge. First, lack of in-depth studies that provided specific strategies for knowledge intensive organizations and different types of industries based on different risk levels. Most of the analysed studies presented high-level strategies that were described in a generalised manner and failed to identify specific strategies for different organizations and risk levels. Second, mobile devices and mobile knowledge management were widely neglected, particularly for the protection of tacit knowledge. And third, the tacit dimension of knowledge remained largely understudied as the majority of the literature focused on formal and informal strategies to protect explicit knowledge.

Therefore, based on these gaps and the research study findings detailed in chapter five and discussed in chapter six, knowledge intensive organizations can mitigate the knowledge leakage risk caused by the use of mobile devices by:

1. **Identifying the type of organizational knowledge asset to protect** (e.g., Individual, process, intellectual property, trade secret) and their associated risks. The findings indicate that high risk level organizations tend to focus primarily on the protection of people and processes, that is, tacit knowledge first, then explicit knowledge.

2. **Understanding the operating environment**. The evidence from this study suggests that knowledge intensive organizations have become increasingly aware of their environmental context and conduct formal processes such as supply chain risk management, market/ competitor analysis and strategic organizational alliances to account for and reduce the uncertainties of their specific conditions.

3. **Identifying the organizational risk profile** (i.e., High, Medium, Low). Although this may be a subjective exercise, the literature and the findings suggest that risk is a multi-dimensional concept, and therefore factors such as the organization's size, environment, type of industry and knowledge assets to protect should be taken into consideration when determining the firm risk profile. The findings also show that high risk level organizations use an active defence approach to knowledge protection. (i.e., deception, active reconnaissance, and counterintelligence).

4. **Selecting the type of specific strategies** (i.e., a combination of formal and informal strategies) that better fit the risk profile of the organization based on the classification scheme proposed by this research study for each one of the risk factors surrounding the organization (i.e., Human, Enterprise, and Technological factors and associated contexts). In addition to this, the results of the study propose that high risk level organizations employ mainly a combination of formal mechanisms such as legal frameworks, mobile policies and cyber security frameworks with informal practices such as ad-hoc processes, deception, misinformation/ disinformation, active defence, and reconnaissance.

5. **Protecting tacit knowledge in the context of mobility**. The research study's findings suggest that applying a risk management approach to the organization's mobile workforce focusing on workers' individual behaviours and risk profiles (i.e., *insider risk management* and *behaviour analytics*) in combination with more traditional approaches such as information security controls and knowledge management processes (i.e., capture, sharing, retention) seem to yield a superior tacit knowledge protection capability. Findings also highlight that high risk level organizations leverage the use of artificial intelligence and machine learning powered technologies such as *User Entity Behaviour Analytics* (UEBA) and *Endpoint Detection and Response* (EDR). In addition to this, the evidence from this study also suggests that while all low and some medium risk level organizations use **trust development** to encourage knowledge sharing and build rapport between individuals and create inter/ intra organizational alliances, by contrast, some other medium and all high risk level organizations have shifted to a **zero trust** paradigm whereby trust is overrated, and workers and their devices are seen as an insider risk. Further strategies derived from the findings to protect tacit knowledge include informal networks, gamification, communities of practice and mentoring programs.

6. **Protecting explicit knowledge in the context of mobility**. The findings from the study propose the use of more conventional strategies such as surveillance, compartmentalization, detection, response, layering and device profiling and data loss / leakage prevention (DLP) to protect explicit knowledge. More innovative strategies include the use of deception (decoy tokens, files, and documentation) to deceive malicious insiders and adversaries. Another important finding in this regard includes conditional and contextual based access for mobile users. Conditional and contextual access are part of the zero trust architecture whereby no user is trusted and access is always verified, access is only granted under particular conditions (i.e., identity, location, time, mobile device fingerprint, application, verification, authorization level, risk level and exceptions) and contextual access (historical data – e.g. has the user logged in from this device in the past? Is this their usual time? Is this their regular behaviour?) is always validated against zero trust and mobile device management (MDM) policies.

## 7.1.2 Answer to Secondary Research Questions

In addition to the overarching research question, the following secondary supporting questions provide the necessary foundation and context required for the main research question:

## 7.1.2.1 Secondary Question 1: Knowledge Leakage Strategies used by Organizations

1. *What strategies are used by knowledge-intensive organizations to mitigate the risk of knowledge leakage (KLR) caused by the use of mobile devices?*

Evidence from the research study suggests that knowledge intensive organizations employ different strategies based on their risk profile, that is, high, medium, and low risk levels. Moreover, this research proposes a classification scheme of the mitigation strategies that was developed based on the evidence and discussed in the previous chapter. This classification scheme provides the specific strategies employed by organizations to mitigate the knowledge leakage risk caused using mobile devices (See Figure 7-1 below).

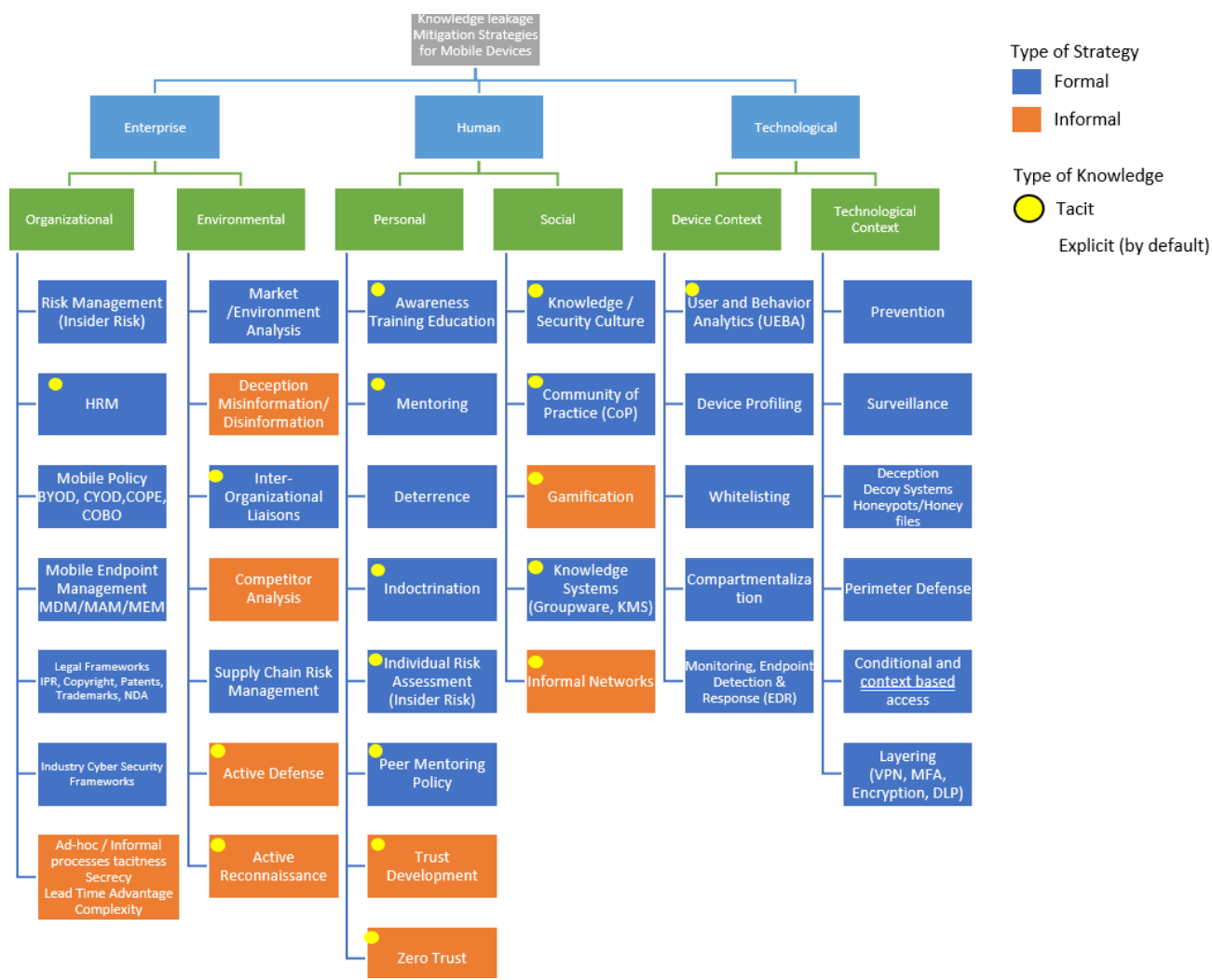

**Figure 7-1. Classification scheme of Knowledge Leakage Mitigation Strategies for Mobile Devices**

In order to answer this research question, the following sub-sections will address the specific strategies used by organizations according to their risk profiles.

### 7.1.2.1.1 High Risk Level Organizations

The findings from the study show that high risk level organizations use a combination of informal and formal strategies and also differ from medium and low risk level firms in their approaches. Formal and informal strategies under each factor are listed below in the following sections.

#### 7.1.2.1.1.1    Formal Strategies

Within the formal strategies the most salient ones are:

- Enterprise Factors
  - **Risk management processes**, namely, insider risk management and supply chain risk management, Mobile Device Management.
  - **Human Risk Management Processes** (onboarding, screening, training, motivation, compensation, culture, compliance, offboarding)
  - **Mobile Policies** (BYOD, CYOD, COPE, COBO)
  - **Mobile Endpoint Management** (Mobile Device Management – MDM, Mobile Threat Defence - MTD)
  - **Legal Frameworks** (IPRs)
  - **Industry Cyber Security Frameworks** (Mitre ATT&CK, Mitre D3fend, NIST)
  - **Organizational Alliances** (Inter /Intra liaisons, partners, government, suppliers, clients)
  - **Market / Environment Analysis** (Operating environment including legislation, competition and governing bodies)
- Human Factors
  - **Awareness Training and Education** (Programs to influence behaviour, culture)
  - **Mentoring** (Peer mentoring policy, rotation, knowledge partners)
  - **Deterrence** (disciplinary action to influence behaviour and attitudes)
  - **Indoctrination** (Defence and Intelligence sectors - code of conduct)
  - **Individual Risk Assessment** (Insider risk, Behavioural Analytics)
  - **Knowledge Security Culture** (Action, practices and habits)
  - **Communities of Practice** (Common groups of interest – Tacit knowledge transfer)
  - **Knowledge Management Systems** (Groupware, Systems and Knowledge experts directory)
- Technological Factors
  - **User Behaviour Entity Analytics** (UEBA)

- o Device Profiling, Whitelisting, Compartmentalization

- o **Endpoint Detection and Response** (EDR)

- o **Deception** (Decoy systems, tokens, files and documentation)

- o Perimeter defence, surveillance, prevention, Layering (MFA, VPN, DLP)

- o **Zero Trust Architecture** (Conditional and Context-based access for mobile devices)

#### 7.1.2.1.1.2    Informal Strategies

Within the informal strategies for high risk level organizations, the following were more prominent:

- • Enterprise Factors

  - o **Ad-hoc processes** (Secrecy, tacitness, fast innovation cycle, complexity , lead-time advantage)

  - o **Deception** (Misinformation/ Disinformation)

  - o **Competitor Analysis** (Strength and Weaknesses analysis - SWOT)

  - o **Active Defence** (Counterintelligence, deception - decoy)

  - o **Active Reconnaissance** (Open Source Intelligence – OSINT,  Dark web monitoring services)

- • Human Factors

  - o **Trust Development** (Rapport, cooperation and alliances)

  - o **Zero Trust** (Insider risk and Insider threat management)

  - o **Gamification** (Motivation to knowledge sharing, capture and protection)

  - o **Informal Networks** (Informal structures of communication irrespective of hierarchies and formal channels)

### 7.1.2.1.2 Medium and Low Risk level Organizations

Medium and low risk level organizations, for the most part, appear to share similar strategies which could be a result of the sample size and this represents a limitation of this research study.

#### 7.1.2.1.2.1 Formal Strategies

Within the formal strategies for medium and low risk level organizations the following represented the most important:

- Enterprise
  - **Risk Management Process** (basic – No Insider risk management or Supply Chain Risk Management)
  - **HRM** (onboarding, screening, training, motivation, compensation, culture, compliance, offboarding)
  - **Mobile Policies** (BYOD, CYOD)
  - **Legal Frameworks** (IPRs)
  - **Cyber Security Frameworks** (some low risk level and some medium risk level organizations -Internal cyber security frameworks – No defence frameworks)

- Human
  - **Awareness, Training and Education**
  - **Mentoring** (Some medium risk level organizations)
  - **Deterrence**
  - **Knowledge Security Culture** (Some medium risk level organizations)
  - **Communities of Practice** (Some medium risk level organizations)
  - **Knowledge Management Systems** (Some medium risk level organizations)

- Technological
  - Device Profiling
  - Whitelisting
  - Compartmentalization, Prevention, Surveillance, Perimeter Defence
  - Detection & Response
  - Layering (VPN, MFA, Encryption)
  - **Zero Trust Architecture** (some medium risk level organizations)

#### 7.1.2.1.2.2 Informal Strategies

Similar to the formal strategies section, low and medium risk level organizations appear to share similar protection mechanisms according to the research study's findings:

- Enterprise
  - **Ad-hoc processes** (Secrecy, tacitness, fast innovation cycle, complexity , lead-time advantage)
  - **Competitor Analysis** (Some medium risk level organizations)
- Human
  - Trust Development
  - **Zero Trust** (Some medium risk level organizations)
  - Gamification
  - Informal Networks

## 7.1.2.2 Secondary Research Question 2: Knowledge Leakage Risk Level

The next question addresses the knowledge leakage risk in knowledge intensive organizations by answering the following research question:

*2. How does the perceived KLR level inform the strategies used by KI organizations?*

As explained during the chapter five — findings— research participants were asked to qualify the knowledge leakage risk within their organizations based on their perceptions of exposure amongst the following risk levels: high, medium, and low. The definition of knowledge risk and risk level was based on the standard definition of risk in the ISO 27001 standard within the risk management section (27005:2011, 2011) which, in turn, extends from the definition of the ISO 31000 standard for Risk Management – Principles and Guidelines. In this ISO 31000 standard, risk is defined as the *effect of uncertainty on objectives* (Australia & New Zealand, 2009). ISO 27001 standard leverages this definition to use a more substantial and specific meaning in the context of information security: *Information security risk is associated with the potential* *that threats will exploit vulnerabilities of an information asset or group of information assets and thereby cause harm to an organization* (27005:2011, 2011, sec. Risk). Furthermore, risk — and by extension knowledge leakage risk — was defined as a function of two dimensions, i.e., likelihood (or probability) and impact (or consequence). Based on this, participants were asked to select their perceived knowledge leakage risk in accordance with the likelihood of a knowledge leakage incident and the impact should such an incident occur within their organizations as indicated in Figure 7-2 below

**Figure 7-2. Knowledge Leakage Risk as a function of Likelihood and Impact**

Overall, out of the twenty participants, 50% deemed their knowledge leakage risk level as high, 35% considered their risk level as medium and the remaining 15% regarded their risk level as low.

The evidence from this study suggests that high risk level organizations seem to be more likely to resort to more aggressive, although proactive, approaches to protect knowledge from internal and external threats. Examples of these approaches include zero trust, user behaviour analytics, deception, misinformation/disinformation, active defence and active reconnaissance of competitors and adversaries.

Furthermore, the findings also propose that high risk level firms assume a different perspective to trust as compared to some medium and low risk level organizations whereby these high risk level firms adopt a zero trust approach towards their operating environment and workers, particularly in the context of mobility and mobile devices. In addition to this, high risk level organizations appear to be more aware of their environment and embrace a risk management approach to manage their supply chain (i.e., supply chain risk management), competitors and alliances.

## 7.1.2.3 Secondary Research Question 3: Knowledge assets Protected by Organizations

The final question addresses the knowledge assets that knowledge intensive organizations protect:

*3. What knowledge assets do knowledge intensive organizations protect from KL?*

The results from the research study imply that knowledge intensive organizations focus on different types of knowledge assets. The significance of this question lies in the fact that different knowledge assets will elicit different risk profiles and therefore will require different strategies.

As mentioned during chapter five —findings, research informants were asked to name the most critical organizational knowledge assets that warranted protection from competitors, customers, partners, and suppliers within their firms. Overall, organizational knowledge assets mentioned by participants varied from organization to organization, but ultimately fell into the following categories ordered from most common to least common: people – 60%, process - 55%, methodology – 55%, product – 50%, intellectual property – 50%, strategy -25%, policy – 15%, client information – 10%, and research findings – 5%. See Table 7-1 below for summarized information on the reported organizational knowledge assets.

**Table 7-1. Reported Organizational Knowledge Assets**

| Identified Organizational Knowledge Asset | Number | Percentage |
|---|---|---|
| People | 12 | 60% |
| Process | 11 | 55% |
| Methodology | 11 | 55% |
| Product | 10 | 50% |
| Intellectual Property | 10 | 50% |
| Strategy | 5 | 25% |
| Policy | 3 | 15% |
| Client Information | 2 | 10% |
| Research Findings | 1 | 5% |

As discussed during the literature review, knowledge assets can be categorized into human capital, that refers to tacit knowledge, i.e., knowledge inside the mind of individuals and organizational knowledge that pertains to relational and structural capital, i.e., explicit knowledge that has been articulated, transferred and captured into knowledge repositories and knowledge management systems. In other words, knowledge in organizations exists in two broad forms explicit knowledge and tacit knowledge. As previously discussed, most of the knowledge protection literature has addressed explicit knowledge protection to the detriment of tacit knowledge. In this study and based on the findings of this research (Table 7-1 above) , knowledge intensive organizations deem their people (i.e., tacit knowledge) as one of, if not, their most valuable resource that warrant protection. Therefore, the mitigation strategies employed to protect an organization's people should subsequently address primarily the tacit knowledge assets in conjunction with the explicit knowledge ones,. In this regard, the findings of this study contribute to addressing the existing gap by providing specific mitigation strategies that organizations may use to protect their tacit knowledge, i.e., people. Some leakage mitigation strategies address accidental leakage due to careless behaviours or habits, other strategies help to contain and limit the impact of intentional tacit knowledge leakage.

Examples of the mitigation strategies to address accidental/intentional tacit knowledge leakage:

1. Enterprise Factors

   a. Human Resource Management (knowledge transfer)

   b. Inter / Intra Organizational Liaisons (knowledge transfer to other organizations)

   c. Active Defence (pre-emptively predict and contain impact of knowledge leakage based on adversarial behavioural analysis)

   d. Active Reconnaissance (pre-emptively predict and contain impact based on counterintelligence)

2. Human Factors

   a. Awareness, Training and Education (influencing behaviour and accidental tacit knowledge leakage)

   b. Mentoring, Indoctrination, Peer mentoring policy (influencing behaviour)

   c. Insider risk / threat management (Individual risk assessment) – helps to prevent accidental leakage and provides risk metrics / analytics to pre-emptively predict potential intentional leakage incidents, helps to contain and limit the impact of intentional knowledge leakage based on behaviour

   d. Trust development (as a way to build rapport and improve knowledge retention)

   e. Zero Trust (by virtue of distrusting, tacit knowledge leakage incidents can be contained and limited in impact)

   f. Gamification (Influences behaviour, improves culture, helps with accidental knowledge leakage)

   g. Informal Networks (Provides alternatives to formal communication channels for tacit knowledge to be transferred within organizations)

   h. Knowledge Security Culture (Influences behaviour, improves culture, helps with accidental knowledge leakage)

   i. Communities of Practice (facilitates knowledge transfer)

3. Technological Factors

    a. User Behaviour and Entity Analytics (Part of a zero trust architecture, it is an Artificial intelligence and machine learning technology used to determine a behavioural baseline based on device usage, triggers notification when abnormal behaviour occurs that prompts immediate policy response, e.g., system suspends access to organizational resources to prevent incident)

## 7.2 Contributions

This research study has a number of theoretical and practical contributions. This section presents the significance of this research to the IS knowledge management and knowledge protection literature.

### 7.2.1 Theoretical Contributions

Current knowledge management and knowledge protection literature has mainly focused on high level protection mechanisms, neglecting in-depth studies that provide specific strategies for knowledge intensive organizations (Ahmad, Bosua, et al., 2014; Manhart & Thalmann, 2015; Oorschot et al., 2018). This research study contributes to this regard and addresses this gap by providing specific mitigation strategies that knowledge intensive organizations can implement.

Furthermore, this research study also provides a synthesis of different knowledge leakage related mechanisms reported by knowledge intensive organizations that are categorized into enterprise, human and technological factors.

In addition to this, the framework used to build the aforementioned categorization was based on the research model discussed in chapter three — research model development. This research model depended on the extant mobile computing literature and extended on different frameworks such as *the social context interaction framework* by Chen & Nath (2008), Bradley and Dunlop's (2005) *model of context* , and *the Integrative model of IT business value framework* developed by Melville et al (2004), then followed the theory taxonomy proposed by Gregor (2006), and underpinned by the theoretical lenses of the knowledge-based view of the firm. Hence, this framework may be extended to account for other contexts and address other knowledge protection phenomena. This research framework contributes to the current body of knowledge protection literature and addresses the gap highlighted by previous researchers (Ilvonen et al., 2018; Kaiser et al., 2020; Manhart & Thalmann, 2015) who have emphasized that knowledge protection is still in its infancy and argued for the need of theoretical frameworks grounded on the knowledge management and information security literature to investigate the knowledge protection phenomena.

A novel contribution of this study that further extends the existing literature of knowledge protection and knowledge management for mobile knowledge workers, pertains to the identification of two new correlations. **1)** A correlation of the **perceived risk** construct and the **type of strategies** organizations seem to implement. And **2)** a relationship of the **perceived risk** with **mobile contexts** (see section *6.3 Discussing the Propositions and Research Questions*). This relationship highlights the dynamic nature of perceived risk associated with the complexities of the environment and interactions of mobile users and contexts.

To explain these associations between **1)** risk and strategies, and **2)** risk and context, the study borrows from the information security management literature, the *Information Warfare Paradigm* proposed by Richard Baskerville (2005) and adapts it to argue that the link between risk and mobile contexts are dynamic in nature, and as such, the relationship between strategies and risk is also dynamic. This highlights the transient and unpredictable nature of risk, as well as the need for control strategies that are agile, exploratory, and able to adapt quickly to changing environments and shifting mobile contexts. The dynamic nature of these strategies is a direct result of the constantly changing risk level, which is in turn affected by the mobile worker's changing contexts. Due to the variability of these contexts and the threat landscape, risk and threats become unpredictable and difficult to measure, necessitating an exploratory approach that allows for the swift deployment of varying strategies based on changing risk levels and contexts.

An additional contribution of this study suggests that organizations should adopt a "hybrid trust" model to strike a balance between the need for knowledge protection with the need for knowledge sharing and innovation whereby a middle way between total trust and zero trust is achieved, through the adoption of a risk-based approach to trust and security.

Another theoretical contribution proposed by this study concerns the lack and underrepresentation of tacit knowledge protection strategies as compared to more formal and explicit knowledge mechanisms in the current knowledge protection literature (Ilvonen et al., 2018; Oorschot et al., 2018), which is addressed by this study's findings in the form of the description and combination of more innovative approaches that address tacit knowledge as highlighted from the evidence.

In line with this, and in response to previous studies (Guo et al., 2020; Manhart & Thalmann, 2015) that have called attention on how to adapt mechanisms from information security to knowledge protection, the results from this study contribute to addressing this gap by providing the adaption of several strategies from the information security literature into the knowledge protection literature, such as, zero trust, deception, active defence, active reconnaissance and behaviour analytics to name a few.

Finally, previous studies (Ahmad, Bosua, et al., 2014; Ilvonen et al., 2018; Manhart & Thalmann, 2015; Olander et al., 2011) have raised concerns with the lack of knowledge protection strategies including mobile technologies which generally facilitate knowledge sharing to the detriment of protection. In this regard, this research study contributes to tackling this concern by presenting protection strategies directly targeting mobility, i.e., mobile workers and mobile devices.

## 7.2.2 Practical Contributions

This research study holds practical significance for information security and knowledge manager practitioners as well as organizations. One of the main contributions of this study to practice is a detailed synthesis, integration, and analysis of multiple streams of IS literature on information security, knowledge management, knowledge protection and mobile computing strategies used by knowledge intensive organizations to protect their knowledge assets, assisting them, in this way, in sustaining their competitive advantage. In order to answer the research aim, the researcher conducted an empirical exploratory study to better understand and identify the managerial practices used by organizations to mitigate the knowledge leakage risk caused by mobile devices. Moreover, by employing a qualitative approach, this study provided key insights into how organizations can mitigate the leakage risk. These insights in combination with the existing literature analysis led to the development of a research model that resulted in a classification scheme tool to assist organizations in improving their knowledge protection capabilities.

The classification scheme contributes to the current practitioner literature on knowledge protection by providing a combination of specific formal and informal strategies that address both tacit and explicit knowledge within knowledge intensive organizations. Additionally, the research study further contributes to practice by finding novel approaches from other fields into the incipient knowledge protection discipline that extend the professional current literature.

Further, the study also contributes to the mobile knowledge management literature by describing specific strategies concerning the use of mobile devices and the challenges posed by mobile workers outside the organizational perimeter. In this regard, the study findings indicate that organizations can employ different and novel mechanisms to deal with individual employees' behaviours that accidentally or intentionally disregard organizational policies (e.g., information security and knowledge management policies). These study results also present options that advance the professional literature in knowledge management by offering innovative approaches, such as, implementing insider risk / threat management programs as part of their knowledge protection efforts to account for incidents occurring as a consequence of internal circumstances.

Additionally, the results from the study contribute to current industry practices in relation to risk and knowledge protection within organizations by suggesting that the organizational risk profile should inform the mitigation strategy and providing specific mechanisms used by high, medium and low risk level organizations accordingly.

Finally, by providing insights on tacit knowledge leakage strategies for information security and knowledge managers to implement within their organizations, the study research contributes to practice and progresses the current industry practices.

## 7.3 Limitations

The qualitative approach of this research study allowed the researcher to gain an in-depth understanding of the knowledge leakage mitigation practices in knowledge intensive organizations. However, given that the researcher was the primary instrument for data collection and analysis, research bias represents the main limitation of this research. Research bias commonly refers to any influence that provides a distortion in the results of a study (Morse et al., 2002). Due to the subjective nature of qualitative research, avoiding bias remains more difficult as compared to quantitative research. This is because qualitative research relies substantially on experience and judgement, as well as the data collected from people which is unique to the person and situation.

 In order to minimize research bias, several measures were conducted. First, the University of Melbourne's Human Ethics Sub-Committee (HESC) reviewed the data collection methods and instruments before approval was granted. Second, to evaluate the rigour in the qualitative study, this research adopted the framework by Carson and colleagues (2001) that suggests a number of techniques that integrates several recommendations from the extant literature and emphasizes on the trustworthiness of the findings in qualitative research (See Chapter 4. Methodology). Third, the researcher also undertook research training to become aware of the most common types of biases relevant to this research study, such as design, selection, sample, interviewer, and response biases, to name a few.

Despite these measures, several limitations remain and should be acknowledged. First, generalizations of this study findings in other contexts should be carried out with caution due to specific context and characteristics of the sample. For this research study, the participating organizations were located in Australia, belonged to broad variety of industries, and included varying sizes. The researcher employed a purposive sample, which is a non-probabilistic type of sampling, subject to his own judgement.

Further, within the selected organizations, only managerial perspectives from knowledge management and information security areas were sought to understand the leakage phenomenon and mitigation strategies used within their organizations.

Second, due to the sensitive nature of the information discussed, the researcher resorted to the use hypothetical scenarios to elicit information from participants, which may have influenced the data collection process. To account for this, several scenarios were used, and the researcher followed up with further questions and was provided with redacted supplementary documentation by participants when available.

Third, the researcher asked participants to assess their organizational risk based on their perceptions of knowledge leakage exposure. This was a self perceived and subjective assessment and was not validated by the researcher.

# 7.4 Future Work

Based on the limitations of this study a number of avenues for future research exist. These areas are primarily derived from the shortcomings of the current research study. First, additional research on knowledge leakage mitigation may include knowledge intensive organizations from other countries different to Australia to identify and contrast mechanisms based on geographical location and culture.

Second, further studies can focus on a particular industry at a time to gain an in-depth understanding of how knowledge intensive organizations address the leakage problem as compared to other industries. For example, the military industry may provide other insights and new strategies. These studies can be used to identify trends and common approaches across industries. Further, additional research can also focus on organizational size to study whether there is an association between size and mitigation strategies (e.g., small, medium and large organizations).

Third, future research should also build on this study and examine ways to objectively validate the subjective perception of organizational risk and its relationship with mitigation strategies.

Fourth, further studies should also include the perspective of mobile workers and analyse the mechanisms to protect knowledge at the operational level and contrast these findings with the higher level strategies from the managerial perspective.

Finally, another avenue for future research may include quantitative studies such as surveys, to confirm or reject the findings in this research, and provide a better generalization of this study by minimizing the inherent research bias pertaining to qualitative studies.

# 7.5 Final Reflections

The primary objective of this research was to examine the complex problem of knowledge leakage in knowledge intensive organizations caused using mobile devices. In doing so, the researcher encountered a wide range of new and challenging ideas, theories, and research methods, through the course of his PhD experience. This process of learning and discovery not only expanded his knowledge and understanding of the field of IS, but also helped him to develop his critical thinking and problem-solving skills.

The researcher also embraced opportunities to collaborate and engage with other professionals in the field, which helped to build a strong network of contacts and opened up new opportunities for professional growth. Numerous challenges were faced during this time, including the recruitment of participants, identifying and addressing the gaps in the literature, the development of the research model and methodology, and at times finding the motivation and discipline to continue.

Through this process, the researcher learned the value of choosing an appropriate research approach and the importance of ethical considerations when conducting research, especially when addressing sensitive topics that could pose potential risks to participants.

In addition to this, as an information security professional, the researcher appreciated the various benefits the PhD program provided such as the advanced knowledge, skills, and network contacts needed to excel in the field and advance his career. The in-depth understanding of the latest research, theories, and methodologies in the field of information security and knowledge management was pivotal to remain updated and current with the latest trends and best practices in the field and apply this knowledge to improve the security of his organization. Similarly, the significant research component of the PhD program, helped him develop skills in areas such as research design, data analysis, and critical thinking. These skills provided the tools and capabilities needed to identify and analyze security threats and vulnerabilities, as well as to develop and implement effective security strategies.

Moreover, the PhD program provided opportunities to connect with other researchers and professionals in the field which enabled the access to a network of experts who offered guidance, support, and opportunities for collaboration.

Additionally, the PhD program prepared the researcher to take on leadership roles in his organization, such as leading security teams, creating security policies, and serving as a subject matter expert on security matters to the executive team and board of directors. The PhD Program has also opened up new career opportunities for the researcher, such as teaching and research roles in academia, consulting, and senior management positions in the Information Security industry.

Most importantly, the experience and the journey of the PhD has allowed the researcher to take a long-term view of the field, and to understand the implications of emerging technologies and trends for the future of information security and developed a well-rounded and integral professional profile that covers various aspects of security including technical, organizational, legal and ethical viewpoints. This holistic perspective has helped him to develop a more comprehensive view of the security challenges faced by his organization and how to mitigate them.

Pursuing a PhD is not only an academic endeavour, but also an opportunity for personal and professional growth. The process of conducting research and writing a thesis requires a great deal of self-discipline, time management, and perseverance. Through this experience, the researcher developed a strong sense of personal motivation and a greater ability to set and achieve goals. Further, the researcher gained a deeper sense of self-awareness and self-confidence, which proved to be beneficial in both his professional and personal life.

Overall, the researcher's PhD experience not only made him a better professional in the field of information security, but also a better person as it has imparted upon him the significance of fully comprehending an issue first, before even trying to address it and refined his abilities in research, communication and dissemination of knowledge. Above all, it ingrained in him the aspiration and the capability to continue researching social phenomena to improve his understanding of the world around him.

The acquired knowledge and skills, as well as the personal growth that comes with it, will be beneficial in his future endeavours and will help him to be more successful in his profession.

Finally, to draw to a close, the researcher would like to re-state the significance and relevance of this study and the contribution to the field of knowledge management and information security. Knowledge leakage in knowledge intensive organizations has become a prevalent issue in recent years and the use of mobile technologies has only made it more pervasive. Traditional strategies and controls have continuously demonstrated to be inefficient against this insidious threat. In this study, the researcher proposed a shift to the current paradigm towards a more active and proactive approach in protecting knowledge. The researcher argued that traditional risk management falls short in addressing the current threat landscape, particularly in the context of mobility and mobile devices. Gone are the days when risks were static, predictable, measurable and persistent. Rather, they have become dynamic, unpredictable, not measurable and transient. A new world order warrants a new paradigm, whereby organizations become agile and employ different and novel strategies that can be swiftly configured and rearranged to deploy dynamic capabilities to meet the changing demands of a complex and chaotic environment. In this study, the author has proposed that some of those prominent novel strategies include zero trust, deception, active defence, insider threat & risk management, and open-source intelligence (OSINT).

# REFERENCES

27005:2011, I. (2011). ISO 27005: 2011. In *Information technology--Security techniques--Information security risk management. ISO.*

Abdoul Aziz Diallo, B. (2012). Mobile and Context-Aware GeoBI Applications: A Multilevel Model for Structuring and Sharing of Contextual Information. In *Journal of Geographic Information System* (Vol. 04, Issue 05, pp. 425–443). https://doi.org/10.4236/jgis.2012.45048

Abdul Molok, N. N., Ahmad, A., & Chang, S. (2010a). Understanding the factors of information leakage through online social networking to safeguard organizational information. *ACIS 2010 Proceedings - 21st Australasian Conference on Information Systems.* http://www.scopus.com/inward/record.url?eid=2-s2.0-84870387706&partnerID=tZOtx3y1

Abdul Molok, N. N., Ahmad, A., & Chang, S. (2011a). Disclosure of Organizational Information by Employees on Facebook: Looking at the Potential for Information Security Risks. *ACIS 2011 Proceedings*, *1*(1), 1–12.

Abdul Molok, N. N., Ahmad, A., & Chang, S. (2011b). Exploring the use of online social networking by employees: Looking at the potential for information leakage. *Pacific Asia Conference on Information Systems - PACIS*, 138.

Abdul Molok, N. N., Ahmad, A., & Chang, S. (2010b). Information leakage through online social networking: Opening the doorway for advanced persistence threats. *Proceedings of the 8th Australian Information Security Management Conference,*

*Perth, Australia: Edith Cowan University. 30 Nov – 2nd Dec, 2010*, 76–88. http://www.scopus.com/inward/record.url?eid=2-s2.0-84864567494&partnerID=40&md5=b95580a446ad0461abaf025fc9477a14

Abedi, E., Shamizanjani, M., Moghadam, F. S., & Bazrafshan, S. (2018). Performance appraisal of knowledge workers in R&D centers using gamification. *Knowledge Management & E-Learning: An International Journal*, *10*(2), 196–216.

Agbo, F. J., & Oyelere, S. S. (2019). Smart mobile learning environment for programming education in Nigeria: adaptivity and context-aware features. *Intelligent Computing-Proceedings of the Computing Conference*, 1061–1077.

Ahlfänger, M., Gemünden, H. G., & Leker, J. (2021). Balancing knowledge sharing with protecting: The efficacy of formal control in open innovation projects. *International Journal of Project Management*, *January*. https://doi.org/10.1016/j.ijproman.2021.09.007

Ahmad, A. (2002). The forensic chain of evidence model: Improving the process of evidence collection in incident handling procedures. *The 6th Pacific Asia Conference on Information Systems*, 1–5.

Ahmad, A., Bosua, R., & Scheepers, R. (2014). Protecting organizational competitive advantage: A knowledge leakage perspective. *Computers and Security*, *42*, 27–39. https://doi.org/10.1016/j.cose.2014.01.001

Ahmad, A., Bosua, R., Scheepers, R., & Tscherning, H. (2015). Guarding Against the Erosion of Competitive Advantage: A Knowledge Leakage Mitigation Model. *International Conference on Information Systems ICIS2015*, 1–13.

Ahmad, A., Desouza, K. C., Maynard, S. B., Naseer, H., & Baskerville, R. L. (2020). How integration of cyber security management and incident response enables organizational learning. *Journal of the Association for Information Science and Technology*, *71*(8), 939–953.

Ahmad, A., & Maynard, S. B. (2014). Teaching information security management: Reflections and experiences. *Information Management & Computer Security*, *22*(5). https://doi.org/10.1108/IMCS-08-2013-0058

Ahmad, A., Maynard, S. B., Desouza, K. C., Kotsias, J., Whitty, M. T., & Baskerville, R. L. (2021). How can organizations develop situation awareness for incident response: A case study of management practice. *Computers & Security*, *101*, 1–15.

Ahmad, A., Maynard, S. B., & Park, S. (2014). Information security strategies: towards an organizational multi-strategy perspective. *Journal of Intelligent Manufacturing*, *25*(2), 357–370. https://doi.org/10.1007/s10845-012-0683-0

Ahmad, A., Ruighaver, A. B., & Teo, W. T. (2005). An information-centric approach to data security in organizations. *IEEE Region 10 Annual International Conference, Proceedings/TENCON*, *2007*, 1–5. https://doi.org/10.1109/TENCON.2005.301322

Ainslie, S., Thompson, D., Maynard, S., & Ahmad, A. (2023). Cyber-Threat Intelligence for Security Decision-Making: A Review and Research Agenda for Practice. *Computers & Security*, 1–16.

Ajzen, I. (1991). The theory of planned behavior. *Organizational Behavior and Human Decision Processes*, *50*(2), 179–211. https://doi.org/10.1016/0749-5978(91)90020-T

Ajzen, I., Netemeyer, R., & Ryn, M. Van. (1991). The theory of planned behavior and its applications. *Orgnizational Behavior and Human Decision Processes*, *55*, 189–221.

Al Saifi, S. A. (2015). Positioning organisational culture in knowledge management research. *Journal of Knowledge Management*, *19*(2), 164–189. https://doi.org/10.1108/JKM-07-2014-0287

Al Sharoufi, H. (2021). Technological context. *The Pragmatics of Adaptability*, *319*, 299.

Alavi, M., Kayworth, T. R., & Leidner, D. E. (2005). An empirical examination of the influence of organizational culture on knowledge management practices. *Journal of Management Information Systems*, *22*(3), 191–224.

Alavi, M., & Leidner, D. E. (1999). Knowledge management systems: issues, challenges, and benefits. *Communications of the Association for Information Systems*, *1*(1), 7.

Alavi, M., & Leidner, D. E. (2001). Knowledge Management and Knowledge Management Systems: Conceptual Foundations and Research Issues. *MIS Quarterly*, *25*(1), 107–136. https://doi.org/10.2307/3250961

Alchian, A. A., & Demsetz, H. (1972). Production, information costs, and economic organization. *The American Economic Review*, *62*(5), 777–795.

Alexandrou, A., & Chen, L.-C. (2022). Perceived security of BYOD devices in medical institutions. *International Journal of Medical Informatics*, *168*, 104882.

Aliannejadi, M., Zamani, H., Crestani, F., & Croft, W. B. (2021). Context-aware target apps selection and recommendation for enhancing personal mobile assistants. *ACM Transactions on Information Systems (TOIS)*, *39*(3), 1–30.

Allam, S., Flowerday, S. V., & Flowerday, E. (2014). Smartphone information security awareness: A victim of operational pressures. *Computers & Security*, *42*, 56–65. https://doi.org/10.1016/j.cose.2014.01.005

Alshaikh, M., Maynard, S. B., & Ahmad, A. (2021). Applying social marketing to evaluate current security education training and awareness programs in organisations. *Computers & Security*, *100*, 1–29.

Alshaikh, M., Maynard, S. B., Ahmad, A., & Chang, S. (2015). Information security policy: a management practice perspective. *26th Australian Conference on Information Systems*, 1–14.

Alshaikh, M., Naseer, H., Ahmad, A., & Maynard, S. B. (2019). Toward sustainable behaviour change: an approach for cyber security education training and awareness. *European Conference on Information Systems*, 1–14.

Amara, N., Landry, R., & Traoré, N. (2008). Managing the protection of innovations in knowledge-intensive business services. *Research Policy*, *37*(9), 1530–1547. https://doi.org/10.1016/j.respol.2008.07.001

Amit, R., & Schoemaker, P. J. H. (1993). Strategic assets and organizational rent. *Strategic Management Journal*, *14*(1), 33–46.

Amoroso, S., & Link, A. N. (2021). Intellectual property protection mechanisms and the characteristics of founding teams. *Scientometrics*, *126*(9), 7329–7350.

Ancori, B., Bureth, A., & Cohendet, P. (2000). The economics of knowledge: The debate about codification and tacit knowledge. *Industrial & Corporate Change*, *9*(2), 255–287. https://doi.org/10.1093/icc/9.2.255

Anderson, R. (2001). Why information security is hard-an economic perspective. *Seventeenth Annual Computer Security Applications Conference*, 358–365.

Annansingh, F. (2006). *Exploring knowledge leakage risks exposure resulting from 3D modellling in organizations: a case study*. University of Sheffield.

Annansingh, F. (2012). *Exploring the Risks of Knowledge Leakage : An Information Systems Case Study Approach*. INTECH Open Access Publisher.

Arias-Pérez, J., Lozada, N., & Henao-García, E. (2020). When it comes to the impact of absorptive capacity on co-innovation, how really harmful is knowledge leakage? *Journal of Knowledge Management*, *24*(8), 1841–1857. https://doi.org/10.1108/JKM-02-2020-0084/FULL/HTML

Armando, A., Costa, G., Verderame, L., & Merlo, A. (2014). Securing the "Bring your own device" paradigm. *Computer*, *47*(6), 48–56. https://doi.org/10.1109/MC.2014.164

Arohan, Y., Yadav, A., Pandey, A., Churi, S., Saxena, M., & Vaghani, A. (2020). *An introduction to context-aware security and User Entity Behavior Analytics*.

Astani, M., Ready, K., & Tessema, M. (2013). BYOD issues and strategies in organizations. *Issues in Information Systems*, *14*(2), 195–201.

Atwood, R. (2019). *Combatting Chinese Cyber Economic Espionage on the Global Stage*. Utica College.

Australia, S., & New Zealand, S. (2009). *AS/NZS ISO 31000: 2009 Risk management-Principles and guidelines*. SAI Global.

Awazu, Y. (2004). Informal network players, knowledge integration, and competitive advantage. *Journal of Knowledge Management*.

Bandura, A. (1978). Self-efficacy: Toward a unifying theory of behavioral change. *Advances in Behaviour Research and Therapy*, *1*(4), 139–161. https://doi.org/10.1016/0146-6402(78)90002-4

Barney, J. (1991). Firm resources and sustained competitive advantage. *Journal of Management*, *17*(1), 99–120. https://doi.org/0803973233

Barney, J. B. (1986). Organizational culture: can it be a source of sustained competitive advantage? *Academy of Management Review*, *11*(3), 656–665.

Baskerville, R. (2005). Information Warfare. *Journal of Information System Security*, *1*(1), 23–50.

Baskerville, R., Kim, J., & Stucke, C. (2022). The cybersecurity risk estimation engine: A tool for possibility based risk analysis. *Computers and Security*, *120*, 102752. https://doi.org/10.1016/j.cose.2022.102752

Baskerville, R., Spagnoletti, P., & Kim, J. (2014). Incident-centered information security: Managing a strategic balance between prevention and response. *Information and Management*, *51*(1), 138–151. https://doi.org/10.1016/j.im.2013.11.004

Becerra, M., Lunnan, R., & Huemer, L. (2008). Trustworthiness, Risk, and the Transfer of Tacit and Explicit Knowledge Between Alliance Partners. *Journal of Management Studies*, *45*(4), 691–713. https://doi.org/10.1111/j.1467-6486.2008.00766.x

Benítez-Guerrero, E., Mezura-Godoy, C., & Montané-Jiménez, L. G. (2012).

Context-aware mobile collaborative systems: Conceptual modeling and case study. *Sensors*, *12*(10), 13491–13507.

Berthon, P., Nairn, A., & Money, A. (2003). Through the paradigm funnel: a conceptual tool for literature analysis. *Marketing Education Review*, *13*(2), 55–66.

Bertino, E., Khan, L. R., Sandhu, R., & Thuraisingham, B. (2006). Secure knowledge management: confidentiality, trust, and privacy. *IEEE Transactions on Systems, Man, and Cybernetics-Part A: Systems and Humans*, *36*(3), 429–438.

Bishop, M., Engle, S., Frincke, D. A., Gates, C., Greitzer, F. L., Peisert, S., & Whalen, S. (2010). A risk management approach to the "insider threat." In *Insider threats in cyber security* (pp. 115–137). Springer.

Bloodgood, J. M., & Chen, A. N. K. (2021). Preventing organizational knowledge leakage: the influence of knowledge seekers' awareness, motivation and capability. *Journal of Knowledge Management*. https://doi.org/10.1108/JKM-12-2020-0894/FULL/HTML

Boisot, M., & Canals, A. (2004). Data, information and knowledge: have we got it right? *Journal of Evolutionary Economics*, *14*(1), 43–67. https://doi.org/10.1007/s00191-003-0181-9

Bolisani, E., Paiola, M., & Scarso, E. (2013). Knowledge protection in knowledge-intensive business services. *Journal of Intellectual Capital Journal of Intellectual Capital Iss Journal of Intellectual Capital Iss Journal of Intellectual Capital*, *14*(2), 192–211. http://dx.doi.org/10.1108/14691931311323841

Boon, C., Den Hartog, D. N., & Lepak, D. P. (2019). A systematic review of human resource management systems and their measurement. *Journal of Management*, *45*(6), 2498–2537.

Bosua, R., & Scheepers, R. (2007). Towards a model to explain knowledge sharing in complex organizational environments. *Knowledge Management Research & Practice*, *5*(March 2006), 93–109.

https://doi.org/10.1057/palgrave.kmrp.8500131

Bouncken, R. B., & Barwinski, R. (2021). Shared digital identity and rich knowledge ties in global 3D printing—A drizzle in the clouds? *Global Strategy Journal*, *11*(1), 81–108. https://doi.org/10.1002/gsj.1370

Bouncken, R. B., Gast, J., Kraus, S., & Bogers, M. (2015). Coopetition: a systematic review, synthesis, and future research directions. *Review of Managerial Science*, *9*(3), 577–601. https://doi.org/10.1007/s11846-015-0168-6

Bradley, N. A., & Dunlop, M. D. (2005). Toward a Multidisciplinary Model of Context to Support Context-Aware Computing. *Human-Computer Interaction*, *20*(4), 403--446. https://doi.org/10.1207/s15327051hci2004

Brown, J. S., & Duguid, P. (2001). Knowledge and Organization: A Social Practice Perspective. *Organization Science*, *12*(2), 198–213. https://doi.org/10.1287/orsc.12.2.198.10116

Buck, C., Olenberger, C., Schweizer, A., Völter, F., & Eymann, T. (2021). Never trust, always verify: A multivocal literature review on current knowledge and research gaps of zero-trust. *Computers & Security*, *110*, 102436.

Campbell, M. (2020). Beyond zero trust: trust is a vulnerability. *Computer*, *53*(10), 110–113.

Carlsson, S.A., El Sawy, O.A., Eriksson, I., and Raven, A. (1996). Gaining Competitive Advantage Through Shared Knowledge Creation: In Search of a New Design Theory For Strategic Information Systems. *4th European Conference on Information Systems*.

Carlsson, S. A. (2003). Knowledge Managing and Knowledge Management Systems in Inter-organizational Networks. *Knowledge and Process Management*, *10*(3), 194–206.

Carson, D., Gilmore, A., Perry, C., & Gronhaug, K. (2001). *Qualitative marketing research*. Sage.

Check Point. (2022). Check Point Cyber Security Report 2022. In *Check Point Research*.

Chen, L., & Nath, R. (2008). A socio-technical perspective of mobile work. *Information, Knowledge, Systems Management*, *7*(1), 41–60.

Chen, T., Ullah, I., Kaafar, M. A., & Boreli, R. (2014). Information leakage through mobile analytics services. *Proceedings of the 15th Workshop on Mobile Computing Systems and Applications, HotMobile 2014*, *1*(1), 15. https://doi.org/10.1145/2565585.2565593

Cheng, S. K., & Kam, B. H. (2008). A conceptual framework for Analysing risk in supply networks. *Journal of Enterprise Information Management*, *22*(4), 345. https://link.springer.com/article/10.1007/s10479-021-04219-5

Choi, J., Sung, W., Choi, C., & Kim, P. (2015). Personal information leakage detection method using the inference-based access control model on the Android platform. *Pervasive and Mobile Computing*, *24*(1), 138–149. https://doi.org/10.1016/j.pmcj.2015.06.005

Cohen, W. M., & Levinthal, D. a. (1990). Absorptive Capacity: A new perspective on learning and innovation. *Administrative Science Quarterly*, *35*(1), 128–152. https://doi.org/10.2307/2393553

Collier, Z. A., & Sarkis, J. (2021). The zero trust supply chain: Managing supply chain risk in the absence of trust. *International Journal of Production Research*, *59*(11), 3430–3445.

Colwill, C. (2009). Human factors in information security: The insider threat - Who can you trust these days? *Information Security Technical Report*, *14*(4), 186–196. https://doi.org/10.1016/j.istr.2010.04.004

Conner, K. R. (1991). A historical comparison of resource-based theory and five schools of thought within industrial organization economics: do we have a new theory of the firm? *Journal of Management*, *17*(1), 121–154.

Corbin, J. M., & Strauss, A. (1990). Grounded theory research: Procedures, canons, and evaluative criteria. *Qualitative Sociology*, *13*(1), 3–21. https://doi.org/10.1007/BF00988593

Corbin, J., & Strauss, A. (2014). *Basics of qualitative research: Techniques and procedures for developing grounded theory*. Sage publications.

Crager, K., Maiti, A., Jadliwala, M., & He, J. (2007). Information Leakage through Mobile Motion Sensors : User Awareness and Concerns. *Internet Society*, *1*(1), 15. https://doi.org/10.14722/eurousec.2017.23013

Cram, W. A., D'arcy, J., & Proudfoot, J. G. (2019). Seeing the forest and the trees: a meta-analysis of the antecedents to information security policy compliance. *MIS Quarterly*, *43*(2), 525–554.

Cram, W. A., Proudfoot, J. G., & D'arcy, J. (2017). Organizational information security policies: a review and research framework. *European Journal of Information Systems*, *26*(6), 605–641.

Creswell, J. W. (2013). *Research design: Qualitative, quantitative, and mixed methods approaches*. Sage publications.

Crossler, R. E., Johnston, A. C., Lowry, P. B., Hu, Q., Warkentin, M., & Baskerville, R. (2013). Future directions for behavioral information security research. *Computers & Security*, *32*, 90–101. https://doi.org/10.1016/j.cose.2012.09.010

D'Arcy, J., & Hovav, A. (2007). Deterring internal information systems misuse. *Communications of the ACM*, *50*(10), 113–117. https://doi.org/10.1145/1290958.1290971

D'Arcy, J., & Hovav, A. (2009). Does one size fit all? Examining the differential effects of IS security countermeasures. *Journal of Business Ethics*, *89*(1), 59–71. https://doi.org/10.1287/isre.1070.0160

D'Arcy, J., Hovav, A., & Galletta, D. (2009). User Awareness of Security Countermeasures and its Impact on Information Systems Misuse: A Deterrence

Approach. *Information Systems Research*, *20*(1), 79–98. https://doi.org/10.1287/isre.l070.0160

Dahlbom, B., & Mathiassen, L. (1993). *Computers in context: The philosophy and practice of systems design.* Blackwell Publishers, Inc.

Dalton, A., Aghaei, E., Al-Shaer, E., Bhatia, A., Castillo, E., Cheng, Z., Dhaduvai, S., Duan, Q., Islam, M. M., Karimi, Y., Masoumzadeh, A., Mather, B., Santhanam, S., Shaikh, S., Strzalkowski, T., & Dorr, B. J. (2020). The panacea threat intelligence and active defense platform. *ArXiv Preprint ArXiv:2004.09662*. http://arxiv.org/abs/2004.09662

Dang-Pham, D., & Pittayachawan, S. (2015). Comparing intention to avoid malware across contexts in a BYOD-enabled Australian university: A Protection Motivation Theory approach. *Computers & Security*, *48*, 281–297. https://doi.org/10.1016/j.cose.2014.11.002

Davenport, T. H., & Prusak, L. (1998). Working knowledge: how organizations manage what they know. *Harvard Business Press*, *31*(4), 1–15. https://doi.org/10.1109/EMR.2003.1267012

David, F. R., & David, F. R. (2011). *Strategic management concepts and cases.* Prentice hall.

De Moya, J.-F., & Pallud, J. (2017). *Quantified Self: A Literature Review Based on the Funnel Paradigm.*

Denning, D. E. (2014). Framework and principles for active cyber defense. *Computers & Security*, *40*, 108–113.

Derballa, V., & Pousttchi, K. (2006a). Mobile Knowledge Management. In David G. Schwartz (Ed.), *Encyclopedia of Knowledge Management* (1st ed., pp. 645–650). Idea Group Reference.

Derballa, V., & Pousttchi, K. (2006b). Mobile Technology for Knowledge Management. In David G. Schwartz (Ed.), *Encyclopedia of Knowledge Management* (1st ed., pp. 651–656).

Desouza, K. (2009). Information and Knowledge Management in Public Sector Networks: The Case of the US Intelligence Community. In *International Journal of Public Administration* (Vol. 32, Issue 14). https://doi.org/10.1080/01900690903344718

Desouza, K. C. (2003a). Knowledge management barriers: Why the technology imperative seldom works. *Business Horizons*, *46*(1), 25–29. https://doi.org/10.1016/S0007-6813(02)00276-8

Desouza, K. C. (2003b). Facilitating tacit knowledge exchange. *Dl.Acm.Org*, *46*(6), 85–88. https://doi.org/10.1145/777313.777317

Desouza, K. C. (2006). Knowledge Security: An Interesting Research Space. *Researchgate.Net.* https://www.researchgate.net/profile/Kevin-Desouza-2/publication/228419064_Knowledge_Security_An_Interesting_Research_Space/links/00b7d5304b15d00391000000/Knowledge-Security-An-Interesting-Research-Space.pdf

Desouza, K. C. (2011). Securing intellectual assets: integrating the knowledge and innovation dimensions. *International Journal of Technology Management*, *54*(2/3), 167. https://doi.org/10.1504/IJTM.2011.039311

Desouza, K. C., Chattaraj, A., & Kraft, G. (2003). Supply chain perspectives to knowledge management: Research propositions. *Journal of Knowledge Management*, *7*(3), 129–138. https://doi.org/10.1108/13673270310485695/FULL/HTML

Desouza, K. C., & Vanapalli, G. K. (2005). Securing knowledge in organizations: Lessons from the defense and intelligence sectors. *International Journal of Information Management*, *25*(1), 85–98. https://doi.org/10.1016/j.ijinfomgt.2004.10.007

Dhillon, G. (2007). *Principles of Information Systems Security: text and cases*. Wiley New York, NY.

Di Vaio, A., Palladino, R., Pezzi, A., & Kalisz, D. E. (2021). The role of digital

innovation in knowledge management systems: A systematic literature review. *Journal of Business Research*, *123*, 220–231.

Diallo, B A A, Badard, T., Hubert, F., & Daniel, S. (2011). Towards context awareness mobile Geospatial BI (GeoBI) applications. *International Cartography Conference (ICC). Paris*.

Diallo, Belko Abdoul Aziz, Badard, T., Hubert, F., & Daniel, S. (2012). Mobile and context-aware GeoBI applications: a multilevel model for structuring and sharing of contextual information. *Journal of Geographic Information System*, *4*(05), 425. https://doi.org/10.1007/s10707-013-0187-x

Diallo, Belko Abdoul Aziz, Badard, T., Hubert, F., & Daniel, S. (2014). Context-based mobile GeoBI: Enhancing business analysis with contextual metrics/statistics and context-based reasoning. *GeoInformatica*, *18*(2), 405–433. https://doi.org/10.1007/s10707-013-0187-x

Dtex. (2019). *The Insider Threat Kill Chain*.

Dtex. (2022). *The 2022 Insider Threat Report*.

Durst, S., Aggestam, L., & Ferenhof, H. A. (2015). Understanding knowledge leakage: a review of previous studies. *Vine*, *45*(4), 568–586. https://doi.org/10.1108/VINE-01-2015-0009

Durst, S., Zięba, M., & Helio, A. F. (2018). *Knowledge risk management in organizations*.

Dwivedi, Y. K., & Kuljis, J. (2008). Profile of IS research published in the European Journal of Information Systems. *European Journal of Information Systems*, *17*(6), 678–693.

Eisenhardt, K. M. (2016). *Building Theories from Case Study. 14*(4), 532–550.

Ezzy, D. (2013). *Qualitative analysis*. Routledge.

Fai, C. M., & Goh, B. (2021). Optimise Defender's Advantage: Practical Approaches for Cybersecurity Defence. *Introduction To Cyber Forensic Psychology: Understanding The Mind Of The Cyber Deviant Perpetrators*, 251.

Figueiredo, R., & de Matos Ferreira, J. J. (2020). Spinner model: prediction of propensity to innovate based on knowledge-intensive business services. *Journal of the Knowledge Economy*, *11*(4), 1316–1335.

Fleisher, C. S., & Bensoussan, B. E. (2003). *Strategic and competitive analysis: methods and techniques for analyzing business competition.*

Foli, S., & Durst, S. (2022). Analysing Drivers of Knowledge Leakage in Collaborative Agreements: A Magnetic Processing Case Firm. *Journal of Risk and Financial Management*, *15*(9), 389.

Ford, D. P. (2004). Trust and knowledge management: the seeds of success. *Handbook on Knowledge Management 1*, 553–575.

Franklin, J., Howell, G., Sritapan, V., Souppaya, M., & Scarfone, K. (2020). *Guidelines for Managing the Security of Mobile Devices in the Enterprise*. National Institute of Standards and Technology.

Fredrich, V., Bouncken, R. B., & Kraus, S. (2019). The race is on: Configurations of absorptive capacity, interdependence and slack resources for interorganizational learning in coopetition alliances. *Journal of Business Research*, *101*, 862–868.

Freeman, R. E. (2010). *Strategic management: A stakeholder approach*. Cambridge university press.

Friedrich, J., Becker, M., Kramer, F., Wirth, M., & Schneider, M. (2020). Incentive design and gamification for knowledge management. *Journal of Business Research*, *106*, 341–352.

Frishammar, J., Ericsson, K., & Patel, P. C. (2015). The dark side of knowledge transfer: Exploring knowledge leakage in joint R&D projects. *Technovation*, *42*, 75–88. https://doi.org/10.1016/j.technovation.2015.01.001

Furnell, S., & Rajendran, A. (2012). Understanding the influences on information security behaviour. *Computer Fraud and Security*, *2012*(3), 12–15. https://doi.org/10.1016/S1361-3723(12)70053-2

Galati, F., Bigliardi, B., Petroni, A., Petroni, G., & Ferraro, G. (2019). A framework for avoiding knowledge leakage: evidence from engineering to order firms. *Knowledge Management Research & Practice*, *17*(3), 340–352.

Georgiadou, A., Mouzakitis, S., & Askounis, D. (2022). Working from home during COVID-19 crisis: a cyber security culture assessment survey. *Security Journal*, *35*(2), 486–505.

Ghosh, A., & Rai, P. K. G. S. (2013). Bring Your Own Device (Byod): Security Risks and Mitigating Strategies. *Journal of Global Research in Computer Science*, *4*(4), 62–70.

Gioia, D. a., Corley, K. G., & Hamilton, A. L. (2012). Seeking Qualitative Rigor in Inductive Research: Notes on the Gioia Methodology. *Organizational Research Methods*, *16*(1), 15–31. https://doi.org/10.1177/1094428112452151

Glaser, B. G., & Strauss, A. L. (1968). *Discovery of grounded theory: Strategies for qualitative research*. Routledge.

Gordon, P. (2007). *Data leakage Threats and mitigations*.

Grant, R. M. (1996a). Prospering in dynamically-competitive environments: Organizational capability as knowledge integration. *Organization Science*, *7*(4), 375–387.

Grant, R. M. (1996b). Toward a Knowledge-Based Theory of the Firm. *Strategic Management Journal*, *17*(s2), 109–122.

Gregor, S. (2006). The Nature Of Theory In Information Systems. *MIS Quarterly*, *30*(3), 611–642. https://doi.org/10.1080/0268396022000017725

Gregor, S., & Hevner, A. R. (2013). Positioning and Presenting Design Science Research for Maximum Impact. *MIS Quarterly*, *37*(2), 337–355. https://doi.org/10.2753/MIS0742-1222240302

Gregor, S., & Jones, D. (2007). The anatomy of a design theory. *Association for Information Systems*, *1*(1). https://doi.org/10.1080/0268396022000017725

Grillitsch, M., & Nilsson, M. (2017). Firm performance in the periphery: on the relation between firm-internal knowledge and local knowledge spillovers. *Regional Studies*, *51*(8), 1219–1231. https://doi.org/10.1080/00343404.2016.1175554

Grimaldi, M., Greco, M., & Cricelli, L. (2021). A framework of intellectual property protection strategies and open innovation. *Journal of Business Research*, *123*, 156–164.

Guba, E. G. (1990). The paradigm dialog. *Alternative Paradigms Conference, Mar, 1989, Indiana u, School of Education, San Francisco, ca, Us*.

Guba, E. G., & Lincoln, Y. S. (1994). Competing paradigms in qualitative research. *Handbook of Qualitative Research*, *2*(163–194), 105.

Guess, A. M., & Lyons, B. A. (2020). Misinformation, disinformation, and online propaganda. *Social Media and Democracy: The State of the Field, Prospects for Reform*, 10–33.

Guo, W., Yang, J., Li, D., & Lyu, C. (2020). Knowledge sharing and knowledge protection in strategic alliances: the effects of trust and formal contracts. *Technology Analysis & Strategic Management*, *32*(11), 1366–1378.

Halcomb, E. J., & Davidson, P. M. (2006). Is verbatim transcription of interview data always necessary? *Applied Nursing Research*, *19*(1), 38–42.

Hasan, H., & Banna, S. (2012). The unit of analysis in IS theory: The case for activity. *Information Systems Foundations*, *191*, 3–33.

Heckman, K. E., Stech, F. J., Schmoker, B. S., & Thomas, R. K. (2015). Denial and deception in cyber defense. *Computer*, *48*(4), 36–44.

Heckman, K. E., Stech, F. J., Thomas, R. K., Schmoker, B., & Tsow, A. W. (2015). Cyber denial, deception and counter deception. *Advances in Information Security*, *64*.

Herath, T., & Rao, H. R. (2009). Encouraging information security behaviors in organizations: Role of penalties, pressures and perceived effectiveness. *Decision Support Systems*, *47*(2), 154–165. https://doi.org/10.1016/j.dss.2009.02.005

Hevner, A. R., March, S. T., Park, J., & Ram, S. (2004). Design Science in Information Systems Research. *MIS Quarterly*, *28*(1), 75–105. https://doi.org/10.2307/25148625

Hitt, M. A., Ireland, R. D., & Hoskisson, R. E. (2017). *Strategic management: Concepts and cases: Competitiveness and globalization*. Cengage Learning.

Hodges, C. (2005). Competitive Intelligence overview feeding the competitive analysis process. *ASQ World Conference on Quality and Improvement Proceedings*, *59*, 441.

Hofer, T., Schwinger, W., Pichler, M., Leonhartsberger, G., Altmann, J., & Retschitzegger, W. (2003). Context-awareness on mobile devices - the hydrogen approach. *Proceedings of The36th Annual Hawaii International Conference on System Sciences, 2003.*, *43*(7236). https://doi.org/10.1109/HICSS.2003.1174831

Homoliak, I., Toffalini, F., Guarnizo, J., Elovici, Y., & Ochoa, M. (2019). Insight into insiders and it: A survey of insider threat taxonomies, analysis, modeling, and countermeasures. *ACM Computing Surveys (CSUR)*, *52*(2), 1–40.

Hopkins, K. D. (1982). The unit of analysis: Group means versus individual observations. *American Educational Research Journal*, *19*(1), 5–18.

Horne, C. A., Maynard, S. B., & Ahmad, A. (2017). Organisational information security strategy: Review, discussion and future research. *Australasian Journal of Information Systems*, *21*, 1–17.

IBM. (2022). *IBM security's cost of a data breach report 2022*.

Ifinedo, P. (2012). Understanding information systems security policy compliance: An integration of the theory of planned behavior and the protection motivation theory. *Computers and Security*, *31*(1), 83–95. https://doi.org/10.1016/j.cose.2011.10.007

Ilvonen, I., Thalmann, S., Manhart, M., & Sillaber, C. (2018). Reconciling digital transformation and knowledge protection: A research agenda. *Knowledge Management Research & Practice*, *16*(2), 235–244.

Inkpen, A., Minbaeva, D., Business, E. T.-J. of I., & 2019, undefined. (2019). Unintentional, unavoidable, and beneficial knowledge leakage from the multinational enterprise. *Springer*, *50*(2), 250–260. https://doi.org/10.1057/s41267-018-0164-6

Intezari, A., Taskin, N., & Pauleen, D. J. (2017). Looking beyond knowledge sharing: an integrative approach to knowledge management culture. *Journal of Knowledge Management*.

Jacob, C., Sanchez-Vazquez, A., & Ivory, C. (2020). Social, organizational, and technological factors impacting clinicians' adoption of mobile health tools: systematic literature review. *JMIR MHealth and UHealth*, *8*(2), e15935.

Jaeger, L., Ament, C., & Eckhardt, A. (2017). The Closer You Get the More Aware You Become – A Case Study about Psychological Distance to Information Security Incidents. *The 38th International Conference on Information Systems*.

Janssen, J., & Spruit, M. (2019). M-RAM: a Mobile Risk Assessment Method for Enterprise Mobile Security. *Technical Report Series*, *UU-CS-2019-009*.

Jarrahi, M. H., & Thomson, L. (2017). The interplay between information practices and information context: The case of mobile knowledge workers. *Journal of the Association for Information Science and Technology*, *68*(5), 1073–1089. https://doi.org/10.1002/asi.23773

Jewels, T., Underwood, A., & Heredero, C. de P. (2003). *The role of informal networks in knowledge sharing.*

Jiang, X., Bao, Y., Xie, Y., & Gao, S. (2016). Partner trustworthiness, knowledge flow in strategic alliances, and firm competitiveness: A contingency perspective. *Journal of Business Research*, *69*(2), 804–814.

https://doi.org/10.1016/j.jbusres.2015.07.009

Jiang, X., Li, M., Gao, S., Bao, Y., & Jiang, F. (2013). Managing knowledge leakage in strategic alliances: The effects of trust and formal contracts. *Industrial Marketing Management*, *42*(6), 983–991. https://doi.org/10.1016/j.indmarman.2013.03.013

Jiang, X., & Li, Y. (2008). The relationship between organizational learning and firms' financial performance in strategic alliances: A contingency approach. *Journal of World Business*, *43*(3), 365–379. https://doi.org/10.1016/j.jwb.2007.11.003

Johnson, G., Scholes, K., & Whittington, R. (1999). Corporate strategy. *Europe: London Prentice Hall.*

Johnston, A. C., & Warkentin, M. (2010). Fear Appeals and Information Security Behaviors: an Empirical Study. *MIS Quarterly*, *34*(3), 549-A4. http://www.uab.edu/cas/thecenter/images/Documents/FEAR-APPEALS-AND-INFORMATION-SECURITY-BEHAVIORS-AN-EMPIRICAL-STUDY.pdf

Kaiser, R., Thalmann, S., & Pammer-Schindler, V. (2020). An investigation of knowledge protection practices in inter-organisational collaboration: protecting specialised engineering knowledge with a practice based on grey-box modelling. *VINE Journal of Information and Knowledge Management Systems.*

Kaloroumakis, P. E., & Smith, M. J. (2021). *Toward a knowledge graph of cybersecurity countermeasures.* Technical report.

Kang, H., & Lee, Y. W. (2017). Innovation Strategies Against Knowledge Leakage: Externality Effects of Non-competes Enforcement. *Academy of Management Proceedings*, *1*(1). http://proceedings.aom.org/content/2017/1/10735.short

Kaplan, S., Schenkel, A., von Krogh, G., & Weber, C. (2001). Knowledge-based theories of the firm in strategic management: A review and extension.

*International Journal of Project Management*, *25*(56), 143–158.

Karabacak, B., & Whittaker, T. (2022). Zero Trust and Advanced Persistent Threats: Who Will Win the War? *International Conference on Cyber Warfare and Security*, *17*(1), 92–101.

Kaster, S. D., & Ensign, P. C. (2022). Privatized espionage: NSO Group Technologies and its Pegasus spyware. *Thunderbird International Business Review*.

Kengatharan, N. (2019). A knowledge-based theory of the firm: Nexus of intellectual capital, productivity and firms' performance. *International Journal of Manpower*.

Khatib, R. A. El. (2021). *A Review Of Knowledge Risk Conception*. *3*(1).

Kianto, A., Ritala, P., Vanhala, M., & Hussinki, H. (2020). Reflections on the criteria for the sound measurement of intellectual capital: A knowledge-based perspective. *Critical Perspectives on Accounting*, *70*, 102046.

Kindervag, J. (2010). Build security into your network's dna: The zero trust network architecture. *Forrester Research Inc*, *27*.

Kindervag, J., Balaouras, S., & Coit, L. (2010). No more chewy centers: Introducing the zero trust model of information security. *Forrester Research*, *3*.

Kitching, J., & Blackburn, R. (1998). Intellectual property management in the small and medium enterprise (SME). *Journal of Small Business and Enterprise Development*.

Kofod-Petersen, A., & Cassens, J. (2006). Using activity theory to model context awareness. In *Modeling and Retrieval of Context* (pp. 1–17). Springer.

Kotsias, J., Ahmad, A., & Scheepers, R. (2022). Adopting and integrating cyber-threat intelligence in a commercial organisation. *European Journal of Information Systems*, 1–17.

Krippendorff, K. (1980). *Validity in content analysis*.

Krishnamurthy, B., & Wills, C. E. (2010). On the leakage of personally identifiable information via online social networks. *ACM SIGCOMM Computer*

*Communication Review*, *40*(1), 112. https://doi.org/10.1145/1672308.1672328

Kumar, S. (2018). Understanding different issues of unit of analysis in a business research. *Journal of General Management Research*, *5*(2), 70–82.

Lallie, H. S., Shepherd, L. A., Nurse, J. R. C., Erola, A., Epiphaniou, G., Maple, C., & Bellekens, X. (2021). Cyber security in the age of COVID-19: A timeline and analysis of cyber-crime and cyber-attacks during the pandemic. *Computers and Security*, *105*, 1–20. https://doi.org/10.1016/j.cose.2021.102248

Langley, A. (1999). Strategies for theorizing from process data. *Academy of Management Review*, *24*(4), 691–710.

Lascaux, A. (2020). Coopetition and trust: What we know, where to go next. *Industrial Marketing Management*, *84*, 2–18.

Lau, A. K. W., Yam, R. C. M., Tang, E. P. Y., & Sun, H. Y. (2010). Factors influencing the relationship between product modularity and supply chain integration. *International Journal of Operations & Production Management*.

Lee, M., Alba, J. D., & Park, D. (2018). Intellectual property rights, informal economy, and FDI into developing countries. *Journal of Policy Modeling*, *40*(5), 1067–1081.

Lee, T. W., Mitchell, T. R., & Sablynski, C. J. (1999). Qualitative research in organizational and vocational psychology, 1979–1999. *Journal of Vocational Behavior*, *55*(2), 161–187.

Leonard-barton, D. (1992). Core Capabilities and Core Rigidities - a Paradox in Managing New Product Development. *Strategic Management Journal*, *13*, 111–125.

Li, Q., Sustainability, Y. K.-, & 2019, U. (2019). Knowledge sharing willingness and leakage risk: an evolutional game model. *Mdpi.Com.* https://www.mdpi.com/400440

Libicki, M. C., Ablon, L., & Webb, T. (2015). *The defender's dilemma: Charting a course toward cybersecurity*. Rand Corporation.

Liebeskind, J. P. (1996). Knowledge, strategy and the theory of the firm. *Knowledge and Strategy*, *17*(Special Issue), 93–107. https://doi.org/10.2307/2486993

Lim, J. S., Ahmad, A., Chang, S., & Maynard, S. (2010). Embedding information security culture emerging concerns and challenges. *Pacific Asia Conference on Information Systems (PACIS)*.

Lim, J. S., Chang, S., Maynard, S., & Ahmad, A. (2009). Exploring the relationship between organizational culture and information security culture. *7th Australian Information Security Management Conference*, 88–97.

Link, A. N., & van Hasselt, M. (2022). The use of intellectual property protection mechanisms by publicly supported firms. *Economics of Innovation and New Technology*, *31*(1–2), 111–121.

Locke, K., Golden-Biddle, K., & Feldman, M. S. (2008). Perspective—Making doubt generative: Rethinking the role of doubt in the research process. *Organization Science*, *19*(6), 907–918.

Luo, C., Goncalves, J., Velloso, E., & Kostakos, V. (2020). A survey of context simulation for testing mobile context-aware applications. *ACM Computing Surveys (CSUR)*, *53*(1), 1–39.

MacDougall, S. L., & Hurst, D. (2005). Identifying tangible costs, benefits and risks of an investment in intellectual capital: Contracting contingent knowledge workers. *Journal of Intellectual Capital*, *6*(1), 53–71. https://doi.org/10.1108/14691930510574663

Manhart, M., & Thalmann, S. (2015). Protecting organizational knowledge: a structured literature review. *Journal of Knowledge Management*, *19*(2), 190–211. https://doi.org/10.1108/JKM-05-2014-0198

Maravilhas, S., & Martins, J. (2019). Strategic knowledge management in a digital environment: Tacit and explicit knowledge in Fab Labs. *Journal of Business Research*, *94*, 353–359.

Marczak, B., Scott-Railton, J., Abdul Razzak, B., Al-Jizawi, N., Anstis, S., Berdan, K., & Deibert, R. (2021). *FORCEDENTRY: NSO Group iMessage Zero-Click Exploit Captured in the Wild.*

Marjanovic, Z. (2013). *Effectiveness of security controls in BYOD environments.* http://dtl.unimelb.edu.au//exlibris/dtl/d3_1/apache_media/L2V4bGlicmlzL2R0bC9kM18xL2FwYWNoZV9tZWRpYS8zMDAzMTU=.pdf

Marsh, S. J., & Stock, G. N. (2006). Creating dynamic capability: The role of intertemporal integration, knowledge retention, and interpretation. *Journal of Product Innovation Management*, *23*(5), 422–436. https://doi.org/10.1111/j.1540-5885.2006.00214.x

Martín, A. G., Beltrán, M., Fernández-Isabel, A., & de Diego, I. M. (2021). An approach to detect user behaviour anomalies within identity federations. *Computers & Security*, *108*, 102356.

Matre, K., Alex, F., Rzasa, A., & Nutting, D. (2021). *The Defender's Advantage.*

Maynard, S. B., Ruighaver, A. B., & Ahmad, A. (2011). Stakeholders in security policy development. *9th Information Security Management Conference*, 182–188.

Maynard, S., Tan, T., Ahmad, A., & Ruighaver, T. (2018). Towards a framework for strategic security context in information security governance. *Pacific Asia Journal of the Association for Information Systems*, *10*(4), 1–24.

Mbonye, M., Nakamanya, S., Nalukenge, W., King, R., Vandepitte, J., & Seeley, J. (2013). 'It is like a tomato stall where someone can pick what he likes': structure and practices of female sex work in Kampala, Uganda. *BMC Public Health*, *13*(1), 1–9.

McLellan, E., MacQueen, K. M., & Neidig, J. L. (2003). Beyond the qualitative interview: Data preparation and transcription. *Field Methods*, *15*(1), 63–84.

McMullin, C. (2021). Transcription and Qualitative Methods: Implications for Third Sector Research. *VOLUNTAS: International Journal of Voluntary and Nonprofit*

*Organizations*, 1–14.

Melnitzky, A. (2011). Defending America against Chinese cyber espionage through the use of active defenses. *Cardozo J. Int'l & Comp. L.*, *20*, 537.

Melville, N., Kraemer, K., & Gurbaxani, V. (2004). Review: Information Technology and Organizational Performance: an Integrative Model of It Business Value. *MIS Quarterly*, *28*(2), 283–322. https://doi.org/10.2307/25148636

Miles, M. B., & Huberman, A. M. (1994). *Qualitative data analysis: An expanded sourcebook*. sage.

Mintzberg, H., & Waters, J. A. (1985). Of strategies, deliberate and emergent. *Strategic Management Journal*, *6*(3), 257–272.

Miric, M., Boudreau, K. J., & Jeppesen, L. B. (2019). Protecting their digital assets: The use of formal & informal appropriability strategies by App developers. *Research Policy*, *48*(8), 103738.

Moein, T., Paalhed, J., & Bengtsson, M. (2015). Dimensions of trust and distrust and their effect on knowledge sharing and knowledge leakage. An Empirical study of Swedish knowledge-intensive firms. *UMEA Universitet Review*, 96.

Mohamed, S., Mynors, D., Andrew, G., Chan, P., Coles, R., & Walsh, K. (2007). Unearthing key drivers of knowledge leakage. *International Journal of Knowledge Management Studies*, *1*(3–4), 456–470.

Mohamed, S., Mynors, D., Grantham, A., Walsh, K., & Chan, P. (2006). Understanding one aspect of the knowledge leakage concept: people. *Proceedings of the European and Mediterranean Conference on Information Systems (EMCIS)*, 2–12. https://doi.org/10.1504/IJEB.2007.012974

Moher, D., Liberati, A., Tetzlaff, J., & Altman, D. G. (2010). Preferred reporting items for systematic reviews and meta-analyses: the PRISMA statement. *Int J Surg*, *8*(5), 336–341.

Morrell, D. (2020). *Continuous Validation and Threat Protection for Mobile Applications*.

LOOKOUT, INC. SAN FRANCISCO United States.

Morrow, B. (2012). BYOD security challenges: Control and protect your most sensitive data. *Network Security*, *2012*(12), 5–8. https://doi.org/10.1016/S1353-4858(12)70111-3

Morse, J. M., Barrett, M., Mayan, M., Olson, K., & Spiers, J. (2002). Verification strategies for establishing reliability and validity in qualitative research. *International Journal of Qualitative Methods*, *1*(2), 13–22.

Mupepi, Mambo Governor Modak, A., Motwani, J., & Mupepi, S. C. (2017). How Can Knowledge Leakage be Stopped: A Socio-Technical System Design Approach to Risk Management. *International Journal of Sociotechnology and Knowledge Development (IJSKD)*, *9*(1), 26--41.

Mupepi, M. (2017). Effective talent management strategies for organizational success. In *igi-global.com*. https://www.igi-global.com/article/how-can-knowledge-leakage-be-stopped/181471

Myers, M. D., & Klein, H. K. (1999). A set of principles for conducting and evaluating interpretive field studies in Information Systems. *MIS Quarterly*, *23*(1), 67–94. https://doi.org/10.2307/249410

Myers, M. D., & Newman, M. (2007). The qualitative interview in IS research: Examining the craft. *Information and Organization*, *17*(1), 2–26. https://doi.org/10.1016/j.infoandorg.2006.11.001

Nelson, S. B., Jarrahi, M. H., & Thomson, L. (2017). Mobility of knowledge work and affordances of digital technologies. *International Journal of Information Management*, *37*(2), 54–62. https://doi.org/10.1016/j.ijinfomgt.2016.11.008

Neuman, W. L. (2006). *Social research methods: Qualitative and quantitative approaches*. (Sixth Edit). Pearson.

Nieto, I., Botía, J. A., & Gómez-Skarmeta, A. F. (2006). Information and hybrid architecture model of the OCP contextual information management system.

*Journal of Universal Computer Science*, *12*(3), 357–366.

Nonaka, Ikujiro. (1991). The knowledge-creating company. *Harvard Business Review*, *85*(7–8). https://doi.org/10.1016/0024-6301(96)81509-3

Nonaka, Ikujiro. (1994). A Dynamic Theory Knowledge of Organizational Creation. *Organization Science*, *5*(1), 14–37. https://doi.org/10.1287/orsc.5.1.14

Nonaka, Ikujiro, & Konno, N. (1998). The concept of" ba": Building a foundation for knowledge creation. *California Management Review*, *40*(3), 40–54.

Nonaka, Ikujirō, o Nonaka, I., Ikujiro, N., & Takeuchi, H. (1995). *The knowledge-creating company: How Japanese companies create the dynamics of innovation* (Vol. 105). OUP USA.

Nonaka, Ikujiro, & Toyama, R. (2003). The knowledge-creating theory revisited: knowledge creation as a synthesizing process. *Knowledge Management Research & Practice*, *1*(1), 2–10. https://doi.org/10.1057/palgrave.kmrp.8500001

Norman, P. M. (2001). Are your secrets safe? Knowledge protection in strategic alliances. *Business Horizons*, *44*(6), 51–60. https://doi.org/10.1016/S0007-6813(01)80073-2

Nunes, M. B., Annansingh, F., Eaglestone, B., & Wakefield, R. (2006). Knowledge management issues in knowledge-intensive SMEs. *Journal of Documentation*, *62*(1), 101–119. https://doi.org/10.1108/00220410610642075

Okoli, C., & Schabram, K. (2010). *A guide to conducting a systematic literature review of information systems research.*

Olander, H., Hurmelinna-Laukkanen, P., & Heilmann, P. (2011). Do SMEs benefit from HRM-related knowledge protection in innovation management? *International Journal of Innovation Management*, *15*(3), 593–616. https://doi.org/10.1142/S1363919611003453

Olander, H., Hurmelinna-Laukkanen, P., & Mähönen, J. (2009). What's small size got to do with it? Protection of intellectual assets in SMEs. *International Journal of*

*Innovation Management*, *13*(03), 349–370.

Olander, H., Hurmelinna-Laukkanen, P., & Vanhala, M. (2014). Mission: Possible But Sensitive — Knowledge Protection Mechanisms Serving Different Purposes. *International Journal of Innovation Management*, *18*(6). https://doi.org/10.1142/S136391961440012X

Oorschot, K. E. Van, Solli-Sæther, H., & Karlsen, J. T. (2018). The knowledge protection paradox: imitation and innovation through knowledge sharing. *International Journal of Technology Management*, *78*(4), 310–342.

Orlikowski, W. J., & Baroudi, J. J. (1991). Studying information technology in organizations: Research approaches and assumptions. *Information Systems Research*, *2*(1), 1–28.

Ortbach, K., Walter, N., & Öksüz, A. (2015). Are You Ready To Lose Control ? a Theory on the Role of Trust and Risk Perception on Bring-Your-Own-Device Policy and Information System Service Quality. *Ecis*, 1–10.

Oxley, J. E., & Sampson, R. C. (2004). The scope and governance of international R&D alliances. *Strategic Management Journal*, *25*(8-9), 723–749.

Oxley, J., & Wada, T. (2009). Alliance structure and the scope of knowledge transfer: Evidence from US-Japan agreements. *Management Science*, *55*(4), 635–649.

Päällysaho, S., & Kuusisto, J. (2011). Informal ways to protect intellectual property (IP) in KIBS businesses. *Innovation*, *13*(1), 62–76.

Palanisamy, R., Norman, A. A., & Kiah, M. L. M. (2020). Compliance with Bring Your Own Device security policies in organizations: A systematic literature review. *Computers & Security*, *98*, 101998.

Park, S., Ruighaver, A. B., Maynard, S. B., & Ahmad, A. (2012). Towards understanding deterrence: Information security managers' perspective. *Proceedings of the International Conference on IT Convergence and Security 2011*, 21–37.

Park, S., & Ruighaver, T. (2008). Strategic approach to information security in organizations. *2008 International Conference on Information Science and Security (ICISS 2008)*, 26–31.

Parker, H. (2012). Knowledge acquisition and leakage in inter-firm relationships involving new technology-based firms. *Management Decision*, *50*(9), 1618–1633. https://doi.org/10.1108/00251741211266714

Patton, M. Q. (1999). Enhancing the quality and credibility of qualitative analysis. *Health Services Research*, *34*(5 Pt 2), 1189.

Pedraza-fariña, L. G. (2017). Spill Your ( Trade ) Secrets : Knowledge Networks as Innovation Drivers. *Notre Dame Law Review*, *92*(4), 1561–1610.

Pemberton, J. D., & Stonehouse, G. H. (2000). Organisational learning and knowledge assets–an essential partnership. *The Learning Organization*.

Penrose, E. T. (1960). The growth of the firm—a case study: the Hercules Powder Company. *Business History Review*, *34*(1), 1–23.

Peyrot, M., Childs, N., Van Doren, D., & Allen, K. (2002). An empirically based model of competitor intelligence use. *Journal of Business Research*, *55*(9), 747–758.

Polanyi, M., & Sen, A. (1997). *The tacit dimension*. University of Chicago press.

Ponemon, I. (2022). Cost of Insider Threats Report. *Ponemon Institute*, *3*(12), 168–178. https://doi.org/10.22184/2227-572x.2022.12.3.168.178

Ponemon Institute. (2018). Ponemon Institute – The 2018 Cost of a Data Breach Study by the Ponemon Institute. In *IBM Security Services* (Issue July). https://www-01.ibm.com/common/ssi/cgi-bin/ssialias?htmlfid=55017055USEN&

Ponemon Institute. (2020). *Cybersecurity in the Remote Work Era.* (Issue October).

Ponemon Institute. (2021a). Cost of a Data Breach Report 2021. *IBM Security*, 1–73. https://www.ibm.com/security/data-breach

Ponemon Institute. (2021b). The State Of Insider Threats 2021 : Publication Date :

September 2021. In *Ponemon Institute* (Issue September).

Pournader, M., Kach, A., & Talluri, S. (2020). A review of the existing and emerging topics in the supply chain risk management literature. *Decision Sciences*, *51*(4), 867–919.

Povarnitsyna, Y. P. (2020). *Interfirm mobility and carreer success of knowledge workers.*

Pratama, A. R., & Scarlatos, L. L. (2020). Ownership and use of mobile devices among adolescents in Indonesia. *Journal of Educational Technology Systems*, *48*(3), 356–384.

Preacher, K. J., Zyphur, M. J., & Zhang, Z. (2010). A general multilevel SEM framework for assessing multilevel mediation. *Psychological Methods*, *15*(3), 209.

Promnick, G. (2017). Cyber economic espionage: Corporate theft and the new Patriot Act. *Hastings Sci. & Tech. LJ*, *9*, 89.

Puhakainen, P., & Siponen, M. (2010). Improving employees' compliance through information systems security training: an action research study. *MIS Quarterly*, 757–778.

Ranft, A. L., & Lord, M. D. (2002). Acquiring new technologies and capabilities: A grounded model of acquisition implementation. *Organization Science*, *13*(4), 420–441.

Rieger, C., & Majchrzak, T. A. (2017). A taxonomy for app-enabled devices: mastering the mobile device jungle. *International Conference on Web Information Systems and Technologies*, 202–220.

Ritala, P., Husted, K., Olander, H., & Michailova, S. (2018). External knowledge sharing and radical innovation: the downsides of uncontrolled openness. *Journal of Knowledge Management*, *1*(1). https://doi.org/10.1016/j.technovation.2014.07.011

Ritala, P., Olander, H., Michailova, S., & Husted, K. (2015). Knowledge sharing, knowledge leaking and relative innovation performance: An empirical study.

*Technovation*, *35*, 22–31. https://doi.org/10.1016/j.technovation.2014.07.011

Rose, S. W., Borchert, O., Mitchell, S., & Connelly, S. (2020). [NIST SP 800-207] Zero Trust Architecture. *NIST Special Publication - 800 Series*, 49. https://nvlpubs.nist.gov/nistpubs/SpecialPublications/NIST.SP.800-207-draft2.pdf

Rutakumwa, R., Mugisha, J. O., Bernays, S., Kabunga, E., Tumwekwase, G., Mbonye, M., & Seeley, J. (2020). Conducting in-depth interviews with and without voice recorders: a comparative analysis. *Qualitative Research*, *20*(5), 565–581.

Sampaio, M. C., Sousa, M. J., & Dionísio, A. (2019). The use of gamification in knowledge management processes: a systematic literature review. *Journal of Reviews on Global Economics*, *8*, 1662–1679.

Scarfo, A. (2012). New security perspectives around BYOD. *Proceedings - 2012 7th International Conference on Broadband, Wireless Computing, Communication and Applications, BWCCA 2012*, 446–451. https://doi.org/10.1109/BWCCA.2012.79

Schilit, B., Adams, N., & Want, R. (1994). Context-aware computing applications. *Mobile Computing Systems and Applications, 1994. WMCSA 1994. First Workshop on (Pp. 85-90). IEEE.* https://doi.org/10.1109/MCSA.1994.512740

Schilit, B. N., & Theimer, M. M. (1994). Disseminating active map information to mobile hosts. *IEEE Network*, *8*(5), 22–32.

Schniederjans, D. G., Curado, C., & Khalajhedayati, M. (2020). Supply chain digitisation trends: An integration of knowledge management. *International Journal of Production Economics*, *220*, 107439.

Schulkind, J., Mbonye, M., Watts, C., & Seeley, J. (2016). The social context of gender-based violence, alcohol use and HIV risk among women involved in high-risk sexual behaviour and their intimate partners in Kampala, Uganda.

*Culture, Health & Sexuality*, *18*(7), 770–784.

Schwartz, D G. (2006). Encyclopedia of knowledge management. In *Encyclopedia of Knowledge Management* (Vol. 1). Idea Group Reference. https://doi.org/10.4018/978-1-59140-573-3

Schwartz, David G., & Te'eni, D. (2011). Preface: The Knowledge Management Pyramid - Unification of a Complex Discipline. In David G. Schwartz & D. Te'eni (Eds.), *Encyclopedia of Knowledge Management* (2nd ed., pp. 2–7). Hershey:IGI Reference.

Schwartz, M., & Hornych, C. (2011). Knowledge sharing through informal networking: an overview and agenda. *International Journal of Knowledge-Based Development*, *2*(3), 282–294.

Sedgwick, P. (2014). Unit of observation versus unit of analysis. *Bmj, 348.*

Shabtai, A., Elovici, Y., & Rokach, L. (2012a). *A survey of data leakage detection and prevention solutions.* Springer. https://doi.org/10.1007/978-1-4614-2053-8

Shabtai, A., Elovici, Y., & Rokach, L. (2012b). A taxonomy of data leakage prevention solutions. In *A survey of data leakage detection and prevention solutions* (pp. 11–15). Springer.

Shanks, G. (2002). Guidelines for conducting positivist case study research in information systems. *Australasian Journal of Information Systems*, *10*(1).

Shanks, G., Arnott, D., & Rouse, A. (1993). *A review of approaches to research and scholarship in information systems.* Department of Information Systems, Faculty of Computing and Information ….

Shedden, P., Scheepers, R., Smith, W., & Ahmad, A. (2011). Incorporating a knowledge perspective into security risk assessments. *Vine*, *41*(2), 152–166. https://doi.org/10.1108/03055721111134790

Shedden, P., Smith, W., & Ahmad, A. (2010). *Information Security Risk Assessment : Towards a Business Practice Perspective. 34*(November), 555–590.

Shedden, P., Smith, W., Scheepers, R., & Ahmad, A. (2012). Towards a knowledge perspective in information security risk assessments–an illustrative case study. *20th Australasian Conference on Information Systems*, 74–84.

Siponen, M. T. (2000). A conceptual foundation for organizational information security awareness. *Information Management & Computer Security*, *8*(1), 31–41. https://doi.org/10.1108/09685220010371394

Snijders, T. A. B., & Bosker, R. J. (2011). *Multilevel analysis: An introduction to basic and advanced multilevel modeling*. sage.

Song, M., & Lee, K. (2014). Proposal of MDM Management Framework for BYOD use of Large Companies. *International Journal of Smart Home*, *8*(1), 123–128.

Sonnenschein, R., Loske, A., & Buxmann, P. (2017). The Role of Top Managers' IT Security Awareness in Organizational IT Security Management. *International Conference on Information Systems (ICIS)*, *1*(1). http://aisel.aisnet.org/cgi/viewcontent.cgi?article=1106&context=icis2017

Spagnoletti, P., & Resca, A. (2007). A Framework for Managing Predictable and Unpredictable Threats: The Duality of Information Security Management. *Proceedings of the Fifteenth European Conference on Information Systems*, June, 1539–1550. 20070161.pdf

Spagnoletti, P., & Resca, A. (2008). The duality of Information Security Management: fighting against predictable and unpredictable threats. *Journal of Information System Security*, *4*(3), 46–62. http://eprints.luiss.it/955/

Spender, J.-C., & Grant, R. M. (1996a). Knowledge and the firm: overview. *Strategic Management Journal*, *17*(s2), 5–9.

Spender, J.-C., & Grant, R. M. (1996b). Making knowledge the basis of a dynamic theory of the firm. *Strategic Management Journal*, *17*(2), 45–62.

Spender, J. (1996). Organizational knowledge, learning and memory: three concepts in search of a theory. *Journal of Organizational Change Management*, *9*(1), 63–78.

https://doi.org/10.1108/09534819610156813

Statistics, A. B. of. (2020). *Australian Industry 2019-2020*. ABS. https://doi.org/10.1002/9781119111931.ch144

Stech, F. J., Heckman, K. E., & Strom, B. E. (2016). Integrating cyber-D&D into adversary modeling for active cyber defense. In *Cyber deception* (pp. 1–22). Springer.

Stiles, W. B. (1993). Quality control in qualitative research. *Elsevier*. https://www.sciencedirect.com/science/article/pii/027273589390048Q

Strom, B. E., Applebaum, A., Miller, D. P., Nickels, K. C., Pennington, A. G., & Thomas, C. B. (2018). Mitre att&ck: Design and philosophy. *Technical Report*.

Styhre, A. (2003). Knowledge management beyond codification: knowing as practice/concept. *Journal of Knowledge Management*, *7*(5), 32–40. https://doi.org/10.1108/13673270310505368

Suddaby, R. (2006). From the editors: What grounded theory is not. In *Academy of management journal* (Vol. 49, Issue 4, pp. 633–642). Academy of Management Briarcliff Manor, NY 10510.

Sumbal, M. S., Tsui, E., See-to, E., & Barendrecht, A. (2017). Knowledge retention and aging workforce in the oil and gas industry: a multi perspective study. *Journal of Knowledge Management*, *21*(4), 907–924. https://doi.org/10.1108/JKM-07-2016-0281

Sveen, F. O., Torres, J. M., & Sarriegi, J. M. (2009). Blind information security strategy. *International Journal of Critical Infrastructure Protection*, *2*(3), 95–109. https://doi.org/10.1016/j.ijcip.2009.07.003

Swacha, J. (2015). Gamification in knowledge management: motivating for knowledge sharing. *Polish Journal of Management Studies*, *12*.

Tamminen, S., Oulasvirta, A., Toiskallio, K., & Kankainen, A. (2004). Understanding mobile contexts. *Personal and Ubiquitous Computing*, *8*(2), 135–143.

Tan, K. H., Wong, W., & Chung, L. (2016). Information and knowledge leakage in supply chain. *Information Systems Frontiers*, *18*(3), 621--638.

Tan, K. H., Wong, W. P., & Chung, L. (2015). Information and Knowledge Leakage in Supply Chain. *Information Systems Frontiers*, 1–18. https://doi.org/10.1007/s10796-015-9553-6

Tavares, M. F. F. (2020). Across establishments, within firms: worker's mobility, knowledge transfer and survival. *Journal for Labour Market Research*, *54*(1), 1–19.

Teece, D J, Pisano, G., & Schuen, A. (1991). *Dynamic Capabilities and Strategic Management. Working Paper, Center for Research in Management, University of California, Berkeley.*

Teece, David J. (2007). Explicating Dynamic Capabilities: The Nature and Microfoundations of (Sustainabile) Enterprise Performance. *Strategic Management Journal*, *298*(13), 1319–1350.

Theriou, N. G., Aggelidia, V., & Theriou, G. (2009). *A theoretical framework contrasting the resource-based perspective and the knowledge-based view.*

Thorleuchter, D., & Van Den Poel, D. (2013). Protecting research and technology from espionage. *Expert Systems with Applications*, *40*(9), 3432–3440. https://doi.org/10.1016/j.eswa.2012.12.051

Timiyo, A. J., & Foli, S. (2023). Knowledge leakage through social networks: a review of existing gaps, strategies for mitigating potential risk factors and future research direction. *VINE Journal of Information and Knowledge Management Systems*.

Toelle, E. (2021). Insider Risk Management. In *Microsoft 365 Compliance* (pp. 289–314). Springer.

Trkman, P., Systems, K. D.-T. J. of S. I., & 2012, U. (2012). Knowledge risks in organizational networks: An exploratory framework. *Elsevier*. https://www.sciencedirect.com/science/article/pii/S0963868711000552

Tsang, H. W. C., Lee, W. B., & Tsui, E. (2016). AHP-Driven Knowledge Leakage

Risk Assessment Model: A Construct-Apply-Control Cycle Approach. *International Journal of Knowledge and Systems Science (IJKSS)*, *7*(3), 1--18.

Tseng, C.-Y., Pai, D. C., & Hung, C.-H. (2011). Knowledge absorptive capacity and innovation performance in KIBS. *Journal of Knowledge Management*, *15*(6), 971–983. https://doi.org/10.1108/13673271111179316

Vafaei-Zadeh, A., Hanifah, H., Foroughi, B., & Salamzadeh, Y. (2019). Knowledge leakage, an Achilles' heel of knowledge sharing. *Eurasian Business Review*, *9*(4), 445–461. https://doi.org/10.1007/S40821-019-00128-7

Van Mierlo, H., Vermunt, J. K., & Rutte, C. G. (2009). Composing group-level constructs from individual-level survey data. *Organizational Research Methods*, *12*(2), 368–392.

Van Wijk, R., Jansen, J. J. P., & Lyles, M. A. (2008). Inter- and intra-organizational knowledge transfer: A meta-analytic review and assessment of its antecedents and consequences. *Journal of Management Studies*, *45*(4), 830–853. https://doi.org/10.1111/j.1467-6486.2008.00771.x

Vishwanath, A. (2016). *Mobile device affordance: Explicating how smartphones influence the outcome of phishing attacks*. https://doi.org/10.1016/j.chb.2016.05.035

Von Solms, R. (1999). Information security management: why standards are important. *Information Management & Computer Security*.

Wakefield, R. L., & Whitten, D. (2006). Mobile computing: a user study on hedonic/utilitarian mobile device usage. *European Journal of Information Systems*, *15*(3), 292–300. https://doi.org/10.1057/palgrave.ejis.3000619

Watson, R. T. (2015). Beyond being systematic in literature reviews in IS. *Journal of Information Technology*, *30*(2), 185–187.

Webster, J., & Watson, R. (2002). Analyzing the Past To Prepare for the Future : Writing a Literature Review. *MIS Quarterly*, *26*(2), 13–23.

Weichbroth, P., & Łysik, Ł. (2020). Mobile Security: Threats and Best Practices.

*Mobile Information Systems*, 2020.

Wernerfelt, B. (1984). A Resource-Based View of the Firm. *Strategic Management Journal*, *5*(2), 171–180. https://doi.org/10.1007/sl0734-011-9485-0

Whitman, Michael and Mattord, H. (2011). *Principles of information security*. Cengage Learning.

Wong, W. P., Tan, K. H., Govindan, K., Li, D., & Kumar, A. (2021). A conceptual framework for information-leakage-resilience. *Annals of Operations Research*. https://doi.org/10.1007/s10479-021-04219-5

Wu, H., Han, Z., & Zhou, Y. (2021). Optimal degree of openness in open innovation: A perspective from knowledge acquisition & knowledge leakage. *Technology in Society*, *67*. https://doi.org/10.1016/j.techsoc.2021.101756

Wynn, D., & Williams, C. K. (2012). Principles for Conducting Critical Realist Case Study Research in Information Systems. *MIS Quarterly*, *36*(3), 787–810. https://doi.org/10.1016/j.ijproman.2012.11.012

Xu, W. (2021). The contingency of neighbourhood diversity: Variation of social context using mobile phone application data. *Urban Studies*, 00420980211019637.

Yahav, I., Schwartz, D. G., & Silverman, G. (2014). Detecting unintentional information leakage in social media news comments. In *2014 IEEE 15th International Conference on Information Reuse and Integration (IRI), Presented at the 2014 IEEE 15th International Conference on Information Reuse and Integration (IRI)* (Issue August, pp. 74–79). https://doi.org/10.1109/IRI.2014.7051874

Yarhi-Milo, K. (2014). Knowing the adversary. In *Knowing the Adversary*. Princeton University Press.

Yin, R. K. (2003). *Case Study Research*. SAGE Publications.

Yin, R. K. (2011). *Applications of case study research*. sage.

Yin, R. K. (2015). *Qualitative research from start to finish*. Guilford publications.

Yin, R. K. (2017). Case study research and applications: Design and methods. Sixth Edition. In *SAGE*.

Yu, X., Tian, Z., Qiu, J., & Jiang, F. (2018). A data leakage prevention method based on the reduction of confidential and context terms for smart mobile devices. *Wireless Communications and Mobile Computing*, *2018*.

Zack, M. H. (1999). Managing Codified Knowledge. (cover story). *Sloan Management Review*, *40*(4), 45–58. https://doi.org/Article

Zahadat, N., Blessner, P., Blackburn, T., & Olson, B. a. (2015). BYOD security engineering: a framework & its analysis. *Computers & Security*. https://doi.org/10.1016/j.cose.2015.06.011

Zaheer, Z., Chang, H., Mukherjee, S., & Van der Merwe, J. (2019). eztrust: Network-independent zero-trust perimeterization for microservices. *Proceedings of the 2019 ACM Symposium on SDN Research*, 49–61.

Zahoor, N., & Al-Tabbaa, O. (2020). Inter-organizational collaboration and SMEs' innovation: A systematic review and future research directions. *Scandinavian Journal of Management*, *36*(2), 101109.

Zaini, M. K., Masrek, M. N., Johari, M. K., Sani, A., & Anwar, N. (2018). Theoretical modeling of information security: organizational agility model based on integrated system theory and resource based view. *International Journal of Academic Research in Progressive Education and Development*, *7*(3).

Zeiringer, J. P., & Thalmann, S. (2022). Knowledge sharing and protection in data-centric collaborations: An exploratory study. *Knowledge Management Research & Practice*, *20*(3), 436–448.

Zhang, D., Li, S., & Zheng, D. (2017). Knowledge search and open innovation performance in an emerging market: Moderating effects of government-enterprise relationship and market focus. *Management Decision*, *55*(4), 634–647. https://doi.org/https://doi.org/10.1108/ MD-04-2016-0211

Zhang, F., Zhou, S., Qin, Z., & Liu, J. (2003). Honeypot: A Supplemented Active Defense System for Network Security. *Parallel and Distributed Computing, Applications and Technologies, PDCAT Proceedings*, 231–235. https://doi.org/10.1109/pdcat.2003.1236295

Zhao, M. (2006). Conducting R&D in countries with weak intellectual property rights protection. *Management Science*, *52*(8), 1185–1199.

Zmud, B. (1998). " Pure" theory manuscripts. *MIS Quarterly*, *22*(2), R29.

# APPENDICES

# Appendix A – Papers and reports analyzed for Literature Review

| N. | Authors | Title | Unit of Analysis | Research Method | Theoretical Lens / Framework | Literature Stream |
|---|---|---|---|---|---|---|
| 1 | Abdoul Aziz Diallo, B. et al. (2012) | Mobile and Context-Aware GeoBI Applications: A Multilevel Model for Structuring and Sharing of Contextual Information | Organization | Mixed | Mobile Contexts | Mobile Computing |
| 2 | Ahmad et al (2014) | Protecting organizational competitive advantage: A knowledge leakage perspective | Organization | Qualitative / Interviews / Field study | VRIN | Knowledge Management / Information Security Management |
| 3 | Ahmad et al. (2015) | Guarding Against the Erosion of Competitive Advantage: A Knowledge Leakage Mitigation Model | Organization | Qualitative | Resource-based View (RBV) | Information Security Management |
| 4 | Ahmad et al. (2012) | Information security strategies: towards an organizational multi-strategy perspective | Organization | Qualitative / Focus Groups | Conceptual | Information Security Management |
| 5 | Alavi and Leidner (2001) | Review: Knowledge Management and Knowledge Management | Organization / Individual | Systematic Review | Knowledge-based View (KBV) | Knowledge Management |

| N. | Authors | Title | Unit of Analysis | Research Method | Theoretical Lens / Framework | Literature Stream |
|---|---|---|---|---|---|---|
| | | Systems: Conceptual Foundations and Research Issues | | | | |
| 6 | Allam et al. (2014) | Review: Knowledge Management and Knowledge Management Systems: Conceptual Foundations and Research Issues | Individual | Mixed | The Awareness Boundary Model | Mobile Computing |
| 7 | Amoroso and Link (2021) | Intellectual property protection mechanisms and the characteristics of founding teams | Groups | Quantitative | Legal Framework | Knowledge Management |
| 8 | Ancori et al. (2000) | The economics of knowledge: The debate about codification and tacit knowledge | Individual | Mixed | Economic Knowledge Theory, Shannon's Theory, Game Theory | Knowledge Management |
| 9 | Anderson (2001) | Why information security is hard- an economic perspective | Organizations / Individual | Conceptual | Microeconomics | Information Security Management |
| 10 | Annasingh (2012) | Exploring the Risks of Knowledge Leakage : An Information Systems Case Study Approach | Organization | Mixed | Practice Based IS Research (PBISR) Framework | Knowledge Management |
| 11 | Annasingh (2006) | Exploring knowledge leakage risks exposure resulting from 3D modellling in organizations: a case study | Organization | Qualitative/ Case Study | Practice Based IS Research (PBISR) Framework | Knowledge Management |
| 12 | Arias-Perez et. Al (2020) | When it comes to the impact of absorptive capacity on co-innovation, how really harmful is knowledge leakage? | Organization | Quantitative | Absorptive capacity and co-innovation Framework (Cohen and Levinthal (1990)) | Knowledge Management |
| 13 | Armando et al (2014) | Formal modeling and automatic enforcement of Bring Your Own Device policies | Organization | Quantitative | Programming Framework (Mathematical Proof) | Mobile Computing / Information Security Management |
| 14 | Arohan et al. (2020) | An introduction to context-aware | Organization / Individual | Quantitative | Markov Model | Mobile Computing / |

| N. | Authors | Title | Unit of Analysis | Research Method | Theoretical Lens / Framework | Literature Stream |
|---|---|---|---|---|---|---|
| | | security and User Entity Behavior Analytics | | | | Information Security Management |
| 15 | Astani et al. (2013) | BYOD issues and strategies in organizations | Organization | Qualitative | Conceptual | Mobile Computing / Information Security Management |
| 16 | Barney (1991) | Firm resources and sustained competitive advantage | Organization | Qualitative | Resource-based View (RBV) | Knowledge Management |
| 17 | Barney (1986) | Organizational culture: can it be a source of sustained competitive advantage? | Organization | Qualitative | Organizational Theory; Strategic Capability framework | Knowledge Management |
| 18 | Baskerville et al. (2012) | Incident-centered information security: Managing a strategic balance between prevention and response | Organization | Qualitative | Incident-centered security framework; Prevention/Respons e paradigm | Information Security Management |
| 19 | Baskerville (2005) | Information Warfare: A Comparative Framework for Business Information Security | Organization | Qualitative | Probability/Possibili ty Theory; Business Information Security Theory/ Information Warfare Theory | Information Security Management |
| 20 | Baskerville and Portougal (2003)) | A possibility theory framework for security evaluation in national infrastructure protection | Organization / Individual | Quantitative | Possibility Theory | Information Security Management |
| 21 | Becerra et al. (2008) | Trustworthiness, Risk, and the Transfer of Tacit and Explicit Knowledge Between Alliance Partners | Organization / Group | Quantitative | Organizational Trust | Knowledge Management |
| 22 | Belsis et al.(2005) | Information systems security from a knowledge management perspective | Organization | Qualitative | SECI Framework | Information Security / Knowledge Management |
| 23 | Benitez et al.(2012) | Context-Aware Mobile Collaborative Systems: | Organization | Qualitative/ Case Study | Activity Theory; Mobile Contexts Framework | Mobile Computing |

| N. | Authors | Title | Unit of Analysis | Research Method | Theoretical Lens / Framework | Literature Stream |
|---|---|---|---|---|---|---|
| | | Conceptual Modeling and Case Study | | | | |
| 24 | Bishop et al.(2010) | A risk management approach to the "insider threat" | Organization / Individual | Qualitative | Conceptual | Information Security Management |
| 25 | Blackler (1995) | Knowledge, knowledge work and organizations: An overview and interpretation | Organization | Qualitative | Activity Theory; Organizational Theory; Org. Competencies; Nonaka's knowledge creating framework | Knowledge Management |
| 26 | Bloodgood and Chen (2021) | Preventing organizational knowledge leakage: the influence of knowledge seekers' awareness, motivation and capability | Organization | Quantitative | Awareness, motivation, capability framework | Knowledge Management |
| 27 | Boisot and Canals (2004) | Data, information and knowledge: have we got it right? | Individual | Quantitative | Information Theory | Knowledge Management |
| 30 | Bolisani et al.(2013) | Knowledge protection in knowledge-intensive business services | Organization | Quantitative | Conceptual | Knowledge Management |
| 31 | Bollinger and Smith (2001) | Managing organizational knowledge as a strategic asset | Organization | Qualitative | Organizational learning | Knowledge Management |
| 32 | Bosua and Scheepers (2007) | Towards a model to explain knowledge sharing in complex organizational environments | Organization | Qualitative | Distributed Cognition Theory; Actor Network Theory | Knowledge Management |
| 33 | Bouncken et al. (2015) | Coopetition: a systematic review, synthesis, and future research directions | Organization / Group | Systematic Review | Resource-based View (RBV); Dynamic Capabilities; Game Theory | Knowledge Management |
| 34 | Cohen and Levinthal (1990) | Absorptive Capacity: A new perspective on learning and innovation | Organization | Quantitative / Survey | Absorptive Capacity | Knowledge Management |
| 35 | Collier and Sarkis (2021) | The zero trust supply chain: Managing supply chain risk in the absence of trust | Organization / Individual | Qualitative | Zero Trust Tenets; NIST 800-207 | Information Security Management |

| N. | Authors | Title | Unit of Analysis | Research Method | Theoretical Lens / Framework | Literature Stream |
|---|---|---|---|---|---|---|
| 36 | Colwill (2009) | Human factors in information security: The insider threat - Who can you trust these days? | Individual | Qualitative | Practitioner view | Information Security Management |
| 37 | Conner (1991) | A historical comparison of resource-based theory and five schools of thought within industrial organization economics: do we have a new theory of the firm? | Organization | Qualitative | Resource-based View (RBV); Industrial Organization Theory | Knowledge Management |
| 38 | Crager (2007) | Information Leakage through Mobile Motion Sensors : User Awareness and Concerns | Individual / Device | Mixed | User Awareness, Perception and Expectations framework | Mobile Computing / Information Security Management |
| 39 | Crossler (2014) | Understanding compliance with bring your own device policies utilizing protection motivation theory: Bridging the intention-behaviour gap | Individual | Quantitative | Protection Motivation Theory; General Deterrence Theory | Mobile Computing/ Information Security Management |
| 40 | Da Pedraza-Farina. (2017) | Spill Your ( Trade ) Secrets : Knowledge Networks as Innovation Drivers | Individual/ Organization | Qualitative | Intellectual Property Rights | Knowledge Management |
| 41 | Dang-Pham and Pittayachawan | Comparing intention to avoid malware across contexts in a BYOD-enabled Australian university: A Protection Motivation Theory approach | Individual | Quantitative | Protection Motivation Theory | Mobile computing/ Information Security Management |
| 42 | D'arcy et al. (2007) | User awareness of security countermeasures and its impact on information systems misuse: a deterrence approach | Individual | Quantitative | General Deterrence Theory | Mobile computing/ Information Security Management |

| N. | Authors | Title | Unit of Analysis | Research Method | Theoretical Lens / Framework | Literature Stream |
|---|---|---|---|---|---|---|
| 43 | Davenport and Prusack (1998) | Working knowledge: how organizations manage what they know | Organization | N/A | Conceptual | Knowledge Management |
| 44 | Dedeche et al. (2013) | Emergent BYOD security challenges and mitigation strategy | Organization | N/A | Information Security Strategy | Information Security Management |
| 45 | Denning | Framework and principles for active cyber defense | Organization | Conceptual | Active and Passive Defence Framework | Information Security Management |
| 46 | Derballa and Poutsttchi (2006) | Mobile Knowledge Management | Individual / Organization | N/A | Knowledge Based View (KBV) | Knowledge Management |
| 47 | Desouza and Evaristo (2004) | Managing knowledge in distributed projects | Individual | N/A | Conceptual | Knowledge Management |
| 48 | Desouza (2009) | Information and Knowledge Management in Public Sector Networks: The Case of the US Intelligence Community | Organization | Qualitative | Semiotics Theory; Desouza Knowledge Management Framework | Knowledge Management |
| 49 | Desouza (2006) | Knowledge Security : An Interesting Research Space | Individual / Organization | N/A | Conceptual | Knowledge Management / Information Security Management |
| 50 | Desouza (2005) | Securing knowledge in organizations: Lessons from the defense and intelligence sectors | Organization | Qualitative / Case study | Conceptual | Knowledge Management |
| 51 | Desouza and Awazu (2006) | Knowledge management at SMEs: Five peculiarities | Organization | Qualitative | SECI model; Semiotic Framework | Knowledge Management |
| 52 | Diallo et al. (2013) | Context-based mobile GeoBI: enhancing business analysis with contextual metrics/statistics and context-based reasoning | Individual | Quantitative | Mobile Contexts | Mobile Computing |
| 53 | Diallo et al. (2011) | Towards Context-Awareness Mobile Geospatial Bi | Individual | Quantitative | Mobile Contexts | Mobile Computing |

| N. | Authors | Title | Unit of Analysis | Research Method | Theoretical Lens / Framework | Literature Stream |
|---|---|---|---|---|---|---|
| | | (Geobi) Applications | | | | |
| 54 | Durst et al. (2015) | Understanding knowledge leakage: a review of previous studies | Individual / Organization | Review | Literature Review | Knowledge Management |
| 55 | Durst et al. (2018) | Knowledge risk management in organizations | Organization | Quantitative | Conceptual | Knowledge Management |
| 56 | Amara et al. (2008) | Managing the protection of innovations in knowledge-intensive business services | Organization | Quantitative / Survey | Legal Framework | Knowledge Management |
| 57 | Emery (2012) | Factors for consideration when developing a bring your own device (BYOD) strategy in higher education | Organization | Qualitative | Technology Acceptance Model (TAM). | Mobile Computing / Information Security Management |
| 58 | Figuereido et al (2020) | Spinner model: prediction of propensity to innovate based on knowledge-intensive business services | Organization | Quantitative | Economic Innovation Theory | Knowledge Management |
| 59 | Ford (2004) | Trust and knowledge management: the seeds of success | Individual | Qualitative | Trust Theory | Knowledge Management |
| 60 | Frishammar et al. (2015) | The dark side of knowledge transfer: Exploring knowledge leakage in joint R&D projects | Organization | Qualitative/ Multiple Case Study | Knowledge Based View (KBV) | Knowledge Management |
| 61 | Galati et al. (2019) | A framework for avoiding knowledge leakage: evidence from engineering to order firms | Organization | Qualitative | Resource based View (RBV); Knowledge Based View (KBV) | Knowledge Management |
| 62 | Gonzalez-Diaz et al. (2021) | Business counterintelligence as a protection strategy for SMEs | Organization | Mixed | Military Strategy | Knowledge Management |
| 63 | Gordon (2007) | Data leakage Threats and mitigations | Organization | Qualitative | Conceptual | Information Security Management |
| 64 | Grant (1996) | Toward a Knowledge-Based Theory of the Firm | Organization | N/A | Resource Based View (RBV); Knowledge Based View (KBV) | Knowledge Management |

| N. | Authors | Title | Unit of Analysis | Research Method | Theoretical Lens / Framework | Literature Stream |
|---|---|---|---|---|---|---|
| 65 | Grimaldi (2021) | A framework of intellectual property protection strategies and open innovation | Organization | Quantitative | Open Innovation Paradigm | Knowledge Management |
| 66 | Guo et al. (2020) | Knowledge sharing and knowledge protection in strategic alliances: the effects of trust and formal contracts | Organization | Quantitative | Transaction Cost Theory; Relational Exchange Theory | Knowledge Management |
| 67 | Hagerdoorn et al. (2005) | Intellectual property rights and the governance of international R&D partnerships | Organization | Quantitative | Legal Framework | Knowledge Management |
| 68 | Herath and Rao (2009) | Protection motivation and deterrence: a framework for security policy compliance in organisations | Individual | Quantitative | Protection Motivation Theory; General Deterrence Theory | Information Security Management |
| 69 | Hofer et al. (2003) | Context-awareness on mobile devices - the hydrogen approach | Individual / Device | Quantitative | Mobile Context | Mobile Computing / Information Security Management |
| 70 | Chen and Nath (2008) | A socio-technical perspective of mobile work | Individual / Device | Quantitative | Mobile Context | Mobile Computing / Knowledge Management |
| 71 | Inkpen et al. (2019) | Unintentional, unavoidable, and beneficial knowledge leakage from the multinational enterprise | Organization / Individual | N/A | Conceptual | Knowledge Management |
| 72 | Janssen and Spruit (2019) | M-RAM: a Mobile Risk Assessment Method for Enterprise Mobile Security | Individual / Device | Qualitative / Case Study | People Process Technology Model | Mobile Computing/ Information Security Management |
| 73 | Carlsson (2003) | Knowledge Managing and Knowledge Management Systems in Inter-organizational Networks | Organization | N/A | Knowledge based view (KBV); Resource based View (RBV); Dynamic Capability; Absorptive capability | Knowledge Management |
| 74 | Shedden et al. (2011) | Incorporating a knowledge | Organization | Qualitative/Case Study | OCTAVE-S methodology | Knowledge Management |

| N. | Authors | Title | Unit of Analysis | Research Method | Theoretical Lens / Framework | Literature Stream |
|---|---|---|---|---|---|---|
| | | perspective into security risk assessments | | | | / Information Security Management |
| 75 | Dey and Abowd (1999) | Towards a Better Understanding of Context and Context-Awareness | Individual / Device | N/A | Mobile Context | Mobile Computing |
| 76 | Schilit et al. (1994) | Context-aware computing applications | Individual / Device | N/A | Mobile Context | Mobile Computing |
| 77 | Siponen (2000) | A conceptual foundation for organizational information security awareness | Individual | N/A | Theory of Reasoned Action; Theory of Planned Behaviour; Intrinsic Motivation; The Technology Acceptance Model | Information Security Management |
| 78 | Homoliak et al. (2019) | Insight into insiders and it: A survey of insider threat taxonomies, analysis, modeling, and countermeasures | Individual | Literature Review | Grounded Theory | Information Security Management |
| 79 | Hurmelinna-Laukkanen | Enabling collaborative innovation – knowledge protection for knowledge sharing | Organization | Quantitative | Knowledge Appropriability regime | Knowledge Management |
| 80 | Jarrahi and Thomson (2017) | The interplay between information practices and information context: The case of mobile knowledge workers | Individual | Qualitative / Interviews | Practice Centric Approach | Knowledge Management / Mobile Computing |
| 81 | Jiang et al. (2016) | Partner trustworthiness, knowledge flow in strategic alliances, and firm competitiveness: A contingency perspective | Organization | Quantitative | Knowledge Appropriability regime | Knowledge Management |
| 82 | Jiang et al. (2013) | Managing knowledge leakage in strategic alliances: The effects of | Organization | Quantitative | Trust; Formal Contracts | Knowledge Management |

| N. | Authors | Title | Unit of Analysis | Research Method | Theoretical Lens / Framework | Literature Stream |
|----|---------|-------|------------------|-----------------|------------------------------|-------------------|
| | | trust and formal contracts | | | | |
| 83 | Kaplan et al. (2001) | Knowledge-based theories of the firm in strategic management: A review and extension | Organization | N/A | Knowledge based View (KBV) | Knowledge Management |
| 84 | Kaplinsky et al. (2006) | Towards a Taxonomy of Knowledge Leakage: Literature and Framework | Organization | Qualitative | Knowledge as a capability | Knowledge Management |
| 85 | Kengatharan (2019) | A knowledge-based theory of the firm: Nexus of intellectual capital, productivity and firms' performance | Organization | Qualitative | Knowledge based View (KBV) | Knowledge Management |
| 86 | Kindervag (2010) | Build security into your network's dna: The zero trust network architecture | Organization | N/A | Conceptual | Information Security Management |
| 87 | Lascaux (2020) | Coopetition and trust: What we know, where to go next | Organization | Literature Review | Trust | Knowledge Management |
| 88 | Leonard-Barton (1992) | Core Capabilities and Core Rigidities - a Paradox in Managing New Product Development | Organization | N/A | Knowledge based view (KBV); Core Capabilities | Knowledge Management |
| 89 | Liebeskind (1996) | Knowledge , Strategy , and the Theory of the Firm | Organization | N/A | Transaction-cost theory | Knowledge Management |
| 90 | MacDougall and Durst (2005) | Identifying tangible costs, benefits and risks of an investment in intellectual capital: Contracting contingent knowledge workers | Individual | Qualitative | Porter's five-forces model | Knowledge Management |
| 91 | Manhart and Thalmann (2015) | Protecting organizational knowledge: a structured literature review | Organizations | Literature Review | Organizational, Legal and Technical framework | Knowledge Management |

| N. | Authors | Title | Unit of Analysis | Research Method | Theoretical Lens / Framework | Literature Stream |
|---|---|---|---|---|---|---|
| 92 | Melville et al. (2004) | Review: Information Technology and Organizational Performance: an Integrative Model of It Business Value | Organization | Literature Review | Resource based view (RBV) | Knowledge Management |
| 93 | Mohamed et al. (2006) | Understanding one aspect of the knowledge leakage concept: people | Individual | Qualitative/ Interview | Thematic Analysis | Knowledge Management |
| 94 | Mohamed et al. (2007) | Unearthing key drivers of knowledge leakage | Organization | Qualitative/ Interview | Thematic Analysis | |
| 95 | Mupepi et al. (2017) | How Can Knowledge Leakage be Stopped: A Socio-Technical System Design Approach to Risk Management | Organization | Qualitative | Risk Management Framework | Knowledge Management |
| 96 | Nelson et al. (2017) | Mobility of knowledge work and affordances of digital technologies | Individual | Qualitative | Affordance Perspective (Orlikowski) | Mobile Computing/ Knowledge Management |
| 97 | Nonaka (1994) | A Dynamic Theory Knowledge of Organizational Creation | Individual / Group / Organization | N/A | SECI model | Knowledge management |
| 98 | Nonaka (1991) | The knowledge-creating company | Individual / Group / Organization | N/A | SECI model | Knowledge management |
| 99 | Nonaka (1995) | The knowledge-creating company: How Japanese companies create the dynamics of innovation | Individual / Group / Organization | N/A | SECI model | Knowledge management |
| 100 | Nunes et al. (2006) | Knowledge management issues in knowledge-intensive SMEs | Organization | Qualitative / Interview | Thematic Analysis | Knowledge Management |
| 101 | Orlander and Hurmelinna-Laukkanen (2011) | Do SMEs benefit from HRM-related knowledge protection in innovation management? | Organization | Qualitative / Interviews | Human-Resource Management | Knowledge Management |
| 102 | Ilvonen and Thalmann | Reconciling digital | Organization | Literature Review | Thematic Analysis | Knowledge Management |

| N. | Authors | Title | Unit of Analysis | Research Method | Theoretical Lens / Framework | Literature Stream |
|---|---|---|---|---|---|---|
| | | transformation and knowledge protection: a research agenda | | | | |
| 103 | Oorschot et al. (2018) | The knowledge protection paradox: imitation and innovation through knowledge sharing | Organization | Quantitative | Theory through Simulation | Knowledge Management |
| 104 | Paallysaho and Kuusisto (2011) | Informal ways to protect intellectual property (IP) in KIBS businesses | Organization | Qualitative | Conceptual | Knowledge Management |
| 105 | Polanyi (2009) | The Tacit Dimension | Individual | N/A | Personal knowledge; embodied knowledge | Knowledge Management |
| 106 | Ritala et al. (2014) | Knowledge sharing, knowledge leaking and relative innovation performance: An empirical study | Organization | Quantitative / Survey | Knowledge Based view (KBV); Resource based View (RBV) | Knowledge Management |
| 107 | Sampaio et al. (2019) | The use of gamification in knowledge management processes: a systematic literature review | Individual / Organization | Literature Review | Thematic Analysis | Knowledge Management |
| 108 | Shabtai et al. (2012) | A taxonomy of data leakage prevention solutions | Organization | Literature Review | Content Analysis | Knowledge Management |
| 109 | Shedden et al. (2010) | Information Security Risk Assessment : Towards a Business Practice Perspective | Organization | Qualitative / Case Study | OCTAVE-S methodology | Information Security Management |
| 110 | Sorensen et al. (2008) | Exploring enterprise mobility: Lessons from the field | Individual /Organization | Qualitative / Interview / Observation | Thematic Analysis | Mobile Computing / Knowledge Management |
| 111 | Spender and Grant (1996) | Knowledge and the firm: overview | Organization | Qualitative | Knowledge Based View (KBV) | Knowledge Management |
| 112 | Teece (2007) | Explicating Dynamic Capabilities: The Nature and Microfoundations of (Sustainabile) | Organization | Literature Review | Dynamic Capabilities | Knowledge Management |

| N. | Authors | Title | Unit of Analysis | Research Method | Theoretical Lens / Framework | Literature Stream |
|---|---|---|---|---|---|---|
| | | Enterprise Performance | | | | |
| 113 | Thalmann et al. (2018) | Balancing Knowledge Protection and Sharing to Create Digital Innovations | Individual Organization | Qualitative | Knowledge Based view (KBV) | Knowledge Management |
| 114 | Trkman and Desouza (2012) | Knowledge risks in organizational networks: An exploratory framework | Organization | Review | Knowledge based view (KBV); Transaction Cost Theory | Knowledge Management |
| 115 | Tsang et al. (2016) | AHP-Driven Knowledge Leakage Risk Assessment Model: A Construct-Apply-Control Cycle Approach | Organization | Quantitative | Analytic Hierarchy Process | Knowledge Management |
| 116 | Vafaei-Zadeh et al. (2019) | Knowledge leakage, an Achilles' heel of knowledge sharing | Organization | Quantitative | Knowledge Based View (KBV) | Knowledge Management |
| 117 | Weichbroth and Lysik (2020) | Mobile Security: Threats and Best Practices | Organization | Quantitative / Survey | Conceptual | Mobile Computing/ Information Security Management |
| 118 | Wu and Zhou (2021) | Optimal degree of openness in open innovation: A perspective from knowledge acquisition & knowledge leakage | Organization | Quantitative | Conceptual | Mobile Computing |
| 119 | Yu et al. (2018) | A data leakage prevention method based on the reduction of confidential and context terms for smart mobile devices | Individual / Device | Quantitative | Set Theory | Mobile Computing |
| 120 | Zahoor and Al-Tabbaa | Inter-organizational collaboration and SMEs' innovation: A systematic review and future research directions | Organization | Literature Review | Thematic Analysis | Knowledge Management |
| 121 | Zaini et al. (2018) | Theoretical modeling of information | Organization | Literature Review | Integrated System Theory; Resource Based View (RBV) | Knowledge Management |

| N. | Authors | Title | Unit of Analysis | Research Method | Theoretical Lens / Framework | Literature Stream |
|---|---|---|---|---|---|---|
| | | security: organizational agility model based on integrated system theory and resource based view | | | | |

# Appendix B − Examples of Scenarios used in the interviews and discussions

## Scenario 1:

John is in the workplace, finishing off a large tender proposal. Work is unusually hectic on a Thursday afternoon as the deadline to submit the tender is Friday. The following events transpire:

A) John is at work, using a laptop connected to the company Wi-Fi to finish the report

B) When work ends, John takes his laptop home. The train ride home was very crowded John was able to get a seat and review documentation on his tablet

C) During the train ride, John receives an urgent call from his boss, asking John to finish the report as soon as possible.

D) John takes out his laptop and starts making changes to the report requested by his boss before he forgets. The passenger next to John moves over to give him extra room to use his laptop and tablet.

E) After arriving home, John makes the final changes to the tender proposal and emails it to his boss.

F) To relieve stress from the week, John goes to his local bar. As the bar is underground with poor network reception, John's phone automatically connects to the bar's Wi-Fi

G) John's boss responds to his email by resending the tender proposal document with added comments for changes. After reading the comments, John finishes his drink and heads home to continue working.

**1. Do any of John's actions relate to you and your working experiences/habits? Please indicate below by circling the corresponding letters**

A       B       C       D       E       F       G

**2. Did John break any of your company's policies in this scenario? If yes, please indicate below by circling the corresponding letters:**

A     B     C     D     E     F     G

**3. Do you think any of John's working habits present a security risk? (If yes, please indicate below by circling the corresponding letters)**

A     B     C     D     E     F     G

## Scenario 2:

Michael, a senior manager, is in an airport waiting for his delayed flight to a business conference. The conference will be attended by many senior professionals within the industry. The following events transpire in the day:

    A) While waiting in the crowded airport lounge, Michael connects his Wi-Fi only tablet to the airport public Wi-Fi. There were multiple public Wi-Fi so Michael chose the one with the strongest signal

    B) Michael receives an email from John regarding a tender proposal and downloads it onto his tablet to review

    C) The airplane has finally arrived and Michael proceeds to the waiting area which had a different which he used while waiting to board.

    D) After finding his seat in the middle of the aisle, Michael noticed that there was in-flight Wi-Fi access

    E) After reviewing and commenting on the tender proposal, Michael sends the revision to John via corporate email

    F) Michael checks into his hotel room and reads tomorrow's conference agenda on hotel Wi-Fi

**1. Do any of Michael's actions relate to you and your working experiences/habits? (Please indicate below by circling the corresponding letters)**

A     B     C     D     E     F

**2. Did Michael's actions break any of your company's policies in this scenario? (If yes, please indicate which actions below by circling the corresponding letters)**

A     B     C     D     E     F

**3. Do you think any of Michael's working habits present a security risk? (If yes, please indicate below by circling the corresponding letters)**

A     B     C     D     E     F

## Scenario 3:

Brigit is working casually for a small company, while completing her Masters degree part-time, and was recently allocated a tablet device. On a typical day,

A) Brigit uses her tablet to take pictures of whiteboards and presentation slides during meetings, as well as audio recordings

B) Uses her tablet to access personal email accounts and browse her favourite websites, such as Facebook, during breaks on the organisation's Wi-Fi

C) Download applications onto her tablet for work as well as entertainment and social networking

D) Review reports and documentation

E) Access the wiki knowledge management portal to upload documentation and update project progress information

F) Brigit also takes the tablet to University to write notes during lectures, complete assignments and study for tests and exams

**1. Do any of Brigit's actions relate to your working habits? Please indicate below by circling the corresponding letters**

A    B    C    D    E    F

**2. Did Brigit break any of your company's policies in this scenario? If yes, please indicate which of her actions by circling the corresponding letters:**

A    B    C    D    E    F

# Appendix C –Example of a Mobile Contexts Table (Transitions)

| User Context | Reading, Writing, Conversations |
|---|---|
| Device Context | Desktop, Tablet, Mobile Phone |
| Technological Context | Corporate network, public network, home network |
| Social Context | Co-workers, strangers in public, family, friends |
| Environmental Context | Workplace, public (train), private (home), public (pub) |

# Scenario 1

| Scenario 1 | User context (what work they are doing) | Device context | Technological context | Social context | Environmental context (location) |
|---|---|---|---|---|---|
| A | *Reading, Writing, Conversations related to confidential document* | Using a desktop | Desktop on corporate network | Co-workers (project team) | Workplace |
| A-> B | *Group work -> individual work* | *Desktop -> tablet* | *Corporate Net -> Cellular Net?* | *Co-workers -> strangers (close enough to read tablet)* | *Private (workplace) -> public (train)* |
| B | *Reading eBooks for leisure, not working* | tablet | *Tablet not connected to any network* | Strangers (close enough to read tablet) | Public (train) |
| B->C | *Leisure -> Individual work Conversation* | *Tablet -> Mobile phone* | *Cellular network* | *Strangers (in earshot)* | - |
| C | Conversation regarding sensitive information | Mobile phone, personally owned | Cellular network | Strangers (in earshot) | Public (train) |
| C->D | *Individual work conversation -> Individual work* | *Mobile phone -> tablet with hotspot connectivity* | *Cellular network* | *Strangers (close enough to read tablet)* | - |
| D | Working on confidential document | Tablet using his phone as a hotspot | Cellular network through hotspot | Strangers | Public (train) |
| D->E | - | *Tablet -> laptop* | *Cellular network -> Home wifi* | *Strangers -> Family* | *Public (train) -> Private (home)* |
| E | Working on confidential document | laptop | Home wifi | Family | Home |
| E->F | - | *Laptop-> tablet* | *Home wifi -> public wifi* | *Family -> friends* | *Private (home) -> Public (pub)* |

| F | Reading a confidential document | tablet | Public wifi | Acquaintances and friends | Public (pub, socialising environment) |

# Scenario 2

| Scenario 2 | User context (what work they are doing) | Device context | Technological context | Social context | Environmental context (location) |
|---|---|---|---|---|---|
| A | *Reading work emails* | Tablet (wifi only) | Public airport wifi (one of many) | Strangers (many moving around him in the lounge) | Airport lounge |
| *A-> B* | *Individual work to individual phone conversation* | *Tablet to mobile* | *Wi-fi Cellular network* | *Busy lounge -> waiting area for boarding* | *Lounge -> Airport waiting area for boarding* |
| B | Conversation regarding sensitive information | Mobile phone | *Cellular* | Strangers (earshot) | *Airport waiting area for boarding* |
| *B->C* | *Individual conversation -> Individual work* | *Mobile to tablet* | *Cellular to wifi* | *Strangers sitting next to him on plane* | *Airport to Airplane* |
| C | Reading and writing regarding confidential proposal | Tablet | Airplane wifi | *Strangers sitting next to him on plane* | Public (Airplane) |
| *C->D* | *Individual work -> working remotely with co-worker (John)* | *-* | *Airplane wifi to hotel wifi* | *Alone* | *Private (hotel)* |
| D | Working on confidential proposal with John remotely | Tablet | Hotel wifi | Alone but in contact with John | Private (hotel) |
| *D->E* | *Working on proposal -> presenting work at strategic meeting* | *-* | *Hotel to corporate connection (implied but not relevant)* | *Alone -> coworkers at headquarters* | *Private to Workplace* |
| E | Presenting confidential proposal during a strategic meeting | Tablet | Corporate | Coworkers | Workplace |

# Appendix D - Interview Protocol

## Background questions:

1. What is your current position at your organization, years of employment, experience, academic training, and industry?

2. In your organization, is knowledge sharing critical for the work of your employees?

## Opening questions

1. What is your general perception of knowledge leakage issues?

2. Have you experienced knowledge leakage issues in you organization?

3. How often does the topic of knowledge leakage arise? In what situations does it happen?

## Scenario Questions:

Considering the scenarios provided to you, please answer the following questions:

1. Have any of the scenarios provided to you resembled a situation in your organization?

2. Do you have policies in place that address this behaviour in your organization? Please explain.

Given the distinction between knowledge and information [Definitions and distinctions given to participants in advanced]:

1. What current mechanisms do you use to protect *information*?

2. What current mechanisms do you use to protect *Knowledge*?

3. What are the main knowledge assets you organization need to protect?

4. Do you have a risk management procedure or strategy in your organization? If yes, how is knowledge managed?

5. What knowledges processes does your organization have in place? (Examples: Capture, retention, transfer, storage)

6. In your opinion, what knowledge is the most critical and therefore should be protected? (Example: Intellectual Property, process knowledge, marketing strategies, client knowledge, etc.)

7. Do you think your knowledge assets are at risk of being appropriated by partners/competitors/clients? Please explain.

8. What would happen if these knowledge assets leaked? What's the potential impact? How likely is it to happen?

9. Based on your experience, how does knowledge leakage occur?

10. How do you discourage/encourage knowledge sharing in your organization?

11. How do you currently share sensitive knowledge among employees, partners and clients?

12. What risk assessments do you undertake when sharing knowledge among employees, partners and clients?

13. Do you protect your knowledge assets in cooperation/transactions/ partnerships with partners/competitors/clients? Please explain.

14. What controls do you have in place to protect these assets?

15. Do you think the use of mobile devices such as smartphones, tablets, and laptops are beneficial to knowledge creation/sharing? Please explain.

16. Do you think the use of mobile devices such as smartphones, tablets, and laptops pose a greater risk? Please explain.

17. Do you have an organizational mobile policy/Strategy? Please explain.

18. How do you secure the mobile devices in your organization when used in knowledge sharing activities? 19. Does your firm have any formal or informal routines for dealing with potential knowledge leakage through mobile devices? Please explain.

20. Based on your experience, what protective actions do you think should be used to address the risk of knowledge leakage caused by the use of mobile devices?

## Qualifying and Selecting Interviewees Process

The selection of participants was a critical step in ensuring the validity and reliability of the study's findings. The participants were chosen based on their expertise in the domains of information security and knowledge management, specifically within knowledge-intensive organizations. This was crucial as these individuals are at the forefront of managing leakage associated risks in their respective organizations.

The qualifying criteria for potential interviewees included their role within the organization, their experience in dealing with knowledge management and information security, and their experience with the organization's sanctioned mobile device policies such as BYOD (Bring Your Own Device), CYOD (Choose Your Own Device), COPE (Company Owned/Personally Enabled), COBO (Company Owned/Business Only), and COSU (Company Owned/Single Use). These criteria ensured that the participants had a deep understanding of the phenomenon under study and could provide rich, detailed, and contextually relevant insights.

The selection process involved initial contact to gauge interest and availability, followed by a screening process to ensure they met the qualifying criteria. This was followed by scheduling and conducting the interviews. The recruitment process was facilitated by various approaches, including personal contacts of the researcher, snowball sampling, perusal of organizational websites, and LinkedIn.

The recruitment process was also guided by ethical considerations. An application for research ethics was submitted to the Melbourne School of Engineering and IT Human Ethics Advisory Group at the University of Melbourne. As part of the research ethics application process, a plain language statement and consent form were developed and sent to the participants, ensuring that the confidentiality and privacy of the participants were preserved at all times.

The selection and recruitment process was rigorous and systematic, ensuring that the participants were well-suited to contribute valuable insights to the study. The added explanation in section 4.4.6 provides a more detailed account of this process, which should help to further clarify the rigorous approach taken in selecting interviewees for this study.

# Appendix E – Example of Supplementary Documentation provided by Organizations

## Decision Map – Person of Security Concern

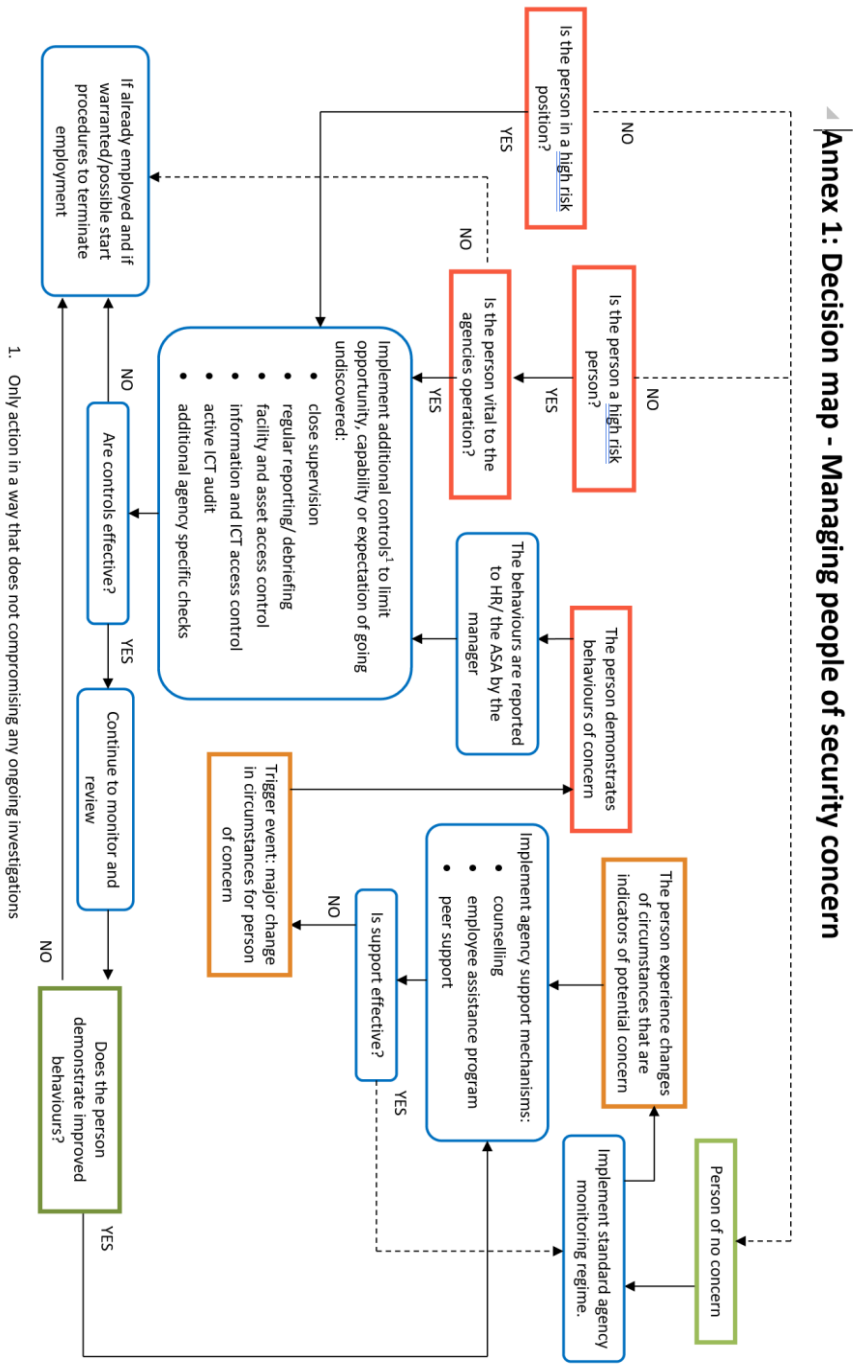

# Personnel Security Framework

| | |
|---|---|
| **Organisational personnel security** | Make sure you: <br>• know your business <br>• have a good security culture <br>• perform a personnel security risk assessment <br>• understand the legal framework <br>• communicate personnel security and the consequences of personnel security breaches to your employees. |
| **Pre-employment personnel security** | Perform the following pre-employment background checks: <br>• identity checks, including overseas applicants or applicants who have spent time overseas <br>• qualification and employment checks <br>• national criminal history checks <br>• financial background checks. <br>All documents for the checks should be secured. Any applicant who fails to meet the standard of your business should be rejected for employment. |
| **Ongoing personnel security** | Make sure you: <br>• have access controls in place <br>• perform protective monitoring <br>• promote a security culture, including <br>  – counter manipulation <br>  – report and investigate, when necessary <br>  – perform ongoing checks <br>  – submit contractors to the same security clearance as in-house personnel <br>• recognise after employment threats. |
| **Information and communications technology security** | Be sure to consider and, if necessary, monitor: <br>• electronic access <br>• shared administrative accounts <br>• account management policies and procedures <br>• the standard operating environment <br>• system logs. |

**Table 1 – A personnel security framework**

# Checklist for Mobile Workers/ Tele-workers

## Annex A—Checklist for mobile computing and communications/tele-working

- ☐ Has the employee been required to read, or been briefed on the requirements for the protection of official resources?

- ☐ What is the security classification or sensitivity of the official resources to be removed?

- ☐ Why are the official resources being removed off-site?

- ☐ How long will the official resources be off-site?

- ☐ Have the details of the official resources being removed been recorded?

- ☐ Do the official resources being removed belong to another agency? If so, has that agency given its approval?

- ☐ How will the official resources be securely transferred or transported?

- ☐ Is the removal of the official resources from the agency a temporary/one off or a permanent/long term arrangement?

- ☐ How will the official resources be securely stored off-site?

- ☐ What is known about the location where the resources are being taken? Is a risk assessment needed in relation to that location?

- ☐ What control does the agency have over the security of the location?

- ☐ Who has access to the location where the official resources are being stored?

- ☐ How will the employee protect his/her work from unwanted scrutiny or unauthorised access?

- ☐ How will the employee protect his/her official conversations from being overheard?

- ☐ Could the resources being carried reasonably expose the employee to targeting by a foreign intelligence service? Has the employee been appropriately briefed? See Contact Reporting Guidelines.

- ☐ Is the employee aware of what action he or she is to take in the event official resources are stolen?

- ☐ Is the employee considering printing, duplication or disposal of official information in a non-secure environment? What measures have been put in place to ensure official information is not compromised by this activity?

- ☐ Has the agency authorised the use of any off-site ICT equipment? If so what equipment and in what circumstances?

- ☐ Does the employee have an authorised email account, or remote ICT access to agency systems, that can be accessed securely?

# Appendix F – Ethics Approval

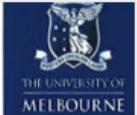

15 April 2016

Dr. R Bosua
Computing and Information Systems
The University of Melbourne

Dear Dr Bosua

I am pleased to advise that the Engineering Human Ethics Advisory Group has approved the following Minimal Risk Project.

Project Title: **Mitigating the risk of knowledge leakage in organisations through mobile devices.**
Researchers: **Dr R Bosua, Dr A Ahmad, Dr S B Maynard, C Agudelo Serna**
Ethics ID: **1646440.1**

The Project has been approved for the period: **15-April-2016 to 31-Dec-2016.**

It is your responsibility to ensure that all people associated with the Project are made aware of what has actually been approved.

Research projects are normally approved to 31 December of the year of approval. Projects may be renewed yearly for up to a total of five years upon receipt of a satisfactory annual report. If a project is to continue beyond five years a new application will normally need to be submitted. Please note that the following conditions apply to your approval. Failure to abide by these conditions may result in suspension or discontinuation of approval and/or disciplinary action.

(a) **Limit of Approval:** Approval is limited strictly to the research as submitted in your Project application.

(b) **Amendments to Project:** Any subsequent variations or modifications you might wish to make to the Project must be notified formally to the Human Ethics Advisory Group for further consideration and approval before the revised Project can commence. If the Human Ethics Advisory Group considers that the proposed amendments are significant, you may be required to submit a new application for approval of the revised Project.

(c) **Incidents or adverse effects:** Researchers must report immediately to the Advisory Group and the relevant Sub-Committee anything which might affect the ethical acceptance of the protocol including adverse effects on participants or unforeseen events that might affect continued ethical acceptability of the Project. Failure to do so may result in suspension or cancellation of approval.

(d) **Monitoring:** All projects are subject to monitoring at any time by the Human Research Ethics Committee.

(e) **Annual Report:** Please be aware that the Human Research Ethics Committee requires that researchers submit an annual report on each of their projects at the end of the year, or at the conclusion of a project if it continues for less than this time. Failure to submit an annual report will mean ethics approval will lapse.

(f) **Auditing:** All projects may be subject to audit by members of the Sub-Committee.

Please quote the ethics registration number and the name of the Project in any future correspondence.

On behalf of the Ethics Committee I wish you well in your research.

Yours sincerely

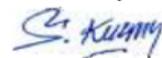

Dr Sherah Kurnia – Chair
Engineering Human Ethics Advisory Group

# Appendix G – Plain Language Statement

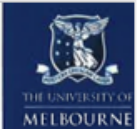

### Project Title: Mitigating the risk of knowledge leakage in organizations through mobile devices

INVESTIGATORS: Dr. Rachelle Bosua (Dept. Computing and Information Systems – Principal researcher)
Dr. Atif Ahmad (Dept. Computing and Information Systems)
Dr. Sean Maynard (Dept. Computing and Information Systems)
Carlos Andres Agudelo (Ph.D. Student. Dept. Computing and Information Systems)

**Introduction**

As someone who has experience in using mobile devices such as laptops, tablets, and smartphones as part of the tools to perform your daily work activities inside and outside your organization, we would like to invite you to participate in our research project. The aim of the study is to investigate to what extent sensitive organizational knowledge used in their daily work activities, can leak through the use of mobile devices. This project has been approved by the Human Research Ethics Committee, the University of Melbourne.

**What will I be asked to do?**

Should you agree to participate, we would ask you to participate in an interview of about 1 hour or a focus group of about 1.5 hours, so that we can get a more detailed picture of your view and perceptions on the use of mobile devices and the potential for sensitive organizational knowledge leakage in different work settings.

With your permission, the interview and focus group will be audio-recorded so that we can ensure that we make an accurate record of what you say. Upon request, you would be provided with a copy of the recorded transcript so that you can verify that the information collected is correct and/or request deletions.

2 |

#### How will my confidentiality be protected?

We intend to protect your anonymity and the confidentiality of your responses to the fullest possible extent, within the limits of the law. The information you provide will be treated as confidential and used for only research purposes connected with this research project. Confidentiality of the information will be ensured and provided subject to any legal limitations. Your name and contact details will be kept in a separate, password-protected computer file separate from any data that you supply. Only the researchers will be able to link your responses with your personal details, for example, in order to know where we should send your interview transcript for verification and final reports if you wish to receive the latter. There are minimal risks associated with this project e.g. you may know the identity of other participants. However we will attempt to protect your identify as best as we can by using pseudonyms and removing any references to personal information that might allow someone to guess your identity. You should note that as the number of people we seek to interview from your organization is small, it may be possible that someone may still be able to identify you. The data will be kept securely in the Department of Information Systems for five years from the last date of publication, before being destroyed.

#### How will I receive feedback?

Once the thesis arising from this research has been completed, a brief summary of the findings is available upon request from the researchers via the email address given below. It is also possible that the results will be presented at academic conferences and be published in academic journals.

#### What happen if I choose not to participate or withdraw?

Please be advised that your participation in this study is completely voluntary. It is your decision whether to participate or not, or to withdraw. Should you wish to withdraw at any stage, or to withdraw any unprocessed data you have supplied, you are free to do so.

#### Where can I get further information?

Should you require any further information, or have any concerns, please do not hesitate to contact any of the researchers on the numbers given above. Should you have any concerns about the conduct of the project, you are welcome to contact the Executive Officer, Human Research Ethics, The University of Melbourne, on Ph: +61 3 8344 2073, or Fax: +61 3 9347 6739.

#### How do I agree to participate?

If you would like to participate, please indicate that you have read and understood this information by signing the accompanying consent form and returning it via email to: cagudelo@student.unimelb.edu.au (Attention: Carlos Andres Agudelo). The researcher will then contact you to arrange a mutually convenient time for you to complete the interview.

# Appendix E – Consent Form

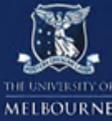

**Consent Form**

**Mitigating the risk of knowledge leakage through mobile devices**

INVESTIGATORS: Dr. Rachelle Bosua (Dept. Computing and Information Systems- Principal Researcher)
Dr. Atif Ahmad (Dept. Computing and Information Systems)
Dr. Sean Maynard (Dept. Computing and Information Systems) and
Carlos Andres Agudelo (Ph.D. Student. Dept. Computing and Information Systems)

You are invited to participate in the above research project, which is being conducted by the above researchers. The Human Research Ethics Committee (HREC) at the University of Melbourne has approved this research project.

1) I consent to participate in this research project, the details of which have been explained to me, and I have been provided with a written plain project description to keep.

2) I understand that after I sign and return this consent form, the researchers will retain it.

3) I understand that my participation will involve one-hour interview or 1.5-hour focus group. In addition the researchers may ask me to share any documents that relate to this project.

4) I acknowledge that:

   a. the possible effects of participating in this research have been explained to my satisfaction,

   b. I understand that my participation is voluntary,

   c. I have been informed that I am free to withdraw from the project at any time without explanation or prejudice, and to withdraw any unprocessed data I have provided,

   d. the project is for research purposes only,

   e. I have been informed that the interview or focus group may be audio-recorded and I understand that audio files and transcripts will be stored securely at The University of Melbourne and will be destroyed after five years following the last date of publication,

   f. my name will be referred to by a pseudonym in any publications or presentations arising from the research,

   g. the sample size for this project is small and the names of participants will only be known to the researchers and will not be revealed at any times,

   h. we intend to protect your anonymity and confidentiality of your responses to the fullest extent possible within limits of the law. However, due to a relatively small number of participants, there will be no guarantee that the identity of the individual persons or organisations can always be protected,

   i. any dependent relationship between me and the researcher(s) will not negatively impact on my participation in this study,

   j. I have been informed that a copy of the research findings will be forwarded to me should I indicate this below.

   *I wish to receive a copy of the project summary at the conclusion of the project*      ❑ *Yes*      ❑ *No*
   *(Please tick)*

   *If yes, please provide your email address: ……………………………………………………………/*

   I consent to participate in this research: Name: …………………………………………………

   Signature: …………………………………………………………

   Date: …………………………………………………………………..

Department of Computing and Information Systems
The University of Melbourne, Victoria 3010 Australia
T: +61 3 8344 1500    F: +61 3 9349 4596
W: http://www.cis.unimelb.edu.au/

THE UNIVERSITY OF
MELBOURNE

unimelb.edu.au